\newcommand{\Gaia}{\emph{Gaia}\xspace}
\newcommand{\Tycho}{\emph{Tycho-2}\xspace}
\newcommand{\Hipparcos}{\textsc{Hipparcos}\xspace}
\newcommand{\TGAS}{\textsc{TGAS}\xspace}

\newcommand{\su}{\textit{su} }
\documentclass[longauth]{aa_gaia}  
\usepackage{graphicx}
\usepackage{txfonts}
\usepackage[draft]{hyperref}
\usepackage{longtable}

\usepackage{float}
\begin{document} 

   \title{Gaia Data Release 1. 
Open cluster astrometry: performance, limitations, and future prospects \footnote{Tables D1 to D19 are available in electronic form at the CDS via anonymous ftp to cdsarc.u-strasbg.fr (130.79.128.5) or via http://cdsweb.u-strasbg.fr/cgi-bin/qcat?J/A+A/}
}
\titlerunning{Gaia Data Release 1. Open cluster astrometry}
\author{
{\it Gaia} Collaboration\relax
\and F.        ~van Leeuwen                   \inst{\ref{inst:0001}}
\and A.        ~Vallenari                     \inst{\ref{inst:0002}}
\and C.        ~Jordi                         \inst{\ref{inst:0003}}
\and L.        ~Lindegren                     \inst{\ref{inst:0004}}
\and U.        ~Bastian                       \inst{\ref{inst:0005}}
\and T.        ~Prusti                        \inst{\ref{inst:0006}}
\and J.H.J.    ~de Bruijne                    \inst{\ref{inst:0006}}
\and A.G.A.    ~Brown                         \inst{\ref{inst:0008}}
\and C.        ~Babusiaux                     \inst{\ref{inst:0009}}
\and C.A.L.    ~Bailer-Jones                  \inst{\ref{inst:0010}}
\and M.        ~Biermann                      \inst{\ref{inst:0005}}
\and D.W.      ~Evans                         \inst{\ref{inst:0001}}
\and L.        ~Eyer                          \inst{\ref{inst:0013}}
\and F.        ~Jansen                        \inst{\ref{inst:0014}}
\and S.A.      ~Klioner                       \inst{\ref{inst:0015}}
\and U.        ~Lammers                       \inst{\ref{inst:0016}}
\and X.        ~Luri                          \inst{\ref{inst:0003}}
\and F.        ~Mignard                       \inst{\ref{inst:0018}}
\and C.        ~Panem                         \inst{\ref{inst:0019}}
\and D.        ~Pourbaix                      \inst{\ref{inst:0020},\ref{inst:0021}}
\and S.        ~Randich                       \inst{\ref{inst:0022}}
\and P.        ~Sartoretti                    \inst{\ref{inst:0009}}
\and H.I.      ~Siddiqui                      \inst{\ref{inst:0024}}
\and C.        ~Soubiran                      \inst{\ref{inst:0025}}
\and V.        ~Valette                       \inst{\ref{inst:0019}}
\and N.A.      ~Walton                        \inst{\ref{inst:0001}}
\and C.        ~Aerts                         \inst{\ref{inst:0028},\ref{inst:0029}}
\and F.        ~Arenou                        \inst{\ref{inst:0009}}
\and M.        ~Cropper                       \inst{\ref{inst:0031}}
\and R.        ~Drimmel                       \inst{\ref{inst:0032}}
\and E.        ~H{\o}g                        \inst{\ref{inst:0033}}
\and D.        ~Katz                          \inst{\ref{inst:0009}}
\and M.G.      ~Lattanzi                      \inst{\ref{inst:0032}}
\and W.        ~O'Mullane                     \inst{\ref{inst:0016}}
\and E.K.      ~Grebel                        \inst{\ref{inst:0005}}
\and A.D.      ~Holland                       \inst{\ref{inst:0038}}
\and C.        ~Huc                           \inst{\ref{inst:0019}}
\and X.        ~Passot                        \inst{\ref{inst:0019}}
\and M.        ~Perryman                      \inst{\ref{inst:0006}}
\and L.        ~Bramante                      \inst{\ref{inst:0042}}
\and C.        ~Cacciari                      \inst{\ref{inst:0043}}
\and J.        ~Casta\~{n}eda                 \inst{\ref{inst:0003}}
\and L.        ~Chaoul                        \inst{\ref{inst:0019}}
\and N.        ~Cheek                         \inst{\ref{inst:0046}}
\and F.        ~De Angeli                     \inst{\ref{inst:0001}}
\and C.        ~Fabricius                     \inst{\ref{inst:0003}}
\and R.        ~Guerra                        \inst{\ref{inst:0016}}
\and J.        ~Hern\'{a}ndez                 \inst{\ref{inst:0016}}
\and A.        ~Jean-Antoine-Piccolo          \inst{\ref{inst:0019}}
\and E.        ~Masana                        \inst{\ref{inst:0003}}
\and R.        ~Messineo                      \inst{\ref{inst:0042}}
\and N.        ~Mowlavi                       \inst{\ref{inst:0013}}
\and K.        ~Nienartowicz                  \inst{\ref{inst:0055}}
\and D.        ~Ord\'{o}\~{n}ez-Blanco        \inst{\ref{inst:0055}}
\and P.        ~Panuzzo                       \inst{\ref{inst:0009}}
\and J.        ~Portell                       \inst{\ref{inst:0003}}
\and P.J.      ~Richards                      \inst{\ref{inst:0059}}
\and M.        ~Riello                        \inst{\ref{inst:0001}}
\and G.M.      ~Seabroke                      \inst{\ref{inst:0031}}
\and P.        ~Tanga                         \inst{\ref{inst:0018}}
\and F.        ~Th\'{e}venin                  \inst{\ref{inst:0018}}
\and J.        ~Torra                         \inst{\ref{inst:0003}}
\and S.G.      ~Els                           \inst{\ref{inst:0065},\ref{inst:0005}}
\and G.        ~Gracia-Abril                  \inst{\ref{inst:0065},\ref{inst:0003}}
\and G.        ~Comoretto                     \inst{\ref{inst:0024}}
\and M.        ~Garcia-Reinaldos              \inst{\ref{inst:0016}}
\and T.        ~Lock                          \inst{\ref{inst:0016}}
\and E.        ~Mercier                       \inst{\ref{inst:0065},\ref{inst:0005}}
\and M.        ~Altmann                       \inst{\ref{inst:0005},\ref{inst:0075}}
\and R.        ~Andrae                        \inst{\ref{inst:0010}}
\and T.L.      ~Astraatmadja                  \inst{\ref{inst:0010}}
\and I.        ~Bellas-Velidis                \inst{\ref{inst:0078}}
\and K.        ~Benson                        \inst{\ref{inst:0031}}
\and J.        ~Berthier                      \inst{\ref{inst:0080}}
\and R.        ~Blomme                        \inst{\ref{inst:0081}}
\and G.        ~Busso                         \inst{\ref{inst:0001}}
\and B.        ~Carry                         \inst{\ref{inst:0018},\ref{inst:0080}}
\and A.        ~Cellino                       \inst{\ref{inst:0032}}
\and G.        ~Clementini                    \inst{\ref{inst:0043}}
\and S.        ~Cowell                        \inst{\ref{inst:0001}}
\and O.        ~Creevey                       \inst{\ref{inst:0018},\ref{inst:0089}}
\and J.        ~Cuypers                       \inst{\ref{inst:0081}}
\and M.        ~Davidson                      \inst{\ref{inst:0091}}
\and J.        ~De Ridder                     \inst{\ref{inst:0028}}
\and A.        ~de Torres                     \inst{\ref{inst:0093}}
\and L.        ~Delchambre                    \inst{\ref{inst:0094}}
\and A.        ~Dell'Oro                      \inst{\ref{inst:0022}}
\and C.        ~Ducourant                     \inst{\ref{inst:0025}}
\and Y.        ~Fr\'{e}mat                    \inst{\ref{inst:0081}}
\and M.        ~Garc\'{i}a-Torres             \inst{\ref{inst:0098}}
\and E.        ~Gosset                        \inst{\ref{inst:0094},\ref{inst:0021}}
\and J.-L.     ~Halbwachs                     \inst{\ref{inst:0101}}
\and N.C.      ~Hambly                        \inst{\ref{inst:0091}}
\and D.L.      ~Harrison                      \inst{\ref{inst:0001},\ref{inst:0104}}
\and M.        ~Hauser                        \inst{\ref{inst:0005}}
\and D.        ~Hestroffer                    \inst{\ref{inst:0080}}
\and S.T.      ~Hodgkin                       \inst{\ref{inst:0001}}
\and H.E.      ~Huckle                        \inst{\ref{inst:0031}}
\and A.        ~Hutton                        \inst{\ref{inst:0109}}
\and G.        ~Jasniewicz                    \inst{\ref{inst:0110}}
\and S.        ~Jordan                        \inst{\ref{inst:0005}}
\and M.        ~Kontizas                      \inst{\ref{inst:0112}}
\and A.J.      ~Korn                          \inst{\ref{inst:0113}}
\and A.C.      ~Lanzafame                     \inst{\ref{inst:0114},\ref{inst:0115}}
\and M.        ~Manteiga                      \inst{\ref{inst:0116}}
\and A.        ~Moitinho                      \inst{\ref{inst:0117}}
\and K.        ~Muinonen                      \inst{\ref{inst:0118},\ref{inst:0119}}
\and J.        ~Osinde                        \inst{\ref{inst:0120}}
\and E.        ~Pancino                       \inst{\ref{inst:0022},\ref{inst:0122}}
\and T.        ~Pauwels                       \inst{\ref{inst:0081}}
\and J.-M.     ~Petit                         \inst{\ref{inst:0124}}
\and A.        ~Recio-Blanco                  \inst{\ref{inst:0018}}
\and A.C.      ~Robin                         \inst{\ref{inst:0124}}
\and L.M.      ~Sarro                         \inst{\ref{inst:0127}}
\and C.        ~Siopis                        \inst{\ref{inst:0020}}
\and M.        ~Smith                         \inst{\ref{inst:0031}}
\and K.W.      ~Smith                         \inst{\ref{inst:0010}}
\and A.        ~Sozzetti                      \inst{\ref{inst:0032}}
\and W.        ~Thuillot                      \inst{\ref{inst:0080}}
\and W.        ~van Reeven                    \inst{\ref{inst:0109}}
\and Y.        ~Viala                         \inst{\ref{inst:0009}}
\and U.        ~Abbas                         \inst{\ref{inst:0032}}
\and A.        ~Abreu Aramburu                \inst{\ref{inst:0136}}
\and S.        ~Accart                        \inst{\ref{inst:0137}}
\and J.J.      ~Aguado                        \inst{\ref{inst:0127}}
\and P.M.      ~Allan                         \inst{\ref{inst:0059}}
\and W.        ~Allasia                       \inst{\ref{inst:0140}}
\and G.        ~Altavilla                     \inst{\ref{inst:0043}}
\and M.A.      ~\'{A}lvarez                   \inst{\ref{inst:0116}}
\and J.        ~Alves                         \inst{\ref{inst:0143}}
\and R.I.      ~Anderson                      \inst{\ref{inst:0144},\ref{inst:0013}}
\and A.H.      ~Andrei                        \inst{\ref{inst:0146},\ref{inst:0147},\ref{inst:0075}}
\and E.        ~Anglada Varela                \inst{\ref{inst:0120},\ref{inst:0046}}
\and E.        ~Antiche                       \inst{\ref{inst:0003}}
\and T.        ~Antoja                        \inst{\ref{inst:0006}}
\and S.        ~Ant\'{o}n                     \inst{\ref{inst:0153},\ref{inst:0154}}
\and B.        ~Arcay                         \inst{\ref{inst:0116}}
\and N.        ~Bach                          \inst{\ref{inst:0109}}
\and S.G.      ~Baker                         \inst{\ref{inst:0031}}
\and L.        ~Balaguer-N\'{u}\~{n}ez        \inst{\ref{inst:0003}}
\and C.        ~Barache                       \inst{\ref{inst:0075}}
\and C.        ~Barata                        \inst{\ref{inst:0117}}
\and A.        ~Barbier                       \inst{\ref{inst:0137}}
\and F.        ~Barblan                       \inst{\ref{inst:0013}}
\and D.        ~Barrado y Navascu\'{e}s       \inst{\ref{inst:0163}}
\and M.        ~Barros                        \inst{\ref{inst:0117}}
\and M.A.      ~Barstow                       \inst{\ref{inst:0165}}
\and U.        ~Becciani                      \inst{\ref{inst:0115}}
\and M.        ~Bellazzini                    \inst{\ref{inst:0043}}
\and A.        ~Bello Garc\'{i}a              \inst{\ref{inst:0168}}
\and V.        ~Belokurov                     \inst{\ref{inst:0001}}
\and P.        ~Bendjoya                      \inst{\ref{inst:0018}}
\and A.        ~Berihuete                     \inst{\ref{inst:0171}}
\and L.        ~Bianchi                       \inst{\ref{inst:0140}}
\and O.        ~Bienaym\'{e}                  \inst{\ref{inst:0101}}
\and F.        ~Billebaud                     \inst{\ref{inst:0025}}
\and N.        ~Blagorodnova                  \inst{\ref{inst:0001}}
\and S.        ~Blanco-Cuaresma               \inst{\ref{inst:0013},\ref{inst:0025}}
\and T.        ~Boch                          \inst{\ref{inst:0101}}
\and A.        ~Bombrun                       \inst{\ref{inst:0093}}
\and R.        ~Borrachero                    \inst{\ref{inst:0003}}
\and S.        ~Bouquillon                    \inst{\ref{inst:0075}}
\and G.        ~Bourda                        \inst{\ref{inst:0025}}
\and H.        ~Bouy                          \inst{\ref{inst:0163}}
\and A.        ~Bragaglia                     \inst{\ref{inst:0043}}
\and M.A.      ~Breddels                      \inst{\ref{inst:0185}}
\and N.        ~Brouillet                     \inst{\ref{inst:0025}}
\and T.        ~Br\"{ u}semeister             \inst{\ref{inst:0005}}
\and B.        ~Bucciarelli                   \inst{\ref{inst:0032}}
\and P.        ~Burgess                       \inst{\ref{inst:0001}}
\and R.        ~Burgon                        \inst{\ref{inst:0038}}
\and A.        ~Burlacu                       \inst{\ref{inst:0019}}
\and D.        ~Busonero                      \inst{\ref{inst:0032}}
\and R.        ~Buzzi                         \inst{\ref{inst:0032}}
\and E.        ~Caffau                        \inst{\ref{inst:0009}}
\and J.        ~Cambras                       \inst{\ref{inst:0195}}
\and H.        ~Campbell                      \inst{\ref{inst:0001}}
\and R.        ~Cancelliere                   \inst{\ref{inst:0197}}
\and T.        ~Cantat-Gaudin                 \inst{\ref{inst:0002}}
\and T.        ~Carlucci                      \inst{\ref{inst:0075}}
\and J.M.      ~Carrasco                      \inst{\ref{inst:0003}}
\and M.        ~Castellani                    \inst{\ref{inst:0201}}
\and P.        ~Charlot                       \inst{\ref{inst:0025}}
\and J.        ~Charnas                       \inst{\ref{inst:0055}}
\and A.        ~Chiavassa                     \inst{\ref{inst:0018}}
\and M.        ~Clotet                        \inst{\ref{inst:0003}}
\and G.        ~Cocozza                       \inst{\ref{inst:0043}}
\and R.S.      ~Collins                       \inst{\ref{inst:0091}}
\and G.        ~Costigan                      \inst{\ref{inst:0008}}
\and F.        ~Crifo                         \inst{\ref{inst:0009}}
\and N.J.G.    ~Cross                         \inst{\ref{inst:0091}}
\and M.        ~Crosta                        \inst{\ref{inst:0032}}
\and C.        ~Crowley                       \inst{\ref{inst:0093}}
\and C.        ~Dafonte                       \inst{\ref{inst:0116}}
\and Y.        ~Damerdji                      \inst{\ref{inst:0094},\ref{inst:0215}}
\and A.        ~Dapergolas                    \inst{\ref{inst:0078}}
\and P.        ~David                         \inst{\ref{inst:0080}}
\and M.        ~David                         \inst{\ref{inst:0218}}
\and P.        ~De Cat                        \inst{\ref{inst:0081}}
\and F.        ~de Felice                     \inst{\ref{inst:0220}}
\and P.        ~de Laverny                    \inst{\ref{inst:0018}}
\and F.        ~De Luise                      \inst{\ref{inst:0222}}
\and R.        ~De March                      \inst{\ref{inst:0042}}
\and D.        ~de Martino                    \inst{\ref{inst:0224}}
\and R.        ~de Souza                      \inst{\ref{inst:0225}}
\and J.        ~Debosscher                    \inst{\ref{inst:0028}}
\and E.        ~del Pozo                      \inst{\ref{inst:0109}}
\and M.        ~Delbo                         \inst{\ref{inst:0018}}
\and A.        ~Delgado                       \inst{\ref{inst:0001}}
\and H.E.      ~Delgado                       \inst{\ref{inst:0127}}
\and P.        ~Di Matteo                     \inst{\ref{inst:0009}}
\and S.        ~Diakite                       \inst{\ref{inst:0124}}
\and E.        ~Distefano                     \inst{\ref{inst:0115}}
\and C.        ~Dolding                       \inst{\ref{inst:0031}}
\and S.        ~Dos Anjos                     \inst{\ref{inst:0225}}
\and P.        ~Drazinos                      \inst{\ref{inst:0112}}
\and J.        ~Dur\'{a}n                     \inst{\ref{inst:0120}}
\and Y.        ~Dzigan                        \inst{\ref{inst:0238},\ref{inst:0239}}
\and B.        ~Edvardsson                    \inst{\ref{inst:0113}}
\and H.        ~Enke                          \inst{\ref{inst:0241}}
\and N.W.      ~Evans                         \inst{\ref{inst:0001}}
\and G.        ~Eynard Bontemps               \inst{\ref{inst:0137}}
\and C.        ~Fabre                         \inst{\ref{inst:0244}}
\and M.        ~Fabrizio                      \inst{\ref{inst:0122},\ref{inst:0222}}
\and S.        ~Faigler                       \inst{\ref{inst:0247}}
\and A.J.      ~Falc\~{a}o                    \inst{\ref{inst:0248}}
\and M.        ~Farr\`{a}s Casas              \inst{\ref{inst:0003}}
\and L.        ~Federici                      \inst{\ref{inst:0043}}
\and G.        ~Fedorets                      \inst{\ref{inst:0118}}
\and J.        ~Fern\'{a}ndez-Hern\'{a}ndez   \inst{\ref{inst:0046}}
\and P.        ~Fernique                      \inst{\ref{inst:0101}}
\and A.        ~Fienga                        \inst{\ref{inst:0254}}
\and F.        ~Figueras                      \inst{\ref{inst:0003}}
\and F.        ~Filippi                       \inst{\ref{inst:0042}}
\and K.        ~Findeisen                     \inst{\ref{inst:0009}}
\and A.        ~Fonti                         \inst{\ref{inst:0042}}
\and M.        ~Fouesneau                     \inst{\ref{inst:0010}}
\and E.        ~Fraile                        \inst{\ref{inst:0260}}
\and M.        ~Fraser                        \inst{\ref{inst:0001}}
\and J.        ~Fuchs                         \inst{\ref{inst:0262}}
\and M.        ~Gai                           \inst{\ref{inst:0032}}
\and S.        ~Galleti                       \inst{\ref{inst:0043}}
\and L.        ~Galluccio                     \inst{\ref{inst:0018}}
\and D.        ~Garabato                      \inst{\ref{inst:0116}}
\and F.        ~Garc\'{i}a-Sedano             \inst{\ref{inst:0127}}
\and A.        ~Garofalo                      \inst{\ref{inst:0043}}
\and N.        ~Garralda                      \inst{\ref{inst:0003}}
\and P.        ~Gavras                        \inst{\ref{inst:0009},\ref{inst:0078},\ref{inst:0112}}
\and J.        ~Gerssen                       \inst{\ref{inst:0241}}
\and R.        ~Geyer                         \inst{\ref{inst:0015}}
\and G.        ~Gilmore                       \inst{\ref{inst:0001}}
\and S.        ~Girona                        \inst{\ref{inst:0276}}
\and G.        ~Giuffrida                     \inst{\ref{inst:0122}}
\and M.        ~Gomes                         \inst{\ref{inst:0117}}
\and A.        ~Gonz\'{a}lez-Marcos           \inst{\ref{inst:0279}}
\and J.        ~Gonz\'{a}lez-N\'{u}\~{n}ez    \inst{\ref{inst:0046},\ref{inst:0281}}
\and J.J.      ~Gonz\'{a}lez-Vidal            \inst{\ref{inst:0003}}
\and M.        ~Granvik                       \inst{\ref{inst:0118}}
\and A.        ~Guerrier                      \inst{\ref{inst:0137}}
\and P.        ~Guillout                      \inst{\ref{inst:0101}}
\and J.        ~Guiraud                       \inst{\ref{inst:0019}}
\and A.        ~G\'{u}rpide                   \inst{\ref{inst:0003}}
\and R.        ~Guti\'{e}rrez-S\'{a}nchez     \inst{\ref{inst:0024}}
\and L.P.      ~Guy                           \inst{\ref{inst:0055}}
\and R.        ~Haigron                       \inst{\ref{inst:0009}}
\and D.        ~Hatzidimitriou                \inst{\ref{inst:0112},\ref{inst:0078}}
\and M.        ~Haywood                       \inst{\ref{inst:0009}}
\and U.        ~Heiter                        \inst{\ref{inst:0113}}
\and A.        ~Helmi                         \inst{\ref{inst:0185}}
\and D.        ~Hobbs                         \inst{\ref{inst:0004}}
\and W.        ~Hofmann                       \inst{\ref{inst:0005}}
\and B.        ~Holl                          \inst{\ref{inst:0013}}
\and G.        ~Holland                       \inst{\ref{inst:0001}}
\and J.A.S.    ~Hunt                          \inst{\ref{inst:0031}}
\and A.        ~Hypki                         \inst{\ref{inst:0008}}
\and V.        ~Icardi                        \inst{\ref{inst:0042}}
\and M.        ~Irwin                         \inst{\ref{inst:0001}}
\and G.        ~Jevardat de Fombelle          \inst{\ref{inst:0055}}
\and P.        ~Jofr\'{e}                     \inst{\ref{inst:0001},\ref{inst:0025}}
\and P.G.      ~Jonker                        \inst{\ref{inst:0307},\ref{inst:0029}}
\and A.        ~Jorissen                      \inst{\ref{inst:0020}}
\and F.        ~Julbe                         \inst{\ref{inst:0003}}
\and A.        ~Karampelas                    \inst{\ref{inst:0112},\ref{inst:0078}}
\and A.        ~Kochoska                      \inst{\ref{inst:0313}}
\and R.        ~Kohley                        \inst{\ref{inst:0016}}
\and K.        ~Kolenberg                     \inst{\ref{inst:0315},\ref{inst:0028},\ref{inst:0317}}
\and E.        ~Kontizas                      \inst{\ref{inst:0078}}
\and S.E.      ~Koposov                       \inst{\ref{inst:0001}}
\and G.        ~Kordopatis                    \inst{\ref{inst:0241},\ref{inst:0018}}
\and P.        ~Koubsky                       \inst{\ref{inst:0262}}
\and A.        ~Krone-Martins                 \inst{\ref{inst:0117}}
\and M.        ~Kudryashova                   \inst{\ref{inst:0080}}
\and I.        ~Kull                          \inst{\ref{inst:0247}}
\and R.K.      ~Bachchan                      \inst{\ref{inst:0004}}
\and F.        ~Lacoste-Seris                 \inst{\ref{inst:0137}}
\and A.F.      ~Lanza                         \inst{\ref{inst:0115}}
\and J.-B.     ~Lavigne                       \inst{\ref{inst:0137}}
\and C.        ~Le Poncin-Lafitte             \inst{\ref{inst:0075}}
\and Y.        ~Lebreton                      \inst{\ref{inst:0009},\ref{inst:0332}}
\and T.        ~Lebzelter                     \inst{\ref{inst:0143}}
\and S.        ~Leccia                        \inst{\ref{inst:0224}}
\and N.        ~Leclerc                       \inst{\ref{inst:0009}}
\and I.        ~Lecoeur-Taibi                 \inst{\ref{inst:0055}}
\and V.        ~Lemaitre                      \inst{\ref{inst:0137}}
\and H.        ~Lenhardt                      \inst{\ref{inst:0005}}
\and F.        ~Leroux                        \inst{\ref{inst:0137}}
\and S.        ~Liao                          \inst{\ref{inst:0032},\ref{inst:0341}}
\and E.        ~Licata                        \inst{\ref{inst:0140}}
\and H.E.P.    ~Lindstr{\o}m                  \inst{\ref{inst:0033},\ref{inst:0344}}
\and T.A.      ~Lister                        \inst{\ref{inst:0345}}
\and E.        ~Livanou                       \inst{\ref{inst:0112}}
\and A.        ~Lobel                         \inst{\ref{inst:0081}}
\and W.        ~L\"{ o}ffler                  \inst{\ref{inst:0005}}
\and M.        ~L\'{o}pez                     \inst{\ref{inst:0163}}
\and D.        ~Lorenz                        \inst{\ref{inst:0143}}
\and I.        ~MacDonald                     \inst{\ref{inst:0091}}
\and T.        ~Magalh\~{a}es Fernandes       \inst{\ref{inst:0248}}
\and S.        ~Managau                       \inst{\ref{inst:0137}}
\and R.G.      ~Mann                          \inst{\ref{inst:0091}}
\and G.        ~Mantelet                      \inst{\ref{inst:0005}}
\and O.        ~Marchal                       \inst{\ref{inst:0009}}
\and J.M.      ~Marchant                      \inst{\ref{inst:0357}}
\and M.        ~Marconi                       \inst{\ref{inst:0224}}
\and S.        ~Marinoni                      \inst{\ref{inst:0201},\ref{inst:0122}}
\and P.M.      ~Marrese                       \inst{\ref{inst:0201},\ref{inst:0122}}
\and G.        ~Marschalk\'{o}                \inst{\ref{inst:0363},\ref{inst:0364}}
\and D.J.      ~Marshall                      \inst{\ref{inst:0365}}
\and J.M.      ~Mart\'{i}n-Fleitas            \inst{\ref{inst:0109}}
\and M.        ~Martino                       \inst{\ref{inst:0042}}
\and N.        ~Mary                          \inst{\ref{inst:0137}}
\and G.        ~Matijevi\v{c}                 \inst{\ref{inst:0241}}
\and T.        ~Mazeh                         \inst{\ref{inst:0247}}
\and P.J.      ~McMillan                      \inst{\ref{inst:0004}}
\and S.        ~Messina                       \inst{\ref{inst:0115}}
\and D.        ~Michalik                      \inst{\ref{inst:0004}}
\and N.R.      ~Millar                        \inst{\ref{inst:0001}}
\and B.M.H.    ~Miranda                       \inst{\ref{inst:0117}}
\and D.        ~Molina                        \inst{\ref{inst:0003}}
\and R.        ~Molinaro                      \inst{\ref{inst:0224}}
\and M.        ~Molinaro                      \inst{\ref{inst:0378}}
\and L.        ~Moln\'{a}r                    \inst{\ref{inst:0363}}
\and M.        ~Moniez                        \inst{\ref{inst:0380}}
\and P.        ~Montegriffo                   \inst{\ref{inst:0043}}
\and R.        ~Mor                           \inst{\ref{inst:0003}}
\and A.        ~Mora                          \inst{\ref{inst:0109}}
\and R.        ~Morbidelli                    \inst{\ref{inst:0032}}
\and T.        ~Morel                         \inst{\ref{inst:0094}}
\and S.        ~Morgenthaler                  \inst{\ref{inst:0386}}
\and D.        ~Morris                        \inst{\ref{inst:0091}}
\and A.F.      ~Mulone                        \inst{\ref{inst:0042}}
\and T.        ~Muraveva                      \inst{\ref{inst:0043}}
\and I.        ~Musella                       \inst{\ref{inst:0224}}
\and J.        ~Narbonne                      \inst{\ref{inst:0137}}
\and G.        ~Nelemans                      \inst{\ref{inst:0029},\ref{inst:0028}}
\and L.        ~Nicastro                      \inst{\ref{inst:0394}}
\and L.        ~Noval                         \inst{\ref{inst:0137}}
\and C.        ~Ord\'{e}novic                 \inst{\ref{inst:0018}}
\and J.        ~Ordieres-Mer\'{e}             \inst{\ref{inst:0397}}
\and P.        ~Osborne                       \inst{\ref{inst:0001}}
\and C.        ~Pagani                        \inst{\ref{inst:0165}}
\and I.        ~Pagano                        \inst{\ref{inst:0115}}
\and F.        ~Pailler                       \inst{\ref{inst:0019}}
\and H.        ~Palacin                       \inst{\ref{inst:0137}}
\and L.        ~Palaversa                     \inst{\ref{inst:0013}}
\and P.        ~Parsons                       \inst{\ref{inst:0024}}
\and M.        ~Pecoraro                      \inst{\ref{inst:0140}}
\and R.        ~Pedrosa                       \inst{\ref{inst:0406}}
\and H.        ~Pentik\"{ a}inen              \inst{\ref{inst:0118}}
\and B.        ~Pichon                        \inst{\ref{inst:0018}}
\and A.M.      ~Piersimoni                    \inst{\ref{inst:0222}}
\and F.-X.     ~Pineau                        \inst{\ref{inst:0101}}
\and E.        ~Plachy                        \inst{\ref{inst:0363}}
\and G.        ~Plum                          \inst{\ref{inst:0009}}
\and E.        ~Poujoulet                     \inst{\ref{inst:0413}}
\and A.        ~Pr\v{s}a                      \inst{\ref{inst:0414}}
\and L.        ~Pulone                        \inst{\ref{inst:0201}}
\and S.        ~Ragaini                       \inst{\ref{inst:0043}}
\and S.        ~Rago                          \inst{\ref{inst:0032}}
\and N.        ~Rambaux                       \inst{\ref{inst:0080}}
\and M.        ~Ramos-Lerate                  \inst{\ref{inst:0419}}
\and P.        ~Ranalli                       \inst{\ref{inst:0004}}
\and G.        ~Rauw                          \inst{\ref{inst:0094}}
\and A.        ~Read                          \inst{\ref{inst:0165}}
\and S.        ~Regibo                        \inst{\ref{inst:0028}}
\and C.        ~Reyl\'{e}                     \inst{\ref{inst:0124}}
\and R.A.      ~Ribeiro                       \inst{\ref{inst:0248}}
\and L.        ~Rimoldini                     \inst{\ref{inst:0055}}
\and V.        ~Ripepi                        \inst{\ref{inst:0224}}
\and A.        ~Riva                          \inst{\ref{inst:0032}}
\and G.        ~Rixon                         \inst{\ref{inst:0001}}
\and M.        ~Roelens                       \inst{\ref{inst:0013}}
\and M.        ~Romero-G\'{o}mez              \inst{\ref{inst:0003}}
\and N.        ~Rowell                        \inst{\ref{inst:0091}}
\and F.        ~Royer                         \inst{\ref{inst:0009}}
\and L.        ~Ruiz-Dern                     \inst{\ref{inst:0009}}
\and G.        ~Sadowski                      \inst{\ref{inst:0020}}
\and T.        ~Sagrist\`{a} Sell\'{e}s       \inst{\ref{inst:0005}}
\and J.        ~Sahlmann                      \inst{\ref{inst:0016}}
\and J.        ~Salgado                       \inst{\ref{inst:0120}}
\and E.        ~Salguero                      \inst{\ref{inst:0120}}
\and M.        ~Sarasso                       \inst{\ref{inst:0032}}
\and H.        ~Savietto                      \inst{\ref{inst:0441}}
\and M.        ~Schultheis                    \inst{\ref{inst:0018}}
\and E.        ~Sciacca                       \inst{\ref{inst:0115}}
\and M.        ~Segol                         \inst{\ref{inst:0444}}
\and J.C.      ~Segovia                       \inst{\ref{inst:0046}}
\and D.        ~Segransan                     \inst{\ref{inst:0013}}
\and I-C.      ~Shih                          \inst{\ref{inst:0009}}
\and R.        ~Smareglia                     \inst{\ref{inst:0378}}
\and R.L.      ~Smart                         \inst{\ref{inst:0032}}
\and E.        ~Solano                        \inst{\ref{inst:0163},\ref{inst:0451}}
\and F.        ~Solitro                       \inst{\ref{inst:0042}}
\and R.        ~Sordo                         \inst{\ref{inst:0002}}
\and S.        ~Soria Nieto                   \inst{\ref{inst:0003}}
\and J.        ~Souchay                       \inst{\ref{inst:0075}}
\and A.        ~Spagna                        \inst{\ref{inst:0032}}
\and F.        ~Spoto                         \inst{\ref{inst:0018}}
\and U.        ~Stampa                        \inst{\ref{inst:0005}}
\and I.A.      ~Steele                        \inst{\ref{inst:0357}}
\and H.        ~Steidelm\"{ u}ller            \inst{\ref{inst:0015}}
\and C.A.      ~Stephenson                    \inst{\ref{inst:0024}}
\and H.        ~Stoev                         \inst{\ref{inst:0462}}
\and F.F.      ~Suess                         \inst{\ref{inst:0001}}
\and M.        ~S\"{ u}veges                  \inst{\ref{inst:0055}}
\and J.        ~Surdej                        \inst{\ref{inst:0094}}
\and L.        ~Szabados                      \inst{\ref{inst:0363}}
\and E.        ~Szegedi-Elek                  \inst{\ref{inst:0363}}
\and D.        ~Tapiador                      \inst{\ref{inst:0468},\ref{inst:0469}}
\and F.        ~Taris                         \inst{\ref{inst:0075}}
\and G.        ~Tauran                        \inst{\ref{inst:0137}}
\and M.B.      ~Taylor                        \inst{\ref{inst:0472}}
\and R.        ~Teixeira                      \inst{\ref{inst:0225}}
\and D.        ~Terrett                       \inst{\ref{inst:0059}}
\and B.        ~Tingley                       \inst{\ref{inst:0475}}
\and S.C.      ~Trager                        \inst{\ref{inst:0185}}
\and C.        ~Turon                         \inst{\ref{inst:0009}}
\and A.        ~Ulla                          \inst{\ref{inst:0478}}
\and E.        ~Utrilla                       \inst{\ref{inst:0109}}
\and G.        ~Valentini                     \inst{\ref{inst:0222}}
\and A.        ~van Elteren                   \inst{\ref{inst:0008}}
\and E.        ~Van Hemelryck                 \inst{\ref{inst:0081}}
\and M.        ~van Leeuwen                   \inst{\ref{inst:0001}}
\and M.        ~Varadi                        \inst{\ref{inst:0013},\ref{inst:0363}}
\and A.        ~Vecchiato                     \inst{\ref{inst:0032}}
\and J.        ~Veljanoski                    \inst{\ref{inst:0185}}
\and T.        ~Via                           \inst{\ref{inst:0195}}
\and D.        ~Vicente                       \inst{\ref{inst:0276}}
\and S.        ~Vogt                          \inst{\ref{inst:0490}}
\and H.        ~Voss                          \inst{\ref{inst:0003}}
\and V.        ~Votruba                       \inst{\ref{inst:0262}}
\and S.        ~Voutsinas                     \inst{\ref{inst:0091}}
\and G.        ~Walmsley                      \inst{\ref{inst:0019}}
\and M.        ~Weiler                        \inst{\ref{inst:0003}}
\and K.        ~Weingrill                     \inst{\ref{inst:0241}}
\and T.        ~Wevers                        \inst{\ref{inst:0029}}
\and \L{}.     ~Wyrzykowski                   \inst{\ref{inst:0001},\ref{inst:0499}}
\and A.        ~Yoldas                        \inst{\ref{inst:0001}}
\and M.        ~\v{Z}erjal                    \inst{\ref{inst:0313}}
\and S.        ~Zucker                        \inst{\ref{inst:0238}}
\and C.        ~Zurbach                       \inst{\ref{inst:0110}}
\and T.        ~Zwitter                       \inst{\ref{inst:0313}}
\and A.        ~Alecu                         \inst{\ref{inst:0001}}
\and M.        ~Allen                         \inst{\ref{inst:0006}}
\and C.        ~Allende Prieto                \inst{\ref{inst:0031},\ref{inst:0508},\ref{inst:0509}}
\and A.        ~Amorim                        \inst{\ref{inst:0117}}
\and G.        ~Anglada-Escud\'{e}            \inst{\ref{inst:0003}}
\and V.        ~Arsenijevic                   \inst{\ref{inst:0117}}
\and S.        ~Azaz                          \inst{\ref{inst:0006}}
\and P.        ~Balm                          \inst{\ref{inst:0024}}
\and M.        ~Beck                          \inst{\ref{inst:0055}}
\and H.-H.     ~Bernstein$^\dagger$           \inst{\ref{inst:0005}}
\and L.        ~Bigot                         \inst{\ref{inst:0018}}
\and A.        ~Bijaoui                       \inst{\ref{inst:0018}}
\and C.        ~Blasco                        \inst{\ref{inst:0519}}
\and M.        ~Bonfigli                      \inst{\ref{inst:0222}}
\and G.        ~Bono                          \inst{\ref{inst:0201}}
\and S.        ~Boudreault                    \inst{\ref{inst:0031},\ref{inst:0523}}
\and A.        ~Bressan                       \inst{\ref{inst:0524}}
\and S.        ~Brown                         \inst{\ref{inst:0001}}
\and P.-M.     ~Brunet                        \inst{\ref{inst:0019}}
\and P.        ~Bunclark$^\dagger$            \inst{\ref{inst:0001}}
\and R.        ~Buonanno                      \inst{\ref{inst:0201}}
\and A.G.      ~Butkevich                     \inst{\ref{inst:0015}}
\and C.        ~Carret                        \inst{\ref{inst:0406}}
\and C.        ~Carrion                       \inst{\ref{inst:0127}}
\and L.        ~Chemin                        \inst{\ref{inst:0025},\ref{inst:0533}}
\and F.        ~Ch\'{e}reau                   \inst{\ref{inst:0009}}
\and L.        ~Corcione                      \inst{\ref{inst:0032}}
\and E.        ~Darmigny                      \inst{\ref{inst:0019}}
\and K.S.      ~de Boer                       \inst{\ref{inst:0537}}
\and P.        ~de Teodoro                    \inst{\ref{inst:0046}}
\and P.T.      ~de Zeeuw                      \inst{\ref{inst:0008},\ref{inst:0540}}
\and C.        ~Delle Luche                   \inst{\ref{inst:0009},\ref{inst:0137}}
\and C.D.      ~Domingues                     \inst{\ref{inst:0543}}
\and P.        ~Dubath                        \inst{\ref{inst:0055}}
\and F.        ~Fodor                         \inst{\ref{inst:0019}}
\and B.        ~Fr\'{e}zouls                  \inst{\ref{inst:0019}}
\and A.        ~Fries                         \inst{\ref{inst:0003}}
\and D.        ~Fustes                        \inst{\ref{inst:0116}}
\and D.        ~Fyfe                          \inst{\ref{inst:0165}}
\and E.        ~Gallardo                      \inst{\ref{inst:0003}}
\and J.        ~Gallegos                      \inst{\ref{inst:0046}}
\and D.        ~Gardiol                       \inst{\ref{inst:0032}}
\and M.        ~Gebran                        \inst{\ref{inst:0003},\ref{inst:0554}}
\and A.        ~Gomboc                        \inst{\ref{inst:0313},\ref{inst:0556}}
\and A.        ~G\'{o}mez                     \inst{\ref{inst:0009}}
\and E.        ~Grux                          \inst{\ref{inst:0124}}
\and A.        ~Gueguen                       \inst{\ref{inst:0009},\ref{inst:0560}}
\and A.        ~Heyrovsky                     \inst{\ref{inst:0091}}
\and J.        ~Hoar                          \inst{\ref{inst:0016}}
\and G.        ~Iannicola                     \inst{\ref{inst:0201}}
\and Y.        ~Isasi Parache                 \inst{\ref{inst:0003}}
\and A.-M.     ~Janotto                       \inst{\ref{inst:0019}}
\and E.        ~Joliet                        \inst{\ref{inst:0093},\ref{inst:0567}}
\and A.        ~Jonckheere                    \inst{\ref{inst:0081}}
\and R.        ~Keil                          \inst{\ref{inst:0569},\ref{inst:0570}}
\and D.-W.     ~Kim                           \inst{\ref{inst:0010}}
\and P.        ~Klagyivik                     \inst{\ref{inst:0363}}
\and J.        ~Klar                          \inst{\ref{inst:0241}}
\and J.        ~Knude                         \inst{\ref{inst:0033}}
\and O.        ~Kochukhov                     \inst{\ref{inst:0113}}
\and I.        ~Kolka                         \inst{\ref{inst:0576}}
\and J.        ~Kos                           \inst{\ref{inst:0313},\ref{inst:0578}}
\and A.        ~Kutka                         \inst{\ref{inst:0262},\ref{inst:0580}}
\and V.        ~Lainey                        \inst{\ref{inst:0080}}
\and D.        ~LeBouquin                     \inst{\ref{inst:0137}}
\and C.        ~Liu                           \inst{\ref{inst:0010},\ref{inst:0584}}
\and D.        ~Loreggia                      \inst{\ref{inst:0032}}
\and V.V.      ~Makarov                       \inst{\ref{inst:0586}}
\and M.G.      ~Marseille                     \inst{\ref{inst:0137}}
\and C.        ~Martayan                      \inst{\ref{inst:0081},\ref{inst:0589}}
\and O.        ~Martinez-Rubi                 \inst{\ref{inst:0003}}
\and B.        ~Massart                       \inst{\ref{inst:0018},\ref{inst:0137},\ref{inst:0593}}
\and F.        ~Meynadier                     \inst{\ref{inst:0009},\ref{inst:0075}}
\and S.        ~Mignot                        \inst{\ref{inst:0009}}
\and U.        ~Munari                        \inst{\ref{inst:0002}}
\and A.-T.     ~Nguyen                        \inst{\ref{inst:0019}}
\and T.        ~Nordlander                    \inst{\ref{inst:0113}}
\and K.S.      ~O'Flaherty                    \inst{\ref{inst:0600}}
\and P.        ~Ocvirk                        \inst{\ref{inst:0241},\ref{inst:0101}}
\and A.        ~Olias Sanz                    \inst{\ref{inst:0603}}
\and P.        ~Ortiz                         \inst{\ref{inst:0165}}
\and J.        ~Osorio                        \inst{\ref{inst:0153}}
\and D.        ~Oszkiewicz                    \inst{\ref{inst:0118},\ref{inst:0607}}
\and A.        ~Ouzounis                      \inst{\ref{inst:0091}}
\and M.        ~Palmer                        \inst{\ref{inst:0003}}
\and P.        ~Park                          \inst{\ref{inst:0013}}
\and E.        ~Pasquato                      \inst{\ref{inst:0020}}
\and C.        ~Peltzer                       \inst{\ref{inst:0001}}
\and J.        ~Peralta                       \inst{\ref{inst:0003}}
\and F.        ~P\'{e}turaud                  \inst{\ref{inst:0009}}
\and T.        ~Pieniluoma                    \inst{\ref{inst:0118}}
\and E.        ~Pigozzi                       \inst{\ref{inst:0042}}
\and J.        ~Poels$^\dagger$               \inst{\ref{inst:0094}}
\and G.        ~Prat                          \inst{\ref{inst:0618}}
\and T.        ~Prod'homme                    \inst{\ref{inst:0008},\ref{inst:0620}}
\and F.        ~Raison                        \inst{\ref{inst:0621},\ref{inst:0560}}
\and J.M.      ~Rebordao                      \inst{\ref{inst:0543}}
\and D.        ~Risquez                       \inst{\ref{inst:0008}}
\and B.        ~Rocca-Volmerange              \inst{\ref{inst:0625}}
\and S.        ~Rosen                         \inst{\ref{inst:0031},\ref{inst:0165}}
\and M.I.      ~Ruiz-Fuertes                  \inst{\ref{inst:0055}}
\and F.        ~Russo                         \inst{\ref{inst:0032}}
\and S.        ~Sembay                        \inst{\ref{inst:0165}}
\and I.        ~Serraller Vizcaino            \inst{\ref{inst:0631}}
\and A.        ~Short                         \inst{\ref{inst:0006}}
\and A.        ~Siebert                       \inst{\ref{inst:0101},\ref{inst:0241}}
\and H.        ~Silva                         \inst{\ref{inst:0248}}
\and D.        ~Sinachopoulos                 \inst{\ref{inst:0078}}
\and E.        ~Slezak                        \inst{\ref{inst:0018}}
\and M.        ~Soffel                        \inst{\ref{inst:0015}}
\and D.        ~Sosnowska                     \inst{\ref{inst:0013}}
\and V.        ~Strai\v{z}ys                  \inst{\ref{inst:0640}}
\and M.        ~ter Linden                    \inst{\ref{inst:0093},\ref{inst:0642}}
\and D.        ~Terrell                       \inst{\ref{inst:0643}}
\and S.        ~Theil                         \inst{\ref{inst:0644}}
\and C.        ~Tiede                         \inst{\ref{inst:0010},\ref{inst:0646}}
\and L.        ~Troisi                        \inst{\ref{inst:0122},\ref{inst:0648}}
\and P.        ~Tsalmantza                    \inst{\ref{inst:0010}}
\and D.        ~Tur                           \inst{\ref{inst:0195}}
\and M.        ~Vaccari                       \inst{\ref{inst:0651},\ref{inst:0652}}
\and F.        ~Vachier                       \inst{\ref{inst:0080}}
\and P.        ~Valles                        \inst{\ref{inst:0003}}
\and W.        ~Van Hamme                     \inst{\ref{inst:0655}}
\and L.        ~Veltz                         \inst{\ref{inst:0241},\ref{inst:0089}}
\and J.        ~Virtanen                      \inst{\ref{inst:0118},\ref{inst:0119}}
\and J.-M.     ~Wallut                        \inst{\ref{inst:0019}}
\and R.        ~Wichmann                      \inst{\ref{inst:0661}}
\and M.I.      ~Wilkinson                     \inst{\ref{inst:0001},\ref{inst:0165}}
\and H.        ~Ziaeepour                     \inst{\ref{inst:0124}}
\and S.        ~Zschocke                      \inst{\ref{inst:0015}}
}
\institute{
     Institute of Astronomy, University of Cambridge, Madingley Road, Cambridge CB3 0HA, United Kingdom\relax                                                                                                \label{inst:0001}
\and INAF - Osservatorio astronomico di Padova, Vicolo Osservatorio 5, 35122 Padova, Italy\relax                                                                                                             \label{inst:0002}
\and Institut de Ci\`{e}ncies del Cosmos, Universitat  de  Barcelona  (IEEC-UB), Mart\'{i}  Franqu\`{e}s  1, E-08028 Barcelona, Spain\relax                                                                  \label{inst:0003}
\and Lund Observatory, Department of Astronomy and Theoretical Physics, Lund University, Box 43, SE-22100 Lund, Sweden\relax                                                                                 \label{inst:0004}
\and Astronomisches Rechen-Institut, Zentrum f\"{ u}r Astronomie der Universit\"{ a}t Heidelberg, M\"{ o}nchhofstr. 12-14, D-69120 Heidelberg, Germany\relax                                                 \label{inst:0005}
\and Scientific Support Office, Directorate of Science, European Space Research and Technology Centre (ESA/ESTEC), Keplerlaan 1, 2201AZ, Noordwijk, The Netherlands\relax                                    \label{inst:0006}
\and Leiden Observatory, Leiden University, Niels Bohrweg 2, 2333 CA Leiden, The Netherlands\relax                                                                                                           \label{inst:0008}
\and GEPI, Observatoire de Paris, PSL Research University, CNRS, Univ. Paris Diderot, Sorbonne Paris Cit{\'e}, 5 Place Jules Janssen, 92190 Meudon, France\relax                                             \label{inst:0009}
\and Max Planck Institute for Astronomy, K\"{ o}nigstuhl 17, 69117 Heidelberg, Germany\relax                                                                                                                 \label{inst:0010}
\and Department of Astronomy, University of Geneva, Chemin des Maillettes 51, CH-1290 Versoix, Switzerland\relax                                                                                             \label{inst:0013}
\and Mission Operations Division, Operations Department, Directorate of Science, European Space Research and Technology Centre (ESA/ESTEC), Keplerlaan 1, 2201 AZ, Noordwijk, The Netherlands\relax          \label{inst:0014}
\and Lohrmann Observatory, Technische Universit\"{ a}t Dresden, Mommsenstra{\ss}e 13, 01062 Dresden, Germany\relax                                                                                           \label{inst:0015}
\and European Space Astronomy Centre (ESA/ESAC), Camino bajo del Castillo, s/n, Urbanizacion Villafranca del Castillo, Villanueva de la Ca\~{n}ada, E-28692 Madrid, Spain\relax                              \label{inst:0016}
\and Laboratoire Lagrange, Universit\'{e} Nice Sophia-Antipolis, Observatoire de la C\^{o}te d'Azur, CNRS, CS 34229, F-06304 Nice Cedex, France\relax                                                        \label{inst:0018}
\and CNES Centre Spatial de Toulouse, 18 avenue Edouard Belin, 31401 Toulouse Cedex 9, France\relax                                                                                                          \label{inst:0019}
\and Institut d'Astronomie et d'Astrophysique, Universit\'{e} Libre de Bruxelles CP 226, Boulevard du Triomphe, 1050 Brussels, Belgium\relax                                                                 \label{inst:0020}
\and F.R.S.-FNRS, Rue d'Egmont 5, 1000 Brussels, Belgium\relax                                                                                                                                               \label{inst:0021}
\and INAF - Osservatorio Astrofisico di Arcetri, Largo Enrico Fermi 5, I-50125 Firenze, Italy\relax                                                                                                          \label{inst:0022}
\and Telespazio Vega UK Ltd for ESA/ESAC, Camino bajo del Castillo, s/n, Urbanizacion Villafranca del Castillo, Villanueva de la Ca\~{n}ada, E-28692 Madrid, Spain\relax                                     \label{inst:0024}
\and Laboratoire d'astrophysique de Bordeaux, Universit\'{e} de Bordeaux, CNRS, B18N, all{\'e}e Geoffroy Saint-Hilaire, 33615 Pessac, France\relax                                                           \label{inst:0025}
\and Instituut voor Sterrenkunde, KU Leuven, Celestijnenlaan 200D, 3001 Leuven, Belgium\relax                                                                                                                \label{inst:0028}
\and Department of Astrophysics/IMAPP, Radboud University Nijmegen, P.O.Box 9010, 6500 GL Nijmegen, The Netherlands\relax                                                                                    \label{inst:0029}
\and Mullard Space Science Laboratory, University College London, Holmbury St Mary, Dorking, Surrey RH5 6NT, United Kingdom\relax                                                                            \label{inst:0031}
\and INAF - Osservatorio Astrofisico di Torino, via Osservatorio 20, 10025 Pino Torinese (TO), Italy\relax                                                                                                   \label{inst:0032}
\and Niels Bohr Institute, University of Copenhagen, Juliane Maries Vej 30, 2100 Copenhagen {\O}, Denmark\relax                                                                                              \label{inst:0033}
\and Centre for Electronic Imaging, Department of Physical Sciences, The Open University, Walton Hall MK7 6AA Milton Keynes, United Kingdom\relax                                                            \label{inst:0038}
\and ALTEC S.p.a, Corso Marche, 79,10146 Torino, Italy\relax                                                                                                                                                 \label{inst:0042}
\and INAF - Osservatorio Astronomico di Bologna, via Ranzani 1, 40127 Bologna,  Italy\relax                                                                                                                  \label{inst:0043}
\and Serco Gesti\'{o}n de Negocios for ESA/ESAC, Camino bajo del Castillo, s/n, Urbanizacion Villafranca del Castillo, Villanueva de la Ca\~{n}ada, E-28692 Madrid, Spain\relax                              \label{inst:0046}
\and Department of Astronomy, University of Geneva, Chemin d'Ecogia 16, CH-1290 Versoix, Switzerland\relax                                                                                                   \label{inst:0055}
\and STFC, Rutherford Appleton Laboratory, Harwell, Didcot, OX11 0QX, United Kingdom\relax                                                                                                                   \label{inst:0059}
\and Gaia DPAC Project Office, ESAC, Camino bajo del Castillo, s/n, Urbanizacion Villafranca del Castillo, Villanueva de la Ca\~{n}ada, E-28692 Madrid, Spain\relax                                          \label{inst:0065}
\and SYRTE, Observatoire de Paris, PSL Research University, CNRS, Sorbonne Universit{\'e}s, UPMC Univ. Paris 06, LNE, 61 avenue de l'Observatoire, 75014 Paris, France\relax                                 \label{inst:0075}
\and National Observatory of Athens, I. Metaxa and Vas. Pavlou, Palaia Penteli, 15236 Athens, Greece\relax                                                                                                   \label{inst:0078}
\and IMCCE, Observatoire de Paris, PSL Research University, CNRS, Sorbonne Universit{\'e}s, UPMC Univ. Paris 06, Univ. Lille, 77 av. Denfert-Rochereau, 75014 Paris, France\relax                            \label{inst:0080}
\and Royal Observatory of Belgium, Ringlaan 3, 1180 Brussels, Belgium\relax                                                                                                                                  \label{inst:0081}
\and Institut d'Astrophysique Spatiale, Universit\'{e} Paris XI, UMR 8617, CNRS, B\^{a}timent 121, 91405, Orsay Cedex, France\relax                                                                          \label{inst:0089}
\and Institute for Astronomy, Royal Observatory, University of Edinburgh, Blackford Hill, Edinburgh EH9 3HJ, United Kingdom\relax                                                                            \label{inst:0091}
\and HE Space Operations BV for ESA/ESAC, Camino bajo del Castillo, s/n, Urbanizacion Villafranca del Castillo, Villanueva de la Ca\~{n}ada, E-28692 Madrid, Spain\relax                                     \label{inst:0093}
\and Institut d'Astrophysique et de G\'{e}ophysique, Universit\'{e} de Li\`{e}ge, 19c, All\'{e}e du 6 Ao\^{u}t, B-4000 Li\`{e}ge, Belgium\relax                                                              \label{inst:0094}
\and \'{A}rea de Lenguajes y Sistemas Inform\'{a}ticos, Universidad Pablo de Olavide, Ctra. de Utrera, km 1. 41013, Sevilla, Spain\relax                                                                     \label{inst:0098}
\and Observatoire Astronomique de Strasbourg, Universit\'{e} de Strasbourg, CNRS, UMR 7550, 11 rue de l'Universit\'{e}, 67000 Strasbourg, France\relax                                                       \label{inst:0101}
\and Kavli Institute for Cosmology, University of Cambridge, Madingley Road, Cambride CB3 0HA, United Kingdom\relax                                                                                          \label{inst:0104}
\and Aurora Technology for ESA/ESAC, Camino bajo del Castillo, s/n, Urbanizacion Villafranca del Castillo, Villanueva de la Ca\~{n}ada, E-28692 Madrid, Spain\relax                                          \label{inst:0109}
\and Laboratoire Univers et Particules de Montpellier, Universit\'{e} Montpellier, Place Eug\`{e}ne Bataillon, CC72, 34095 Montpellier Cedex 05, France\relax                                                \label{inst:0110}
\and Department of Astrophysics, Astronomy and Mechanics, National and Kapodistrian University of Athens, Panepistimiopolis, Zografos, 15783 Athens, Greece\relax                                            \label{inst:0112}
\and Department of Physics and Astronomy, Division of Astronomy and Space Physics, Uppsala University, Box 516, 75120 Uppsala, Sweden\relax                                                                  \label{inst:0113}
\and Universit\`{a} di Catania, Dipartimento di Fisica e Astronomia, Sezione Astrofisica, Via S. Sofia 78, I-95123 Catania, Italy\relax                                                                      \label{inst:0114}
\and INAF - Osservatorio Astrofisico di Catania, via S. Sofia 78, 95123 Catania, Italy\relax                                                                                                                 \label{inst:0115}
\and Universidade da Coru\~{n}a, Facultade de Inform\'{a}tica, Campus de Elvi\~{n}a S/N, 15071, A Coru\~{n}a, Spain\relax                                                                                    \label{inst:0116}
\and CENTRA, Universidade de Lisboa, FCUL, Campo Grande, Edif. C8, 1749-016 Lisboa, Portugal\relax                                                                                                           \label{inst:0117}
\and University of Helsinki, Department of Physics, P.O. Box 64, FI-00014 University of Helsinki, Finland\relax                                                                                              \label{inst:0118}
\and Finnish Geospatial Research Institute FGI, Geodeetinrinne 2, FI-02430 Masala, Finland\relax                                                                                                             \label{inst:0119}
\and Isdefe for ESA/ESAC, Camino bajo del Castillo, s/n, Urbanizacion Villafranca del Castillo, Villanueva de la Ca\~{n}ada, E-28692 Madrid, Spain\relax                                                     \label{inst:0120}
\and ASI Science Data Center, via del Politecnico SNC, 00133 Roma, Italy\relax                                                                                                                               \label{inst:0122}
\and Institut UTINAM UMR6213, CNRS, OSU THETA Franche-Comt\'{e} Bourgogne, Universit\'{e} Bourgogne Franche-Comt\'{e}, F-25000 Besan\c{c}on, France\relax                                                    \label{inst:0124}
\and Dpto. de Inteligencia Artificial, UNED, c/ Juan del Rosal 16, 28040 Madrid, Spain\relax                                                                                                                 \label{inst:0127}
\and Elecnor Deimos Space for ESA/ESAC, Camino bajo del Castillo, s/n, Urbanizacion Villafranca del Castillo, Villanueva de la Ca\~{n}ada, E-28692 Madrid, Spain\relax                                       \label{inst:0136}
\and Thales Services for CNES Centre Spatial de Toulouse, 18 avenue Edouard Belin, 31401 Toulouse Cedex 9, France\relax                                                                                      \label{inst:0137}
\and EURIX S.r.l., via Carcano 26, 10153, Torino, Italy\relax                                                                                                                                                \label{inst:0140}
\and University of Vienna, Department of Astrophysics, T\"{ u}rkenschanzstra{\ss}e 17, A1180 Vienna, Austria\relax                                                                                           \label{inst:0143}
\and Department of Physics and Astronomy, The Johns Hopkins University, 3400 N Charles St, Baltimore, MD 21218, USA\relax                                                                                    \label{inst:0144}
\and ON/MCTI-BR, Rua Gal. Jos\'{e} Cristino 77, Rio de Janeiro, CEP 20921-400, RJ,  Brazil\relax                                                                                                             \label{inst:0146}
\and OV/UFRJ-BR, Ladeira Pedro Ant\^{o}nio 43, Rio de Janeiro, CEP 20080-090, RJ, Brazil\relax                                                                                                               \label{inst:0147}
\and Faculdade Ciencias, Universidade do Porto, Departamento Matematica Aplicada, Rua do Campo Alegre, 687 4169-007 Porto, Portugal\relax                                                                    \label{inst:0153}
\and Instituto de Astrof\'{\i}sica e Ci\^encias do Espa\,co, Universidade de Lisboa Faculdade de Ci\^encias, Campo Grande, PT1749-016 Lisboa, Portugal\relax                                                 \label{inst:0154}
\and Departamento de Astrof\'{i}sica, Centro de Astrobiolog\'{i}a (CSIC-INTA), ESA-ESAC. Camino Bajo del Castillo s/n. 28692 Villanueva de la Ca\~{n}ada, Madrid, Spain\relax                                \label{inst:0163}
\and Department of Physics and Astronomy, University of Leicester, University Road, Leicester LE1 7RH, United Kingdom\relax                                                                                  \label{inst:0165}
\and University of Oviedo, Campus Universitario, 33203 Gij\'{o}n, Spain\relax                                                                                                                                \label{inst:0168}
\and University of C\'{a}diz, Avd. De la universidad, Jerez de la Frontera, C\'{a}diz, Spain\relax                                                                                                           \label{inst:0171}
\and Kapteyn Astronomical Institute, University of Groningen, Landleven 12, 9747 AD Groningen, The Netherlands\relax                                                                                         \label{inst:0185}
\and Consorci de Serveis Universitaris de Catalunya, C/ Gran Capit\`{a}, 2-4 3rd floor, 08034 Barcelona, Spain\relax                                                                                         \label{inst:0195}
\and University of Turin, Department of Computer Sciences, Corso Svizzera 185, 10149 Torino, Italy\relax                                                                                                     \label{inst:0197}
\and INAF - Osservatorio Astronomico di Roma, Via di Frascati 33, 00078 Monte Porzio Catone (Roma), Italy\relax                                                                                              \label{inst:0201}
\and CRAAG - Centre de Recherche en Astronomie, Astrophysique et G\'{e}ophysique, Route de l'Observatoire Bp 63 Bouzareah 16340 Algiers, Algeria\relax                                                       \label{inst:0215}
\and Universiteit Antwerpen, Onderzoeksgroep Toegepaste Wiskunde, Middelheimlaan 1, 2020 Antwerpen, Belgium\relax                                                                                            \label{inst:0218}
\and Department of Physics and Astronomy, University of Padova, Via Marzolo 8, I-35131 Padova, Italy\relax                                                                                                   \label{inst:0220}
\and INAF - Osservatorio Astronomico di Teramo, Via Mentore Maggini, 64100 Teramo, Italy\relax                                                                                                               \label{inst:0222}
\and INAF - Osservatorio Astronomico di Capodimonte, Via Moiariello 16, 80131, Napoli, Italy\relax                                                                                                           \label{inst:0224}
\and Instituto de Astronomia, Geof\`{i}sica e Ci\^{e}ncias Atmosf\'{e}ricas, Universidade de S\~{a}o Paulo, Rua do Mat\~{a}o, 1226, Cidade Universitaria, 05508-900 S\~{a}o Paulo, SP, Brazil\relax          \label{inst:0225}
\and Department of Geosciences, Tel Aviv University, Tel Aviv 6997801, Israel\relax                                                                                                                          \label{inst:0238}
\and Astronomical Institute Anton Pannekoek, University of Amsterdam, PO Box 94249, 1090 GE, Amsterdam, The Netherlands\relax                                                                                \label{inst:0239}
\and Leibniz Institute for Astrophysics Potsdam (AIP), An der Sternwarte 16, 14482 Potsdam, Germany\relax                                                                                                    \label{inst:0241}
\and ATOS for CNES Centre Spatial de Toulouse, 18 avenue Edouard Belin, 31401 Toulouse Cedex 9, France\relax                                                                                                 \label{inst:0244}
\and School of Physics and Astronomy, Tel Aviv University, Tel Aviv 6997801, Israel\relax                                                                                                                    \label{inst:0247}
\and UNINOVA - CTS, Campus FCT-UNL, Monte da Caparica, 2829-516 Caparica, Portugal\relax                                                                                                                     \label{inst:0248}
\and Laboratoire G\'{e}oazur, Universit\'{e} Nice Sophia-Antipolis, UMR 7329, CNRS, Observatoire de la C\^{o}te d'Azur, 250 rue A. Einstein, F-06560 Valbonne, France\relax                                  \label{inst:0254}
\and RHEA for ESA/ESAC, Camino bajo del Castillo, s/n, Urbanizacion Villafranca del Castillo, Villanueva de la Ca\~{n}ada, E-28692 Madrid, Spain\relax                                                       \label{inst:0260}
\and Astronomical Institute, Academy of Sciences of the Czech Republic, Fri\v{c}ova 298, 25165 Ond\v{r}ejov, Czech Republic\relax                                                                            \label{inst:0262}
\and Barcelona Supercomputing Center - Centro Nacional de Supercomputaci\'{o}n, c/ Jordi Girona 29, Ed. Nexus II, 08034 Barcelona, Spain\relax                                                               \label{inst:0276}
\and Department of Mechanical Engineering, University of La Rioja, c/ San Jos\'{e} de Calasanz, 31, 26004 Logro\~{n}o, La Rioja, Spain\relax                                                                 \label{inst:0279}
\and ETSE Telecomunicaci\'{o}n, Universidade de Vigo, Campus Lagoas-Marcosende, 36310 Vigo, Galicia, Spain\relax                                                                                             \label{inst:0281}
\and SRON, Netherlands Institute for Space Research, Sorbonnelaan 2, 3584CA, Utrecht, The Netherlands\relax                                                                                                  \label{inst:0307}
\and Faculty of Mathematics and Physics, University of Ljubljana, Jadranska ulica 19, 1000 Ljubljana, Slovenia\relax                                                                                         \label{inst:0313}
\and Physics Department, University of Antwerp, Groenenborgerlaan 171, 2020 Antwerp, Belgium\relax                                                                                                           \label{inst:0315}
\and Harvard-Smithsonian Center for Astrophysics, 60 Garden Street, Cambridge MA 02138, USA\relax                                                                                                            \label{inst:0317}
\and Institut de Physique de Rennes, Universit{\'e} de Rennes 1, F-35042 Rennes, France\relax                                                                                                                \label{inst:0332}
\and Shanghai Astronomical Observatory, Chinese Academy of Sciences, 80 Nandan Rd, 200030 Shanghai, China\relax                                                                                              \label{inst:0341}
\and CSC Danmark A/S, Retortvej 8, 2500 Valby, Denmark\relax                                                                                                                                                 \label{inst:0344}
\and Las Cumbres Observatory Global Telescope Network, Inc., 6740 Cortona Drive, Suite 102, Goleta, CA  93117, USA\relax                                                                                     \label{inst:0345}
\and Astrophysics Research Institute, Liverpool John Moores University, L3 5RF, United Kingdom\relax                                                                                                         \label{inst:0357}
\and Konkoly Observatory, Research Centre for Astronomy and Earth Sciences, Hungarian Academy of Sciences, Konkoly Thege Mikl\'{o}s \'{u}t 15-17, 1121 Budapest, Hungary\relax                               \label{inst:0363}
\and Baja Observatory of University of Szeged, Szegedi \'{u}t III/70, 6500 Baja, Hungary\relax                                                                                                               \label{inst:0364}
\and Laboratoire AIM, IRFU/Service d'Astrophysique - CEA/DSM - CNRS - Universit\'{e} Paris Diderot, B\^{a}t 709, CEA-Saclay, F-91191 Gif-sur-Yvette Cedex, France\relax                                      \label{inst:0365}
\and INAF - Osservatorio Astronomico di Trieste, Via G.B. Tiepolo 11, 34143, Trieste, Italy\relax                                                                                                            \label{inst:0378}
\and Laboratoire de l'Acc\'{e}l\'{e}rateur Lin\'{e}aire, Universit\'{e} Paris-Sud, CNRS/IN2P3, Universit\'{e} Paris-Saclay, 91898 Orsay Cedex, France\relax                                                  \label{inst:0380}
\and \'{E}cole polytechnique f\'{e}d\'{e}rale de Lausanne, SB MATHAA STAP, MA B1 473 (B\^{a}timent MA), Station 8, CH-1015 Lausanne, Switzerland\relax                                                       \label{inst:0386}
\and INAF/IASF-Bologna, Via P. Gobetti 101, 40129 Bologna, Italy\relax                                                                                                                                       \label{inst:0394}
\and Technical University of Madrid, Jos\'{e} Guti\'{e}rrez Abascal 2, 28006 Madrid, Spain\relax                                                                                                             \label{inst:0397}
\and EQUERT International for CNES Centre Spatial de Toulouse, 18 avenue Edouard Belin, 31401 Toulouse Cedex 9, France\relax                                                                                 \label{inst:0406}
\and AKKA for CNES Centre Spatial de Toulouse, 18 avenue Edouard Belin, 31401 Toulouse Cedex 9, France\relax                                                                                                 \label{inst:0413}
\and Villanova University, Dept. of Astrophysics and Planetary Science, 800 E Lancaster Ave, Villanova PA 19085, USA\relax                                                                                   \label{inst:0414}
\and Vitrociset Belgium for ESA/ESAC, Camino bajo del Castillo, s/n, Urbanizacion Villafranca del Castillo, Villanueva de la Ca\~{n}ada, E-28692 Madrid, Spain\relax                                         \label{inst:0419}
\and Fork Research, Rua do Cruzado Osberno, Lt. 1, 9 esq., Lisboa, Portugal\relax                                                                                                                            \label{inst:0441}
\and APAVE SUDEUROPE SAS for CNES Centre Spatial de Toulouse, 18 avenue Edouard Belin, 31401 Toulouse Cedex 9, France\relax                                                                                  \label{inst:0444}
\and Spanish Virtual Observatory\relax                                                                                                                                                                       \label{inst:0451}
\and Fundaci\'{o}n Galileo Galilei - INAF, Rambla Jos\'{e} Ana Fern\'{a}ndez P\'{e}rez 7, E-38712 Bre\~{n}a Baja, Santa Cruz de Tenerife, Spain\relax                                                        \label{inst:0462}
\and INSA for ESA/ESAC, Camino bajo del Castillo, s/n, Urbanizacion Villafranca del Castillo, Villanueva de la Ca\~{n}ada, E-28692 Madrid, Spain\relax                                                       \label{inst:0468}
\and Dpto. Arquitectura de Computadores y Autom\'{a}tica, Facultad de Inform\'{a}tica, Universidad Complutense de Madrid, C/ Prof. Jos\'{e} Garc\'{i}a Santesmases s/n, 28040 Madrid, Spain\relax            \label{inst:0469}
\and H H Wills Physics Laboratory, University of Bristol, Tyndall Avenue, Bristol BS8 1TL, United Kingdom\relax                                                                                              \label{inst:0472}
\and Stellar Astrophysics Centre, Aarhus University, Department of Physics and Astronomy, 120 Ny Munkegade, Building 1520, DK-8000 Aarhus C, Denmark\relax                                                   \label{inst:0475}
\and Applied Physics Department, University of Vigo, E-36310 Vigo, Spain\relax                                                                                                                               \label{inst:0478}
\and HE Space Operations BV for ESA/ESTEC, Keplerlaan 1, 2201AZ, Noordwijk, The Netherlands\relax                                                                                                            \label{inst:0490}
\and Warsaw University Observatory, Al. Ujazdowskie 4, 00-478 Warszawa, Poland\relax                                                                                                                         \label{inst:0499}
\and Instituto de Astrof\'{\i}sica de Canarias, E-38205 La Laguna, Tenerife, Spain\relax                                                                                                                     \label{inst:0508}
\and Universidad de La Laguna, Departamento de Astrof\'{\i}sica, E-38206 La Laguna, Tenerife, Spain\relax                                                                                                    \label{inst:0509}
\and RHEA for ESA/ESTEC, Keplerlaan 1, 2201AZ, Noordwijk, The Netherlands\relax                                                                                                                              \label{inst:0519}
\and Max Planck Institute for Solar System Research, Justus-von-Liebig-Weg 3, 37077 G\"{ o}ttingen, Germany\relax                                                                                            \label{inst:0523}
\and SISSA (Scuola Internazionale Superiore di Studi Avanzati), via Bonomea 265, 34136 Trieste, Italy\relax                                                                                                  \label{inst:0524}
\and Instituto Nacional de Pesquisas Espaciais/Minist\'{e}rio da Ciencia Tecnologia, Avenida dos Astronautas 1758, S\~{a}o Jos\'{e} Dos Campos, SP 12227-010, Brazil\relax                                   \label{inst:0533}
\and Argelander Institut f\"{ u}r Astronomie der Universit\"{ a}t Bonn, Auf dem H\"{ u}gel 71, 53121 Bonn, Germany\relax                                                                                     \label{inst:0537}
\and European Southern Observatory (ESO), Karl-Schwarzschild-Stra{\ss}e 2, 85748 Garching bei M\"{ u}nchen, Germany\relax                                                                                    \label{inst:0540}
\and Laboratory of Optics, Lasers and Systems, Faculty of Sciences, University of Lisbon, Campus do Lumiar, Estrada do Pa\c{c}o do Lumiar, 22, 1649-038 Lisboa, Portugal\relax                               \label{inst:0543}
\and Department of Physics and Astronomy, Notre Dame University, Louaize, PO Box 72, Zouk Mika\"{ e}l, Lebanon\relax                                                                                         \label{inst:0554}
\and University of Nova Gorica, Vipavska 13, 5000 Nova Gorica, Slovenia\relax                                                                                                                                \label{inst:0556}
\and Max Planck Institute for Extraterrestrial Physics, OPINAS, Gie{\ss}enbachstra{\ss}e, 85741 Garching, Germany\relax                                                                                      \label{inst:0560}
\and NASA/IPAC Infrared Science Archive, California Institute of Technology, Mail Code 100-22, 770 South Wilson Avenue, Pasadena, CA, 91125, USA\relax                                                       \label{inst:0567}
\and Center of Applied Space Technology and Microgravity (ZARM), c/o Universit\"{ a}t Bremen, Am Fallturm 1, 28359 Bremen, Germany\relax                                                                     \label{inst:0569}
\and RHEA System for ESA/ESOC, Robert Bosch Stra{\ss}e 5, 64293 Darmstadt, Germany\relax                                                                                                                     \label{inst:0570}
\and Tartu Observatory, 61602 T\~{o}ravere, Estonia\relax                                                                                                                                                    \label{inst:0576}
\and Sydney Institute for Astronomy, School of Physics A28, The University of Sydney, NSW 2006, Australia\relax                                                                                              \label{inst:0578}
\and Slovak Organisation for Space Activities, Zamocka 18, 85101 Bratislava, Slovak Republic\relax                                                                                                           \label{inst:0580}
\and National Astronomical Observatories, CAS, 100012 Beijing, China\relax                                                                                                                                   \label{inst:0584}
\and US Naval Observatory, Astrometry Department, 3450 Massachusetts Ave. NW, Washington DC 20392-5420 D.C., USA\relax                                                                                       \label{inst:0586}
\and European Southern Observatory (ESO), Alonso de C\'{o}rdova 3107, Vitacura, Casilla 19001, Santiago de Chile, Chile\relax                                                                                \label{inst:0589}
\and Airbus Defence and Space SAS, 31 Rue des Cosmonautes, 31402 Toulouse Cedex 4, France\relax                                                                                                              \label{inst:0593}
\and EJR-Quartz BV for ESA/ESTEC, Keplerlaan 1, 2201AZ, Noordwijk, The Netherlands\relax                                                                                                                     \label{inst:0600}
\and The Server Labs for ESA/ESAC, Camino bajo del Castillo, s/n, Urbanizacion Villafranca del Castillo, Villanueva de la Ca\~{n}ada, E-28692 Madrid, Spain\relax                                            \label{inst:0603}
\and Astronomical Observatory Institute, Faculty of Physics, A. Mickiewicz University, ul. S\l{}oneczna 36, 60-286 Pozna\'{n}, Poland\relax                                                                  \label{inst:0607}
\and CS Syst\`{e}mes d'Information for CNES Centre Spatial de Toulouse, 18 avenue Edouard Belin, 31401 Toulouse Cedex 9, France\relax                                                                        \label{inst:0618}
\and Directorate of Science, European Space Research and Technology Centre (ESA/ESTEC), Keplerlaan 1, 2201AZ, Noordwijk, The Netherlands\relax                                                               \label{inst:0620}
\and Praesepe BV for ESA/ESAC, Camino bajo del Castillo, s/n, Urbanizacion Villafranca del Castillo, Villanueva de la Ca\~{n}ada, E-28692 Madrid, Spain\relax                                                \label{inst:0621}
\and Sorbonne Universit\'{e}s UPMC et CNRS, UMR7095, Institut d'Astrophysique de Paris, F75014, Paris, France\relax                                                                                          \label{inst:0625}
\and GMV for ESA/ESAC, Camino bajo del Castillo, s/n, Urbanizacion Villafranca del Castillo, Villanueva de la Ca\~{n}ada, E-28692 Madrid, Spain\relax                                                        \label{inst:0631}
\and Institute of Theoretical Physics and Astronomy, Vilnius University, Sauletekio al. 3, Vilnius, LT-10222, Lithuania\relax                                                                                \label{inst:0640}
\and S[\&]T Corporation, PO Box 608, 2600 AP, Delft, The Netherlands\relax                                                                                                                                   \label{inst:0642}
\and Department of Space Studies, Southwest Research Institute (SwRI), 1050 Walnut Street, Suite 300, Boulder, Colorado 80302, USA\relax                                                                     \label{inst:0643}
\and Deutsches Zentrum f\"{ u}r Luft- und Raumfahrt, Institute of Space Systems, Am Fallturm 1, D-28359 Bremen, Germany\relax                                                                                \label{inst:0644}
\and University of Applied Sciences Munich, Karlstr. 6, 80333 Munich, Germany\relax                                                                                                                          \label{inst:0646}
\and Dipartimento di Fisica, Universit\`{a} di Roma Tor Vergata, via della Ricerca Scientifica 1, 00133 Rome, Italy\relax                                                                                    \label{inst:0648}
\and Department of Physics and Astronomy, University of the Western Cape, Robert Sobukwe Road, 7535 Bellville, Cape Town, South Africa\relax                                                                 \label{inst:0651}
\and INAF - Istituto di Radioastronomia, via Gobetti 101, 40129 Bologna, Italy\relax                                                                                                                         \label{inst:0652}
\and Department of Physics, Florida International University, 11200 SW 8th Street, Miami, FL 33199, USA\relax                                                                                                \label{inst:0655}
\and Hamburger Sternwarte, Gojenbergsweg 112, D-21029 Hamburg, Germany\relax                                                                                                                                 \label{inst:0661}
}

\date{Received \textbf{Febr. 3, 2017}; accepted \textbf{Febr. 25, 2017}}

 
\abstract
{The first \Gaia Data Release contains the Tycho-Gaia Astrometric Solution (\TGAS). This is a subset of about 2 million stars for which, besides the position and photometry, the proper motion and parallax are calculated using \Hipparcos and \Tycho positions in 1991.25 as prior information.}
{We investigate the scientific potential and limitations of the \TGAS component by means of the astrometric data for open clusters.}
{Mean cluster parallax and proper motion values are derived taking into account the error correlations within the astrometric solutions for individual stars, an estimate of the internal velocity dispersion in the cluster, and, where relevant, the effects of the depth of the cluster along the line of sight. Internal consistency of the \TGAS data is assessed.}
{Values given for standard uncertainties are still inaccurate and may lead to unrealistic unit-weight standard deviations of least squares solutions for cluster parameters. Reconstructed mean cluster parallax and proper motion values are generally in very good agreement with earlier \Hipparcos-based determination, although the \Gaia mean parallax for the Pleiades is a significant exception. We have no current explanation for that discrepancy. Most clusters are observed to extend to nearly 15 pc from the cluster centre, and it will be up to future \Gaia releases to establish whether those potential cluster-member stars are still dynamically bound to the clusters.}
{The \Gaia DR1 provides the means to examine open clusters far beyond their more easily visible cores, and can provide membership assessments based on proper motions and parallaxes. A combined HR diagram shows the same features as observed before using the \Hipparcos data, with clearly increased luminosities for older A and F dwarfs.}

 


\maketitle
%


\section{Introduction}

The homogeneity in age and composition of stars in open clusters makes them unique and very valuable potential tracers of stellar evolution and galactic structure. However, to reach this potential it is essential that cluster membership and absolute distances are determined fully independent of assumptions on luminosities. Photometric and spectroscopic data should be obtained on a single accurate and full-sky-coverage system. To determine distances for open clusters, a sizeable fraction of the members need to be covered, and for the nearby clusters the variations along the line of sight, and direction on the sky, in parallax and proper motion need to be fully accounted for. This is the kind of task that is only possible to achieve with a dedicated satellite mission, and was first done using the \Hipparcos astrometric data in conjunction with the Geneva photometric surveys \citep[fvl09 from hereon]{2009A&A...497..209V}.

The \TGAS catalogue in the first \Gaia data release \citep{DPACP-18} (DR1 from hereon) provides an order of magnitude more data than the \Hipparcos catalogue did, but at the same time, because of the limitations in its construction, it is more problematic and complicated in its use and interpretation \citep{DPACP-14, DPACP-8, DPACP-16}. The combination with the first epoch from the new reduction of the \Hipparcos data \citep{1997ESASP1200.....E, 2007A&A...474..653V} and \Tycho \citep{Tycho2} data, as well as the still very limited scan coverage of the \Gaia data in this first data release, creates locally strong and systematic correlations between the astrometric parameters as determined for individual stars. Error-correlation coefficients between the five astrometric parameters still frequently exceed values as high as 0.8, and need to be taken into account when determining both mean parallax and mean proper motion data for a cluster. Many details on this can be found in \cite{DPACP-14}.

The way the data had to be processed also plays an important role. In particular simplifications in the attitude reconstruction (because of low numbers of reference stars) meant that the effects of clanks\footnote{discrete adjustments of the satellite structure, and thus telescope pointing, to temperature changes} and minor hits\footnote{impacts of external particles, causing discrete rate changes}  were smoothed over, leading to locally correlated errors on the epoch astrometric data, a problem that should be largely resolved in future releases. This first release on the \Gaia star cluster data is therefore a taste of things to come, and provides some ideas on how to handle the \Gaia astrometric data for a star cluster. The data derived for the clusters can still be affected by local systematics in the \TGAS catalogue, claimed to be at a level of 0.3~mas \citep{DPACP-8}, and, as we will show, comparisons with the \Hipparcos 
astrometric data for clusters are consistent with a slightly lower level of 
systematics, at 0.25~mas.

The homogeneity of the astrometric data for members of an open cluster offers possibilities to study some aspects of the proper motions and parallaxes as presented in the \TGAS section of the \Gaia DR1. In particular the reliability of the standard uncertainties (\su from hereon) as quoted in DR1 can be checked, and localized correlated errors may show up. Different roles are there for the nearest cluster (Hyades), eight medium distance clusters (within 300 pc: Coma Berenices, Pleiades, IC2391, IC2602, $\alpha$~Per cluster, Praesepe, Blanco 1, NGC2451A) and ten more distant clusters (between 300 and 500 pc: NGC6475, NGC7092, NGC2516, NGC2232, IC4665, NGC6633, Coll140, NGC2422, NGC3532 and NGC2547). Table~\ref{tab:clusterid} provides further identifiers of the clusters presented in this paper. The Hyades permits a consistency comparison between proper motions and parallaxes over an area up to 36 degrees in diameter on the sky. The second group is used to assess consistency of the \su on the astrometric parameters of individual stars. The third group, for which the density on the sky of potential cluster members is higher, can be used to assess the effects of error correlations between neighbouring stars. Most of these tests are ultimately limited by the uncertainty in the estimate of the internal velocity dispersion in the clusters, and in particular its dependence on the 3D position within the cluster. For the more distant clusters there is the additional limitation of ascertaining membership of a cluster.

\begin{table}[t]
\caption{Open cluster names and identifiers}
\centering
\begin{tabular}{lrr}
\hline
\hline
Name &  Lynga & Melotte \\
\hline
Hyades & C0423+157 & 25 \\
Coma Ber &  C1222+263 & 111 \\
Praesepe &  C0837+201 & 88 \\
Pleiades &  C0344+239 & 22 \\
$\alpha$~Per &  C0318+484 & 20 \\
IC2391 &  C0838$-$528 & \\
IC2602 &  C1041$-$641 & \\
Blanco 1 &  C0001$-$302 & \\ 
NGC2451A & C0743$-$378 & \\
NGC6475 &  C1750$-$348 & \\
NGC7092 &  C2130+482 & \\
NGC2516 & C0757$-$607 & \\
NGC2232 &  C0624$-$047 & \\
IC4665 &  C1743+057 & 179 \\
NGC6633 & C1825+065 & \\
Coll140 & C0722$-$321 & \\
NGC2422 &  C0734$-$143 & \\
NGC3532 &  C1104$-$584 & \\
NGC2547 &  C0809$-$491 & \\
\hline
\end{tabular}
\label{tab:clusterid}
\end{table}

\begin{table}[t]
\caption{Supplementary data}
\centering
\begin{tabular}{l|rrr}
\hline
\hline
Name & Fe/H & E(B$-$V) & log(age) \\
\hline
Hyades &       $0.15\pm 0.004$ & 0.00 & 8.90 \\
Coma Ber &     $0.00\pm  0.08$ & 0.00 & 8.75 \\
Praesepe &     $0.16\pm 0.004$ & 0.01 & 8.90 \\
Pleiades &     $-0.01\pm 0.05$ & 0.04 & 8.08\\
$\alpha$~Per & $0.14\pm  0.11$ & 0.09 & 7.55 \\
IC2391 &       $-0.01\pm 0.03$ & 0.05 & 7.55\\
IC2606 &       $-0.02\pm 0.02$ & 0.03 & 7.88\\
Blanco 1 &     $0.03\pm  0.07$ & 0.01 & 8.32\\
NGC2451A &     -0.08           & 0.00 & 7.76 \\
NGC6475 &      $0.02\pm 0.02$  & 0.21 & 8.22\\
NGC7092 &       0.00           & 0.01 & 8.57\\
NGC2516 &      $+0.05\pm 0.11$ & 0.07 & 8.08\\
NGC2232 &       0.11           & 0.03 & 7.49\\
IC4665 &       $-0.03\pm 0.04$ & 0.17 & 7.63 \\
NGC6633 &      $-0.08\pm 0.12$ & 0.17 & 8.76\\
Coll140 &      $0.01\pm 0.04$  & 0.05 & 7.57\\
NGC2422 &      $0.09\pm 0.03$  & 0.10 & 8.12 \\
NGC3532 &      $0.00\pm 0.07$  & 0.04 & 8.45\\
NGC2547 &      $-0.14\pm 0.10$ & 0.04 & 7.70 \\
\hline
\end{tabular}
\tablefoot{Metallicities for Hyades and Praesepe are from \cite{2017arXiv170203936C}. For the other clusters are from \cite{2016A&A...585A.150N}. $E_{B-V}$ are from \cite{2016A&A...585A.101K}}.
\label{tab:suppldata}
\end{table}

Comparisons of the astrometric data are generally kept limited to fvl09, based on the re-reduction of the \Hipparcos data \citep{2007A&A...474..653V}, and which superseded the earlier analysis of the \Hipparcos astrometry for open clusters in \cite{1999A&A...341L..71V} and \cite{1999A&A...345..471R}. The paper fvl09 provides more extensive references to earlier studies of the clusters selected for the current study. Table~\ref{tab:suppldata} summarizes, where available, external data on the clusters.

In order to appreciate the possibilities as well as the limitations inherent to the \TGAS component of the \Gaia DR1, and in particular where these affect our analysis of cluster data, we provide some background information on the data in Sect.~\ref{sec:inputdata}. This includes a discussion of the not-published epoch astrometry data in order to assess the potential level of error correlations between neighbouring stars.

A summary of the methods used to derive cluster astrometry is presented in Sect.~\ref{sec:genappr}, with more details provided in App.~\ref{app:combine}. This is followed by the analysis of the Hyades (Sect.~\ref{sec:hyades}) and the nearby clusters (Sect.~\ref{sec:nearby}). The distant clusters (Sect.~\ref{sec:distant}) pose their own specific problems, and are only briefly discussed here. A summary of the results is presented in Sect.~\ref{sect:summ}.

\Gaia source identifiers are based on the HEALPix pixelization (Nested, level 12) of the sky \citep{2005ApJ...622..759G}, and all-sky maps shown in the current paper use this pixelization, usually at level 5 or 6, where level 6 has pixel-size of just under one square degree. An integer division of the source identifier by $2^{35}$ gives the level 12 HEALPix pixel for the source location on the sky. Source identifiers may change in future releases. The positions, magnitudes and HD numbers are therefore the more relevant source identifiers.

The additional photometric data used here comes primarily from the Geneva photometric catalogue \citep{genevaphot}, which provides multi-colour intermediate bandwidth photometry for a wide range of open clusters. Where possible the photometric data as presented is for cluster members confirmed by \Gaia or \Hipparcos astrometric data only. 

\section{The input data \label{sec:inputdata}}
The \Gaia data is obtained from an array of CCDs in the focal plane, operating in Time-Delayed Integration (TDI) mode. The CCD charges are following the images as these move across the CCDs, taking about 4.5~s to cross a single CCD. In order to extend the brightness range for sources to be observed, gates are applied to shorten the integration time for the brighter stars. For more details see \cite{DPACP-8, DPACP-18}.

\begin{figure}[t]
\centering
\includegraphics[width=8.5cm]{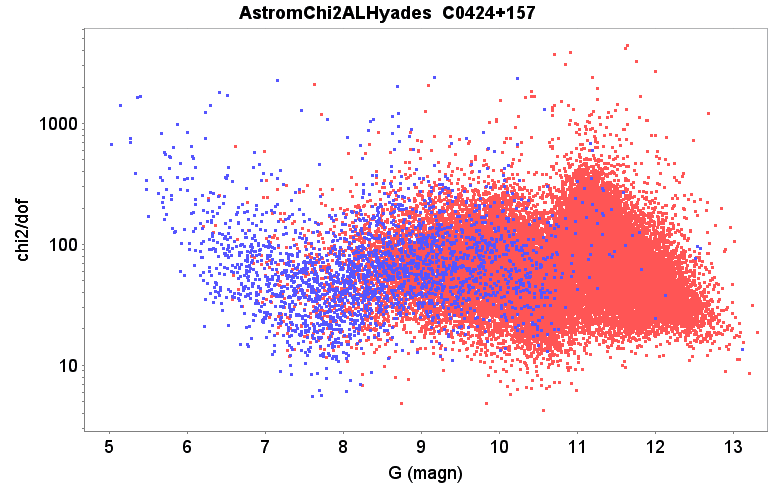}
\caption{This diagram shows the logarithm of the square root of the normalized $\chi^2$ values for the astrometric solutions as a function of the $G$-band magnitude. The data come from an 18 degrees radius field, centred on the Hyades cluster. The blue dots used first epoch from the \Hipparcos catalogue, the red dots from the \Tycho catalogue.}
\label{fig:logchi2}
\end{figure}
The \TGAS astrometric data, forming part of the \Gaia DR1 \citep{DPACP-8, DPACP-18}, are based on first-epoch positions from the new reduction of the \Hipparcos catalogue \citep{2007A&A...474..653V} and the \Tycho \citep{Tycho2} catalogue (when a star was not included in the \Hipparcos catalogue) and overall 14 months of \Gaia data, though locally the coverage will often be significantly less than 14 months. The \Gaia survey nominally covers the sky in at least two scan directions every six months. Having been collected at the start of the mission, this is not the best data \Gaia will obtain. There have been a range of issues that affected the data and the data processing, most of it leading to some form of (temporary) data loss and still poorly defined \su values on extracted parameters. In particular the transit time \su estimates were still inaccurate due to early limitations on the modelling of the point-spread functions, leading to large $\chi^2$ values for astrometric solutions (see Fig.~\ref{fig:logchi2}). When the normalized $\chi^2$ values are as large as observed here it means that there are quite significant modelling errors still present. Naturally, the worst affected are the brightest stars (brighter than $G\approx 6$), of which, as a result, a large fraction is not included in the \Gaia DR1. Modelling errors tend to be non-Gaussian, and can hide a range of systematic errors in the data.

\begin{figure}[t]
\centering
\includegraphics[width=8cm]{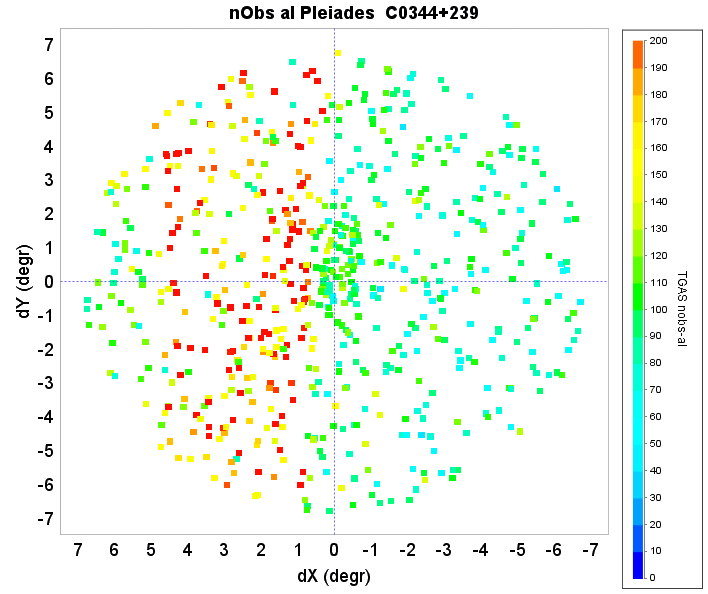}
\caption{The number of observations (CCD transits) per source in the \TGAS catalogue for the Pleiades field. At this early stage there are still large local variations in the coverage. Positions are relative to the assumed centre of the Pleiades cluster. Each point represents a cluster or field star in the area.}
\label{fig:pleinobs}
\end{figure}
\begin{figure}[t]
\centering
\includegraphics[width=8cm]{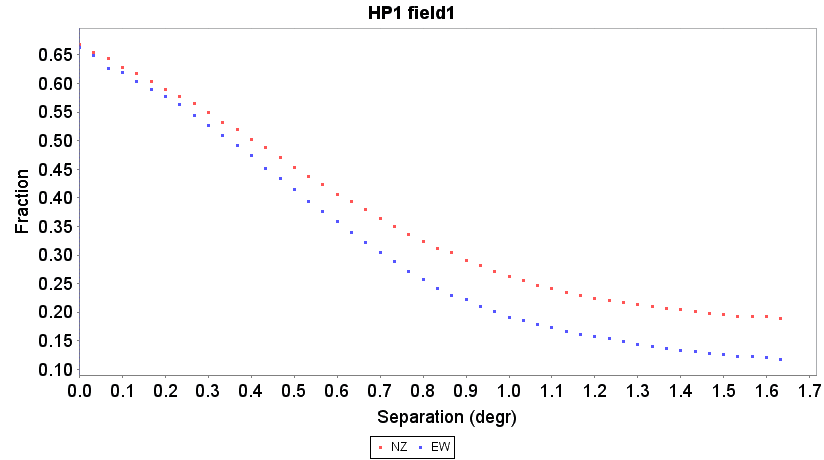}
\caption{An example of the scan-coincidence fraction as a function of separation on the sky. The coincidence level should be approaching 1 at separations much smaller than the size of the field of view (as it is for the \Hipparcos data), but is found to be between 0.55 and 0.7 in the \TGAS data. The red dots represent correlations (in ecliptic coordinates) in the North and South quadrants (between $\pm 45\degr$ from the North or South directions), the blue dots in the East and West quadrants (between $\pm 45\degr$ from the East or West directions). The scan coverage is significantly different between North-South and East-West directions. The data are for all stars in HEALPix (level 2), pixel 1, an area of about 833 square degrees (equatorial coordinates).
\label{fig:scancoinc}}
\end{figure}
\begin{figure}[t]
\centering
\includegraphics[width=8cm]{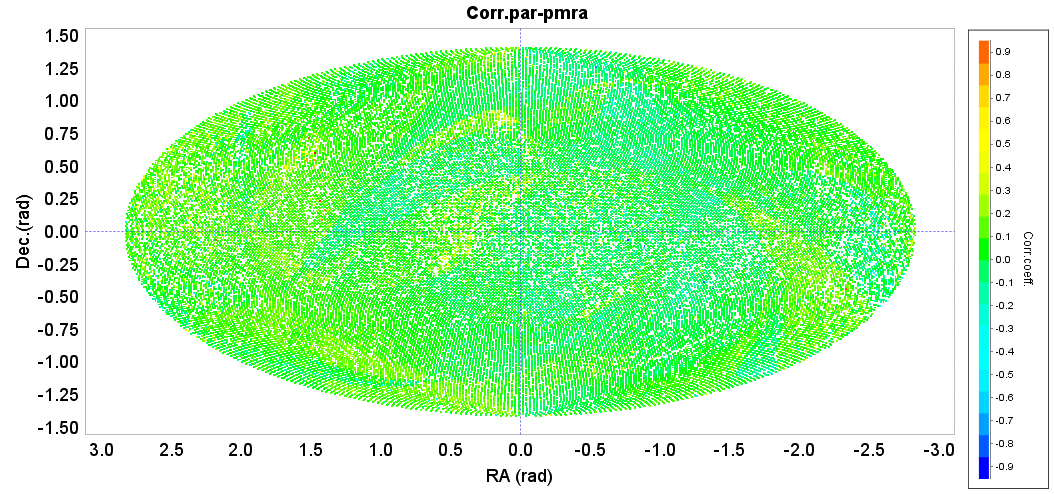}
\includegraphics[width=8cm]{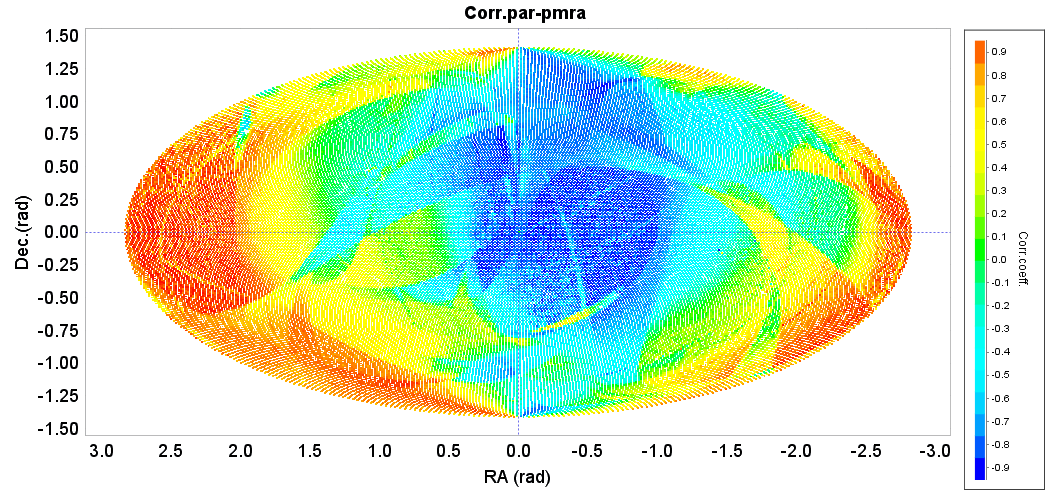}
\caption{Correlation coefficients between the proper motion in Right Ascension and the parallax, averaged over HEALPix level 5 pixels. Top: for stars with first-epoch \Hipparcos positions; bottom: for stars with first-epoch \Tycho positions. Correlations in the bottom graph are clearly systematic over the sky (linked to scan coverage) and can reach values over $\pm$ 0.9. Similar correlations, but differently distributed, are observed between all the astrometric parameters for data with first epoch \Tycho positions.}
\label{fig:pmraparcorr}
\end{figure}

Quite large variations in the number of transits per star across the field of a cluster do often occur. This is an early-mission feature and is due to the scanning law and data gaps when it concerns large-scale features, such as shown in Fig.~\ref{fig:pleinobs} for the Pleiades field. For small-scale, local variations this is probably due to a variety of source-identification problems which, at this early stage, still appears to cause a significant loss of data. The approximate level of data loss can be derived from the epoch astrometric data (see below), which shows typically a coincidence of scans between stars at relatively short separations on the sky (much shorter than the size of the field of view) to be around 55 to 70 ~per~cent (an example is shown in Fig.~\ref{fig:scancoinc}), when values close to 100~per~cent would be expected, as has been observed for the \Hipparcos data  \citep[see][Fig.~9]{2007A&A...474..653V}. Assuming that neighbouring stars are affected by the same percentage of data loss, then a 60~per~cent coincidence of scans would indicate that this loss amounts to 22.5~per~cent. These differences in coverage may explain the large local variations observed in the covariance matrices for the individual stellar astrometric solutions, which have to be taken properly into account. This is shown to affect stars with first-epoch \Tycho data much more severely than those with first-epoch \Hipparcos data. An example of the correlation coefficients, and the variations thereof, between the derived astrometric parameters is shown in Fig.~\ref{fig:pmraparcorr}. 

The modulation of the basic angle, though corrected for with great care, adds uncertainty about the local parallax zero point, which is reflected in the assumed additional noise on the parallax determinations. The orientation of the payload as a function of time, which is referred to as the satellite attitude, is controlled by micro-propulsion thrusters, and affected by numerous clanks and hits \citep{DPACP-14, 2013A&A...551A..19R}. The on-ground reconstruction of the attitude provides an estimate of the orientation of the telescope reference frame as a function of time, and as such is the reference against which the observed transit times are converted to positions, creating the so-called one-dimensional epoch astrometric data. These are the measurements used in the astrometric solutions. Inaccuracies in the modelling of the reconstructed attitude will reflect in the epoch astrometric data as correlated errors for neighbouring stars. Simplifications in the attitude reconstruction model as used in GDR1 concern:
\begin{enumerate}
\item use of gated observations in the attitude reconstructions;
\item smoothing over clanks and hits.
\end{enumerate}
A gated observation is one for which the integration was done over a fraction of the CCD to avoid saturation for very bright stars. The effects on the attitude reconstruction are described in \cite{2013A&A...551A..19R}. In simple terms, the different integration times affect the way clanks are `seen' by transits. 

Against this background, one has to be careful in deriving conclusions on, for example, open cluster astrometric data, which relies on combining data as obtained for individual member stars contained within a small area on the sky, within which the data may be affected by correlated errors. 

\begin{figure}[t]
\centering
\includegraphics[width=8cm]{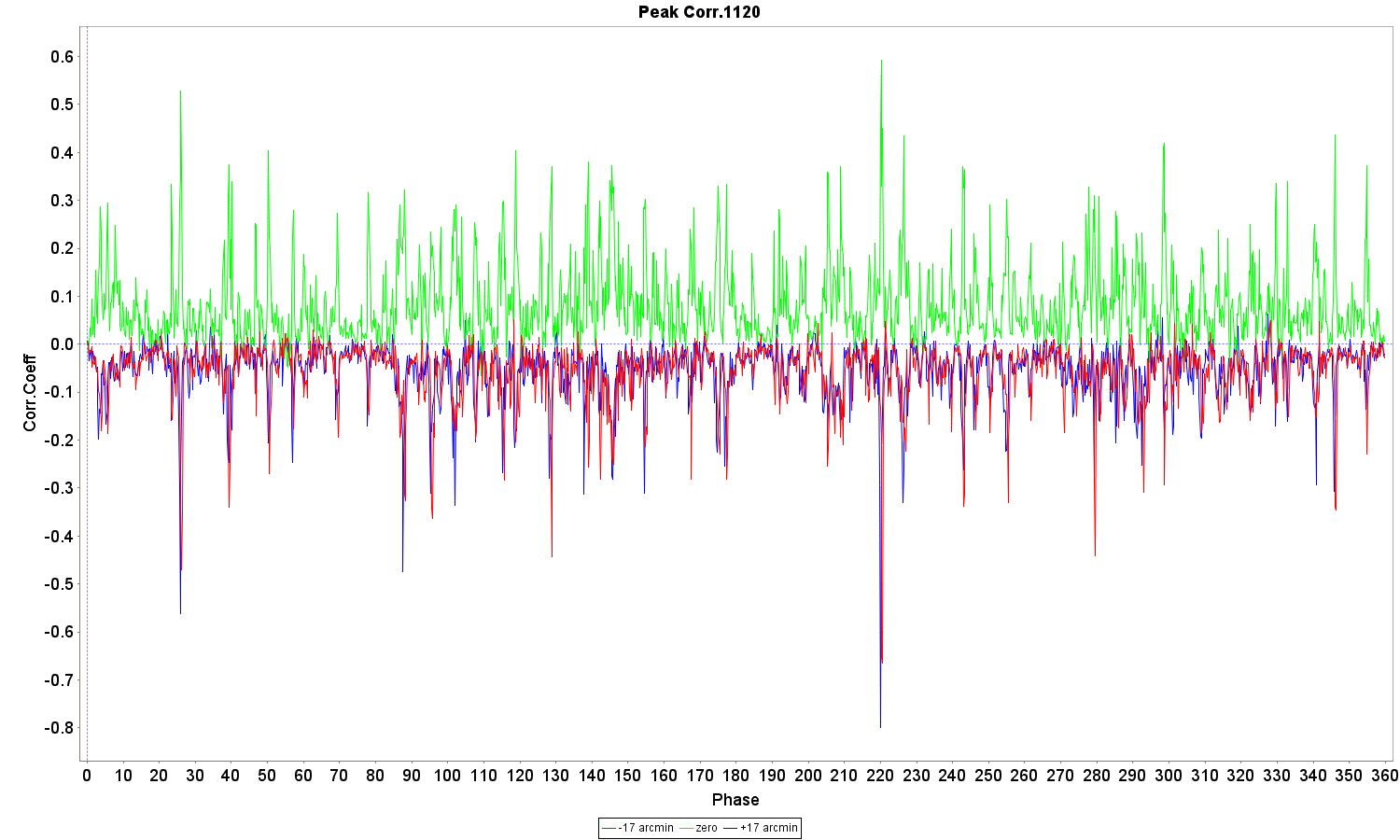}
\caption{An extract from the along-scan error correlations averaged over 20 satellite revolutions, against the rotation phase of the satellite. The green line shows the positive error correlations for sources separated by no more than 1 arcmin in transit phase (1~s in transit time). The red and blue lines show the error correlations for sources separated by $\pm$ 17 arcmin respectively. All of the larger peaks can be related to clanks, and can be observed as such in a reconstruction of the satellite spin rate.}
\label{fig:errorcorr}
\end{figure}

For the current study we had access to the \TGAS epoch astrometric data to study the error correlation levels, and to see if these effects are significant and sufficiently predictable to be compensated for. Systematics and correlation levels for residuals were, as expected, found to be strongly correlated with the occurrence of clanks. Most of the clanks are linked to the rotation phase of the satellite over period of days to weeks,  where the rotation phase is defined with respect to the direction of the Sun as seen from the satellite. This created significant error-correlation patterns as a function of the rotation phase of the satellite (Fig.~\ref{fig:errorcorr}),

An error correlation pattern such as this is very complicated and cannot reasonably be corrected for in the data reductions. It must be left to the next \Gaia data release, where the clanks are planned to be incorporated in the attitude model, to derive astrometric solutions from data much less seriously affected by this type disturbances. In the current \TGAS data these events have to be accepted as unresolved and contributing to the overall astrometric noise.

\begin{figure}[t]
\centering
\includegraphics[width=8.5cm]{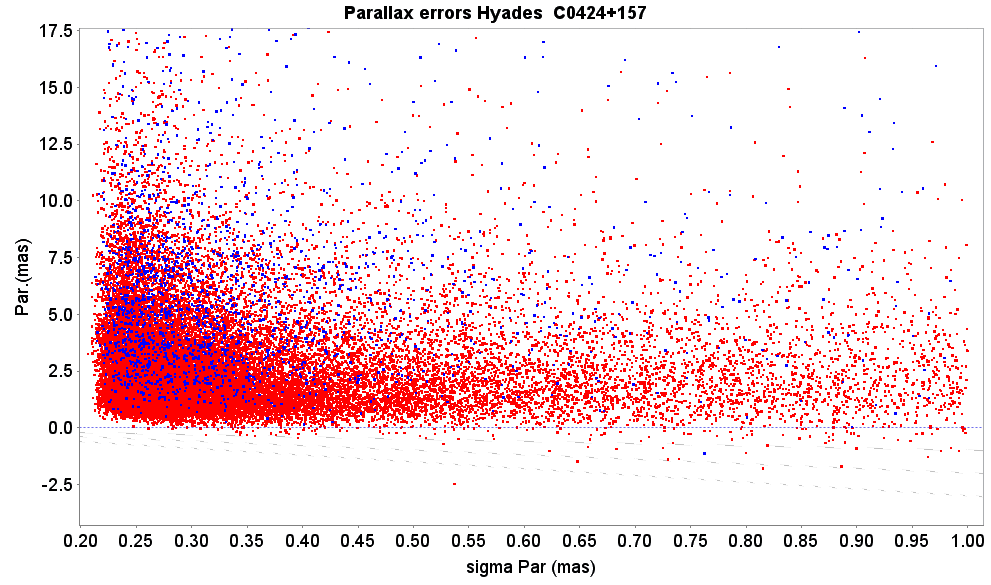}
\caption{The distribution of parallaxes as a function of \su for stars in a field of 18 degrees radius centred on the Hyades cluster. The red and blue points as in Fig.~\ref{fig:logchi2}. The three grey lines show the 1, 2 and 3$\sigma$ \su levels.}
\label{fig:parerrors}
\end{figure}
\begin{figure}[t]
\centering
\includegraphics[width=8.5cm]{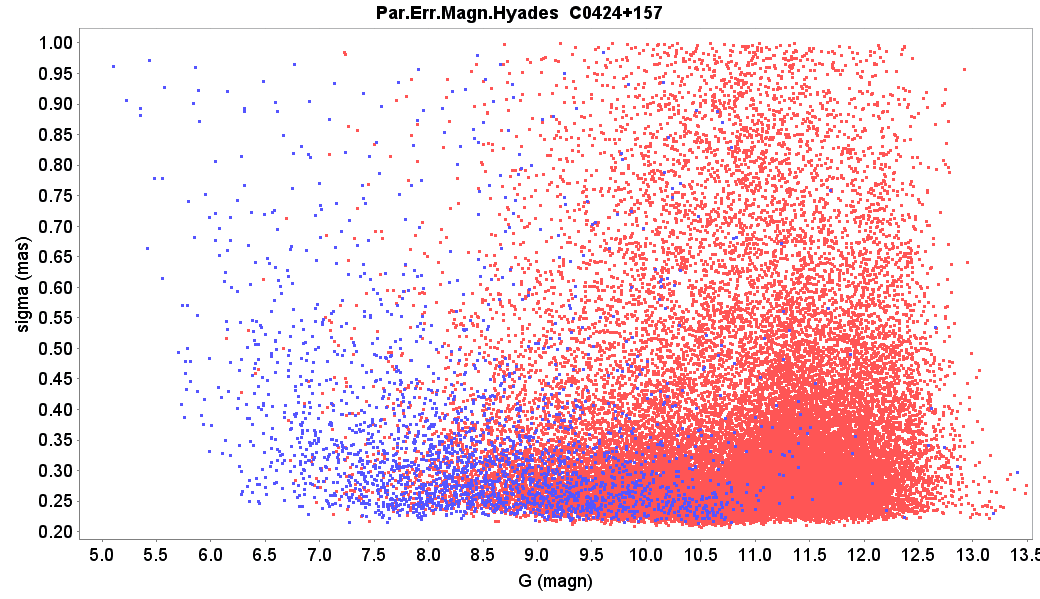}
\caption{Standard uncertainties for parallax measurements in \TGAS, as a function of the G magnitude, for stars in a field of 18 degrees radius centred on the Hyades cluster. The red and blue points as in Fig.~\ref{fig:logchi2}.}
\label{fig:parerrmagn}
\end{figure}
The distribution of $\sigma_\varpi$ (\su) for the \TGAS parallax measurements is furthermore affected by post-processing adjustments and filtering. The effect of the applied filter cutoff at 1 mas can be seen in Fig.~\ref{fig:parerrors} and \ref{fig:parerrmagn}. The majority of values for $\sigma_\varpi$ is found in the range 0.22 to 0.35~mas. 

\begin{figure}[t]
\centering
\includegraphics[width=8.cm]{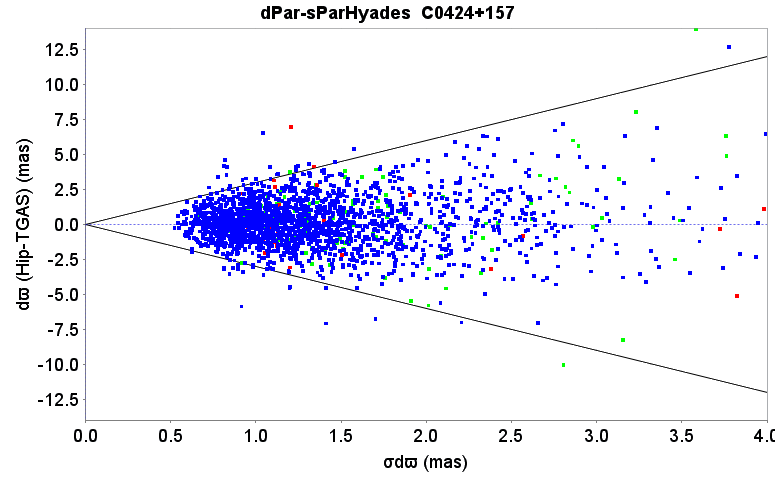}
\includegraphics[width=8.cm]{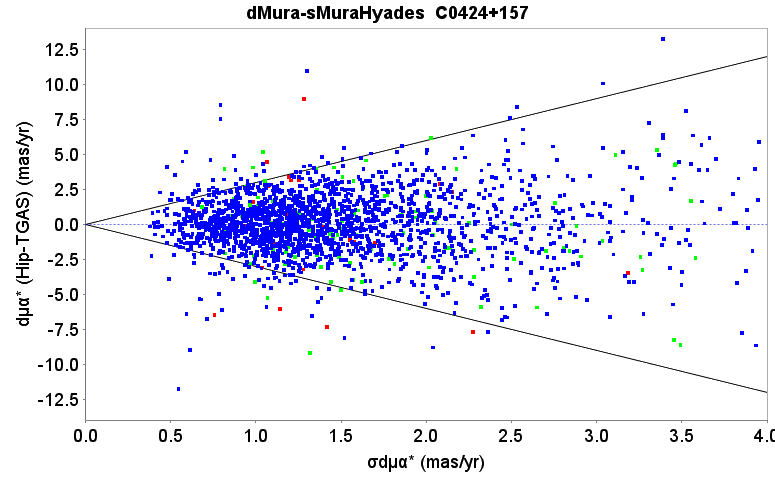}
\includegraphics[width=8.cm]{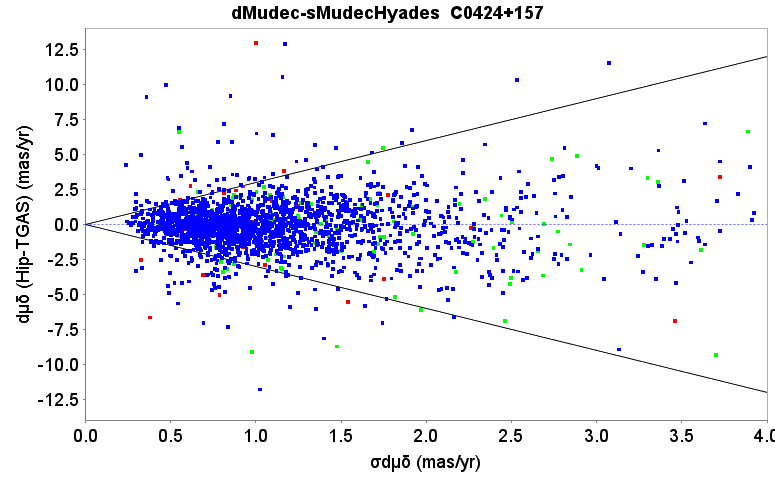}
\caption{Differences in astrometric parameters as a function of the \su of the differences between the \Hipparcos and \TGAS solutions for stars, as measured in a field of 18\degr\ radius centred on the Hyades cluster. The blue dots represent clean 5-parameter solutions in the \Hipparcos data. The red dots represent primarily accelerated solutions (so-called 7 and 9 parameter solutions). The green dots were solved as double stars in the \Hipparcos solution. The two black lines show the $\pm 3\sigma$ \su levels. From top to bottom: Parallaxes, proper motions in right ascension, proper motion in declination.}
\label{fig:parerr}
\end{figure}
Differences with the \Hipparcos parallaxes and their \su values show generally small systematics and underestimates of the combined \su values of the parallax differences. For the an area of 18\degr\ radius field centred on the Hyades the differences for 2059 stars in common with the \Hipparcos catalogue showed a difference of $0.14\pm 0.03$~mas and a unit weight standard deviation of 1.25 (see also Fig.~\ref{fig:parerr}). The situation for the differences in proper motions is different. Because of the much longer epoch span for the \TGAS data compared to the \Hipparcos data, these differences will start to show the presence of long-period orbital effects on the \Hipparcos proper motions of some stars, leading to more outliers than observed for the parallax differences. 

\begin{figure}[t]
\centering
\includegraphics[width=8cm]{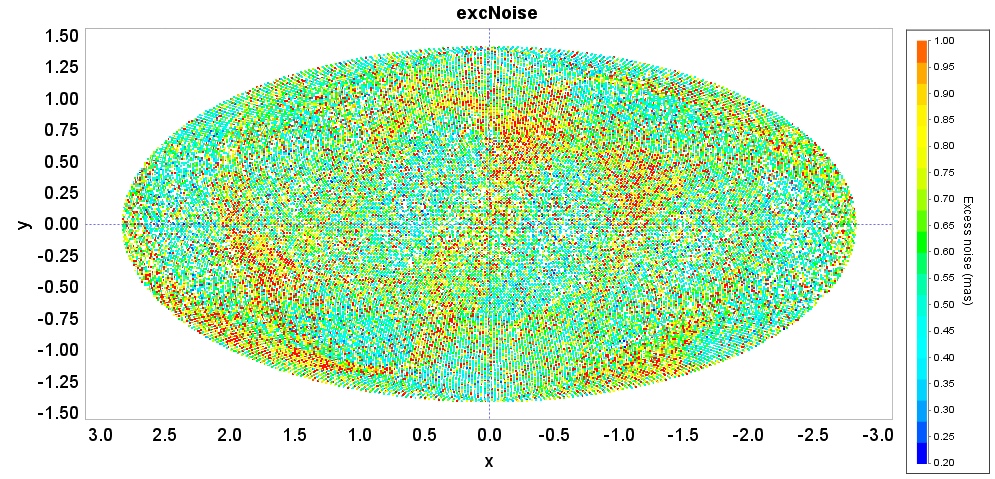}
\includegraphics[width=8cm]{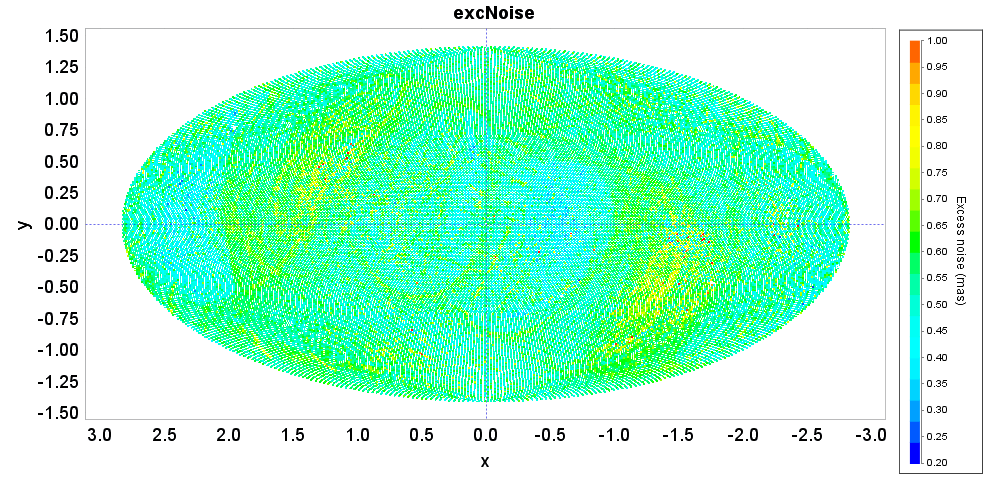}
\caption{Excess noise levels as applied to astrometric solutions. Top: for stars with \Hipparcos first epoch positions; bottom: for stars with \Tycho first epoch positions.}
\label{fig:excessnoise}
\end{figure}
There is at least one further aspect in which the data differ depending on the origin of the first epoch positions, and that is the addition of excess noise. Here the stars with first-epoch \Hipparcos data are much more affected than those using \Tycho data (Fig.~\ref{fig:excessnoise}). In addition, the application of excess noise, which effectively compensates the astrometric solution for imperfections in the data model, is predominantly found there where the number of observations is highest. These imperfections may be caused by the unresolved issues in the along-scan attitude reconstruction, such as clanks and hits, in which case the astrometric parameters can partly absorb these effects when relatively few observations are available. But it may also be caused by a very small mis-alignment between the \Hipparcos first-epoch positions and the \TGAS proper motion reference frame. Stars with first epoch \Tycho positions are much less affected, as those positions had assigned significantly larger \su values than the \Hipparcos positions. In both cases, it would affect the astrometric solutions more severely when more \Gaia data is available and relatively more weight in the astrometric solution comes from the \Gaia data, as appears to be the case. 

For the field of each cluster that we analyzed, the weighted mean differences, with \su and unit-weight standard deviation, between the \Hipparcos and \TGAS data are provided in Table~\ref{tab:hipcomp} for the parallaxes and proper motions. The unit-weight standard deviation is obtained by normalizing the error on each observation by its estimated \su. In these comparisons only those stars are used which have simple 5-parameter astrometric solutions in the \Hipparcos catalogue, while initial selection of stars in the field of a cluster was done independent of solution type.

\section{General approach to the cluster data analysis \label{sec:genappr}}

\subsection{Cluster membership selection \label{ssec:clustmemb}}

Different approaches to cluster membership were used for the selection of cluster members, depending on the distance of the cluster. For the nearest clusters, the Hyades and initially also Coma Ber, the cluster membership has been determined based on first of all the coincidence in space within a volume around the assumed 3D position $\vec{R}_c$ of the cluster centre
\begin{equation}
\vec{R}_c = R_c\cdot\left[\begin{array}{r}
\cos\alpha\cos\delta_c\\\sin\alpha_c\cos\delta_c\\\sin\delta_c\\
\end{array}
\right],
\end{equation}
where $R_c=1/\varpi_c$ is the assumed distance of the cluster, and $(\alpha_c,\delta_c)$ are the equatorial coordinates of the projected cluster centre. It is further based on the assumed space motion $\dot{\vec{R}}_c$, and an assumed outer radius $r$ of the cluster. The position in space of a potential cluster member is derived from its position and parallax, where the main uncertainty comes from the measured parallax. The observed proper motion and its standard errors are compared with the projection of the space motion of the cluster at the coordinates of the star. Determining the 3D positions of individual stars limits this method for the \TGAS data to the nearest clusters. Details on the calculations and associated accuracies are presented in App.~\ref{appdist}, where it is shown that the uncertainties in the estimates of the 3D positions of individual stars increases with the square of the distance of the cluster. Thus, in future releases, with potentially a ten-fold improvement in parallax and proper motion accuracies, we may expect this method to be applicable up to about 100 to 150 pc distance.

For the more distant clusters, which in the case of the \TGAS is any cluster more distant than about 50 to 75 pc, an iteration between membership selection and mean parallax and proper motion is performed. As first approximation for the parallax and proper motion (or space velocity) the astrometric data for open clusters from the \Hipparcos data presented in fvl09 are used. Margins around these initial values are set generously to avoid introducing a bias on the \Gaia solution. 

\subsection{Radial velocity projection}
The radial velocity values used in these solutions play only a minor role through projection on the sky away from the cluster centre. Only in the analysis of the Hyades data this is an important quantity. The projection of the radial velocity $V_{rad}$ onto the proper motion at a distance $\rho$ from the projected cluster centre for a cluster with a mean parallax $\varpi_c$ is given by:
\begin{equation}
\Delta\mu = \varpi_c \sin\rho V_{rad}/\kappa,
\end{equation} 
where the parallax and proper motion are expressed in mas and mas~yr$^{-1}$ respectively, and the radial velocity in km~s$^{-1}$. The constant $\kappa=4.74047$ provides the scaling factor between the proper motions and radial velocities. For example, the Pleiades cluster has a radial velocity of 8.6 km~s$^{-1}$ and most members are found within about 4.5 degrees on the sky from the cluster centre. At a parallax of about 8 mas this gives a maximum projection of the radial velocity of 1 mas~yr$^{-1}$. This is at the same level as the internal velocity dispersion in the cluster \citep{1979A&AS...37..333V}. For the Hyades the radial velocity is, at 39 km~s$^{-1}$, much higher. The spread over the sky and the parallax are three times larger. This leads to projection effects as large as 41 mas~yr$^{-1}$. The projection effects for the tangential component of the space motion on the proper motions are still smaller, being proportional to $\cos\rho$. This amounts to 3 to 4 per~cent for the Hyades (about 5 mas~yr$^{-1}$) and less than 0.5 per~cent for the Pleiades (less than 0.2~mas~yr$^{-1}$). The observed proper motions are also affected by a systematic scaling of the cluster proper motion, depending on the offset along the line-of-sight for an individual cluster member, relative to the cluster centre. It can be observed as an increased dispersion in the proper motions of the cluster members along the direction of the cluster proper motion, an effect also known as the relative secular parallax. In the analysis of the cluster data this can be treated as an individual correction per star, based on the observed proper motion and parallax and their standard errors, and using the latest estimate of the cluster parallax and space velocity vector. Within the constraints of the current data set this is still only possible for the Hyades cluster.

\section{The Hyades \label{sec:hyades}}

The \Hipparcos data for the Hyades cluster have been covered extensively by \cite{1998A&A...331...81P, madsen99, 2001A&A...367..111D} for the 1997 reduction, and in fvl09 for the new reduction. The \Hipparcos input catalogue \citep{1989hmps....2.....P} contained a selection of around 150 stars considered from earlier studies to be members of the Hyades cluster. Many of these are relatively bright and are not included in the \TGAS catalogue. Because of the pre-selection done for the \Hipparcos catalogue, there is only a small number of members found among the additional \Tycho stars, and the total number of members, with \Hipparcos first epoch data, available for the current study is just over half the number that was available for the \Hipparcos studies. 

Starting with the cluster centre and parallax as derived in fvl09, 285 stars are found within the \Gaia DR1 \TGAS catalogue for which the position is likely to be within 16~pc from the assumed cluster centre in space, taking into account the \su on the parallaxes of the individual stars and their positions as projected on the sky, relative to the projected cluster centre. The data selection has to be limited to relative errors on the parallaxes of at most 20~per~cent, else distances to the individual stars become effectively undetermined. In Appendix~\ref{appdist} further details are presented on deriving the relative distance and its \su for a star from the assumed cluster centre. 

The next selection step calculates predicted proper motions from the space velocity of the cluster as projected perpendicular to the line of sight, and scaled according to the observed parallax. The details for the projection calculations are given in Appendix~\ref{app:projeff}. These predicted proper motions only account for the projection of the space motion of the entire cluster at the position on the sky and the observed parallax of the star. When comparing these predicted proper motions with the observed values there are three types of error contributions that need to be considered:
\begin{enumerate}
\item the \su on the observed proper motions;
\item the \su of the predicted proper motions, mainly resulting from the errors on the observed parallaxes;
\item the internal velocity dispersion and possible systematic motions in the cluster, estimated to be at a level of about 0.6~km~s$^{-1}$.
\end{enumerate}
\begin{figure}[t]
\centering
\includegraphics[width=8.5cm]{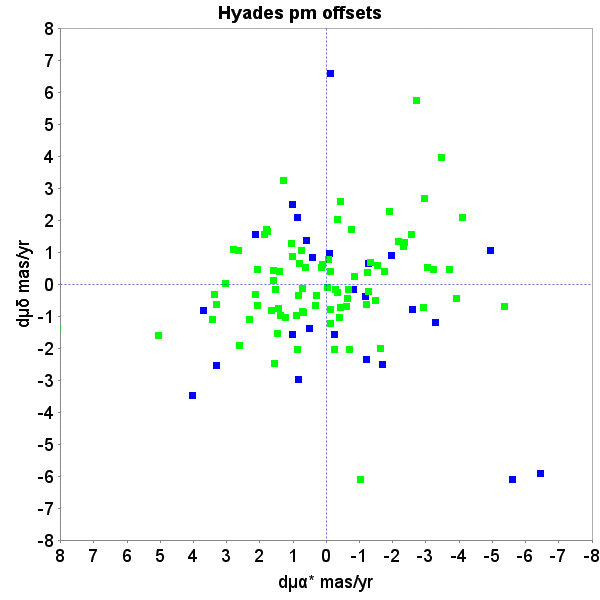}
\caption{Differences between predicted and observed proper motions for stars within the space volume of the Hyades cluster, showing the results for 112 possible members. Green dots: First epoch \Hipparcos; blue dots: first epoch \Tycho.}
\label{fig:hyadespmdiff}
\end{figure}
In the \Gaia DR1 \TGAS data the first item is by far the smallest contribution, while the second and third items give comparable error contributions, at a level of 1 to 2 mas~yr$^{-1}$. Added in quadrature, these three contributions provide the estimated uncertainty on the differences between predicted and observed proper motions. Applying this to the initial selection of 285 stars within the space of the Hyades cluster leaves 112 stars for which the observed proper motions are in both coordinates within 3 sigma from the predicted proper motions. Of the original 150 Hyades members found in the \Hipparcos data only 85 are included here, primarily because of the problems still experienced with the calibrations for bright stars and filters applied to the \TGAS data. A further 27 possible members with first epoch \Tycho data are included. Figure~\ref{fig:hyadespmdiff} shows the observed differences in proper motions, and the membership selection based on this. It is clear that there is a generally very good agreement with the cluster distance and space motion as derived in fvl09.

A new value for the space motion $\dot{\vec{R}}_c$ of the cluster can be derived from applying Eq.~\ref{equ:spacVel3} in a least squares solution with the observed proper motions and parallaxes for the cluster members, indicated with index $i$:
\begin{equation}
\left[\begin{array}{rrr}
-\sin\alpha_i & \cos\alpha_i & 0 \\
-\cos\alpha_i\sin\delta_i & -\sin\alpha_i\sin\delta_i & \cos\delta_i\\
\end{array}\right] \cdot \dot{\vec{R}_c} = \left[\begin{array}{r}
\kappa\mu_{\alpha \cdot, i}/\varpi_i \\ \kappa\mu_{\delta,i}/\varpi_i \\
\end{array}\right]
\end{equation}
The standard errors on the observations are derived from the errors on the proper motions and parallaxes and a contribution from the internal velocity dispersion. The value for the latter was determined at 0.58 km~s$^{-1}$, which should be interpreted as the weighted-average velocity dispersion over the whole cluster, where most of the weight comes from the projected centre of the cluster. Two solutions were obtained, the first solution is based on the proper motions only, and in the second solution an additional observation of the radial velocity of the cluster was added. For this second solution a value of $V_\mathrm{rad} = 39.1\pm 0.2$~km~s$^{-1}$ was used, as derived by \cite{1984AJ.....89.1038D} based on radial velocity measurements for 17 non-variable cluster members. The two solutions gave the following results (first without, second with radial velocity constraint):
\begin{equation}
\dot{\vec{R}} = \left[\begin{array}{rr}
-6.03\pm 0.08 & -6.14\pm 0.03 \\45.56\pm 0.18 & 45.28\pm  0.02\\ 5.57   \pm 0.06 & 5.48\pm 0.02\\
\end{array}\right] \mathrm{km~s}^{-1}.
\label{equ:v0hyades}
\end{equation}
A standard deviation of 1.00 was obtained by adjusting the internal velocity dispersion to the value of 0.58 km~s$^{-1}$ given above. Of the 112 possible members entering the solution, initially 6, and later (in the fitting of the kinematically improved parallaxes, see App.~\ref{app:reduced}) still 3 more were rejected in the iterations, leaving 103 probable members, for which identifiers are presented in Table~\ref{tab:hyades}, and a map is shown in Fig~\ref{fig:maphyades}.

The following data apply to the second solution in Eq.~\ref{equ:v0hyades}, i.e.\ including the mean radial velocity measurement for the cluster as an observation. The position of the convergent point is
\begin{align}
\alpha_\mathrm{conv} &= 97\fdg 73\pm 0\fdg 04\phantom{0} =  6^\mathrm{h}30.92^m,\nonumber \\
\delta_\mathrm{conv} &= \phantom{1}6\fdg 83\pm 0\fdg 03 =  6\degr 
49.8\arcmin.
\end{align}
The result in Eq.~\ref{equ:v0hyades} can be transformed back to a radial velocity and proper motion for the cluster centre:
\begin{align}
\mathrm{v_{rad, c}} &= \phantom{-}39.10\pm 0.02~~\mathrm{km~s^{-1}}, \nonumber\\
\mu_{\alpha*, c} &= 104.92\pm 0.12~~\mathrm{mas~yr^{-1}}, \nonumber \\
\mu_{\delta, c} &= -28.00 \pm 0.09~~\mathrm{mas~yr^{-1}}. 
\end{align}
From the first solution, using only proper motion data, the radial velocity of the cluster is recovered at a value of $39.38\pm 0.16$~km~s$^{-1}$, not significantly different from the spectroscopic value, considering the different size and composition of the radial velocity sample. 
 
The weighted mean parallax for the 103 probable member stars (as projected on the line of sight towards the cluster centre) is 
\begin{equation}
\varpi_c = 21.39\pm 0.21 \mathrm{mas},
\end{equation} 
in good agreement with earlier \Hipparcos-based determination in fvl09, which gave a value of $21.53\pm 0.23$~mas. The parallax is equivalent to a distance of $46.75\pm 0.46$~pc and a distance modulus of $3.349\pm 0.021$~mag. The error given is the \su on the mean. The standard deviation is much larger, at about 8~mas, due to the size of the cluster relative to its distance. The mean position on the sky of the 103 selected stars is
\begin{align}
\alpha_c &= 66\fdg 85 = 4^\mathrm{h}27.4^m, \nonumber \\
\delta_c &= 17\fdg 04 = 17\degr 2.4\arcmin.
\end{align}
The largest separation on the sky for a cluster member as found here is 17.2 degrees from the cluster centre, equivalent to 14.5~pc. There is an indication of more cluster members found at still larger distances from the centre, but whether these are actually bound to the cluster is unlikely and unclear from the data at this stage. For all numbers given above it should be realized that they are dependent on the initial values and criteria used for member selection, such as the maximum radius of the field and the internal velocity dispersion. However, those dependencies are small, as the figures shown here are the result of a converged iterative process, in which the assumed parallax, cluster centre position and space velocity vector were adjusted. 

\begin{figure}[t]
\centering
\includegraphics[width=7.5cm]{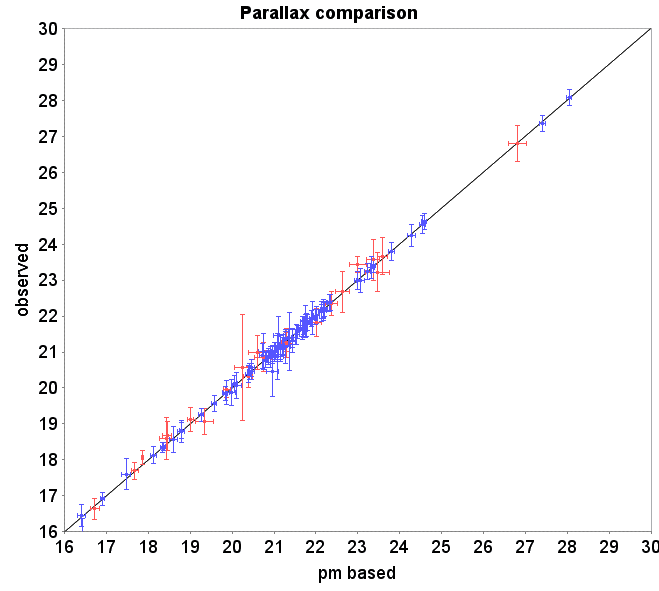}
\caption{Comparison between the parallaxes as measured and kinematically improved by means of the proper motion data. The blue data points use \Hipparcos data as first epoch, the red data points use \Tycho data instead.}
\label{fig:hyadparcomp}
\end{figure}
\begin{figure}[t]
\centering
\includegraphics[width=7.5cm]{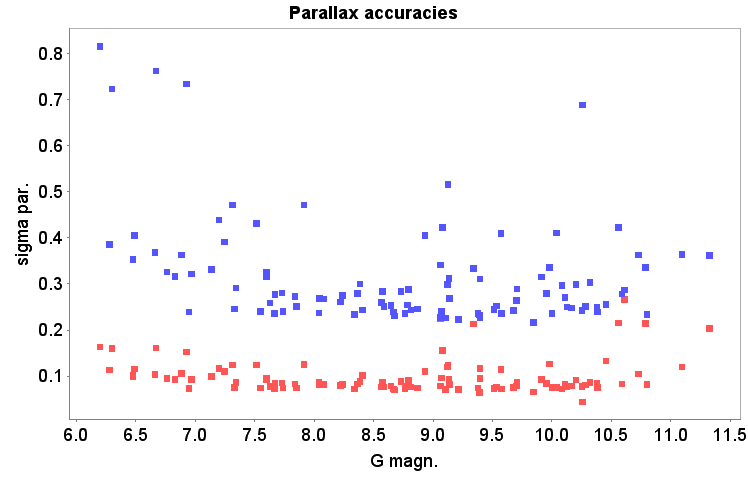}
\caption{Standard uncertainties on the parallax determinations. Blue dots: as derived from the \TGAS catalogue; red dots: kinematically improved parallaxes using the cluster space velocity vector.}
\label{fig:paracchyad}
\end{figure}
\begin{figure}[t]
\centering
\includegraphics[width=8.5cm]{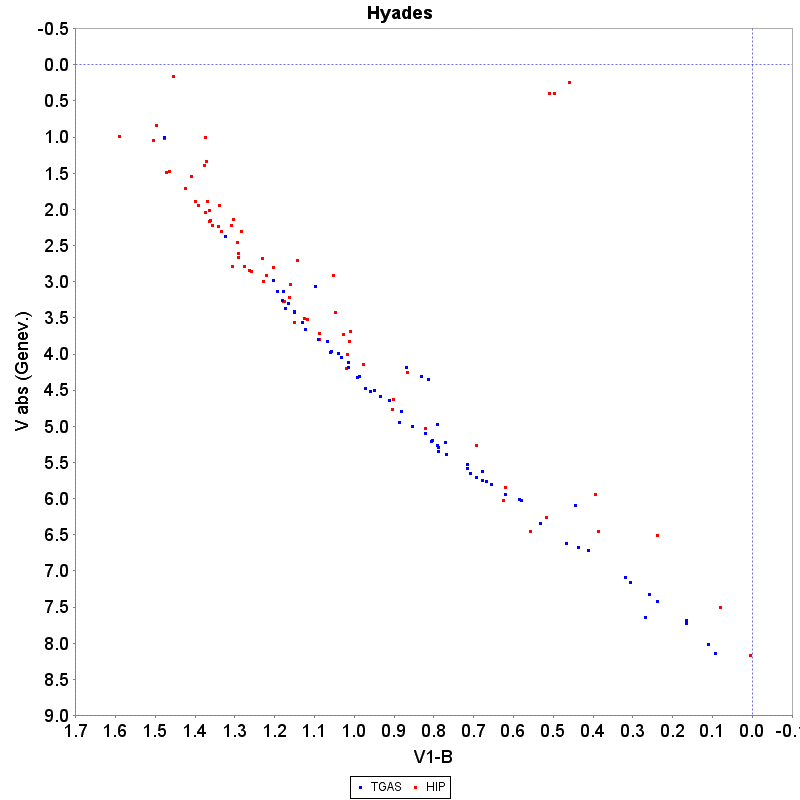}
\caption{The absolute magnitudes and colour indices in the Geneva photometry for cluster members, after applying distance moduli based on individual kinematically improved parallaxes. The red dots represent stars not present in the \TGAS catalogue, but with similarly treated data in fvl09.}
\label{fig:hrdiagrhyades}
\end{figure}

For the next step the reduced proper motions are derived as described in Appendix~\ref{app:reduced}. This allows to extract the differential parallax information from the proper motions, the so-called kinematically improved parallaxes, a process first described by \cite{madsen99}. Figure ~\ref{fig:hyadparcomp} shows the comparison between the parallaxes as published in the \TGAS catalogue and the kinematically improved parallaxes, with the \su error bars for both determinations. Including the proper motion data reduces the standard errors on the parallaxes by about a factor two to three, down to a level of 0.1 to 0.2 mas (Fig.~\ref{fig:paracchyad}), equivalent to relative errors below 1~per~cent. A relative error on the parallax of 1 per~cent is almost equivalent to an uncertainty in the distance modulus of 0.02 magnitude. This shows in the HR diagram for the cluster in the form of a very narrow main sequence (Fig.~\ref{fig:hrdiagrhyades}). It is the reconstruction of a multitude of such sequences, for clusters of different age and composition, that will provide the detailed observational isochrones that may provide further insights into the many processes that are involved in producing theoretical isochrones. 

The process of improving parallaxes by means of proper motion data in the Hyades  is ultimately limited by the internal velocity dispersion in the cluster. This contributes on average an uncertainty at about the same 1~per~cent level as the current determination of the parallaxes. In the calculations of the standard uncertainties on the kinematically improved parallaxes this has been taken into account in as far as possible. For future releases of the \Gaia data, with improved accuracies for the parallaxes and proper motions, the process presented here can be inverted, and used to reconstruct the internal velocity dispersion throughout the cluster.

Table~\ref{tab:Spacialdens} gives the spatial densities for the 106 stars used in the current analysis. Considering the low number of stars and various selections that have been applied to the \TGAS data, it seems a bit premature to further interpret and analyse the space density profile.
 
A full list of the source identifiers, cross matches with HD identifiers and the  kinematically improved distance moduli is presented in Table~\ref{tab:hyades}. It is these individual kinematically improved distance moduli that should be used in the construction of the Hyades HR diagram.
\begin{table}[t]
\caption{Spatial densities in the Hyades cluster for 106 selected stars (before the final elimination of 3 possible members).\label{tab:Spacialdens}}
\centering
\begin{tabular}{rrrr}
\hline\hline
r$_1$ &r$_2$ & stars & $\log d$ \\
\hline
 0 &  1 &  1 &  -0.62\\
 1 &  2 & 10 &  -0.47\\
 2 &  3 & 15 &  -0.72\\
 3 &  4 & 14 &  -1.04\\
 4 &  5 & 13 &  -1.29\\
 5 &  6 & 10 &  -1.58\\
 6 &  8 & 16 &  -1.89\\
 8 & 11 & 12 &  -2.45\\
11 & 16 & 15 &  -2.89\\
\hline
\end{tabular}
\tablefoot{r$_1$ and r$_2$ are the inner and outer radius in pc. $d$ gives the density in number of stars per cubic parsec.}
\end{table}

\section{The nearby clusters \label{sec:nearby}}
\subsection{General considerations\label{sec:gencons}}
For the following clusters, the mean parallax and proper motions have been determined while taking into account the local projection effects and the full covariance matrix for the astrometric solution of each member star. Membership selection was based on position, proper motion and parallax information, but will always be slightly ambiguous, and in particular for most of the younger clusters that are still close to, or even embedded into, an OB association. Because of the high levels of error correlations present in the astrometric parameters of the individual stars, the solution for the mean proper motion and parallax have to be done simultaneously, solving Eq.\ref{equ:obsequcluster} after deconvolving with the square root of the inverse of the noise matrix. The noise matrix takes account of the correlations and standard uncertainties on the astrometric parameters as well as the internal velocity and parallax dispersions, all as described in Appendix~\ref{app:combine}. Here we use a velocity dispersion of 0.6~km~s$^{-1}$ and a position dispersion along the line of sight of 5 pc was used. The outer radius of the cluster has been set at 15~pc. All results have a slight dependency on these assumptions, mostly where it affects membership selection. 

Mean positions for the member stars, as an estimate for the projected position of the cluster centre, have been determined from the tangential projection of the member-star positions on the sky relative to an assumed position of the cluster centre (see Appendix~\ref{app:tangproj}). The new centre was then obtained through de-projection on the sky. As corrections tend to be very small, this process generally converged rapidly through the iterations.   

\subsection{The Ursa Major moving group and the Coma Berenices cluster \label{sec:umajcorbor}}
Very little can be said here about the Ursa Major moving group. The brightest members of the group are not included in the \TGAS catalogue, and a search for fainter members coinciding (in proper motion) with the local projection of the space velocity of the group showed no more than about three possible candidates.
\begin{figure}[t]
\centering
\includegraphics[width=8.5cm]{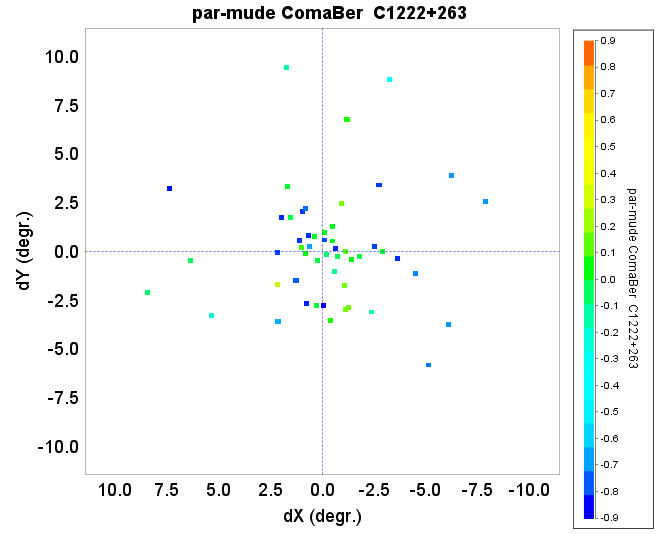}
\caption{The stars in the Coma Berenice cluster, colour coded according to the error-correlations between the parallax and proper motion in declination. The dark blue dots, representing strong negative error correlations, have first epoch \Tycho data, the green dots, representing near-zero correlations, have first epoch \Hipparcos data.}
\label{fig:parcompcoma}
\end{figure}
The Coma Berenices cluster is more interesting at this stage. It has first been analysed in the same way as the Hyades cluster. Starting with the \Hipparcos solution for the cluster, a volume of 15~pc radius at a distance of 86.7~pc was initially searched. Likely cluster members were found to be restricted to within a radius of 13~pc only, and the distance had to be adjusted to 85.5~pc. Within that volume 142 stars are found. 

In determining of the space velocity of the cluster, an additional `observation' was added for the mean radial velocity at the cluster centre in order to stabilize the solution, similar to the processing of the Hyades cluster. Assuming a radial velocity of -1.2~km~s$^{-1}$, the space motion is found to be
\begin{equation}
\dot{\vec{R}} = \left[\begin{array}{r}
-0.41\pm 0.85 \\4.86\pm  0.11\\ -4.11\pm 0.42\\
\end{array}\right] \mathrm{km~s}^{-1},
\end{equation}
as based on 44 stars identified as probable members. Of these, 25 have \Hipparcos and 19 have \Tycho first epoch data. The space motion is equivalent to the following values at the centre of the cluster:
\begin{align}
\mathrm{v_{rad, c}} &= -\phantom{1}1.89\pm 0.10~~\mathrm{km~s^{-1}}, \nonumber\\
\mu_{\alpha *, c} &= -12.04\pm 0.15~~\mathrm{mas~yr^{-1}}, \nonumber \\
\mu_{\delta, c} &= -\phantom{1}8.97 \pm 0.19~~\mathrm{mas~yr^{-1}}. 
\end{align}

The weighted mean parallax for these stars is $11.69\pm 0.06$~mas, which differs by 1.2 $\sigma$ from the determination in fvl09 ($11.53\pm 0.12$ mas). The parallax is equivalent to a distance of $85.5\pm 0.4$~pc, and a distance modulus of $4.66\pm 0.01$~mag. The distance moduli for individual stars in the cluster range from about 4.47 to 4.84, and individual parallaxes need to be taken into account when reconstructing absolute magnitudes. Compared with isochrone fitting by \cite{pinso98}, who derived a distance modulus of $4.54\pm 0.04$, there is still a difference of nearly 3$\sigma$. Even more discrepant is the MAP-based trigonometric parallax for the cluster by \cite{1995ApJ...445..712G}, which gave a parallax of $13.54\pm 0.54$~mas, a difference in distance modulus ($4.34\pm 0.09$) of 0.3 magnitudes. Also the parallax derived by \cite{2003AJ....126.2408M} is, at a value of $12.40\pm 0.17$~mas off by 4$\sigma$, and in distance modulus by 0.13 magnitude. 
\begin{figure}[t]
\centering
\includegraphics[width=8cm]{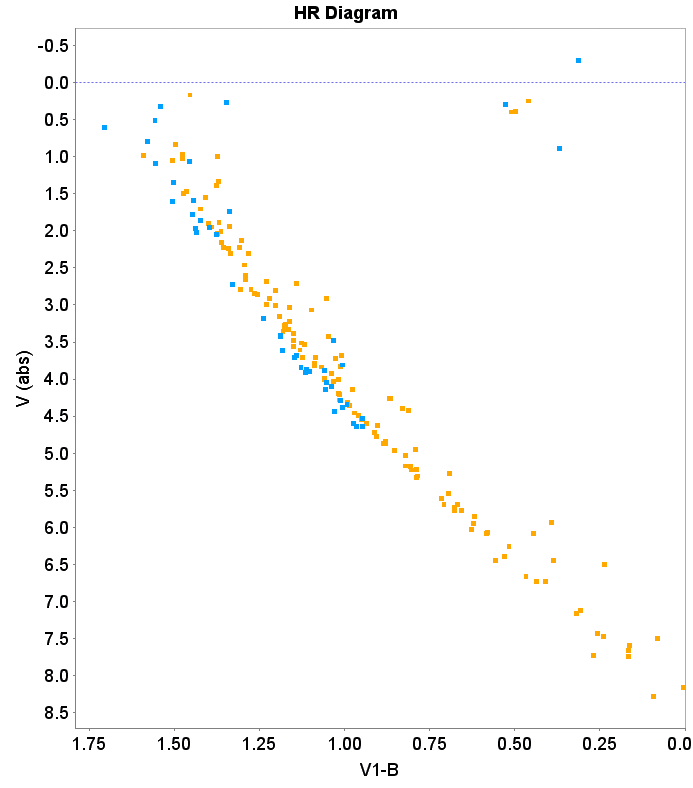}
\caption{The HR diagram for the Coma Berenices cluster (light blue) compared with the Hyades cluster (orange-red). Geneva photometry.}
\label{fig:hrdcomab}
\end{figure}

The cluster centre is confirmed to be at 
\begin{align}
\alpha_c &= 186\fdg 02 = 12^\mathrm{h}24.08^m, \nonumber \\
\delta_c &= \phantom{0}25\fdg 95 = 25\degr 57\arcmin.
\end{align}
All values are subject to minor adjustments depending on the exact selection criteria. They can be compared with the data presented in Table~\ref{tab:overview}, which have been obtained with the weighted mean parallax and proper motion method as described in App.~\ref{app:combine}. For this solution a field with a 10.4 degrees radius was investigated, containing 6717 stars, 52 of which were considered possible members of the Coma Ber cluster. Two of the 52 possible members were eliminated during the iterative solutions for the astrometric parameters of the cluster. For 786 stars and 28 cluster members in common with the \Hipparcos catalogue, the weighted mean differences in the astrometric parameters are shown in Table~\ref{tab:hipcomp}.
\begin{figure}[t]
\centering
\includegraphics[width=8cm]{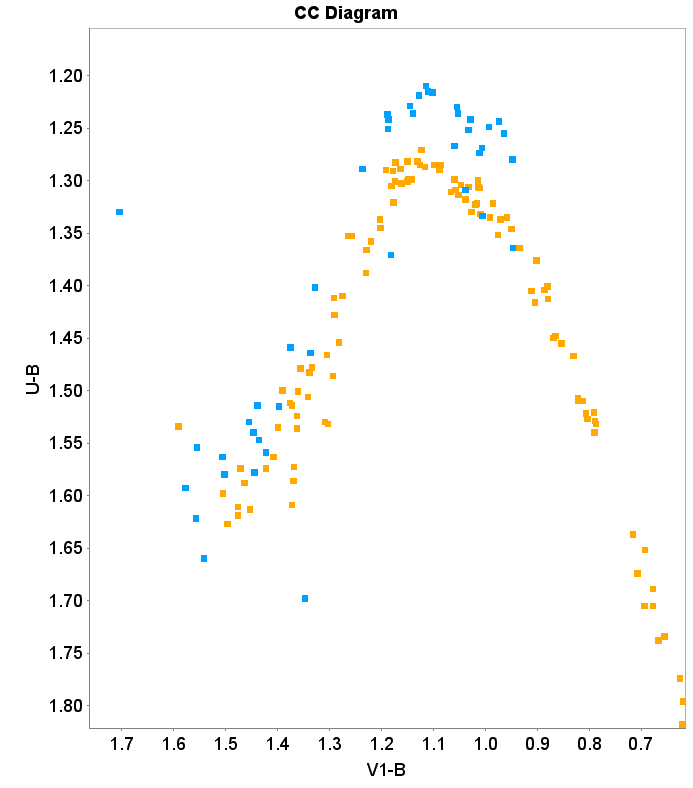}
\caption{The two colour diagram for the Coma Berenices cluster (light blue) compared with the Hyades (orange-red), showing the so-called Hyades anomaly.}
\label{fig:bv1comab}
\end{figure}
The differences are all well within the range of the formal \su values, and are primarily due to slight differences in member selection. A full list of the 50 probable member stars is presented in Table~\ref{tab:comaber} and shown as a map in Fig.~\ref{fig:mapcomaber}. The HR diagram in Geneva photometry is shown in Fig.~\ref{fig:hrdcomab}. As has been assessed before, these clusters are of closely the same age, with the impression of Coma Berenices cluster being slightly younger. Also in chemical composition they appear to be very similar \citep{2014A&A...561A..93H}. However, there is a marked difference in the two-colour diagrams (Fig.~\ref{fig:bv1comab}) for late F and G stars, a difference which in field stars is directly related to luminosity differences in the sense that it would imply the Hyades stars to be more luminous at the same temperature than those in the Coma Berenices cluster. This is the so-called Hyades anomaly, first noted half a century ago by \cite{1966AJ.....71..482V}.

\subsection{The Praesepe cluster\label{sec:praesepe}}
\begin{figure}[t]
\centering
\includegraphics[width=8cm]{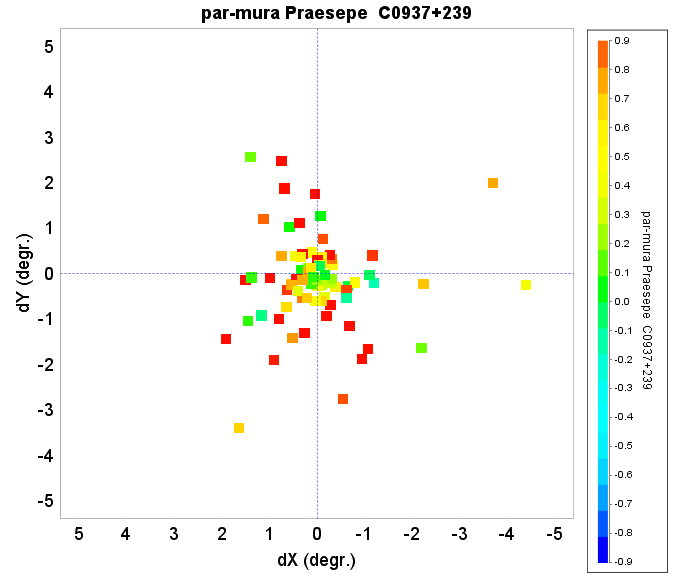}
\caption{Distribution of stars in the Praesepe cluster, colour coded according to the error-correlation factor between the parallax and the proper motion in right ascension for the individual solutions. The red dots, representing the highest correlations, belong to stars with \Tycho first epoch positions.}
\label{fig:praesepecorr}
\end{figure}
The Praesepe cluster has been investigated over a 5.47 degrees radius field, an area for which 2082 stars are contained in the \TGAS catalogue, 156 of which are also contained in the \Hipparcos catalogue. 84 stars were selected as possible members of Praesepe, of which 5 were later eliminated in the iterative solutions for the astrometric parameters of the cluster. The weighted mean differences in this field between the \TGAS and \Hipparcos astrometric parameters for 146 stars (with a simple 5-parameter solution, excluding 10 stars with complex solutions), of which 23 are identified as probable cluster members,  are summarized in Table~\ref{tab:hipcomp}. 

There is a significant increase in the number of member stars with respect to the solution in fvl09, from 24 to 79 stars. Probable members are found projected up to 4.4 degrees from the cluster centre, equivalent to a distance of about 14~pc, much like what is observed in the Hyades. The Praesepe field shows generally strong to very strong correlations for the astrometric parameters of individual stars, in particular for stars for which \Tycho first epoch data was used. These are the red points in Fig.~\ref{fig:praesepecorr}. 

\begin{figure}[t]
\centering
\includegraphics[width=8.5cm]{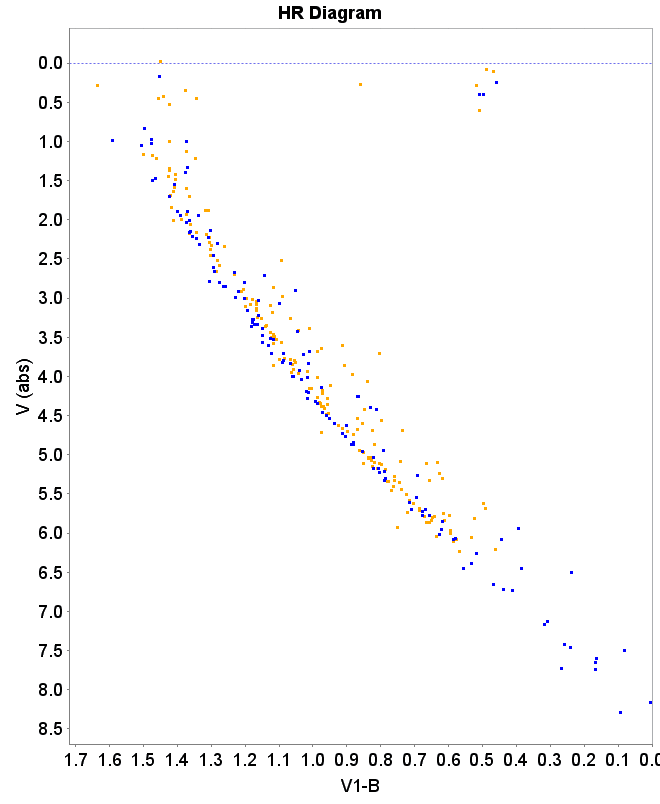}
\caption{The HR diagram in Geneva photometry for the Hyades (blue dots) and Praesepe (orange dots) clusters. For the Hyades stars individual kinematically improved parallaxes were used, for the Praesepe stars the common cluster parallax. As was also observed in fvl09, the two main sequences accurately coincide.}
\label{fig:hrhyadprae}
\end{figure}

As was noticed before in fvl09, the Praesepe and Hyades clusters appear to be very similar in composition and age. Figure~\ref{fig:hrhyadprae} shows the combined HR diagram for the two clusters and the closely coinciding main sequences. In contrast, the main sequence of the Coma Ber cluster, considered to be of the same age as the Hyades, appears to be sub-luminous by about 0.1 to 0.15 magnitudes with respect to the Hyades and Praesepe. This difference has increased slightly (by 0.06 mag.) in the current analysis with respect to fvl09. 

\subsection{The Pleiades cluster \label{sec:pleiades}}
\begin{figure}[t]
\centering
\includegraphics[width=80mm]{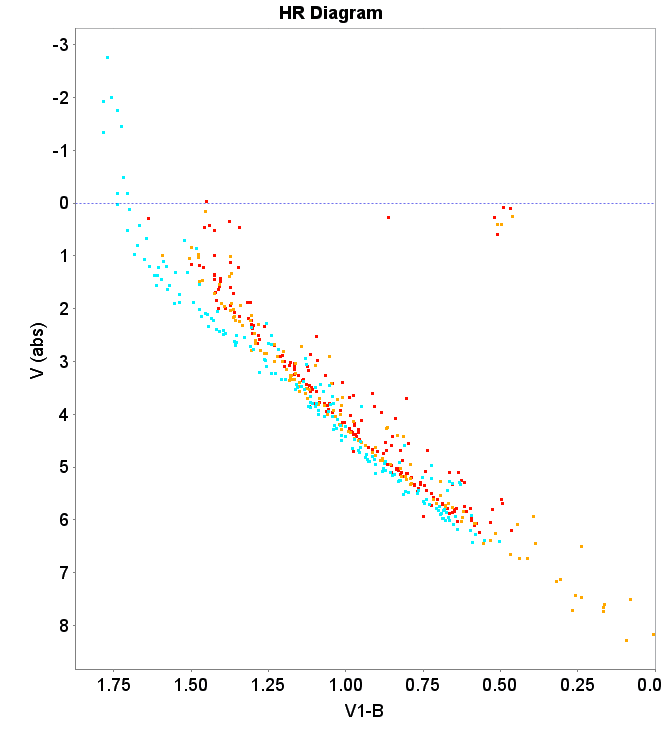}
\caption{diagram for the Pleiades (light blue) compared with the Hyades and Praesepe clusters (orange, red) in Geneva photometry.}
\label{fig:hrpleiades}
\end{figure}
The Pleiades cluster has been investigated over a 6.7 degrees radius field, for which the \TGAS catalogue contains 4996 stars, 160 of which were marked as possible cluster members. Within that area 325 stars are in common with the \Hipparcos catalogue, and of these 285 have single-star 5-parameter astrometric solutions (44 of which are probable cluster members), and a \su on the difference in parallax between the \Hipparcos and TGAS solutions that is below 3~mas. For those stars the mean difference in the astrometric parameters between the \Hipparcos and \TGAS solutions are shown in Table~\ref{tab:hipcomp}.
In a smaller field, at 4.5 degrees radius more compatible with the area of the sky used in the \Hipparcos determination of the Pleiades parallax, the differences are as shown in Table~\ref{tab:systdiffpleiades}. 
\begin{table}[t]
\caption{Systematic differences (\Hipparcos - \TGAS) between the astrometric parameters for 134 stars in the Pleiades field.}
\label{tab:systdiffpleiades}
\centering
\begin{tabular}{l|rr|l}
\hline\hline
Diff. &  Mean & Stand.dev. & units \\ 
\hline
$\mathrm{d}\varpi_c$ & $0.60\pm 0.12$ & 1.27 &  mas \\
$\mathrm{d}\mu_{\alpha *, c}$ & $0.22\pm0.14$ & 1.58 & mas~yr$^{-1}$ \\
$\mathrm{d}\mu_\delta, c$ & $0.01\pm0.15$ & 2.06 & mas~yr$^{-1}$ \\
\hline
\end{tabular}
\end{table} 

From the \TGAS catalogue we can identify 155 possible members, based on their proper motions, parallaxes and confirmed by consistency in the HR diagram. The mean parallax for 152 stars confirmed as Pleiades members in the subsequent iterations for the cluster astrometric solution is 
\begin{equation}
\varpi_c = 7.48\pm 0.03~\mathrm{mas}.
\end{equation}
Details on the 152 probable Pleiades members are presented in Table~\ref{tab:pleiades} in App.~\ref{app:pleiades}.

The difference with the Pleiades parallax as derived in fvl09 is part of an overall parallax difference in that part of the sky between the \Hipparcos and \TGAS catalogues, for which there is currently no explanation. No such differences were observed between the three independent reductions of the \Hipparcos data (the two reductions from which the first catalogue was constructed, and the new reduction). The current \TGAS parallax for the Pleiades, dominated by fainter cluster members, agrees with other studies of the cluster distance that are also based on the fainter members of the cluster.

The HR diagram for the Pleiades is shown in comparison with the Hyades and Praesepe clusters in Fig.~\ref{fig:hrpleiades}.

\begin{figure}[t]
\centering
\includegraphics[width=7.5cm]{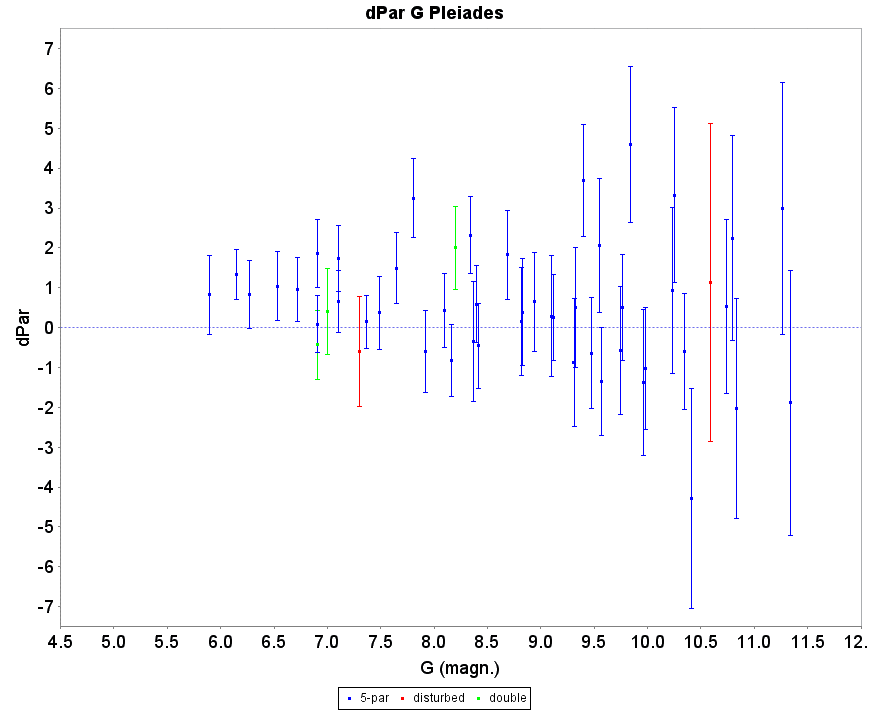}
\caption{Parallax differences between the \Hipparcos and \TGAS solutions as a function of the Gaia $G$ magnitude for members of the Pleiades cluster.}
\label{fig:dparmaggplei}
\end{figure}

\begin{figure}[t]
\centering
\includegraphics[width=7.5cm]{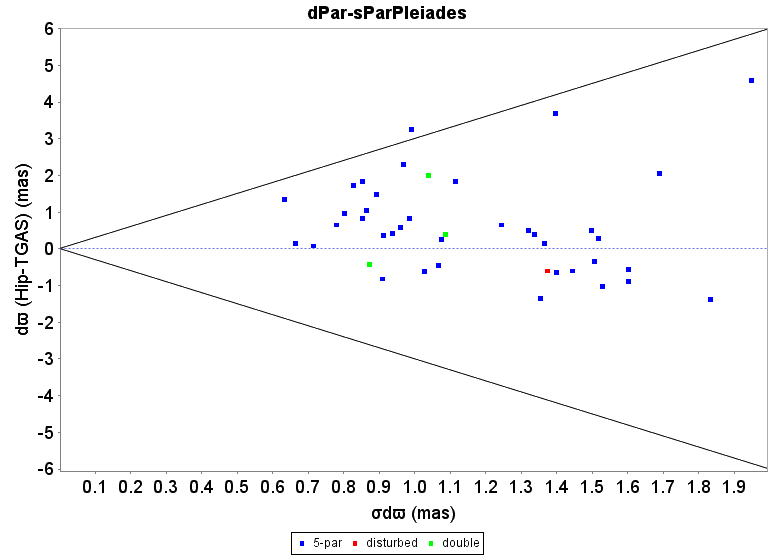}
\includegraphics[width=7.5cm]{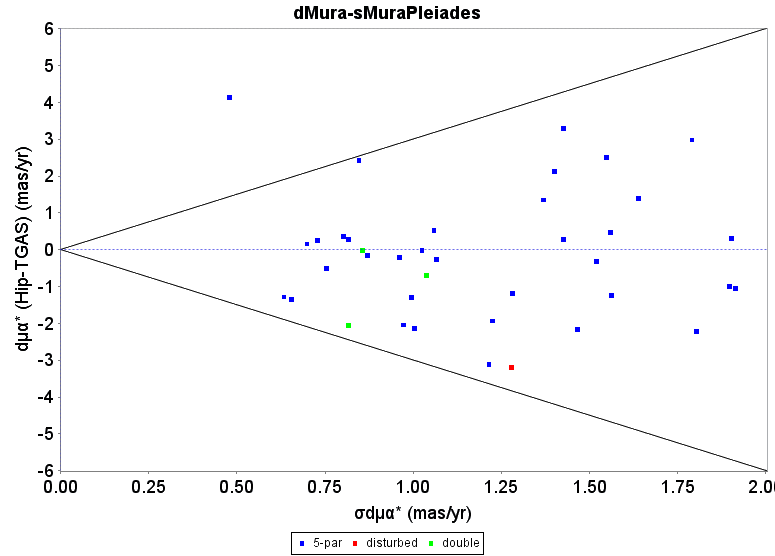}
\caption{Differences in parallax (top) and proper motion in Right Ascension (bottom) between the \Hipparcos and \TGAS solutions, for members of the Pleiades cluster, and as a function of the errors on the differences.}
\label{fig:pleidpar}
\end{figure}

The differences in parallax for this field between the \Hipparcos and \TGAS solutions are not entirely random, but show correlations with brightness or colour (Fig.~\ref{fig:dparmaggplei}). From the small volume of data and the strong correlation between brightness and colour it is not possible to distinguish which of these is the actual source of the correlation. This also affects comparisons between differences and their \su values, as the latter are, for the \Hipparcos solution in particular, strongly correlated with brightness.

\begin{figure}[t]
\centering
\includegraphics[width=6.5cm]{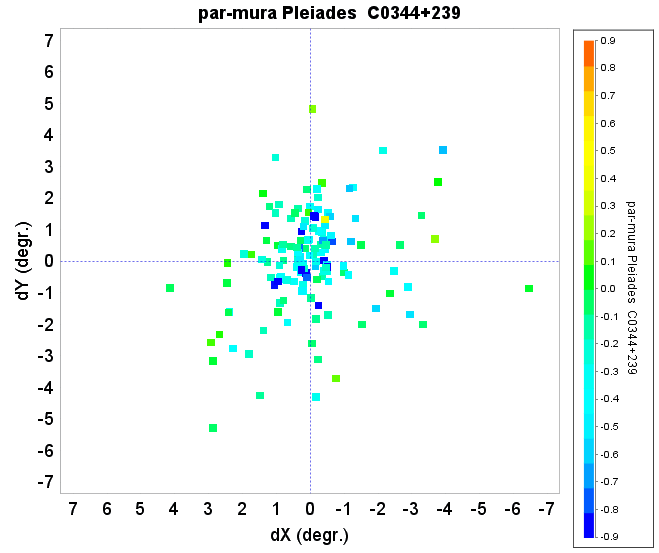}
\caption{Error correlation levels for Pleiades members as a function of position relative to the cluster centre. For reasons not understood, correlations appear to be stronger towards the cluster centre. The blue points, representing the negative correlation coefficients, all have \Tycho first-epoch data. For stars with \Hipparcos first epoch data (green points) the correlations are almost zero.}
\label{fig:parpmracorrplei}
\end{figure}
It is noted that there is a similar difference in the proper motion in Right Ascension (Fig.~\ref{fig:pleidpar}), and that in the \TGAS solution there is strong negative error correlation between the parallax and that component of the proper motion for stars with first epoch data from \Tycho (which dominate the parallax determination for the Pleiades), in particular towards the centre of the cluster (Fig.~\ref{fig:parpmracorrplei}).
 
\begin{figure}[t]
\centering
\includegraphics[width=8cm]{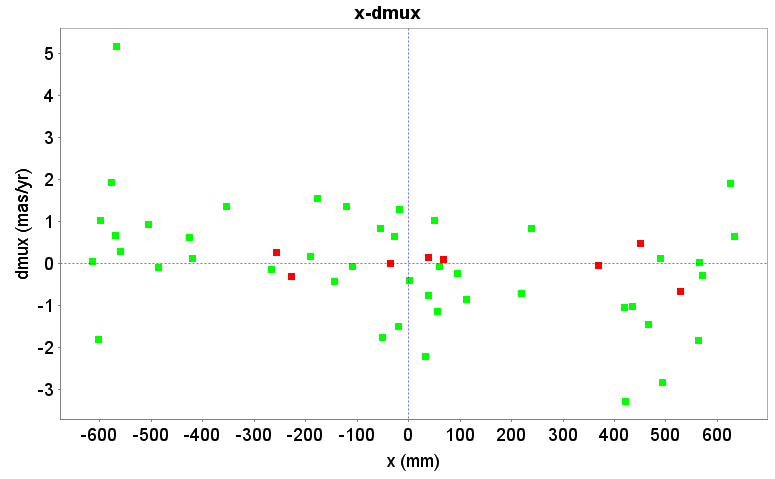}
\caption{A comparision between the proper motions in Right Ascension as measured from photographic plates \citep{1979A&AS...37..333V} and as published in \TGAS. The red dots represent stars with first epoch \Hipparcos data, the green dots have first epoch \Tycho data.}
\label{fig:propmotcompplei}
\end{figure}
Another potentially interesting comparison is that with the high-accuracy differential, ground-based proper motion studies of the Pleiades, such as \cite{1979A&AS...37..333V}, and the proper motions found in the \TGAS catalogue. In both cases accuracies significantly better than 1 mas~yr$^{-1}$ are claimed. The epoch coverage in this ground-based study is 77 years (1899 to 1976, all taken with the same telescope at the same site), with good coverage up to the mid 1940s (which were used by \citet{1947AnLei..19A...1H} in his study of the Pleiades) and a large volume of data in 1975/76. Of the 146 stars in this study, 52 are contained in the \TGAS catalogue, of which 8 have first epoch \Hipparcos data. Scale corrections to the proper motions in Right Ascension ($\times 0.90$) and Declination ($\times 1.05$) as well as a colour dependence in Declination (-0.43 (B$-$V)) had to be applied, after which a unit-weight standard deviation of 1.14 was obtained for the differences in proper motion, largely confirming the accuracies claimed in both \cite{1979A&AS...37..333V} and \TGAS. When also considering the 44 stars with \Tycho first epoch data the unit-weight standard deviation increases to 1.29, which may indicate a slight underestimate of the proper motion uncertainties for those stars in the \TGAS catalogue. 

\begin{figure}[t]
\centering
\includegraphics[width=8.5cm]{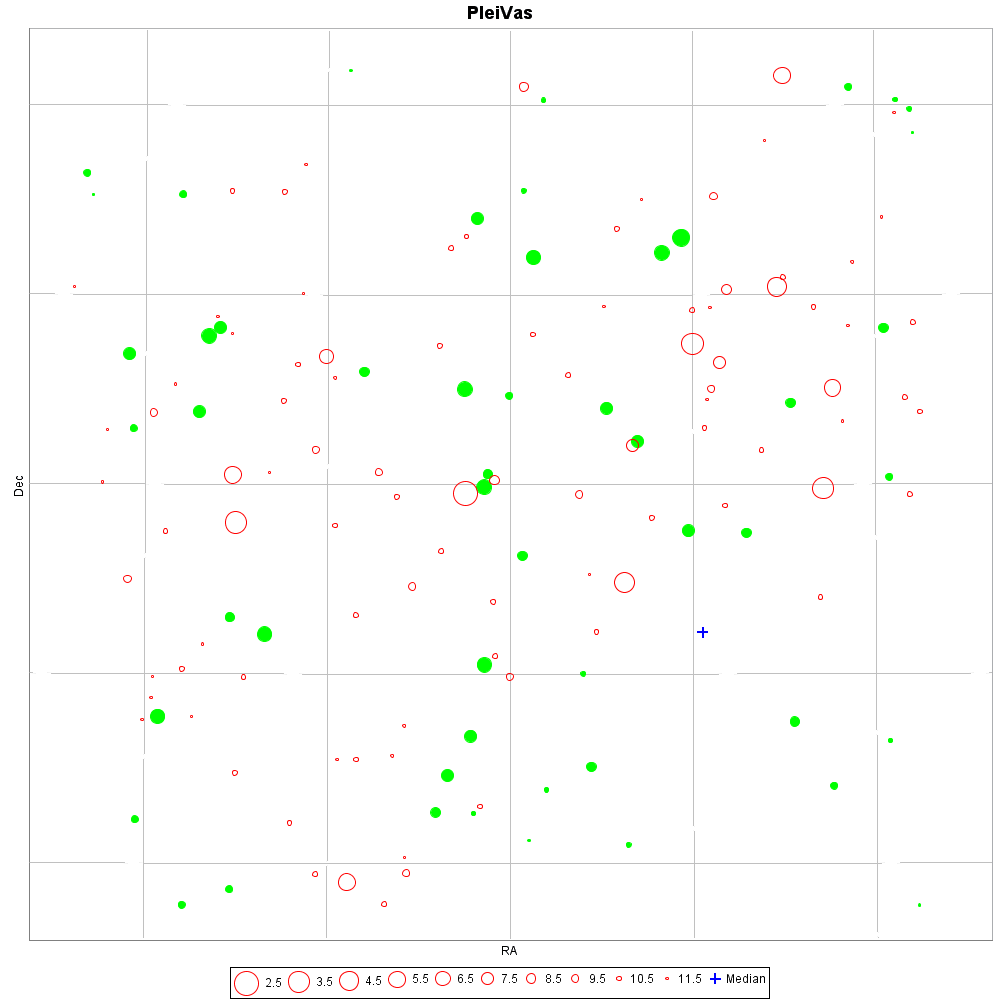}
\caption{The central area of the Pleiades cluster as defined by the brightest stars, showing as filled green circles all stars in this field that are included in \TGAS  and as open red circles those not included. The grid size is about 0.33 degrees. The cluster centre as defined by the cluster members in the TGAS catalogue, covering a much larger field, is indicated by the cross.}
\label{fig:pleivas}
\end{figure}
Figure~\ref{fig:pleivas} shows the central field of the Pleiades cluster, as defined by the brightest stars, indicating which stars are included in \TGAS. All the brightest stars are missing. These are the same stars that dominated by their weight the \Hipparcos parallax and proper motion solution for the Pleiades. It is also noted that the cluster centre as determined based on the fainter stars in the cluster, is markedly offset from the mean position of the bright stars. However, a similar offset is not observed in a sample of 333 probable Pleiades cluster members as extracted from the URAT1 catalogue \citep{2015AJ....150..101Z}.

Although it may seem tempting to suggest that this has resolved the so-called Pleiades issue, there are still some unexplained, and quite serious, issues left. The systematic parallax difference at a level of 0.6 mas in the Pleiades field affects all stars in that field, not just those of the Pleiades cluster. This is relevant, as field stars in the same part of the sky have been observed to show no anomalous luminosities when applying \Hipparcos parallaxes \citep{2016ApJS..222...19K}. It is a difference of which there has been no sign in comparisons between the three independent \Hipparcos reductions (the two reductions that contributed to the 1997 catalogue, and the new reduction). Strongly correlated errors over an area of more than a degree in diameter are very difficult to explain because of the rapidly decreasing fraction of shared scans for pairs of stars with increasing separations on the sky. Differences between the 1997 and 2007 reductions only show localized features on a scale of 0.5 to 1.0 degrees on the sky. Those features could be attributed to smoothing over clanks and hits in the 1997 publication. It should be noted too that, unlike for \Gaia, the basic angle for \Hipparcos was observed to be only slowly evolving, and stable at the sub-mas level over 24 hour periods, for almost the entire duration of the mission. Hits and clanks were very much less frequent for \Hipparcos than they are for \Gaia, and were in addition in the attitude reconstruction for the new reduction fully accounted for. For the \Gaia GDR1 this is not yet the case. On the other hand, the apparent internal consistency of the \TGAS data, such as shown for example by the distribution of negative parallaxes with respect to their formal errors, also does not leave much room for a discrepancy at the level observed for the Pleiades solutions. 

\subsubsection{HR Diagrams of Pleiades and Praesepe} 
\begin{figure}[t]
\centering
\includegraphics[width=8.cm]{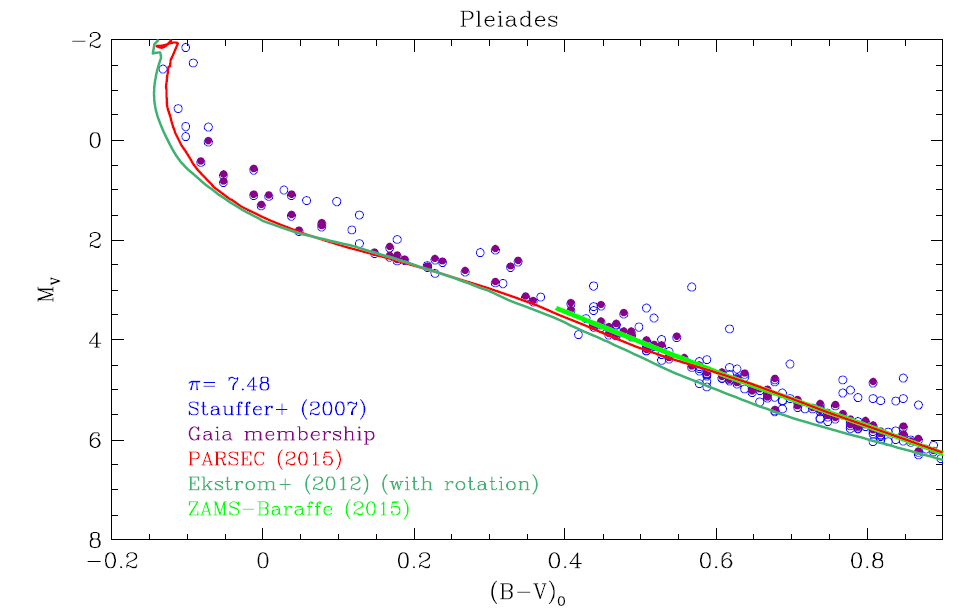}
\includegraphics[width=8.cm]{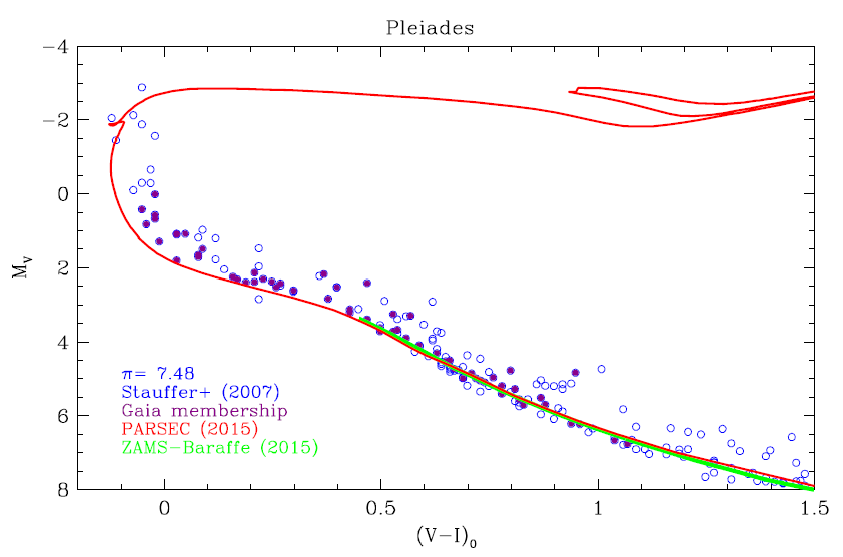}
\caption{ $M_V$, $(B{-}V)_0$ HR diagram of the Pleiades, with several sets of commonly used isochrones (top). Filled dots: members confirmed with Gaia data; open dots: other cluster members. Bottom panel is the analogous in the $M_V$, $(V{-}I)_0$. We assume an age of 130 Myr, solar metallicity, $A_V$=0.1.}
\label{fig:PleiadesCMD}
\end{figure}

\begin{figure}[t]
\centering
\includegraphics[width=8.cm]{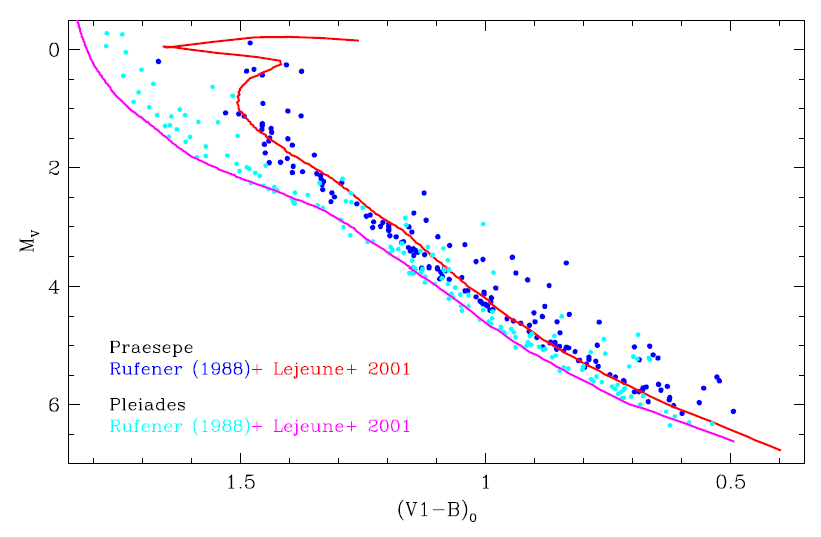}
\caption{ $M_V$, $(V1{-}B)_0$ HR diagram of the Pleiades (cyan dots) and Praesepe (blue dots), compared  with  Geneva stellar models.}
\label{fig:PleiadesV1B}
\end{figure}

Main sequence fitting has long being considered a powerful tool to derive distances. In the \Gaia era, when distances are known by direct measurements, it  provides a powerful test-bed for stellar models. Having this goal in mind, we compare the HR diagrams of two of the most studied clusters, Pleiades and Praesepe with stellar models, focusing on the main sequence fitting. We make use of literature values for the cluster ages and extinctions that are well constrained and have been derived using independent methods.  

In the case of the Pleiades, we assume an age  of about 130 Myr that is derived using the lithium depletion boundary \citep{2004ApJ...614..386B}. We point out that the error budget is quite large, going from 120 to 150 Myr, depending on differences in the stellar models and on adopted photometry. The extinction $A_V=0.1$ is by \cite{1998ApJ...504..805S} and the metal content is derived by high resolution spectroscopy,  [Fe/H]=$+0.03$ \citep{2009AJ....138.1292S}.

Using a similar approach for Praesepe,  we assume a metallicity from recent high resolution spectral analyses that have pointed in favour of super-solar values, going from [Fe/H]=$+0.27 \pm 0.10$ \citep{2008A&A...489..403P}, to [Fe/H]=$+0.12 \pm 0.04$ \citep{2013ApJ...775...58B}. We adopt an extinction of $A_V=0.1$ \citep{2006AJ....132.2453T}. The age of Praesepe is less well constraint, since techniques such as lithium boundary depletion are not applicable to intermediate-age clusters.  Stellar isochrones seem to suggest
an age range of several hundred Myr, with the main-sequence turnoff giving an age of about 600 to 650 Myr for
the most massive members \citep{2008A&A...483..891F}. Applying rotating stellar models, \cite{2015ApJ...807...24B} derive a best-fit age of about 800 Myrs, in agreement with fvl09. Here we assume a conservative estimate  of 600 Myr.

Figure~\ref{fig:PleiadesCMD} presents the HR diagram of the Pleiades in the $(B{-}V)$-$M_V$ and $(V{-}I)$-$M_V$ planes, using \cite{2007ApJS..172..663S} data corrected by  the \Gaia distance modulus and interstellar absorption. Only about 100 stars in common between \Gaia and  \cite{2007ApJS..172..663S} photometry were found. We  compare the data with several sets of commonly used stellar models, either including stellar rotation \citep{2012A&A...537A.146E} or  without \citep{ 2015A&A...577A..42B, 2015MNRAS.452.1068C}.

Figure~\ref{fig:PleiadesV1B} presents the HR diagram of the Pleiades and Praesepe in the Geneva photometry \citep{genevaphot} compared with \cite{2001A&A...366..538L} Geneva isochrone data base. This data set includes \cite{1992A&AS...96..269S} stellar tracks for solar and super-solar metallicity that are of interest here. Although these stellar models make use of quite old prescriptions, we note  that, concerning the main sequence, the  combined effects  of no rotational mixing and a stronger overshoot parameter $d_\mathrm{over}/H_P=0.2$ (used in the '92 models)  mimic the effect obtained in the more recent models \citep{2012A&A...537A.146E} including rotational mixing and  an overshoot parameter of 0.1. 

A discussion on the age of the Pleiades and Praesepe is outside the scope of the paper. Here we point out that broadly speaking the HR diagrams of Pleiades and Praesepe are reasonably fitted. The new Pleiades parallax seems to solve the discrepancy between \Hipparcos distance and those estimated via HRD fitting \citep{2007ApJ...655..233A}. However, it is clear that even in the zero age main sequence region (in the magnitude range $M_V \sim 3-6$), the fit critically depends on the ingredients of the stellar models and is often far from optimal as already noticed by \citet{2012MNRAS.424.3178B}. 

\subsection{The $\alpha$~Per cluster \label{sec:alpper}}

\begin{figure}[t]
\centering
\includegraphics[width=8.5cm]{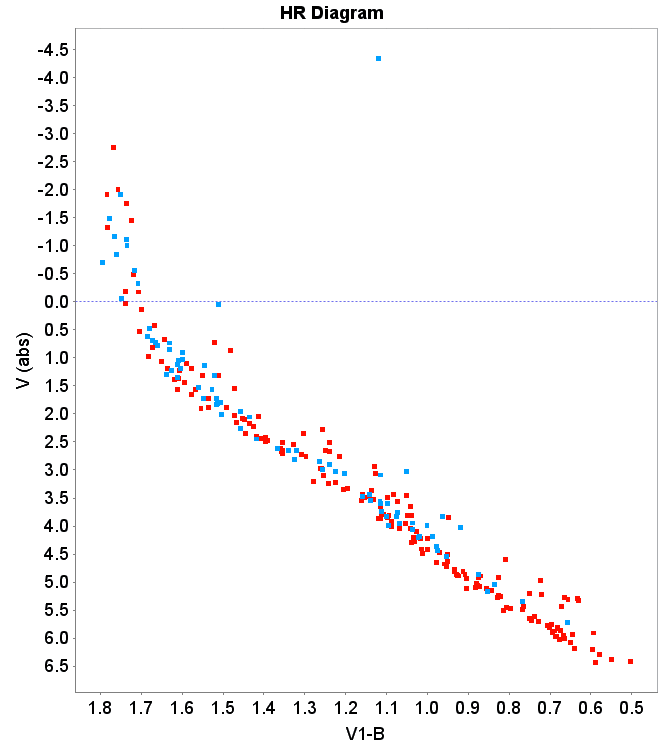}
\caption{The Geneva photometry HR diagram for the $\alpha$~Per cluster (blue dots) compared with the data on the Pleiades cluster (red dots). Reddening corrections were applied for both clusters. Geneva photometry.}
\label{fig:hralphper}
\end{figure}

The $\alpha$~Per cluster has been investigated over a 5.3 degrees radius field, an area for which 5475 stars are contained in the \TGAS catalogue, 323 of which are also contained in the \Hipparcos catalogue. The weighted mean differences in this field between the \TGAS and \Hipparcos astrometric parameters for 295 stars (only those with a basic 5-parameter \Hipparcos solution) of which 50 are identified as probable cluster members are summarized in Table~\ref{tab:hipcomp}. 

The parallax as determined for the $\alpha$~Per cluster corresponds to a distance modulus of $6.17\pm 0.01$, which is within 1$\sigma$ of the distance modulus give in \cite{pinso98}. The difference with the parallax determination in fvl09 is also within one sigma. A list of member stars and a map of the cluster are presented in Appendix~\ref{app:alphaper}. Figure~\ref{fig:hralphper} shows the Geneva photometry for stars in the $\alpha$~Per cluster that have been confirmed as cluster members from the \TGAS or the \Hipparcos astrometric data. The data is shown in comparison with the Pleiades cluster photometry.

There is no indication of increased scatter on the main sequence, at least compared to what is observed for Pleiades. This may contradict the suggested relatively high fraction of binary stars in the  $\alpha$~Per cluster, as reported to by \cite{2016MNRAS.457.1028S}.

\subsection{The cluster IC2391 \label{sec:ic2391}}

The cluster IC2391 was examined over a 6.3 degrees radius field, in which 13999 stars are contained in the \TGAS catalogue, 45 of which were indicated as possible cluster members. Only a small fraction of those stars have \Hipparcos first epoch data, 444 stars of which 8 are possible cluster members. The mean parallax and proper motion for the cluster are presented in Table~\ref{tab:overview}. The list and maps of cluster members shown in Appendix~\ref{app:ic2391}. 

Figure~\ref{fig:ic2391errcorr} shows the error correlations for stars in the field of the cluster that have \Tycho first epoch positions. There are substantial and systematic differences in error correlations between the astrometric parameters over the field of the cluster. 
\begin{figure}[t]
\centering
\includegraphics[width=7cm]{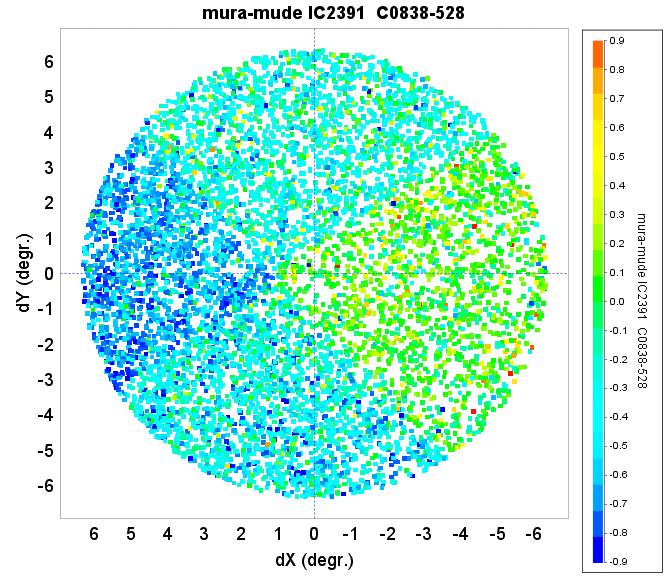}
\caption{Error correlations for the proper motion components in IC2391, for stars with \Tycho first epoch positions. The contributions of scans in different directions are clearly visible.}
\label{fig:ic2391errcorr}
\end{figure}
Of particular interest here is that the brightest star in the field of IC2391 is not a cluster member (fvl09). Three more stars indicated as members by \cite{1969AJ.....74..899P} are also unlikely to be members as based on the parallax determinations in \TGAS. They are HD 74582, 74955 and 75066. In proper motion these three stars do not deviate significantly from the cluster proper motion though. All three stars were also indicated as non-members in a spectroscopic follow-up study by \cite{1969PASP...81..629P}. Compared to that paper, there are four stars for which the current astrometric solution reaches a different conclusion on membership. These are HD 74009 and 74195, which now do not appear to be cluster members based on their parallaxes, and HD 74169 and 74535 (rejected on spectral type criterion) which do appear to be members of IC2391, as based on their proper motion and parallax. There are in addition 6 stars indicated as members in the photometric study that are not included in either the \Hipparcos or the \TGAS catalogue.

\subsection{The cluster IC2602 \label{sec:ic2602}}

\begin{figure}[t]
\centering
\includegraphics[width=7cm]{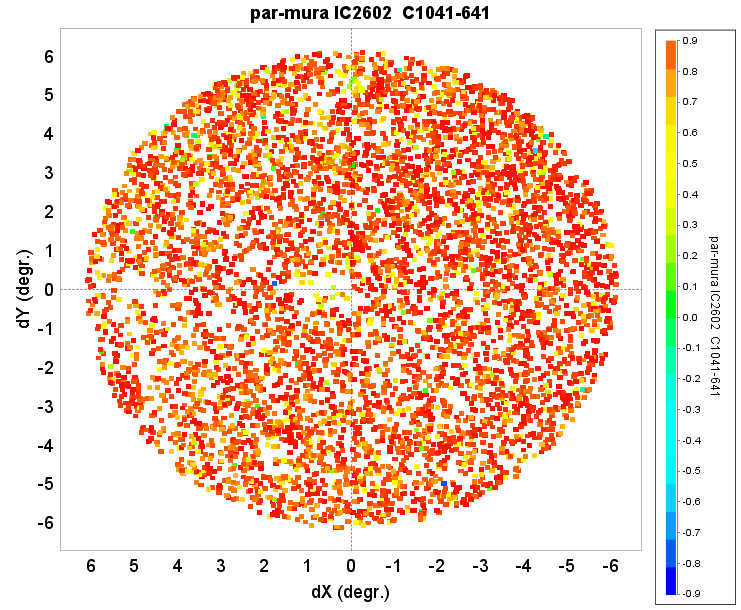}
\caption{Error correlations between the parallax and proper motion in Right Ascension in the field of IC2602, for stars with first epoch positions from the \Tycho catalogue.}
\label{fig:ic2602errcorr}
\end{figure}
A field of 6.1 degrees radius was investigated, containing 20762 stars, of which 70 were found to be possible cluster members. Of these stars 479 and 23 respectively have first epoch positions from the \Hipparcos catalogue. The result of the astrometric solution for the cluster proper motion and parallax led to 4 more rejections and 66 probable members, the details for which are presented in Appendix~\ref{app:ic2602}. Compared to the photometric study of \cite{1969AJ.....74.1011H} there is only one star now rejected as a cluster member, HD 93012. However, 6 of the member stars mentioned in that paper are not contained in the \TGAS catalogue. 
\begin{figure}[t]
\centering
\includegraphics[width=8cm]{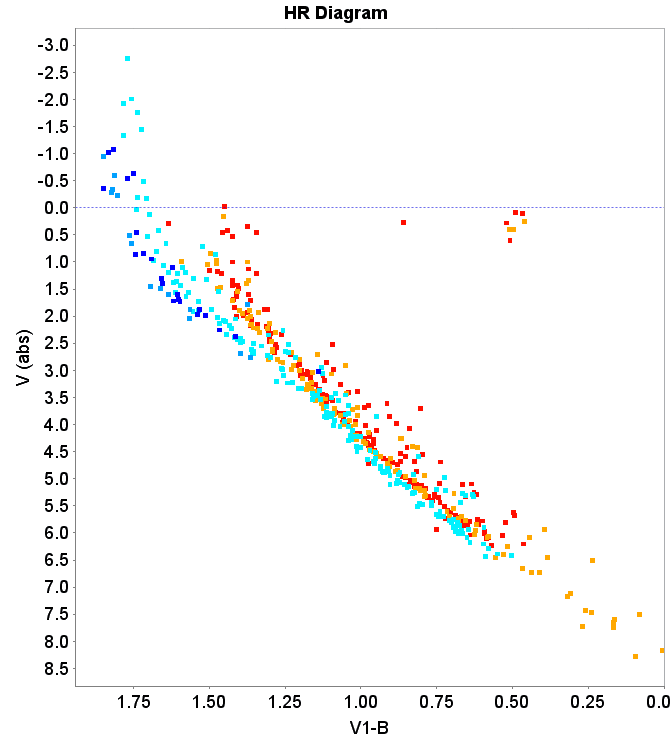}
\caption{The combined HR diagram for the Hyades and Praesepe (orange, red dots), Pleiades (light blue), IC2391 (blue) and IC 2602 (dark blue dots). Only data for astrometrically confirmed cluster members is shown. Geneva photometric data.}
\label{fig:hr23912602}
\end{figure}
Error correlations are particularly strong between parallax and proper motion in Right Ascension for stars with \Tycho first-epoch positions. The field coverage shows some holes where bright stars are found (Fig.~\ref{fig:ic2602errcorr}). 

The HR diagram for IC2391 and IC2602, compared with the combined main sequence for the Hyades and Praesepe, is shown in Fig.~\ref{fig:hr23912602}. The main sequences for the two clusters coincide very well, confirming their very similar age.

\subsection{The cluster Blanco 1 \label{sec:blanco1}}
\begin{figure}[t]
\centering
\includegraphics[width=7.8cm]{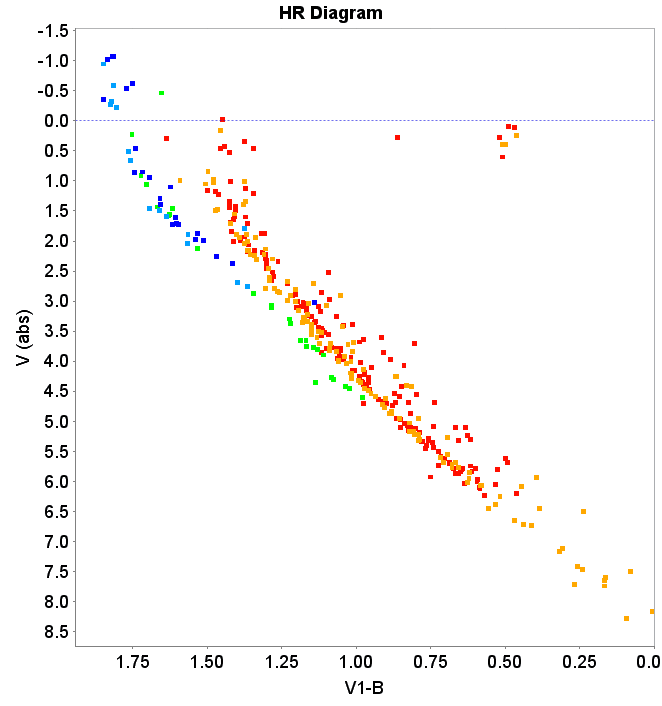}
\caption{The HR diagram of Blanco 1 (green dots, only confirmed members) compared with the Hyades and Praesepe (orange, red dots) and IC2391 and IC2602 (blue dots). Geneva photometry.}
\label{fig:hrblanco1}
\end{figure}
A field of 3.9 degrees radius was investigated for which 1169 stars are contained in the \TGAS catalogue, 121 of which have first-epoch \Hipparcos positions. Of these stars, 46 were marked as possible cluster members, of which 8 also have \Hipparcos data. The astrometric solution resulted in two further rejections, and the final selection details are presented in App.~\ref{app:blanco1}. The parallax is just under $2\sigma$ less than what was found in fvl09, putting the cluster at around 232~pc, close to a recent estimate based on isochrone fitting \citep{2015AAS...22524726K}. The small number of members available in fvl09 led to a relatively large \su on the parallax estimate. 

The Geneva photometry for Blanco 1 contains 64 entries, of which 26 could be identified as cluster members in the \TGAS or else the \Hipparcos data, using information from \cite{1988A&AS...76..101W}. Twelve non-members were found in the list, and the remainder of sources has not been identified as insufficient information was available. The HR diagram for Blanco 1, compared with other clusters, is shown in Fig.~\ref{fig:hrblanco1}. 

\subsection{The cluster NGC2451A \label{sec:ngc2451a}}
\begin{figure}[t]
\centering
\includegraphics[width=7.8cm]{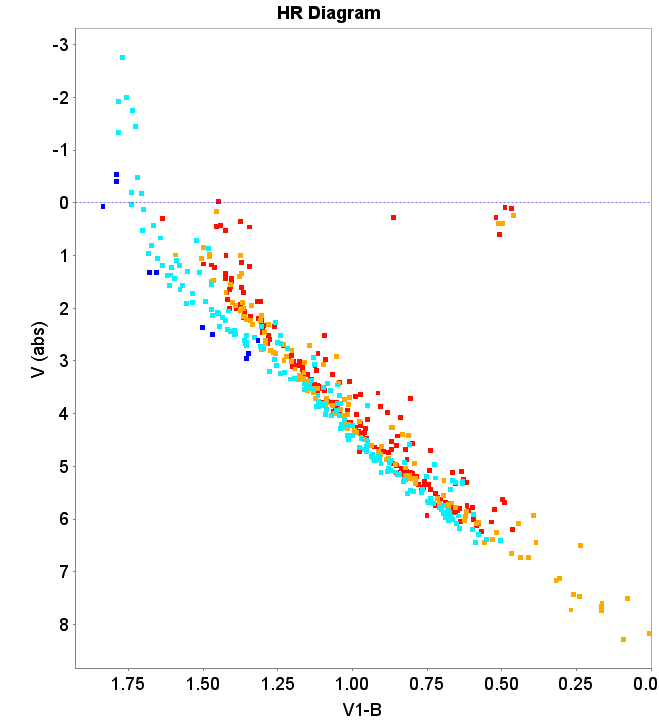}
\caption{The combined HR diagrams for the Hyades  and Praesepe (orange, red), Pleiades (light blue) and confirmed members of NGC2451A (dark blue), in Geneva photometry.}
\label{fig:hrngc2451}
\end{figure}
An extensive ground-based proper motion study of NC2451A was presented by \cite{2001AJ....122.1486P}. The \TGAS results cover only the brightest 5 magnitudes of that study, where membership is close to unambiguous. Only NGC2451A is covered, NGC2451B, if it exists, is not an obvious feature in the proper motion or the parallax distributions in the field, and with its assumed distance, will anyway be eliminated from the analysis of NGC2451A on the basis of the parallax selection criterion.

A field of 5 degrees radius was investigated, for which 7815 stars are contained in the \TGAS catalogue. Of these, 247 have \Hipparcos first epoch data. 39 stars were selected as possible cluster members, and of these 4 have \Hipparcos first epoch data. Two of the possible members were later rejected in the cluster solution.

The parallax found is at $5.99\pm 0.11$~mas slightly more than the $5.54\pm 0.11$~mas determined from the \Hipparcos data in fvl09. Taking into account a possible local calibration error of order 0.25~mas (see below), these results are in good agreement. On the other hand, the \TGAS result for this cluster is closer to earlier results based on the first \Hipparcos data publication, $5.30\pm 0.20$~mas \citep{1999A&A...341L..71V}. This shows how vulnerable these determinations can be to relatively large variations when only small numbers of stars are involved. The details for NGC2451A are presented in Appendix~\ref{app:ngc2451}.

Figure~\ref{fig:hrngc2451} shows the HR diagram for NGC2451A with respect to the Hyades, Praesepe and Pleiades clusters. The Geneva photometry gives 49 entries for NGC2451A. Of these, only 10 could be confirmed as members of the cluster based on either \TGAS or, for the brighter stars, \Hipparcos astrometry, using HD identifiers or positions as given in \cite{1967MNSSA..26...30W}. Around 12 stars could not be identified in either catalogue, and may still be members. The HR diagrams of the Pleiades and NGC2451A have moved further apart with the \TGAS parallaxes compared to fvl09.

\section{The more distant clusters\label{sec:distant}} 

\begin{table}[t]
\caption{Star selection numbers in the fields of the more distant clusters.}
\label{tab:distclust}
\centering
\begin{tabular}{l|rrrrrr}
\hline\hline
Cluster & Rad. & N-T & N-H & n-T & n-H & Rej. \\
\hline
NGC6475 & 3.2 & 2554 & 79 & 81 & 11 & 2 \\
NGC7092 & 2.7 & 3044 & 81 & 24 & 5 & 1 \\
NGC2516 & 2.6 & 1751 & 82 & 76 & 5 & 1 \\
NGC2232 & 3.7 & 1785 & 45 & 32 & 4 & 1 \\
IC4665 & 2.5 & 959 & 38 & 16 & 6 & 0 \\
NGC6633 & 2.2 & 1278 & 31 & 51 & 4 & 2 \\
Coll140 & 2.6 & 2984 & 59 & 32 & 3 & 1 \\
NGC2422 & 2.3 & 3204 & 46 & 39 & 2 & 2 \\
NGC3532 & 2.2 & 4304 & 76 & 140 & 4 & 8 \\
NGC2547 & 1.9 & 1152 & 42 & 36 & 8 & 2 \\
\hline
\end{tabular}
\tablefoot{Columns as follows, 1: Cluster identifier; 2: field radius in degrees; 3, N-T: \TGAS stars within the radius; 4, N-H: \Hipparcos stars within the radius; 5, n-T: Possible cluster members found in TGAS; 6, n-H: \Hipparcos stars among the possible cluster members; 7, rej.: number of possible members rejected in astrometric parameter solution.}
\end{table}

\begin{figure}[t]
\centering
\includegraphics[width=7.5cm]{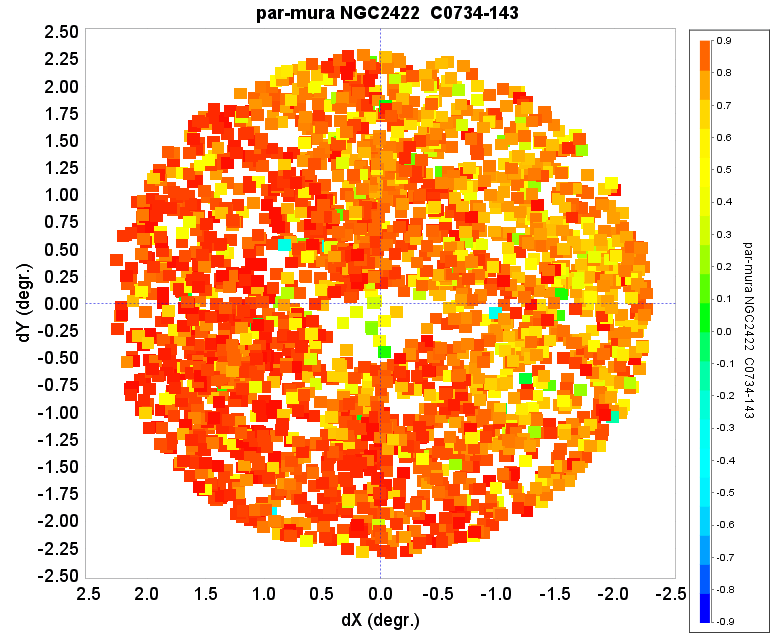}
\caption{Correlation levels between parallax and proper motion in Right Ascension for the field of NGC2422, showing the hole in the centre where the cluster core is situated. Data points are for stars with \Tycho first epoch positions.}
\label{fig:ngc2422corr}
\end{figure}
Table~\ref{tab:distclust} gives an overview of the fields and their contents as these have been investigated for more distant clusters. The detailed astrometric solutions are presented in Table~\ref{tab:overview}, while the details for each cluster can be found in Appendix~\ref{appselected}. A few clusters need special attention. For NGC2422 the core of the cluster was essentially missing from the \TGAS catalogue (see Fig.~\ref{fig:ngc2422corr}).

The field of NGC6633 is crossed by 4 diagonal `empty' lines, which may have affected the selection. A similar situation, though less severe, is found for NGC3532. 

The \Hipparcos data for the same clusters has mostly been obtained with a significantly smaller sample of stars (see Table~\ref{tab:hipcomp}), but also often using the brightest stars that are not included in \TGAS, leaving a generally small overlap between the two solutions. Despite that, there is in most cases a good agreement. The main exception is NGC2547.

In addition to these clusters, the possible existence of a cluster associated with $\delta$~Cep \citep{1999AJ....117..354D, 2012ApJ...747..145M} was looked into. Although there are around 18 stars detected within a 5 degrees radius around $\delta$~Cep, with similar distances and proper motions, these stars do not show any noticeable clustering in their distribution on the sky or the distribution of proper motions. The average parallax for these 18 stars is slightly larger than the measured parallax for $\delta$~Cep in fvl09.

\section{Summary of results\label{sect:summ}}

\begin{figure}[t]
\centering
\includegraphics[width=8cm]{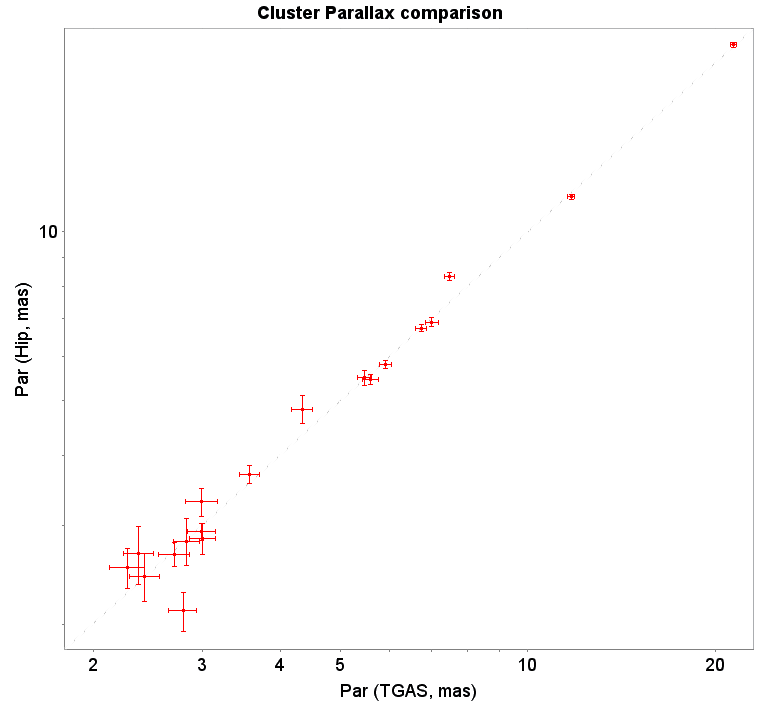}
\caption{Comparison between the cluster parallaxes as determined by the \Hipparcos and \TGAS analyses.}
\label{fig:clustcomp}
\end{figure}

We have determined and examined the astrometric data for 19 open clusters, ranging from the Hyades at just under 47~pc to IC2422 at nearly 440~pc. The results are summarized in Table~\ref{tab:overview}. Overall the agreement with a similar study using the \Hipparcos data is better than expected. There is one exception which remains unexplained, which is the Pleiades cluster. Whether the difference originates in the \TGAS data or in the \Hipparcos data, it remains at this stage unresolved. The differences between the current solution and fvl09 are shown in Table~\ref{tab:parcomp} and Fig.~\ref{fig:clustcomp}.  Without taking into account as additional noise local parallax zeropoint variations of 0.3 mas, as suggested in \cite{DPACP-14}, the unit weight standard deviation of the differences of the differences between the two solutions is 1.45. An additional noise at a level of 0.25~mas brings this down to 1.01. When excluding the Pleiades determinations, a much lower additional noise of 0.14~mas is required, which would make the Pleiades result stand out by 4.4 times the \su of the parallax differences between the \TGAS and \Hipparcos solutions. 

The main result, and unique to the \Gaia data, is that we seem to detect cluster members, bound or escaped, often still at nearly 15~pc from the cluster centre. With its complete survey, the \Gaia mission can detect these potential cluster members from the combined parallax and proper motion data, and future releases will further supplement this with radial velocity measurements. Without the parallax as an additional distinction the contrast of the cluster members from the field stars is much more difficult and uncertain.

There were assumptions still to be made for the current reductions. The one most affecting the results concerns properties of the internal velocities. Once proper motion and parallax accuracies for a significant group of stars are down to the 0.01 mas level it will become possible to examine for example the internal structure of the Pleiades cluster, and describe the distribution of positions and velocities of stars in the cluster. Being able to do so for clusters of different ages, such as Hyades, Coma Ber and the Pleiades, can then provide data that can be directly compared with N-body simulations.

The results for the Hyades confirmed what had earlier been observed in the Pleiades too, that the main sequence for a population that is homogeneous in age and composition can in fact be very narrow. This contrasts sharply with the width of the main sequence for field stars, in particular in the region of late G to early B stars. With future releases of the \Gaia data and its application to star clusters of different ages and chemical composition, it should become possible to reach a better understanding of the broad distribution of the field stars.

\begin{table*}[t]
\caption{Overview of the results\label{tab:overview}}
\centering
\begin{tabular}{l|rrrrrrr}
\hline\hline
Name & $\alpha$ (degr) &$\varpi$ & $\mu_{\alpha *}$ & $\mu_\delta$ & $c_{12}$ & $c_{23}$ & nMemb \\
ClustId & $\delta$ (degr) & $\sigma_\varpi$ & $\sigma_{\mu\alpha *}$ & 
$\sigma_{\mu\delta}$ & $c_{13}$ & r(max)\degr & st.dev. \\ 
\hline
Hyades & 66.85 & 21.39 & 104.92 & -28.00 & 0. & 0. & 103 \\
\vspace{2mm}
C0424+157 & 17.04 & 0.21 & 0.12 & 0.09 & 0. & 17.2 & 1.00 \\
ComaBer & 185.983 &   11.73 & -12.14 &  -8.90 &  0.14 &  0.32  &  50 \\
\vspace{2mm}
C1222+263 &  26.093 &   0.05 &   0.14 &   0.16 &  -0.07 &  9.42 &  0.86 \\
Pleiades &  56.438 &    7.48 &  20.38 & -45.39 & -0.17 & -0.03  & 152 \\
\vspace{2mm}
C0344+239 &  23.844 &   0.03 &   0.07 &   0.08 &   0.02 &  6.49 &  1.10 \\
Praesepe & 130.081 &    5.47 & -36.06 & -13.15 &  0.40 &  0.06  &  79 \\
\vspace{2mm}
C0937+201 &  19.675 &   0.05 &   0.07 &   0.08 &  -0.05 &  4.36 &  1.09 \\
alphaPer &  52.069 &    5.91 &  23.06 & -25.36 & -0.29 & -0.01  & 116 \\
\vspace{2mm}
C0318+484 &  49.060 &   0.03 &   0.06 &   0.07 &   0.09 &  5.11 &  1.04 \\
IC2391 & 130.064 &    7.01 & -24.35 &  23.76 &  0.53 & -0.04  &  44 \\
\vspace{2mm}
C0838-528 & -52.919 &   0.11 &   0.14 &   0.22 &  -0.01 &  6.07 &  1.60 \\
IC2602 & 159.809 &    6.74 & -17.67 &  11.06 &  0.36 & -0.19  &  66 \\
\vspace{2mm}
C1041-641 & -64.496 &   0.05 &   0.09 &   0.13 &   0.05 &  5.35 &  1.18 \\
Blanco 1 &   0.855 &    4.34 &  18.20 &   2.66 & -0.46 &  0.07  &  43 \\
\vspace{2mm}
C0001-302 & -30.079 &   0.11 &   0.12 &   0.11 &  -0.44 &  3.77 &  1.41 \\
NGC2451A & 115.799 &    5.59 & -21.82 &  15.59 &  0.63 &  0.13  &  37 \\
\vspace{2mm}
C0743-378 & -38.579 &   0.11 &   0.11 &   0.16 &  -0.01 &  4.70 &  1.38 \\
NGC6475 & 268.530 &    3.57 &   3.10 &  -5.32 &  0.01 &  0.37  &  78 \\
\vspace{2mm}
C1750-348 & -34.849 &   0.02 &   0.06 &   0.04 &   0.10 &  2.37 &  0.62 \\
NGC7092 & 323.437 &    2.99 &  -7.34 & -19.94 & -0.58 &  0.09  &  23 \\
\vspace{2mm}
C2130+482 &  48.438 &   0.12 &   0.11 &   0.13 &   0.21 &  2.14 &  1.30 \\
NGC2516 & 119.438 &    2.99 &  -4.07 &  11.21 &  0.77 & -0.08  &  87 \\
\vspace{2mm}
C0757-607 & -60.688 &   0.08 &   0.06 &   0.06 &   0.05 &  2.66 &  1.27 \\
NGC2232 &  97.149 &    3.00 &  -4.34 &  -1.71 &  0.32 & -0.32  &  31 \\
\vspace{2mm}
C0624-047 &  -5.111 &   0.06 &   0.10 &   0.08 &  -0.06 &  2.64 &  0.86 \\
IC4665 & 266.618 &    2.83 &  -0.78 &  -8.37 & -0.12 &  0.19  &  16 \\
\vspace{2mm}
C1743+057 &   5.583 &   0.05 &   0.07 &   0.06 &   0.09 &  1.92 &  0.60 \\
NGC6633 & 276.900 &    2.37 &   1.45 &  -1.77 & -0.36 &  0.44  &  47 \\
\vspace{2mm}
C1825+065 &   6.698 &   0.03 &   0.05 &   0.04 &  -0.03 &  1.83 &  0.63 \\
Coll140 & 111.127 &    2.70 &  -8.02 &   4.88 &  0.69 & -0.14  &  30 \\
\vspace{2mm}
C0722-321 & -32.183 &   0.08 &   0.07 &   0.07 &  -0.23 &  2.61 &  0.84 \\
NGC2422 & 114.139 &    2.27 &  -6.89 &   0.90 &  0.59 & -0.25  &  34 \\
\vspace{2mm}
C0734-143 & -14.156 &   0.07 &   0.07 &   0.07 &  -0.33 &  1.83 &  0.83 \\
NGC3532 & 166.270 &    2.42 & -10.54 &   5.19 &  0.61 & -0.04  & 128 \\
\vspace{2mm}
C1104-584 & -58.753 &   0.03 &   0.03 &   0.04 &   0.26 &  2.19 &  0.78 \\
NGC2547 & 123.079 &    2.79 &  -8.70 &   4.16 &  0.63 & -0.01  &  40 \\
C0809-491 & -49.230 &   0.06 &   0.06 &   0.07 &  -0.11 &  2.38 &  0.91 \\
\hline
\end{tabular}
\end{table*}

\begin{table*}[t]
\caption{Comparison between the \Hipparcos and \TGAS parallax determinations}
\label{tab:parcomp}
\centering
\begin{tabular}{l|rrrrr|rrr|rr}
\hline\hline
Cluster & N(TH) & N(TT) & $\varpi_T$ & $\sigma\varpi$ & $\Sigma\varpi$ &
N(Hip) &  $\varpi_H$ & $\sigma\varpi$ & $\Delta\varpi$ & $\sigma\Delta\varpi$ \\
\hline
Hyades & 88 & 22 & 21.39 & 0.21 & 0.33 & 150 & 21.53 & 0.23 & -0.14 & 0.40 \\ 
Coma Ber & 28 & 22 & 11.73 & 0.05 & 0.25 & 27 & 11.53 & 0.12 & +0.20 & 0.28 \\
Pleiades & 51 & 101 & 7.48 & 0.03 & 0.25 & 53 & 8.32 & 0.13 & -0.85 & 0.28 \\
Praesepe & 24 & 55 & 5.47 & 0.05 & 0.25 & 24 & 5.49 & 0.18 & -0.02 & 0.31 \\
$\alpha$ Per & 51 & 65 & 5.91 & 0.03 & 0.25 & 50 & 5.80 & 0.10 & 0.11 & 0.27 \\
IC2391 & 8 & 36 & 7.01 & 0.11 & 0.28 & 11 & 6.90 & 0.12 & 0.11 & 0.30 \\
IC2602 & 23 & 43 & 6.74 & 0.05 & 0.25 & 15 & 6.73 & 0.09 & 0.01 & 0.27 \\
Blanco 1 & 8 & 35 & 4.34 & 0.11 & 0.27 & 13 & 4.83 & 0.27 & -0.49 & 0.38 \\
NGC2451A & 4 & 33 & 5.59 & 0.11 & 0.27 & 14 & 5.45 & 0.11 & 0.14 & 0.29 \\
NGC6475 & 11 & 67 & 3.57 &0.02 & 0.25 & 20 & 3.70 & 0.14 & -0.13 & 0.29 \\
NGC7092 & 5 & 18 & 2.99 & 0.12 & 0.28 & 7 & 3.30 & 0.19 & -0.31 & 0.34 \\
NGC2516 & 7 & 80 & 2.99 & 0.08 & 0.26 & 11 & 2.92 & 0.10 & 0.07 & 0.29 \\
NGC2232 & 4 & 27 & 3.00 & 0.06 & 0.26 & 6 & 2.84 & 0.18 & 0.16 & 0.32 \\
IC4665 & 6 & 10 & 2.83 & 0.05 & 0.25 & 7 & 2.81 & 0.27 & 0.02 & 0.37 \\ 
NGC6633 & 4 &43 & 2.37 & 0.03 & 0.25 & 6 & 2.67 & 0.32 & -0.30 & 0.41 \\
Coll140 & 3 & 28 & 2.70 & 0.08 & 0.27 & 9 & 2.66 & 0.13 & 0.04 & 0.30 \\
NGC2422 & 2 & 33 & 2.27 & 0.07 & 0.27 & 7 & 2.52 & 0.21 & -0.25 & 0.34 \\
NGC3532 & 4 & 128 & 2.42 & 0.03 &  0.25 & 6 & 2.43 & 0.24 & -0.01 & 0.35 \\
NGC2547 & 8 & 26 & 2.79 & 0.06 &  0.26 & 8 & 2.11 & 0.17 & 0.68 & 0.31 \\
\hline
\end{tabular}
\tablefoot{The meaning of the columns is as follows: N(TH): number of stars in the \TGAS solution with \Hipparcos first epoch data; N(TT): number of stars in the \TGAS solution with \Tycho first epoch data; $\varpi_T$: \TGAS parallax for the cluster; $\sigma\varpi$: Formal \su on $\varpi_T$; $\Sigma\varpi$: \su including calibration uncertainty of 0.25 mas; N(Hip): Number of stars in the \Hipparcos solution; $\varpi_H$: \Hipparcos parallax for the cluster; $\sigma\varpi$: formal \su on $\varpi_H$; $\Delta\varpi=\varpi_T - \varpi_H$;  $\sigma\Delta\varpi$: \su on the parallax difference. }
\end{table*}

\begin{table*}[t]
\caption{Comparisons between \TGAS and \Hipparcos astrometric parameters in cluster fields.}
\label{tab:hipcomp}
\centering
\begin{tabular}{l|r|rr|rr|rr}
\hline\hline
Cluster & N & $\Delta\varpi$ & UWSD & $\Delta\mu_{\alpha *}$ & 
UWSD & $\Delta\mu_\delta$ & UWSD \\
\hline
 ComaBer &   28 &  $-0.14\pm 0.22$ &1.44 &  $-0.08\pm 0.40$ &2.77 &   $0.17\pm 0.34$ &3.69\\
 ComaBer &  758 & $ -0.01\pm 0.05$ &1.22 & $  0.14\pm 0.05$ &1.42 & $  0.21\pm 0.04$ &1.73\\
Pleiades &   44 & $  0.72\pm 0.17$ &1.05 & $  0.28\pm 0.30$ &1.80 & $  0.28\pm 0.19$ &1.36\\
Pleiades &  241 & $  0.36\pm 0.09$ &1.30 & $  0.25\pm 0.11$ &1.65 & $ -0.22\pm 0.11$ &2.08\\
Praesepe &   23 &   $0.19\pm 0.21$ &1.18 &   $0.77\pm 0.32$ &1.87 &   $0.02\pm 0.21$ &1.77\\
Praesepe &  123 & $  0.30\pm 0.13$ &1.35 & $ -0.07\pm 0.13$ &1.36 & $ -0.32\pm 0.11$ &1.79\\
alphaPer &   50 &  $-0.41\pm 0.14$ &1.20 &  $-0.24\pm 0.15$ &1.60 &  $-0.72\pm 0.18$ &2.06\\
alphaPer &  245 & $ -0.05\pm 0.07$ &1.24 & $ -0.08\pm 0.08$ &1.63 & $ -0.38\pm 0.08$ &1.72\\
  IC2391 &    7 &   $0.03\pm 0.33$ &1.50 &   $0.29\pm 0.25$ &1.66 &  $-0.18\pm 0.33$ &2.29\\
  IC2391 &  390 & $ -0.05\pm 0.04$ &1.19 & $  0.18\pm 0.05$ &1.76 & $ -0.06\pm 0.04$ &1.67\\
  IC2602 &   19 & $ -0.01\pm 0.15$ &1.17 & $  0.12\pm 0.12$ &1.11 & $  0.28\pm 0.19$ &2.13\\
  IC2602 &  394 & $  0.03\pm 0.04$ &1.29 & $  0.18\pm 0.06$ &2.05 & $  0.01\pm 0.05$ &1.77\\
 Blanco 1 &    8 & $ -0.26\pm 0.39$ &1.15 & $  0.16\pm 0.35$ &1.03 & $ -0.11\pm 0.34$ &1.60\\
 Blanco1 &   98 & $  0.14\pm 0.16$ &1.41 & $  0.10\pm 0.15$ &1.46 & $  0.08\pm 0.28$ &4.18\\
 NGC2451A &    4 & $ -0.20\pm 0.24$ &0.69 & $ -0.28\pm 0.31$ &2.70 & $ -0.41\pm 0.37$ &2.68\\
 NGC2451A &  224 & $  0.00\pm 0.05$ &1.16 & $  0.05\pm 0.04$ &1.54 & $  0.11\pm 0.05$ &1.41\\
 NGC6475 &   11 & $  0.23\pm 0.22$ &0.93 & $ -1.02\pm 0.31$ &1.33 & $  0.35\pm 0.15$ &1.09\\
 NGC6475 &   64 & $  0.13\pm 0.14$ &1.04 & $ -0.54\pm 0.26$ &1.78 & $  0.23\pm 0.11$ &1.44\\
 NGC7092 &    5 & $  0.05\pm 0.26$ &1.03 & $  0.08\pm 0.17$ &0.90 & $  0.15\pm 0.51$ &2.87\\
 NGC7092 &   63 & $ -0.01\pm 0.10$ &1.22 & $ -0.01\pm 0.12$ &1.69 & $ -0.10\pm 0.14$ &2.20\\
 NGC2516 &    6 & $  0.16\pm 0.21$ &1.01 & $  0.47\pm 0.25$ &1.65 & $  0.43\pm 0.17$ &1.03\\
 NGC2516 &   63 & $  0.12\pm 0.09$ &1.19 & $  0.47\pm 0.12$ &1.58 & $ -0.04\pm 0.13$ &1.76\\
 NGC2232 &    3 & $ -0.56\pm 0.23$ &0.55 & $  0.18\pm 0.20$ &0.55 & $ -0.11\pm 0.19$ &0.62\\
 NGC2232 &   41 & $  0.19\pm 0.17$ &1.36 & $ -0.04\pm 0.33$ &3.19 & $  0.12\pm 0.19$ &2.20\\
  IC4665 &    6 & $  0.35\pm 0.35$ &1.00 & $ -0.46\pm 0.98$ &3.36 & $  0.13\pm 0.36$ &1.83\\
  IC4665 &   28 & $ -0.48\pm 0.18$ &0.96 & $  0.46\pm 0.25$ &1.85 & $  0.16\pm 0.16$ &1.77\\
 NGC6633 &    4 & $  0.50\pm 0.43$ &0.93 & $ -0.56\pm 0.62$ &1.54 & $  0.13\pm 0.07$ &0.18\\
 NGC6633 &   24 & $ -0.30\pm 0.23$ &1.34 & $ -0.43\pm 0.31$ &2.17 & $  0.58\pm 0.21$ &1.62\\
 Coll140 &    3 & $  0.24\pm 0.83$ &2.10 & $ -0.31\pm 0.31$ &1.23 & $ -0.32\pm 0.20$ &0.67\\
 Coll140 &   52 & $  0.12\pm 0.12$ &1.09 & $  0.11\pm 0.20$ &2.99 & $  0.30\pm 0.19$ &2.16\\
 NGC2422 &    2 & $  1.46\pm 0.55$ &0.72 & $  0.29\pm 0.24$ &0.38 & $  0.10\pm 0.66$ &1.16\\
 NGC2422 &   33 & $  0.24\pm 0.20$ &1.31 & $  0.09\pm 0.18$ &1.66 & $  0.26\pm 0.14$ &1.54\\
 NGC3532 &    4 & $  0.32\pm 0.50$ &1.40 & $  0.47\pm 0.13$ &0.38 & $ -0.29\pm 0.43$ &1.49\\
 NGC3532 &   63 & $ -0.04\pm 0.11$ &1.15 & $  0.37\pm 0.17$ &2.23 & $ -0.07\pm 0.16$ &2.33\\
 NGC2547 &    9 & $ -0.65\pm 0.19$ &0.80 & $ -0.07\pm 0.25$ &1.26 & $  0.06\pm 0.25$ &1.35\\
 NGC2547 &   48 & $ -0.26\pm 0.14$ &1.50 & $  0.49\pm 0.18$ &2.39 & $ -0.09\pm 0.10$ &1.52\\
\hline
\end{tabular}
\tablefoot{For each cluster: first line for cluster members, second line for the remaining stars in the field of the cluster. Only stars with clean 5-parameter solutions in the \Hipparcos catalogue were used. For each value is given the mean, error on the mean and unit-weight standard deviation of the differences.}
\end{table*}

\section{Conclusions}

The \Gaia data, like the \Hipparcos data before, can not be validated or invalidated by results derived for the open clusters. A limited set of conclusions can be drawn from internal consistency of the data, and the most important one is the agreement between the parallaxes of the Hyades stars as measured and as derived from the proper motions. This agreement is, however, limited by the internal velocity dispersion of the cluster. The proper motion comparison with ground-based differential data in the Pleiades field is also reassuring. The overall agreement for the parallaxes of the 19 clusters investigated here with the earlier study (fvl09)  based on the new reduction of the \Hipparcos data is more than satisfactory, and an indication that earlier estimates for an additional local noise on the \TGAS parallaxes of 0.3 mas may have been slightly overestimated. 

Although questions remain on the one discrepancy between the \Hipparcos and \TGAS results, as well as on the \su levels of the current determination, the overall results are very promising for future releases, when parallaxes and proper motions at similar and higher accuracies will come available for much larger numbers of stars, extending over a wider range of magnitudes. Future releases should also gradually become less complicated to use, with error correlation levels between astrometric parameters reduced, and also modelling errors in the attitude solution becoming much less significant.

\begin{acknowledgements}
This work has made use of results from the European Space Agency (ESA) space
mission {\it Gaia}, the data from which were processed by the {\it Gaia}
Data Processing and Analysis Consortium (DPAC). Funding for the DPAC has
been provided by national institutions, in particular the institutions
participating in the {\it Gaia} Multilateral Agreement. The {\it Gaia}
mission website is \url{http://www.cosmos.esa.int/gaia}. The authors are
current or past members of the ESA {\it Gaia} mission team and of the {\it
Gaia} DPAC.
This work has financially been supported by:
the Algerian Centre de Recherche en Astronomie, Astrophysique et
G\'{e}ophysique of Bouzareah Observatory;
the Austrian FWF Hertha Firnberg Programme through grants T359, P20046, and
P23737;
the BELgian federal Science Policy Office (BELSPO) through various PROgramme
de D\'eveloppement d'Exp\'eriences scientifiques (PRODEX) grants;
the Brazil-France exchange programmes FAPESP-COFECUB and CAPES-COFECUB;
the Chinese National Science Foundation through grant NSFC 11573054;
the Czech-Republic Ministry of Education, Youth, and Sports through grant LG
15010;
the Danish Ministry of Science;
the Estonian Ministry of Education and Research through grant IUT40-1;
the European Commission's Sixth Framework Programme through the European
Leadership in Space Astrometry (ELSA) Marie Curie Research Training Network
(MRTN-CT-2006-033481), through Marie Curie project PIOF-GA-2009-255267
(SAS-RRL), and through a Marie Curie Transfer-of-Knowledge (ToK) fellowship
(MTKD-CT-2004-014188); the European Commission's Seventh Framework Programme
through grant FP7-606740 (FP7-SPACE-2013-1) for the {\it Gaia} European
Network for Improved data User Services (GENIUS) and through grant 264895
for the {\it Gaia} Research for European Astronomy Training (GREAT-ITN)
network;
the European Research Council (ERC) through grant 320360 and through the
European Union's Horizon 2020 research and innovation programme through
grant agreement 670519 (Mixing and Angular Momentum tranSport of massIvE
stars -- MAMSIE);
the European Science Foundation (ESF), in the framework of the {\it Gaia}
Research for European Astronomy Training Research Network Programme
(GREAT-ESF);
the European Space Agency in the framework of the {\it Gaia} project;
the European Space Agency Plan for European Cooperating States (PECS)
programme through grants for Slovenia; the Czech Space Office through ESA
PECS contract 98058;
the Academy of Finland; the Magnus Ehrnrooth Foundation;
the French Centre National de la Recherche Scientifique (CNRS) through
action `D\'efi MASTODONS';
the French Centre National d'Etudes Spatiales (CNES);
the French L'Agence Nationale de la Recherche (ANR) investissements d'avenir
Initiatives D'EXcellence (IDEX) programme PSL$\ast$ through grant
ANR-10-IDEX-0001-02;
the R\'egion Aquitaine;
the Universit\'e de Bordeaux;
the French Utinam Institute of the Universit\'e de Franche-Comt\'e,
supported by the R\'egion de Franche-Comt\'e and the Institut des Sciences
de l'Univers (INSU);
the German Aerospace Agency (Deutsches Zentrum f\"{u}r Luft- und Raumfahrt
e.V., DLR) through grants 50QG0501, 50QG0601, 50QG0602, 50QG0701, 50QG0901,
50QG1001, 50QG1101, 50QG140, 50QG1401, 50QG1402, and 50QG1404;
the Hungarian Academy of Sciences through Lend\"ulet Programme LP2014-17;
the Hungarian National Research, Development, and Innovation Office through
grants NKFIH K-115709 and PD-116175;
the Israel Ministry of Science and Technology through grant 3-9082;
the Agenzia Spaziale Italiana (ASI) through grants I/037/08/0, I/058/10/0,
2014-025-R.0, and 2014-025-R.1.2015 to INAF and contracts I/008/10/0 and
2013/030/I.0 to ALTEC S.p.A.;
the Italian Istituto Nazionale di Astrofisica (INAF);
the Netherlands Organisation for Scientific Research (NWO) through grant
NWO-M-614.061.414 and through a VICI grant to A.~Helmi;
the Netherlands Research School for Astronomy (NOVA);
the Polish National Science Centre through HARMONIA grant
2015/18/M/ST9/00544;
the Portugese Funda\c{c}\~ao para a Ci\^{e}ncia e a Tecnologia (FCT) through
grants PTDC/CTE-SPA/118692/2010, PDCTE/CTE-AST/81711/2003, and
SFRH/BPD/74697/2010; the Strategic Programmes PEst-OE/AMB/UI4006/2011 for
SIM, UID/FIS/00099/2013 for CENTRA, and UID/EEA/00066/2013 for UNINOVA;
the Slovenian Research Agency;
the Spanish Ministry of Economy MINECO-FEDER through grants AyA2014-55216,
AyA2011-24052, ESP2013-48318-C2-R, and ESP2014-55996-C2-R and MDM-2014-0369
of ICCUB (Unidad de Excelencia Mar\'{\i}a de Maeztu);
the Swedish National Space Board (SNSB/Rymdstyrelsen);
the Swiss State Secretariat for Education, Research, and Innovation through
the ESA PRODEX programme, the Mesures d'Accompagnement, and the Activit\'es
Nationales Compl\'ementaires;
the Swiss National Science Foundation, including an Early Postdoc.Mobility
fellowship;
the United Kingdom Rutherford Appleton Laboratory;
the United Kingdom Science and Technology Facilities Council (STFC) through
grants PP/C506756/1 and ST/I00047X/1; and
the United Kingdom Space Agency (UKSA) through grants ST/K000578/1 and
ST/N000978/1.
\end{acknowledgements}

\appendix

\section{Combined astrometric solutions \label{app:combine}}

\subsection{Observations and noise contributions}
In the combined astrometric solution the observed parallaxes and proper motions are compared with predicted ones, based on the assumed parallax and space motion of the cluster centre, and the position of the star on the sky relative to the projection of the cluster centre. This forms the common solution which provides an update to the proper motion of the cluster. 

The correction for the parallax offset $\mathrm{d}\varpi_i$ along the line of sight of the observed parallax is reflected in the proper motion for each star $i$:
\begin{equation}
\left[\begin{array}{r}
 1 \\ \mu_{\alpha *, c}/\varpi_c \\ \mu_{\delta, c}/\varpi_c\\ 
\end{array}\right]\cdot \mathrm{d}\varpi_i.
\label{equ:distcorr}
\end{equation}
In reality this contribution is only significant for the nearby clusters. The complete observation equations for the cluster parallax and proper motion corrections are as follows:
\begin{equation}
\left[\begin{array}{rrrr}
1 & 0 & 0 & 1 \\
0 & 1 & 0 & \mu_{\alpha *, c}/\varpi_c \\
0 & 0 & 1 & \mu_{\delta, c}/\varpi_c\\
\end{array}\right]\cdot
\left[\begin{array}{r}
\mathrm{d}\varpi_c \\ \mathrm{d}\mu_{\alpha *, c} \\ \mathrm{d}\mu_{\delta, c} \\ \mathrm{d}\varpi_i \\
\end{array}\right] = 
\left[\begin{array}{r}
\delta\varpi_i \\ \delta\mu_{\alpha *, i} \\ \delta\mu_{\delta, i}\\
\end{array}\right] + \sqrt{\tens{N}_i}~\vec{\epsilon}.
\label{equ:obsequ}
\end{equation}
where the index $c$  refers to the cluster  parameters, $\tens{N}_i$ is the noise covariance matrix for the astrometric parameters of star $i$ (see below), and each element of the vector $\epsilon$ has expectation value 0 and sigma of one. The value of $\delta\varpi_i$ is the difference between the assumed cluster parallax and the observed parallax for star $i$. The values of $\delta\mu_{\alpha *, i}$ and $\delta\mu_{\delta, i}$ are the differences between the observed and predicted proper motion assuming the parallax to be the same as the cluster. This way, the expression in Eq.~\ref{equ:distcorr} allows for a compensation of the relative distance of a star, as based on the parallax and proper motion measurements. However, due to the still fairly limited accuracies of both proper motions and parallaxes for individual stars, the inclusion of the relative parallax corrections (Eq.~\ref{equ:distcorr}) creates a near-singularity in the solution when also the cluster parallax is solved for. There are therefore, at this stage, two types of solutions, one for the cluster parallax, 
\begin{equation}
\left[\begin{array}{r}
\tens{I}_3\\
\end{array}\right]\cdot
\left[\begin{array}{r}
\mathrm{d}\varpi_c \\ \mathrm{d}\mu_{\alpha *, c} \\ \mathrm{d}\mu_{\delta, c} \\
\end{array}\right] = 
\left[\begin{array}{r}
\delta\varpi_i \\ \delta\mu_{\alpha *, i} \\ \delta\mu_{\delta, i}\\
\end{array}\right] + \sqrt{\tens{N}_i}~\vec{\epsilon}.
\label{equ:obsequcluster},
\end{equation}
and one for the differential parallaxes within the cluster,
\begin{equation}
\left[\begin{array}{rrr}
0 & 0 & 1 \\
1 & 0 &  \mu_{\alpha *, c}/\varpi_c \\
0 & 1 & \mu_{\delta, c}/\varpi_c \\
\end{array}\right]\cdot
\left[\begin{array}{r}
\mathrm{d}\mu_{\alpha *, c} \\ \mathrm{d}\mu_{\delta, c} \\ \mathrm{d}\varpi_i \\
\end{array}\right] = 
\left[\begin{array}{r}
\delta\varpi_i \\ \delta\mu_{\alpha *, i} \\ \delta\mu_{\delta, i}\\
\end{array}\right] + \sqrt{\tens{N}_i}~\vec{\epsilon}.
\label{equ:obsequind}
\end{equation}
Only when both the parallaxes and proper motions reach a higher accuracy it may become possible to combine the two solutions.

Two noise matrices are associated with the observations. The first is the covariance matrix $\tens{N}_a$ for the astrometric parameter determination as applicable to each individual member. The second is the noise on the proper motions introduced by the internal velocity dispersion, $\tens{N}_v$. The sum of these two contributions is given by $\tens{N}_i$. If the matrix $\tens{U}_i$ is an upper-triangular square root of $\tens{N}_i$, then we can normalize the noise on the observation equations by multiplying both sides of Eq.~\ref{equ:obsequcluster} by the upper-triangular inverse of $\tens{U}_i$:
\begin{equation}
\left[\begin{array}{r}
\tens{U}^{-1}_i \\ 
\end{array}\right]\cdot
\left[\begin{array}{r}
\mathrm{d}\varpi_c \\ \mathrm{d}\mu_{\alpha *, c} \\ \mathrm{d}\mu_{\delta, c} \\
\end{array}\right] = \tens{U}^{-1}_i\cdot\left[\begin{array}{r}\delta\varpi_i\\ 
\delta\mu_{\alpha *}\\ \delta\mu_\delta \end{array}\right] +\vec{\epsilon}. 
\label{equ:normobs}
\end{equation}
Equations of the type Eq.~\ref{equ:normobs} are the input observation equations for the cluster astrometric parameters solution. The matrix $\tens{U}^{-1}$ is referred to as the weight matrix, and is a square root of the normal equations. It has the same dimensions as the observation equations. A similar procedure can be applied to Eq.~\ref{equ:obsequind}.

The first component of $\tens{N}_i$ can be reconstructed from the data provided in the \Gaia DR1 \TGAS records, where the standard errors $\sigma$ and correlation coefficients $c$ for the astrometric parameter solution are given. Here we are only concerned about the parallax and proper motion determinations. The 3 by 3 matrix $\tens{N}_a$ is then given as:
\begin{equation}
\tens{N}_a = \left[\begin{array}{rrr}
\sigma_1^2 & c_{12}\sigma_1\sigma_2 & c_{13}\sigma_1\sigma_3 \\
c_{12}\sigma_1\sigma_2 & \sigma_2^2 & c_{23}\sigma_2\sigma_3 \\
c_{13}\sigma_1\sigma_3 & c_{23}\sigma_2\sigma_3 & \sigma_3^2 \\
\end{array}\right],
\end{equation} 
where the indices 1, 2, 3 stand for parallax, proper motion in right ascension and proper motion in declination respectively. The values for $c_{i,j}$ are provided with the astrometric data as the correlation coefficients for the astrometric parameters.

The noise matrix for the internal velocity dispersion is given by:
\begin{equation}
\tens{N}_v = \left[\begin{array}{rrr}
0 & 0 & 0 \\
0 & \sigma_v^2 & 0 \\
0 & 0 & \sigma_v^2 \\
\end{array}\right],
\end{equation}
where $\sigma_v$ is equivalent to an internal velocity dispersion of 0.6 km~s$^{-1}$ ($\kappa=4.74047$, the transformation factor from mas~yr$^{-1}$ to km~s$^{-1}$):
\begin{equation}
\sigma_v = 0.6 \cdot \varpi_c/\kappa~~\mathrm{mas~s^{-1}},
\end{equation}
which is roughly equivalent to what has been observed in the Pleiades and Hyades. The assumptions concerning this internal velocity dispersion can significantly affect the outcome of the cluster parallax due to the strong correlation coefficients in $\tens{N}_a$. It is a value that is going to be dependent on stellar mass and distance from the cluster centre, but these are considerations that become possible to implement with future releases of the \Gaia data. The current data is still too complicated to determine and implement such dependencies. 

When solving Eq.~\ref{equ:obsequcluster} the parallax dispersion has to be taken into account. This again is a somewhat uncertain quantity, that will differ from cluster to cluster. The dispersion in actual distance for an `average cluster' is assumed to be 0.003 kpc. In first approximation this will give a parallax dispersion $\sigma_\varpi \approx \varpi^2\sigma_r$. At a parallax of, say, 8 mas, this implies a parallax dispersion of just under 0.2~mas. The noise matrix contribution is simply 
\begin{equation}
\tens{N}_\varpi = \left[\begin{array}{rrr}
\sigma_\varpi^2 & 0 & 0 \\
0 & 0 & 0 \\
0 & 0 & 0 \\
\end{array}\right].
\end{equation}

For any individual star the measured proper motion may be further disturbed by unresolved orbital motion, but these have to be resolved with increase in the data volume and epoch coverage.

\subsection{Projection effects \label{app:projeff}}

If we consider the centre of the cluster to be represented by the vector $\vec{R}$, then the cluster space velocity is given by the derivative of this vector, $\dot{\vec{R}}$. Expressed in equatorial coordinates, these vectors have the following familiar expressions:
\begin{equation}
\vec{R} = R\cdot \left[\begin{array}{r}\cos\alpha\cos\delta\\\sin\alpha\cos\delta\\\sin\delta\\ \end{array}\right]
\end{equation} 
and
\begin{equation}
\dot{\vec{R}} = \left[\begin{array}{rrr}\cos\alpha\cos\delta & -\sin\alpha &
-\cos\alpha\sin\delta \\ \sin\alpha\cos\delta & \cos\alpha & -\sin\alpha\sin\delta \\
\sin\delta & 0 & \cos\delta \\ \end{array}\right]\,\cdot\,\left[\begin{array}{l}
\dot{R} \\ R\,\dot{\alpha}\cos\delta \\ R\,\dot{\delta}\\ \end{array}\right].
\label{equ:spacVel1}
\end{equation}
The vector on the right-hand side of Eq.~\ref{equ:spacVel1} relates directly to the proper motion and radial velocity of the cluster:
\begin{equation}
\left[\begin{array}{r}
\dot{R} \\ R\,\dot{\alpha}\cos\delta \\ R\,\dot{\delta}\\ \end{array}\right] = 
\left[\begin{array}{l} V_\mathrm{rad} \\ \kappa\mu_{\alpha *}/\varpi \\
\kappa\mu_\delta/\varpi\\ \end{array}\right].
\end{equation}
Similarly, the projection of the cluster space motion on the observed parameters of a cluster member (index $i$) can be expressed as:
\begin{equation}
\left[\begin{array}{l}V_{\mathrm{rad},i}\\
\kappa\,\mu_{\alpha *,i}/\varpi_i \\ 
\kappa\,\mu_{\delta,i}/\varpi_i \\\end{array}\right] = 
\left[\begin{array}{rrr} 
\cos\alpha_i\cos\delta_i & \sin\alpha_i\cos\delta_i & \sin\delta_i \\
-\sin\alpha_i & \cos\alpha_i & 0 \\ 
-\cos\alpha_i\sin\delta_i & -\sin\alpha_i\sin\delta_i & \cos\delta_i \\
\end{array}\right]
\,\cdot\,\dot{\vec{R}},
\label{equ:spacVel3}
\end{equation}

An approximation of these equations for distant clusters, with small differences between the position of the cluster centre and those of the member stars, can be found in fvl09. Equation~\ref{equ:spacVel3} is used to provide predicted values for the proper motions.

\subsection{Transformation to reduced proper motions \label{app:reduced}}

For further analysis the reference system can be rotated such that one component of the proper motion is aligned with the cluster proper motion, while the other is perpendicular to it. This system of reduced proper motions is particularly useful for analysing the Hyades and other nearby systems. The shared cluster motion for any cluster member is in the direction of the convergent point $(\alpha_t, \delta_t)$, the position of which is set by the direction of the space motion vector $\dot{\vec{R}}$ of the cluster:
\begin{equation}
\hat{\dot{\vec{R}}} = \left[\begin{array}{r}
\cos\alpha_t\cos\delta_t \\ \sin\alpha_t\cos\delta_t \\ \sin\delta_t \\
\end{array}\right]
\end{equation}
The transformation of the positional reference system to the new coordinates $(\rho, \tau)$, with the pole at the convergent point, is given by:
\begin{equation}
\left[\begin{array}{r}
\cos\rho\cos\tau \\ \sin\rho\cos\tau \\ \sin\tau\\
\end{array}\right] = \left[\begin{array}{rrr}
\cos\alpha_t\sin\delta_t & \sin\alpha_t\sin\delta_t & -\cos\delta_t \\
-\sin\alpha_t & \cos\alpha_t & 0 \\
\cos\alpha_t\cos\delta_t & \sin\alpha_t\cos\delta_t & \sin\delta_t \\
\end{array} \right] \cdot \hat{\vec{R}}
\end{equation}
For the rotation of the proper motions to the new system the local orientation of the equatorial coordinates needs to be reconstructed. 

The vector product of the direction to the source, $\hat{\vec{R}}$ and the direction of the convergent point is a vector $\vec{u}$ in the plane tangential to the direction of source and perpendicular to the direction of the convergent point as seen from the source. In that same plane the vectors
\begin{equation}
\vec{p} = \left[\begin{array}{r}
-\sin\alpha \\ \cos\alpha \\ 0 \\ 
\end{array}\right]
\end{equation}
and
\begin{equation}
\vec{q} = \left[\begin{array}{r}
-\cos\alpha\sin\delta \\ -\sin\alpha\sin\delta \\ \cos\delta \\ 
\end{array}\right]
\end{equation}
describe the direction, from the source, of right ascension and declination respectively (see also Eq.~\ref{equ:spacVel1}). Thus, the inner products
\begin{equation}
\cos\phi = \hat{\vec{u}}\cdot\vec{p}
\end{equation}
and 
\begin{equation}
\cos(90-\phi) = \sin\phi = \hat{\vec{u}}\cdot\vec{q}
\end{equation}
define the orientation angle $\phi$ needed for the transformation of the proper motions. The proper motions in the new coordinate system are
\begin{align}
\mu_{\rho\cos\tau} & =  \cos\phi~\mu_{\alpha\cos\delta} - \sin\phi~\mu_\delta \nonumber \\
\mu_\tau & =  \sin\phi~\mu_{\alpha\cos\delta} + \cos\phi~\mu_\delta. 
\end{align}
The transformation of the weight matrix $\tens{U}^{-1}$ is having the same form:
\begin{equation}
\tens{W} = \tens{U}^{-1}\cdot\left[\begin{array}{rrr}
1 & 0 & 0 \\ 0 & \cos\phi & \sin\phi \\ 0 & -\sin\phi & \cos\phi \\
\end{array} \right],
\end{equation}
which transforms the application of the standard errors and correlations to the new system. Note that the weight matrix $\tens{W}$ is no longer upper triangular. The predicted projected cluster proper motion has only one component, in the $\tau$ direction. It is zero in the $\rho$ direction. Thus, the vector $\vec{s}$ in Eq.~\ref{equ:obsequind} is simplified to
\begin{equation}
\vec{s}'\cdot\mathrm{d}\varpi_i = \left[\begin{array}{r}
 1 \\ 0 \\ \mu_{\tau, c}/\varpi_c\\ 
\end{array}\right]\cdot \mathrm{d}\varpi_i
\label{equ:distcorrtau}
\end{equation}
and the observation equations become
\begin{equation}
\tens{W}_i\cdot\left[\begin{array}{rrr}
0 & 0 & 1\\ 1 & 0 & 0 \\ 0 & 1 & \mu_{\tau, c}/\varpi_c \\ 
\end{array}\right]\cdot\left[\begin{array}{r}
\mathrm{d}\mu_{\rho\cos\tau, c} \\ \mathrm{d}\mu_{\tau, c} \\ \mathrm{d}\varpi_i\\
\end{array}\right] = \tens{W}_i\cdot\left[\begin{array}{r}\delta\varpi_i\\ 
\delta\mu_{\rho \cdot}\\ \delta\mu_\tau \end{array}\right] +\vec{\epsilon}. 
\label{equ:normobs2}
\end{equation}
The proper motion in the $\tau$ direction is primarily a function of the parallax of the star (relative to the mean cluster parallax) and the angular separation from the convergent point, and can as such be used to derive differential parallaxes of cluster members \citep{madsen99}. These are referred to as the kinematically improved parallaxes. The added uncertainty is in the internal velocity dispersion of the cluster members.

The observed proper motion dispersion in the $\rho$ direction, after correcting for observational standard errors, provides a potential measure for the internal velocity dispersion in the cluster. This reduced proper motion solution, which can be seen as the inverse of the convergent point cluster parallax determination (see also see \cite{madsen99, 2009A&A...497..209V}), is only useful in that context. For solving the cluster parallax and proper motion it is better to use Eq.~\ref{equ:obsequcluster} and staying that way closer to the original observations.

\section{Tangential projection and de-projection \label{app:tangproj}}
The tangential projection is used here as a simple tool to determine cluster centre positions, based on the average of the positions of all selected member stars. Just for reference, the equations are given here. Using the subscripts $i$ and $c$ for the star and the cluster centre respectively, and $\Delta\alpha_i\equiv(\alpha_i-\alpha_c)$, the projection is:
\begin{align}
x_i & =  \frac{\sin\Delta\alpha_i\cos\delta_i}{\sin\delta_i\sin\delta_c + \cos\delta_i\cos\delta_c\cos\Delta\alpha_i}, \nonumber \\
y_i & =  \frac{\sin\delta_i\cos\delta_c - \cos\delta_i\sin\delta_c \cos\Delta\alpha_i}{\sin\delta_i\sin\delta_c + \cos\delta_i \cos(\delta_c) \cos\Delta\alpha_i}.
\end{align}

For the inverse derivation, first derive
\begin{align}
w_i &=  \sin\delta_i\sin\delta_c + \cos\delta_i\cos\delta_c\cos\Delta\alpha_i \nonumber \\
&=  \frac{1}{\sqrt{1+x_i^2+y_i^2}}
\label{equ:w}
\end{align}
and similarly
\begin{align}
u_i & = \sin\delta_i\cos\delta_c - \cos\delta_i\sin\delta_c\cos\Delta\alpha_i \nonumber \\
 & =  y_i\cdot w_i
\label{equ:u}
\end{align}
and
\begin{align}
v_i &= \sin\Delta\alpha_i\cos\delta_i \nonumber\\
 &= x_i w_i.
\label{equ:v}
\end{align}
Combine Eq.~\ref{equ:w} and \ref{equ:u} to give
\begin{align}
\sin\delta_i &= \phantom{-}u_i\cos\delta_c + w_i\sin\delta_c \nonumber\\
\cos\delta_i\cos\Delta\alpha_i &= -u_i\sin\delta_c + w_i\cos\delta_c
\label{equ:uvw}
\end{align}
Equations \ref{equ:v} and \ref{equ:uvw} are all that is needed to recover $(\alpha_c, \delta_c)$.

\section{Three-dimensional distance from cluster centre \label{appdist}}

If the vector towards the star is given by $\vec{R}_s$ and for the cluster centre as $\vec{R}_c$, then the position of the star within the cluster is given by
\begin{equation}
\vec{r} = \vec{R}_s - \vec{R}_c.
\end{equation}
The angular separation $\rho$ of the star from the centre of the cluster is given by
\begin{equation}
\cos\rho = \varpi_c\varpi_s\vec{R}_c\cdot\vec{R}_s,
\end{equation}
where $\varpi_c$ is the assumed parallax for the cluster centre, and $\varpi_s$ the observed parallax for the star. The length of $\vec{r}$ is given by
\begin{equation}
r = ||\vec{r}|| = \sqrt{\frac{1}{\varpi_c^2} + \frac{1}{\varpi_s^2}-\frac{2\cos\rho}{\varpi_c\varpi_s}}.
\end{equation} 
Along the line of sight, the \su on $r$ is dominated by the relative error on the stellar parallax:
\begin{equation}
\sigma_r = \frac{\partial r}{\partial \varpi_s}\sigma_{\varpi,s} = 
\frac{|\varpi_s\cos\rho/\varpi_c - 1|}{r\varpi_s^3}\sigma_{\varpi,s}.
\label{equ:erronr}
\end{equation}
The \su $\sigma_r$ leads to a `stretched out' appearance of the cluster along the line of sight. For clusters much more distant than the Hyades, the parallax of the star can be expressed as $\varpi_s = \varpi_c + \Delta\varpi_s$, with $\Delta\varpi_s\ll\varpi_c$. In addition, $\cos\rho\approx 1$, which gives in first approximation (expressed in the parallax of the cluster):
\begin{equation}
\sigma_r\approx\frac{\sigma_{\varpi,s}}{r\varpi_c^3}\frac{|\Delta\varpi_s|}{\varpi_c}.
\end{equation}
Also, $\Delta\varpi_s\approx \varpi_c^2 r\cos\theta$, where $\theta$ is measured from the line of sight through the cluster centre. Substituting gives
\begin{equation}
\sigma_r\approx\frac{\sigma_{\varpi,s}\cos\theta}{\varpi_c^2}.
\end{equation}
Thus, in Eq.~\ref{equ:erronr} the error $\sigma_r$ effectively scales with the distance of the cluster squared, which makes it at this stage only just applicable to the Hyades.

\onecolumn
\section{Selected stars \label{appselected}}
Here we present the tables with the selected members for the different clusters, and other information that may be of interest. Cross identifications with HD numbers were obtained from the \Hipparcos or \Tycho identifiers in the \TGAS records, and the cross matches of those identifiers with the HD catalogue as provided by \cite{1997ESASP1200.....E} and \cite{tycho2hd}. Positions in the tables are in the ICRS, at epoch 2015.0. Further \Gaia data on the sources in the tables can be extracted from the \Gaia archive at https://gea.esac.esa.int/archive/, using the option "file", providing a file with source identifiers. The option "Tycho-Gaia Astrometric Solution" should be selected. 
\subsection{The Hyades cluster}
For the Hyades cluster the individual distance moduli, as based on the combined information from the parallax measurement and the proper motion, are included in Table~\ref{tab:hyades}. Figure~\ref{fig:maphyades} shows the distribution of the members as projected on the sky.
\begin{table*}
\centering
\caption{Identifiers, positions and distance moduli for members of the Hyades cluster.}
\label{tab:hyades} 
\scriptsize{
\begin{tabular}{rrrrrr|rrrrrr}
\hline\hline 
SourceId & HD & $\alpha$ (degr) & $\delta$ (degr) & G & dm & SourceId & HD & $\alpha$ (degr) & $\delta$ (degr) & G & dm\\ \hline 
   68000018174329600 &  &  53.2094 &  23.6920 &    8.568 &   3.12 &  3307645127438373888 & 286789 &  66.7269 &  13.1381 &    9.999 &   3.41 \\
   71487325460694912 &  &  54.7836 &  28.3821 &   10.259 &   3.73 &  3312709374919349248 &  28205 &  66.9000 &  15.5891 &    7.247 &   3.37 \\
 3277270534605393920 &  &  57.0502 &   7.1463 &   10.129 &   2.81 &  3306922954457367936 &  28237 &  66.9424 &  11.7364 &    7.331 &   3.32 \\
   43789768566924416 &  &  57.6046 &  17.2464 &    9.126 &   3.41 &  3310903736305456512 & 285830 &  66.9464 &  14.4177 &    9.138 &   3.45 \\
   66482348530642176 & 283066 &  57.7637 &  23.9035 &    9.705 &   3.05 &  3307844860597241088 &  28258 &  67.0189 &  13.8679 &    8.764 &   3.40 \\
   67351752990540544 & 283044 &  58.1715 &  25.8042 &   10.457 &   3.27 &  3312921374502681984 & 285804 &  67.0458 &  16.4708 &   10.381 &   3.15 \\
   38354676428572288 & 286363 &  58.7566 &  12.4855 &    9.676 &   3.31 &    48061405596787712 &  28291 &  67.1555 &  19.7405 &    8.384 &   3.39 \\
   43538289638888064 & 285252 &  58.7777 &  16.9984 &    8.674 &   3.05 &  3314109912215994112 &  28344 &  67.2017 &  17.2853 &    7.671 &   3.32 \\
  170457596891797760 & 281459 &  60.2819 &  33.1958 &    9.394 &   3.30 &  3314212063714381056 &  28406 &  67.3769 &  17.8630 &    6.764 &   3.33 \\
   50327292903510144 &  &  60.9132 &  19.4549 &    9.701 &   3.40 &   151379146007107200 & 283704 &  67.3786 &  26.6713 &    8.933 &   3.79 \\
   50298121485861120 &  &  61.3575 &  19.4420 &   10.728 &   3.37 &  3314213025787054592 & 285773 &  67.3822 &  17.8930 &    8.669 &   3.29 \\
   45367052352895360 &  25825 &  61.5677 &  15.6980 &    7.666 &   3.36 &  3312951748510907648 &  28462 &  67.4910 &  16.6727 &    8.786 &   3.26 \\
   45159897490770816 & 285507 &  61.7556 &  15.3349 &    9.954 &   3.26 &  3305871821341047808 &  28608 &  67.7387 &  10.7517 &    6.886 &   3.36 \\
   45567507066546048 & 285482 &  61.9305 &  16.5187 &    9.512 &   3.30 &   144171228809559808 &  28593 &  67.8159 &  20.1330 &    8.341 &   3.31 \\
 3304337452864501120 & 286554 &  62.1116 &  12.1918 &   10.589 &   3.32 &  3307815001984777088 &  28635 &  67.8727 &  13.9034 &    7.600 &   3.65 \\
   53942246617146240 & 284155 &  62.1514 &  23.7684 &    9.120 &   3.36 &  3312564033223630720 & 285876 &  67.9691 &  15.4994 &   10.381 &   3.29 \\
 3300315439330018304 &  &  62.4563 &   9.3055 &    9.533 &   2.76 &  3307504218151520256 & 286839 &  68.1073 &  13.1132 &   10.382 &   3.34 \\
   46975431705914112 &  26345 &  62.6770 &  18.4231 &    6.477 &   3.34 &  3410640882737635200 & 285836 &  68.1710 &  19.1133 &   10.039 &   3.61 \\
 3311514205777562496 &  26756 &  63.6073 &  14.6250 &    8.238 &   3.45 &  3312644881687518976 &  28805 &  68.2481 &  15.8189 &    8.407 &   3.36 \\
 3304412597612195328 &  26767 &  63.6141 &  12.4353 &    7.840 &   3.39 &   144377799556207488 & 284552 &  68.4054 &  21.1507 &   10.110 &   3.19 \\
   52813460492850304 &  26737 &  63.6272 &  22.4517 &    6.925 &   3.92 &  3313259165090609280 &  28878 &  68.4086 &  16.7624 &    9.080 &   3.45 \\
  149005266040519808 &  26736 &  63.6352 &  23.5747 &    7.851 &   3.27 &  3410453484725565312 & 285837 &  68.4251 &  19.0139 &   10.166 &   3.39 \\
 3300934223858467072 &  26784 &  63.6436 &  10.7014 &    6.945 &   3.27 &  3307528029449757056 &  28911 &  68.4448 &  13.2518 &    6.489 &   3.32 \\
 3311492799660064384 & 285625 &  63.7939 &  14.3984 &   10.802 &   3.39 &  3312602344331419136 &  28977 &  68.6345 &  15.8275 &    9.342 &   3.57 \\
 3312197930211158784 & 285590 &  63.8909 &  15.7062 &   10.320 &   3.27 &  3312575681175439616 &  28992 &  68.6476 &  15.5045 &    7.739 &   3.31 \\
   49365082792386816 &  26874 &  63.9273 &  20.8197 &    7.600 &   3.50 &  3309956850635519488 &  29159 &  69.0224 &  15.6839 &    9.071 &   3.51 \\
 3312136494998639872 &  26911 &  63.9434 &  15.4006 &    6.199 &   3.32 &  3282171745125201792 &  26911 &  69.4931 &   4.6698 &   10.788 &   3.42 \\
   52548237672091392 & 284253 &  64.1400 &  21.9073 &    8.867 &   3.54 &   146677874804442240 &  29419 &  69.7142 &  23.1497 &    7.343 &   3.19 \\
   45789299177700352 &  27130 &  64.4128 &  16.9477 &    8.047 &   3.34 &  3307992332594320640 & 286929 &  69.9628 &  12.7284 &    9.577 &   3.16 \\
   47620260916592384 &  27149 &  64.5082 &  18.2567 &    7.313 &   3.35 &   146698078328904064 & 284574 &  70.0247 &  23.3044 &    9.107 &   3.86 \\
  148946064212226944 & 284303 &  64.5454 &  23.2845 &    9.063 &   3.65 &   148183862135533952 & 283810 &  70.0389 &  25.5921 &    9.978 &   3.25 \\
 3312281664893305728 & 285690 &  64.5808 &  16.0882 &    9.215 &   3.27 &  3281064262038614912 &  &  71.5787 &   3.6364 &   10.282 &   3.28 \\
   49005576847854080 &  27250 &  64.7421 &  19.9065 &    8.363 &   3.32 &  3309006597711379328 &  30246 &  71.6270 &  15.4719 &    8.082 &   3.36 \\
   47345005052090880 &  27282 &  64.7839 &  17.5246 &    8.220 &   3.37 &  3412605297699792512 & 284785 &  71.7877 &  20.8821 &    9.392 &   3.14 \\
 3283285790922135424 &  &  65.2677 &   3.2688 &    9.064 &   2.86 &  3413146910255989248 & 284653 &  71.8595 &  23.0508 &   10.203 &   3.68 \\
   49231663928585344 &  27524 &  65.3823 &  21.0397 &    6.661 &   3.42 &  3405127244241184256 &  27524 &  72.2156 &  15.9475 &    9.397 &   3.88 \\
   47541096078933376 &  27534 &  65.3849 &  18.4174 &    6.670 &   3.39 &   147182172683187712 & 283882 &  72.3045 &  24.8026 &    9.134 &   3.45 \\
 3311024785663873920 &  27561 &  65.3954 &  14.4097 &    6.477 &   3.45 &  3405113740864365440 &  30589 &  72.3842 &  15.8886 &    7.550 &   3.40 \\
  145325544220443904 &  27732 &  65.8435 &  21.3789 &    8.586 &   3.49 &  3405988677241799040 & 286085 &  72.5033 &  16.4119 &   10.087 &   3.49 \\
 3312025581763840512 & 285749 &  65.8559 &  15.7630 &    9.914 &   3.07 &  3404812680839290368 &  30712 &  72.6413 &  15.0833 &    7.517 &   3.36 \\
 3311148824319241472 &  27771 &  65.8852 &  14.6704 &    8.814 &   3.33 &  3405220084257276416 &  30738 &  72.7026 &  16.2103 &    7.135 &   3.48 \\
 3310820620098473728 & 286734 &  65.9772 &  14.0520 &   10.252 &   3.17 &  3404850785786832512 &  30809 &  72.8470 &  15.4334 &    7.728 &   3.92 \\
 3313630078465745280 &  27835 &  66.0537 &  16.3788 &    8.039 &   3.76 &  3406943087694799744 & 284930 &  73.0984 &  18.9968 &    9.846 &   3.51 \\
  145373372976256512 &  27808 &  66.0613 &  21.7361 &    6.969 &   3.16 &  3408463506117452544 &  31236 &  73.7435 &  19.4853 &    6.278 &   3.99 \\
 3314079503847287424 & 285720 &  66.0711 &  18.0028 &    9.571 &   3.31 &  3392446817156214784 &  31609 &  74.4566 &  14.0021 &    8.647 &   3.68 \\
 3313947699887831808 &  27848 &  66.0932 &  17.0788 &    6.828 &   3.51 &  3239389678968988288 &  &  75.2040 &   4.7332 &    9.377 &   3.58 \\
 3313662892016181504 &  27859 &  66.1185 &  16.8861 &    7.627 &   3.27 &  3391728561185367168 &  27859 &  75.4005 &  13.9329 &   10.612 &   3.12 \\
 3312783557591565440 &  27991 &  66.4060 &  15.9409 &    6.297 &   3.38 &  3407518510233429248 &  27991 &  75.7799 &  19.0178 &   11.096 &   3.66 \\
 3311179335766914944 &  &  66.5200 &  15.0413 &   11.326 &   3.42 &  3391712034151625984 &  32347 &  75.7822 &  13.7306 &    8.729 &   3.63 \\
  145293177350363264 &  28033 &  66.5775 &  21.4703 &    7.201 &   3.37 &  3407121827053483776 & 240648 &  76.5752 &  17.8163 &    8.574 &   3.63 \\
 3313689417734366720 &  28099 &  66.6676 &  16.7468 &    7.916 &   3.30 &  3387381641964995712 & 242780 &  80.1062 &  11.6098 &    8.795 &   3.71 \\
  144534720481849856 & 284455 &  66.6989 &  21.2347 &   10.560 &   3.19  & & & & \\
\hline
\end{tabular}
}
\end{table*}
\begin{figure}[t]
\centering
\includegraphics[width=7cm]{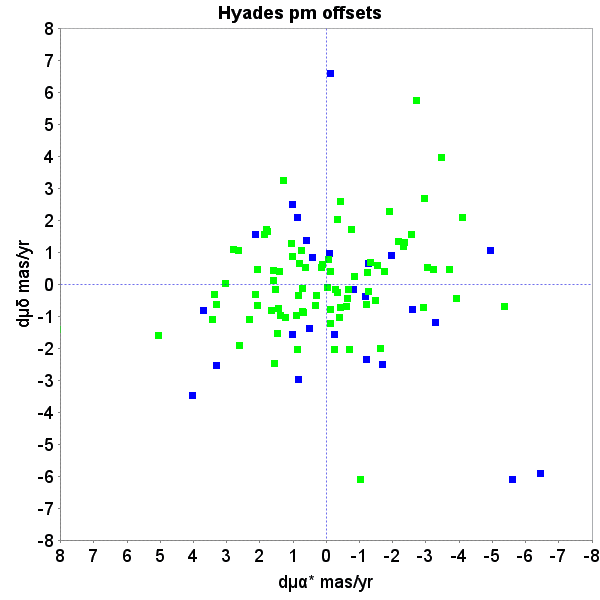}
\caption{Offsets between measured proper motions and the predicted values as based on the measured parallax, position on the sky and space velocity vector of the cluster. The main noise contribution is likely to be the internal velocity dispersion.}
\label{fig:hyadespmres}
\end{figure}
\begin{figure*}
\centering
\includegraphics[width=12cm]{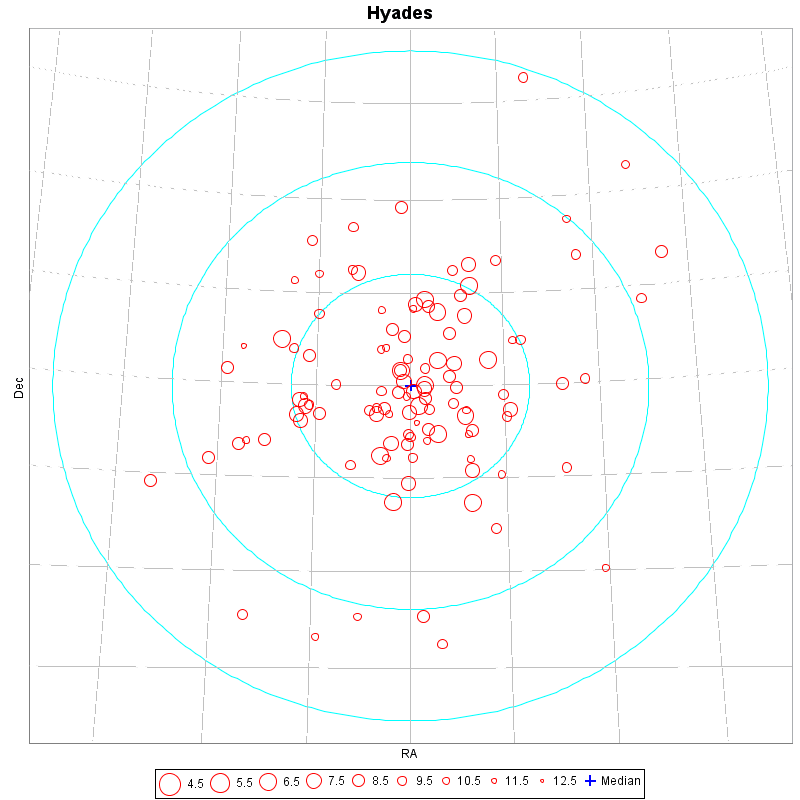}
\caption{A map of the Hyades members as identified from the TGAS catalogue. The coordinate grid is at 5 degrees intervals, the three concentric circles are at 5, 10 and 15 pc from the cluster centre at the cluster distance.}
\label{fig:maphyades}
\end{figure*}

\subsection{Coma Berenices}
\begin{table*}
\centering
\caption{Identifiers and positions for members of the Coma Berenices cluster.}
\label{tab:comaber}
\scriptsize{
\begin{tabular}{rrrrr|rrrrr}
\hline\hline 
SourceId & HD & $\alpha$ (degr) & $\delta$ (degr) & G & SourceId & HD &  $\alpha$ (degr) & $\delta$ (degr) & G \\ \hline 
 4019588595869297792 &  &177.1570 &  28.2751 &  10.243&  4008681509241392256 &  & 185.9242 &  26.6015 &   9.187 \\
 4020237891845231872 &  &178.8889 &  29.7282 &  11.285&  3953816566210308864 &  & 185.9467 &  23.2457 &  11.242 \\
 3999244366580288512 &  &180.6109 &  20.1230 &   9.711&  4008364334496501504 & 108102 & 186.2593 &  25.5606 &   8.073 \\
 4002921030384951936 &  &181.0969 &  24.8206 &   9.892&  3953772070349122816 & 108154 & 186.3437 &  23.2290 &   8.505 \\
 4003439518837094272 &  &181.9904 &  25.5865 &  11.070&  4009051048227419520 & 108226 & 186.4664 &  26.7766 &   8.276 \\
 4003559674841400320 & 105863 &182.7807 &  25.9901 &   9.442&  4008790017295714688 &  & 186.7126 &  26.2671 &  11.514 \\
 4013188785360910336 &  &182.8965 &  29.3790 &  11.019&  4008867670304423424 &  & 186.7760 &  26.8457 &   9.732 \\
 4003406533487752832 &  &183.2218 &  26.2504 &  11.091&  3953787429152384256 &  & 186.8361 &  23.3298 &  10.118 \\
 4001595500398414848 & 106293 &183.4329 &  22.8880 &   8.029&  4008390928933851264 & 108486 & 186.9098 &  25.9121 &   6.724 \\
 4002565304013314944 & 106691 &184.0348 &  25.7603 &   8.041&  4009300946604526208 &  & 186.9512 &  28.1944 &   9.536 \\
 4002550288811032832 & 106946 &184.4621 &  25.5713 &   7.821&  4009295139808743936 &  & 187.0879 &  28.0405 &  10.164 \\
 4016309611677805568 & 107053 &184.6211 &  32.7489 &   6.702&  4008777029313889152 & 108642 & 187.1589 &  26.2269 &   6.556 \\
 3953951874859947904 & 107067 &184.6507 &  23.1200 &   8.602&  3960008294143258240 &  & 187.4205 &  24.5207 &   9.501 \\
 4008571351920493440 & 107131 &184.7584 &  26.0083 &   6.492&  4009041083903247488 & 108976 & 187.7628 &  27.7303 &   8.489 \\
 3953900747569804928 & 107168 &184.8299 &  23.0346 &   6.307&  4010969146262166144 & 109069 & 187.9606 &  29.3141 &   7.527 \\
 4002172366045267072 & 107214 &184.8681 &  24.2842 &   8.886&  1518264342066234368 &  & 188.1294 &  35.3312 &   9.467 \\
 4010203989248653184 & 107276 &184.9609 &  28.4643 &   6.681&  4010481375416160000 &  & 188.2525 &  27.7124 &  10.925 \\
 4008525241151501824 & 107399 &185.1898 &  25.7658 &   8.956&  3958685993971916672 &  & 188.3333 &  22.4065 &  10.099 \\
 4008553484856786048 &  &185.3150 &  26.1539 &  11.144&  3959212282084332544 & 109307 & 188.3925 &  24.2829 &   6.328 \\
 4008227720176880384 & 107513 &185.3614 &  24.9970 &   7.393&  3960681298339257728 &  & 188.4254 &  25.9427 &  10.312 \\
 4008706729289355520 & 107583 &185.4542 &  26.5491 &   9.177&  3955895055503655552 & 111154 & 191.7780 &  22.6168 &   8.284 \\
 4009518100151157248 & 107611 &185.4839 &  27.3095 &   8.454&  3958022919740939904 & 111878 & 193.0483 &  25.3735 &   8.787 \\
 3953625835302703488 & 107685 &185.6031 &  22.4641 &   8.461&  1464103808031348480 &  & 194.4035 &  28.9791 &  10.071 \\
 4008433603729046784 & 107793 &185.7849 &  25.8513 &   8.995&  3956998690300563200 & 113037 & 195.1465 &  23.6517 &   8.195 \\
 4008744250123486720 & 107877 &185.9208 &  26.9799 &   8.312&&&&\\
\hline
\end{tabular}
}
\end{table*}
\begin{figure}[t]
\centering
\includegraphics[width=7cm]{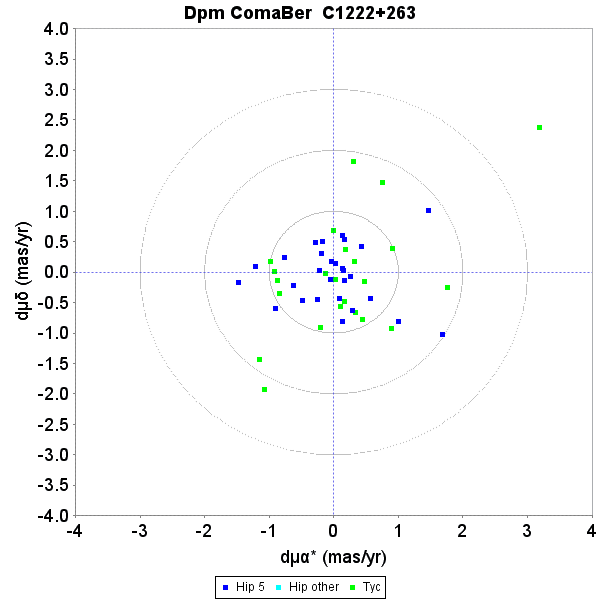}
\includegraphics[width=7cm]{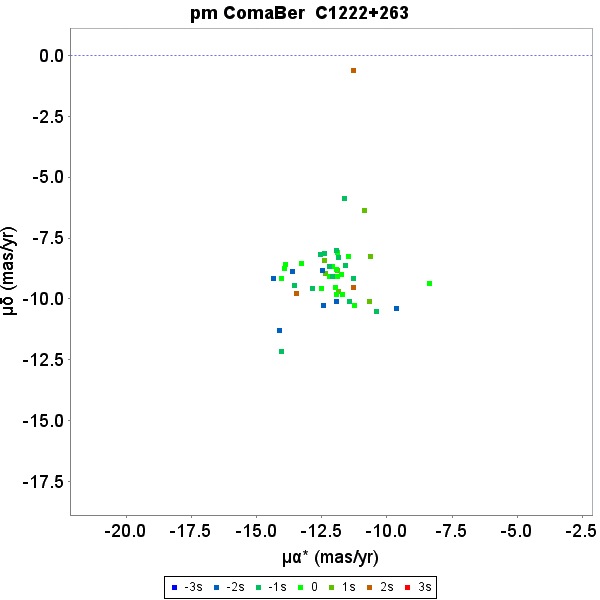}
\caption{Proper motion charts for the Coma Ber cluster. Left: unit weight residual proper motions. Green dots have first epoch \Tycho data, the dark blue dots have \Hipparcos first epoch 5-parameter solutions. The concentric circles represent 1, 2, and 3$\sigma$ \su levels. Right: actual proper motion distribution, where the colour indicate the difference from the cluster parallax in \su units.}
\label{fig:pmcomaber}
\end{figure}
\begin{figure}[t]
\centering
\includegraphics[width=12cm]{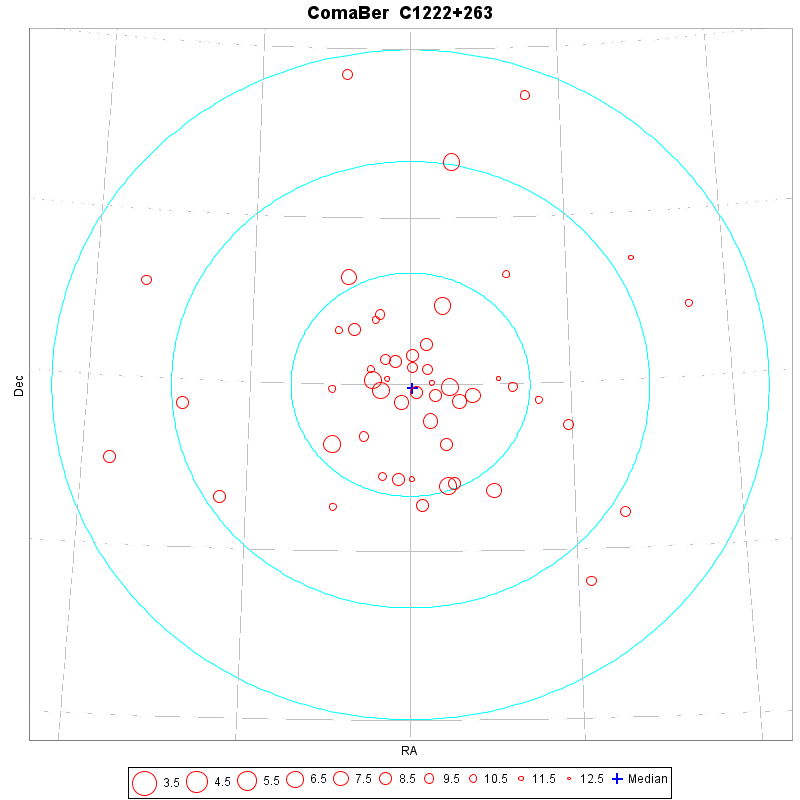}
\caption{A map of members of the Coma Ber cluster as identified from the TGAS catalogue. The coordinate grid is at 5 degrees intervals, the three concentric circles are at 5, 10 and 15 pc from the cluster centre at the cluster distance.}
\label{fig:mapcomaber}
\end{figure}

\subsection{The Pleiades \label{app:pleiades}}
\begin{table*}
\centering
\caption{Identifiers and positions for members of the Pleiades cluster.}
\label{tab:pleiades}
\scriptsize{
\begin{tabular}{rrrrr|rrrrr}
\hline\hline 
SourceId & HD & $\alpha$ (degr) & $\delta$ (degr) & G & SourceId & HD & $\alpha$ (degr) & $\delta$ (degr) & G \\ \hline 
   62413983709539584 &  20420 & 49.4574 &  22.8320 &   7.567&    65221483571888128 &  23409 &  56.4652 &  24.0387 &   7.840 \\
  118078787089883392 &  & 52.0463 &  27.3070 &  10.529&    66798492484311936 &  23432 &  56.4771 &  24.5543 &   5.815 \\
  117672070866974976 &  21510 & 52.2364 &  26.3084 &   8.315&    69872039800655744 &  23430 &  56.4965 &  25.3984 &   8.022 \\
   68897838137902208 &  & 52.4100 &  24.5103 &  11.809&    66786500935624320 &  23441 &  56.5122 &  24.5277 &   6.456 \\
   69335615566555904 &  21744 & 52.8166 &  25.2553 &   8.084&    70049473488842112 &  &  56.5397 &  26.1387 &  11.423 \\
   61519668439604992 &  & 52.8682 &  21.8217 &   8.745&    64981927476000128 &  &  56.5751 &  23.4865 &  10.275 \\
   69643753698133376 &  & 52.8902 &  26.2653 &  10.512&    63289916519862656 &  &  56.5807 &  20.8796 &  11.261 \\
   61554646652580736 &  & 53.2744 &  22.1341 &  11.136&    65231482255770112 &  23489 &  56.6138 &  24.2548 &   7.365 \\
   67680850564641792 &  & 53.3080 &  23.0062 &  11.586&    69847610026687104 & 282952 &  56.6141 &  25.1353 &  10.342 \\
   69250231614573056 &  & 53.5076 &  24.8807 &  10.663&    65008693712182656 &  23512 &  56.6426 &  23.6238 &   8.013 \\
   68097015715726208 &  & 53.5306 &  24.3443 &   9.394&    65199974375696512 & 282975 &  56.6571 &  23.7875 &   9.997 \\
   67799185503558016 &  22146 & 53.7445 &  23.5300 &   8.764&    64808204638390912 &  23514 &  56.6601 &  22.9196 &   9.265 \\
   67618281484716544 &  & 53.8822 &  22.8234 &   9.688&    66838452861270272 & 282954 &  56.6617 &  24.9595 &  10.051 \\
   71296663272200320 &  & 54.0353 &  27.3428 &   9.168&    64952824777149440 &  23513 &  56.6667 &  23.1101 &   9.260 \\
   64575726648969600 &  22444 & 54.3503 &  22.3508 &   9.034&    64899017426872960 &  &  56.6962 &  22.9144 &  10.291 \\
   64671899556831360 &  & 54.5942 &  22.4995 &   9.884&    66789868189090816 &  23568 &  56.7476 &  24.5199 &   6.828 \\
   68444873706967808 &  & 54.7370 &  24.5696 &  10.331&    64979728452744192 &  &  56.7560 &  23.4948 &  11.209 \\
   68250260148846720 &  22627 & 54.7881 &  24.3676 &   9.692&    65212103363335296 &  23585 &  56.7677 &  23.9950 &   8.313 \\
   64313218247427456 &  22637 & 54.8051 &  21.8431 &   7.291&    66729257611496704 &  23584 &  56.7920 &  24.2765 &   9.334 \\
   68254245878512384 &  & 54.8062 &  24.4663 &  10.172&    66715273197982848 &  23607 &  56.8307 &  24.1389 &   8.238 \\
   65113559634339200 &  22680 & 54.9217 &  23.2907 &   9.763&    69964879813013248 &  23598 &  56.8371 &  25.5257 &   9.663 \\
   68593342136568448 &  22702 & 54.9633 &  25.1947 &   8.728&    65017627244152064 &  23632 &  56.8375 &  23.8032 &   7.025 \\
   71371258264471424 &  & 55.0129 &  27.7403 &   9.468&    66715101399291392 &  23629 &  56.8378 &  24.1161 &   6.328 \\
   70941383577307392 &  & 55.0241 &  26.1962 &  10.942&    64898364591843712 &  23610 &  56.8455 &  22.9219 &   8.118 \\
   68334235349446528 &  & 55.1281 &  24.4871 &  11.420&    65207705316826752 &  23631 &  56.8518 &  23.9145 &   7.306 \\
   70190245337962368 &  & 55.1495 &  26.1512 &  10.897&    65007078804476928 &  23643 &  56.8619 &  23.6781 &   7.753 \\
   65119160271381248 &  & 55.2103 &  23.4183 &  10.591&    66729876086786944 &  23642 &  56.8728 &  24.2881 &   6.833 \\
   65120465942548864 &  22887 & 55.3458 &  23.4867 &   9.037&    69948249699646720 &  23664 &  56.9452 &  25.3855 &   8.259 \\
   65150943028579200 &  & 55.3660 &  23.7081 &  10.894&    64913448517448192 &  &  56.9506 &  23.2179 &  10.875 \\
   70108469159560448 &  & 55.4008 &  25.6191 &   9.899&    66724447247218048 &  23733 &  57.0566 &  24.3182 &   8.188 \\
   65027591568709632 &  22977 & 55.5198 &  22.8584 &   9.023&    66939848447027584 &  23732 &  57.0704 &  25.2149 &   9.066 \\
   64317994252099840 &  & 55.6001 &  21.4733 &   9.664&    63948214747182848 &  23792 &  57.1640 &  21.9248 &   8.280 \\
   64449729487990912 &  & 55.6002 &  22.4210 &  10.157&    64933755122821120 &  23791 &  57.1830 &  23.2596 &   8.320 \\
   63144712265008512 &  23028 & 55.6245 &  20.1498 &   8.340&    64114241002810496 &  23852 &  57.2970 &  22.6093 &   7.713 \\
   68310561489710336 &  23061 & 55.7297 &  24.4929 &   9.330&    66506331628024832 &  23863 &  57.3009 &  23.8866 &   8.113 \\
   68317605236065408 &  & 55.7624 &  24.6696 &  11.588&    66746437479801088 &  23872 &  57.3201 &  24.3959 &   7.537 \\
   69896469573892224 &  & 55.8068 &  25.2699 &  11.850&    66745612851296256 &  23873 &  57.3407 &  24.3808 &   6.637 \\
   65063703653090176 &  & 55.8523 &  23.2258 &  10.281&    66555981449928832 &  23886 &  57.3584 &  24.2475 &   7.957 \\
   69904097435804672 &  & 55.8631 &  25.3874 &  11.110&    64930490947667840 &  23912 &  57.3864 &  23.3802 &   9.010 \\
   64380632054140416 &  & 55.8798 &  22.1582 &  10.171&    64109911675780224 &  23913 &  57.4092 &  22.5333 &   7.028 \\
   65232100731054592 &  & 55.8832 &  23.6739 &  10.574&    64928601162063744 &  &  57.4143 &  23.2899 &  10.938 \\
   65233784358231168 &  & 55.8935 &  23.7617 &  11.121&    64929391436043264 &  23924 &  57.4206 &  23.3414 &   8.093 \\
   65089336018173440 &  & 55.9073 &  23.5358 &  10.514&    66453486350431232 &  23923 &  57.4315 &  23.7117 &   6.196 \\
   65184924810275968 &  23157 & 55.9231 &  23.6490 &   7.844&    66980358578521856 &  23935 &  57.4706 &  25.6473 &   9.388 \\
   69876712724339456 &  & 55.9515 &  25.0042 &  11.358&    64053561704835584 &  23950 &  57.4796 &  22.2440 &   6.073 \\
   65289275335247872 &  & 55.9614 &  24.2473 &  11.035&    64924409273987712 &  &  57.4855 &  23.2184 &  10.091 \\
   68306094723718528 &  23194 & 56.0012 &  24.5568 &   8.037&    66558249192653952 &  23948 &  57.4859 &  24.3488 &   7.562 \\
   65086587239103616 &  & 56.0026 &  23.5438 &  10.766&    66507465499396224 &  23964 &  57.4920 &  23.8485 &   6.828 \\
   68296886313838848 &  & 56.0149 &  24.5041 &  10.458&    66960258131598720 &  23975 &  57.5737 &  25.3794 &   9.483 \\
   65072431026899712 &  & 56.0468 &  23.3791 &  11.133&    64172034082472448 &  &  57.5889 &  23.0962 &  11.102 \\
   68322140721525760 &  & 56.0581 &  24.7792 &  10.528&    70506870326054656 &  &  57.5889 &  27.1442 &  11.088 \\
   70252676982016768 &  & 56.0641 &  26.3310 &  10.798&    65677368580114560 &  &  57.7144 &  23.3289 &  12.084 \\
   69811635379765760 &  & 56.0838 &  24.7960 &  10.407&    66969672699893248 &  24086 &  57.7765 &  25.5945 &   8.983 \\
   65272817021006848 &  23247 & 56.0982 &  24.1325 &   8.932&    66480939781378048 &  &  57.8204 &  23.8264 &  11.976 \\
   65292230272755200 &  23246 & 56.1073 &  24.3945 &   8.133&    66657411397616128 &  24132 &  57.8635 &  24.5185 &   8.744 \\
   69811944617410688 &  23269 & 56.1699 &  24.8183 &   9.693&    66863054431798272 &  &  57.9186 &  24.9830 &  10.643 \\
   69945810158924672 &  & 56.1800 &  25.8753 &  11.347&    63730305286697600 &  &  57.9255 &  21.6682 &  11.220 \\
   69917635172718720 &  & 56.1832 &  25.4991 &  11.744&    66471215975411200 &  24194 &  57.9870 &  23.9018 &   9.910 \\
   63378976961208704 &  23290 & 56.1870 &  20.7478 &   8.606&    67369654414221952 &  24178 &  57.9893 &  25.9987 &   7.653 \\
   64879398017459072 &  23289 & 56.2136 &  23.2687 &   8.858&    50905051903831680 &  &  58.0034 &  19.5967 &  10.265 \\
   70234878637524480 &  & 56.2210 &  26.1418 &  12.012&    66570549979009280 &  &  58.3489 &  24.0648 &  11.150 \\
   64739244643463552 &  23312 & 56.2456 &  22.0323 &   9.322&    51619115986889472 &  &  58.3703 &  20.9072 &  11.553 \\
   63052044051306112 &  & 56.2570 &  19.5592 &   9.264&    65819961494790400 &  24463 &  58.5900 &  24.0755 &   9.554 \\
   65194648616227840 &  & 56.2637 &  23.8392 &  11.597&    66581957412169728 &  &  58.6052 &  24.3599 &  10.758 \\
   69840875517213184 &  & 56.2668 &  25.2577 &  10.482&    51742467447748224 &  &  58.6161 &  21.3895 &  10.713 \\
   65188085906203520 &  23326 & 56.2721 &  23.7025 &   8.859&    51674916201705344 &  &  58.8832 &  21.0793 &  10.843 \\
   65275497080596480 &  23325 & 56.2773 &  24.2633 &   8.527&    65308413709777664 &  24655 &  59.0163 &  22.2268 &   8.944 \\
   70242781377368704 &  & 56.2843 &  26.2923 &  11.183&    65309100904545280 &  &  59.0464 &  22.2210 &  11.241 \\
   69864308858778112 &  & 56.3141 &  25.2893 &  10.939&    65754128236100096 &  &  59.1093 &  23.7841 &  11.049 \\
   69819400680743808 &  23351 & 56.3370 &  24.9219 &   8.901&    65437640685532288 &  24711 &  59.1173 &  23.1501 &   8.289 \\
   69819022723623296 &  23352 & 56.3506 &  24.8858 &   9.524&    51717109960976896 &  &  59.3132 &  21.5156 &  11.420 \\
   65222205126393984 &  23361 & 56.3590 &  24.0350 &   8.013&    49809491645958528 &  &  59.4573 &  18.5622 &  11.160 \\
   71729527256889216 &  23336 & 56.3648 &  28.6685 &   7.405&    51452746133437696 &  &  59.5072 &  20.6766 &   9.249 \\
   63502259702709888 &  23388 & 56.3834 &  21.2465 &   7.731&    65776736943479808 &  24899 &  59.5872 &  24.0809 &   7.224 \\
   64798686990854400 &  23402 & 56.4163 &  22.6943 &   7.802&    51694741770737152 &  &  59.5902 &  21.2575 &  11.180 \\
   69922239378409088 &  & 56.4381 &  25.5956 &   9.938&    53783848223326976 &  &  60.9341 &  22.9441 &   9.491 \\
   64956123312029952 &  23410 & 56.4535 &  23.1470 &   6.920&    51861420861864448 & 284215 &  62.2309 &  20.3858 &   9.232 \\
\hline
\end{tabular}
}
\end{table*}
\begin{figure}[t]
\centering
\includegraphics[width=7cm]{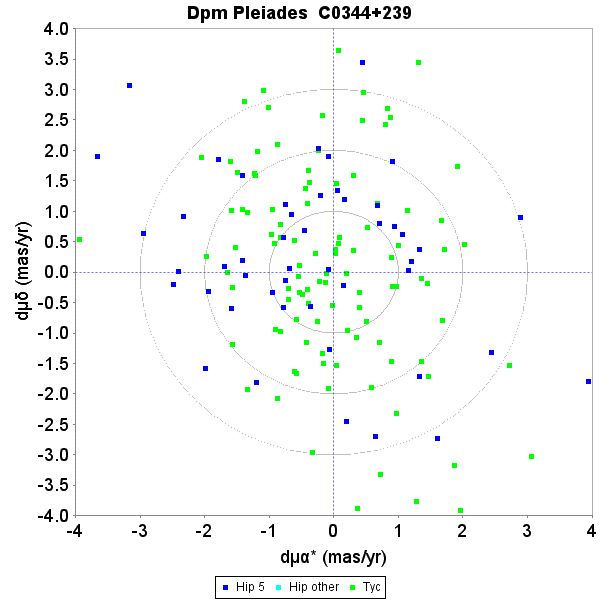}
\includegraphics[width=7cm]{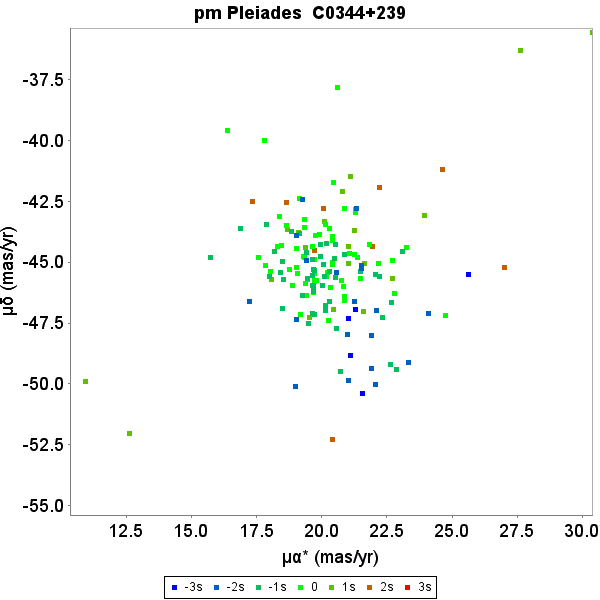}
\caption{Proper motion charts for the Pleiades cluster. Left: unit weight residual proper motions. Green dots have first epoch \Tycho data, the dark blue dots have \Hipparcos first epoch 5-parameter solutions. The concentric circles represent 1, 2, and 3$\sigma$ \su levels. Right: actual proper motion distribution, where the colour indicate the difference from the cluster parallax in \su units.}
\label{fig:pmpleiades}
\end{figure}
\begin{figure}[t]
\centering
\includegraphics[width=12cm]{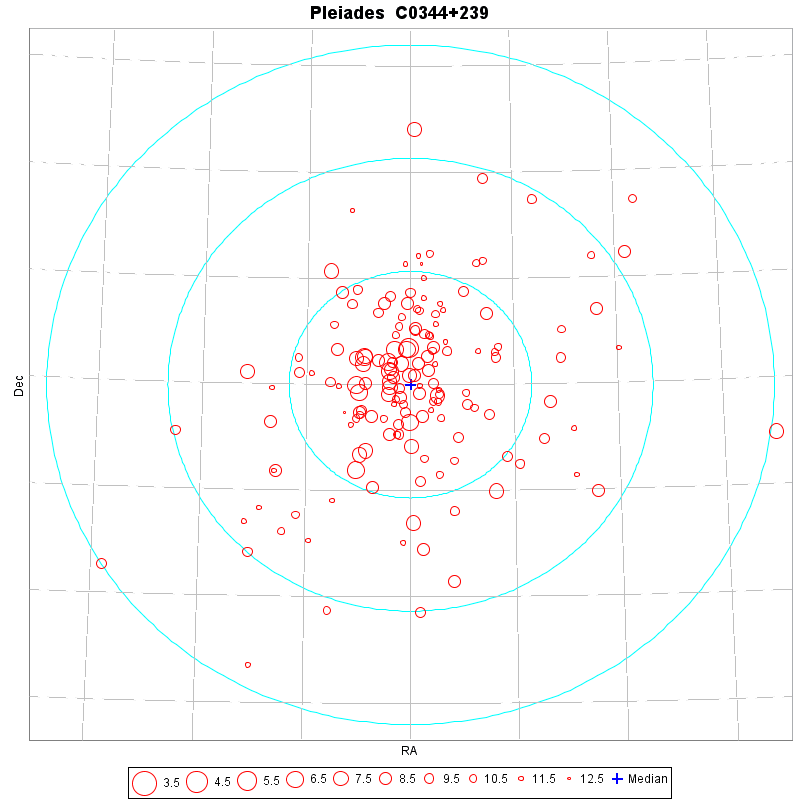}
\caption{A map of members of the Pleiades cluster as identified from the TGAS catalogue. The coordinate grid is at 2 degrees intervals, the three concentric circles are at 5, 10 and 15 pc from the cluster centre at the cluster distance.}
\label{fig:mappleiades}
\end{figure}

\subsection{The Praesepe cluster}

\begin{table*}
\centering
\caption{Identifiers and positions for members of the Praesepe cluster.}
\label{tab:praesepe}
\scriptsize{
\begin{tabular}{rrrrr|rrrrr}
\hline\hline 
SourceId & HD & $\alpha$ (degr) & $\delta$ (degr) & G & SourceId & HD & $\alpha$ (degr) & $\delta$ (degr) & G \\ \hline 
  663386973364333056 &  70297 &125.4596 &  19.4890 &   8.681&   661291132403077760 &  73731 & 130.1124 &  19.5448 &   6.272 \\
  676256482090436480 &  &126.1375 &  21.7394 &   9.823&   661199907297724416 &  73746 & 130.1372 &  19.1943 &   8.601 \\
  662925625157936256 &  &127.7309 &  19.5554 &  10.606&   661300546971382016 &  73785 & 130.1799 &  19.7193 &   6.800 \\
  662215546804136064 &  &127.8039 &  18.1536 &   9.718&   664424637463005824 &  73798 & 130.2185 &  20.2665 &   8.420 \\
  664683125774748032 &  &128.7483 &  21.0969 &  10.988&   661246018066598656 &  73819 & 130.2344 &  19.5803 &   6.741 \\
  662794164798230016 &  72779 &128.8309 &  19.5900 &   6.362&   661319547906689024 &  73818 & 130.2371 &  19.9348 &   8.633 \\
  664376430751104512 &  &128.8667 &  20.1963 &  10.141&   661305838371084288 &  73854 & 130.2943 &  19.8295 &   8.930 \\
  662849896293859712 &  72846 &128.9376 &  19.7711 &   7.461&   661401358443742720 &  73872 & 130.3072 &  19.9219 &   8.348 \\
  659539232422905472 &  &128.9771 &  18.1493 &  10.676&   661224577591160448 &  & 130.3267 &  19.2604 &  10.074 \\
  659343622432367360 &  &129.1157 &  17.9148 &  10.610&   659472230933525760 &  & 130.3812 &  18.5006 &  10.221 \\
  664286476955590016 &  73081 &129.2583 &  19.6047 &   9.035&   661252993093483392 &  & 130.4261 &  19.6605 &   9.445 \\
  664282663024631808 &  &129.3663 &  19.5625 &   9.556&   661424070230823040 &  & 130.4324 &  20.2268 &  10.302 \\
  664281494793527936 &  73175 &129.4195 &  19.5184 &   8.214&   661221760091324544 &  & 130.4394 &  19.2672 &  10.006 \\
  661277457228117504 &  &129.4441 &  19.4383 &  10.473&   661396754238802816 &  73974 & 130.4585 &  19.8741 &   6.583 \\
  659771023218259072 &  73210 &129.4447 &  19.2672 &   6.696&   661422764560767360 &  73993 & 130.4713 &  20.1594 &   8.462 \\
  658465868556261888 &  &129.5320 &  17.0506 &   9.640&   664486794229860864 &  & 130.4967 &  20.9186 &  11.075 \\
  661281752195411328 &  73397 &129.6955 &  19.5009 &   8.906&   661233133164719616 &  74028 & 130.5269 &  19.4112 &   7.924 \\
  664323276234817664 &  73430 &129.7648 &  19.9997 &   8.282&   661344183839101952 &  & 130.5644 &  19.6876 &   9.738 \\
  664330835377270912 &  73429 &129.7716 &  20.1171 &   9.265&   661412490998994176 &  74058 & 130.5899 &  20.1816 &   9.109 \\
  661288624142170368 &  73449 &129.7753 &  19.6768 &   7.374&   660909189551857664 &  & 130.6532 &  18.3888 &   9.951 \\
  661284019938140032 &  73450 &129.7877 &  19.5923 &   8.448&   661235985023012864 &  & 130.6695 &  19.5431 &   9.636 \\
  659439073785562240 &  &129.7955 &  18.1759 &  10.091&   661329958907430272 &  & 130.6849 &  19.5799 &   9.555 \\
  661207019764473344 &  &129.8006 &  19.1156 &  10.396&   664845303740260992 &  74135 & 130.7209 &  20.8192 &   8.753 \\
  664344819790782592 &  &129.8123 &  20.2107 &  11.074&   661043845366041984 &  & 130.7746 &  19.4375 &   9.738 \\
  659687494694306688 &  &129.8979 &  18.8768 &  10.542&   661015910898910080 &  74186 & 130.7792 &  19.0683 &   9.398 \\
  661311439008441600 &  73575 &129.9276 &  19.7784 &   6.601&   664963020203495552 &  & 130.8439 &  21.6716 &  10.349 \\
  661210730616213248 &  73576 &129.9359 &  19.2752 &   7.630&   661386240158893568 &  & 130.8979 &  20.1895 &   9.968 \\
  664452949887595136 &  73597 &129.9762 &  20.5602 &   9.214&   665276415377154688 &  & 130.9115 &  22.2692 &  11.963 \\
  661268248817325312 &  73619 &129.9906 &  19.5414 &   7.493&   658628905514597376 &  & 131.0497 &  17.9021 &   9.780 \\
  661206573087872256 &  73641 &129.9918 &  19.2016 &   9.346&   661872464816550144 &  74547 & 131.3110 &  20.9975 &   9.366 \\
  664329186109826944 &  73616 &129.9931 &  20.1582 &   8.818&   660953994650380416 &  74589 & 131.3354 &  18.8753 &   8.382 \\
  661324701867480576 &  73617 &129.9960 &  20.0314 &   9.114&   665004698566151168 &  & 131.3663 &  21.6535 &  10.364 \\
  664328911231920896 &  73640 &130.0052 &  20.1356 &   9.536&   661779796602531712 &  & 131.3769 &  20.5901 &   9.690 \\
  664547885844673792 &  73639 &130.0320 &  21.0627 &   9.212&   661156854545540480 &  74718 & 131.5643 &  19.7091 &   8.324 \\
  661322056167588736 &  73666 &130.0476 &  19.9711 &   6.619&   665104341807795584 &  74740 & 131.6203 &  22.3521 &   8.165 \\
  661323636715553152 &  &130.0638 &  19.9942 &   8.741&   660288790115813632 &  74780 & 131.6385 &  18.7609 &   9.056 \\
  661290754445957248 &  73711 &130.0752 &  19.5319 &   7.516&   610194902914961664 &  & 131.8087 &  16.3964 &  10.461 \\
  661217911800624384 &  73712 &130.0838 &  19.3489 &   6.698&   660225911794604416 &  & 132.0071 &  18.6771 &  10.185 \\
  661297076637809024 &  73710 &130.0919 &  19.6699 &   6.097&   660204402598397312 &  & 132.1158 &  18.3455 &  11.207 \\
  661419259867455488 &  &130.0928 &  20.1067 &   9.919&&&&\\
\hline
\end{tabular}
}
\end{table*}
\begin{figure}[t]
\centering
\includegraphics[width=7cm]{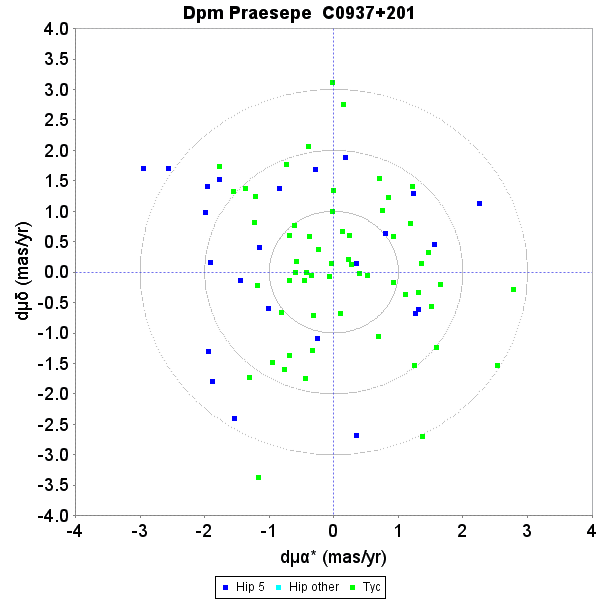}
\includegraphics[width=7cm]{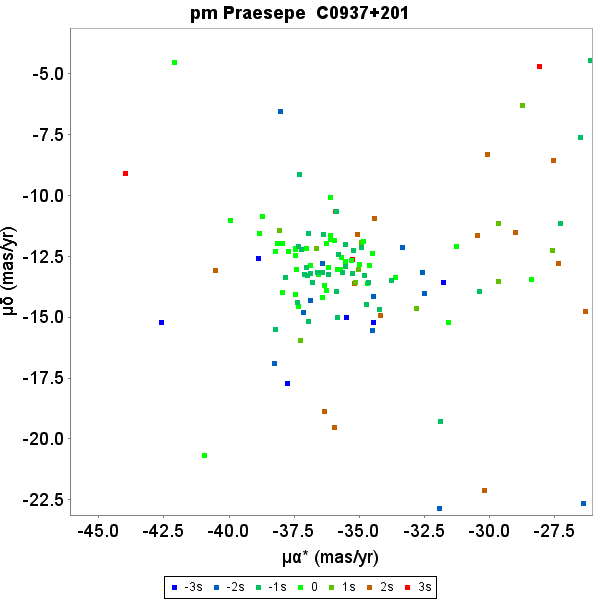}
\caption{Proper motion charts for the Praesepe cluster. Left: unit weight residual proper motions. Green dots have first epoch \Tycho data, the dark blue dots have \Hipparcos first epoch 5-parameter solutions. The concentric circles represent 1, 2, and 3$\sigma$ \su levels. Right: actual proper motion distribution, where the colour indicate the difference from the cluster parallax in \su units.}
\label{fig:praesepe}
\end{figure}
\begin{figure}[t]
\centering
\includegraphics[width=12cm]{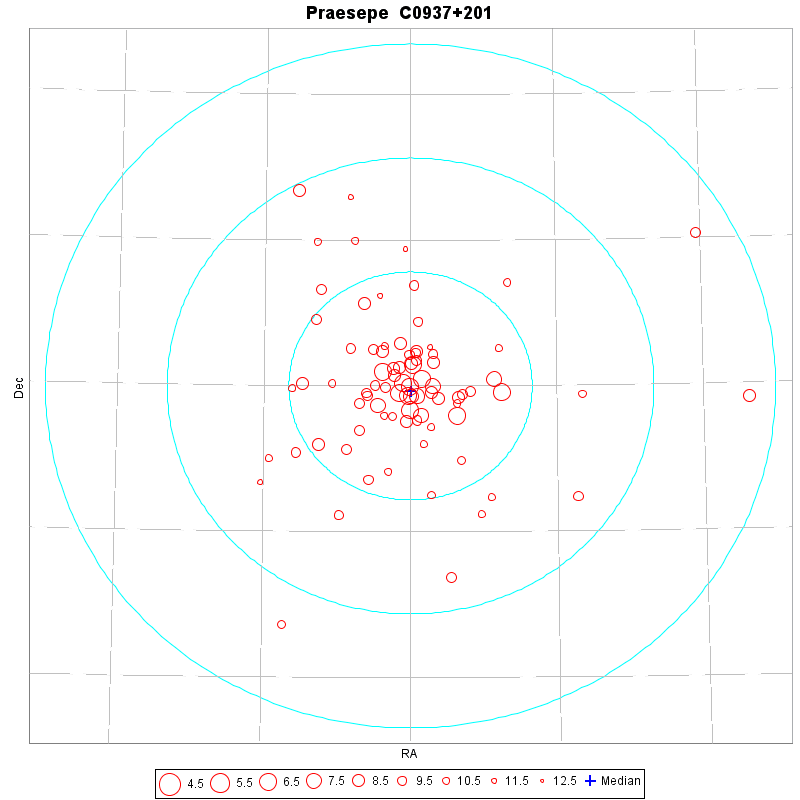}
\caption{A map of members of the Praesepe cluster as identified from the TGAS catalogue. The coordinate grid is at 2 degrees intervals, the three concentric circles are at 5, 10 and 15 pc from the cluster centre at the cluster distance.}
\label{fig:mappraesepe}
\end{figure}

\subsection{The $\alpha$ Per cluster \label{app:alphaper}}
\begin{table*}
\centering
\caption{Identifiers and positions for members of the $\alpha$ Per cluster.}
\label{tab:alphaper}
\scriptsize{
\begin{tabular}{rrrrr|rrrrr}
\hline\hline 
SourceId & HD & $\alpha$ (degr) & $\delta$ (degr) & G & SourceId & HD & $\alpha$ (degr) & $\delta$ (degr) & G \\ \hline 
  439191432859770624 &  18280 & 44.4756 &  48.8786 &   8.676&   249203176564080000 &  21239 &  51.9068 &  48.2728 &   8.344 \\
  436211962508592000 &  & 46.9603 &  49.1082 &   9.953&   441415504364607360 &  &  51.9077 &  48.9912 &  10.370 \\
  440153471174320768 &  19268 & 47.0162 &  52.2134 &   6.337&   441582114738266368 &  21238 &  51.9125 &  49.5998 &   6.963 \\
  436802640772120960 &  & 47.6742 &  50.5255 &  10.872&   441585722510785408 &  &  51.9794 &  49.7602 &  10.474 \\
  445951986261569280 &  19624 & 47.9289 &  52.1634 &   6.899&   249126519989327872 &  21279 &  51.9826 &  47.7358 &   7.269 \\
  436786663491188096 &  & 47.9585 &  50.3796 &   9.586&   441597954577636608 &  21302 &  52.0776 &  49.9528 &   8.100 \\
  435152789212540800 &  19767 & 48.1779 &  47.8385 &   8.896&   441401932267956096 &  &  52.1314 &  48.9408 &   9.623 \\
  436392282414665344 &  19805 & 48.2720 &  49.0093 &   7.966&   441815142481538048 &  &  52.1446 &  50.2668 &   9.904 \\
  436345965487346560 &  & 48.4522 &  48.9864 &  11.932&   441519064618374656 &  21345 &  52.1584 &  49.3875 &   8.409 \\
  436422244105716480 &  19893 & 48.4599 &  49.5688 &   7.156&   249212797292342400 &  21398 &  52.2820 &  48.3028 &   7.390 \\
  436499553517681024 &  & 48.8483 &  49.4402 &   9.975&   441493775848573440 &  &  52.3541 &  48.9624 &  10.111 \\
  436461517287294464 &  & 49.0967 &  49.6258 &   9.863&   249164418780724224 &  &  52.3595 &  48.2031 &   9.793 \\
  436455126375970304 &  & 49.1625 &  49.4523 &  11.573&   242888956163684096 &  21455 &  52.3597 &  46.9378 &   6.186 \\
  442841914544333696 &  20191 & 49.2047 &  51.2181 &   7.195&   441494188165433216 &  &  52.4454 &  49.0094 &   9.047 \\
  436536249718223744 &  & 49.2479 &  49.9265 &  10.190&   249222383659342336 &  21527 &  52.5807 &  48.4992 &   8.754 \\
  436482064410835968 &  & 49.5074 &  49.6440 &   9.583&   248924965763985280 &  21553 &  52.6418 &  47.6281 &   8.723 \\
  436493849800482048 &  & 49.5221 &  49.9060 &   9.778&   249149025617941248 &  21551 &  52.6541 &  48.1035 &   5.859 \\
  442576794801933312 &  20344 & 49.5998 &  50.5556 &   8.014&   441550125819632768 &  21600 &  52.8113 &  49.7061 &   8.582 \\
  435242914806781184 &  & 49.6143 &  47.3542 &  11.253&   249331785065507968 &  &  52.8709 &  48.9911 &   9.919 \\
  442556556916042496 &  & 49.6822 &  50.3862 &  10.105&   441560399381392768 &  21619 &  52.8759 &  49.9020 &   8.729 \\
  436477460205283584 &  20391 & 49.6867 &  49.7699 &   7.952&   248960218855540736 &  21641 &  52.8882 &  47.8623 &   6.790 \\
  436477013528685184 &  & 49.7098 &  49.7311 &  10.920&   441532499273859840 &  &  52.9358 &  49.5367 &   9.493 \\
  435641144174237568 &  20475 & 49.9241 &  48.9135 &   9.093&   249278355672330624 &  21672 &  52.9749 &  48.7350 &   6.635 \\
  435609670653875328 &  20487 & 49.9470 &  48.6278 &   7.659&   249267120037890432 &  &  52.9760 &  48.5272 &  10.272 \\
  442609127315775744 &  20510 & 50.0262 &  50.9687 &   7.070&   249267875952132736 &  &  52.9827 &  48.5837 &   8.207 \\
  442921800934903424 &  20537 & 50.0988 &  51.6183 &   7.299&   441558440876307584 &  &  52.9948 &  49.8701 &   9.089 \\
  435429007149649920 &  & 50.3761 &  48.4938 &   9.060&   442265083255962624 & 232804 &  53.1330 &  51.4895 &  10.037 \\
  435647638163950848 &  & 50.4177 &  49.1201 &   9.443&   248077998211731840 &  &  53.1944 &  46.7008 &   9.705 \\
  441653239394337024 &  & 50.4944 &  49.2148 &   9.081&   248886001820685184 &  21855 &  53.3429 &  47.4219 &   8.199 \\
  242950116497972608 &  & 50.7646 &  46.3341 &  11.393&   241699525101944576 &  &  53.3994 &  44.8708 &  10.210 \\
  435341561615535360 &  & 50.9181 &  47.9581 &   9.806&   442041985477060480 &  &  53.4957 &  50.8821 &   9.747 \\
  442750826879003648 &  20842 & 50.9300 &  51.7703 &   7.871&   249282100885527040 &  21931 &  53.5542 &  48.6174 &   7.392 \\
  435422581877640704 &  20863 & 50.9474 &  48.6043 &   7.002&   247787692782638976 &  &  53.5747 &  45.7300 &  11.278 \\
  441645233577061248 &  20919 & 51.0801 &  49.2212 &   8.903&   249282341403694592 &  &  53.5901 &  48.6599 &   8.756 \\
  441454914984502656 &  20931 & 51.1253 &  49.1398 &   7.891&   442036934595519872 &  &  53.7711 &  50.9124 &  11.047 \\
  441371523899475840 &  & 51.1964 &  48.4116 &   9.742&   249765198807892736 &  &  53.7866 &  49.7442 &  10.947 \\
  441436601243976448 &  & 51.2073 &  48.8716 &  11.361&   248192106902849408 &  22136 &  53.9938 &  47.0909 &   6.900 \\
  243137067835529344 &  & 51.2298 &  47.4149 &   9.598&   249087281167662464 &  &  54.1329 &  48.6545 &   9.944 \\
  441696326507978752 &  20969 & 51.2683 &  49.7953 &   8.953&   249475649291238784 &  &  54.2297 &  48.8285 &   9.916 \\
  441480409912850432 &  20986 & 51.2919 &  49.2514 &   8.157&   444862507681846016 &  22222 &  54.2836 &  53.9832 &   8.456 \\
  441481165827089024 &  21005 & 51.3365 &  49.3161 &   8.403&   248229524660290688 &  22401 &  54.5651 &  47.5769 &   7.472 \\
  243190978263909632 &  & 51.3729 &  47.9672 &   9.140&   249408682161136000 &  22440 &  54.6464 &  48.5934 &   8.585 \\
  441849158625581568 &  & 51.4070 &  50.3215 &   9.913&   443683693773493504 &  22603 &  55.1440 &  52.0083 &   9.076 \\
  243068657596900224 &  21046 & 51.4071 &  47.0205 &   8.903&   443627412522052992 &  &  55.1938 &  51.6344 &  11.372 \\
  441463607998311808 &  21071 & 51.4892 &  49.1206 &   6.106&   244833442477429888 &  &  55.2741 &  45.7937 &  10.070 \\
  249199878029198208 &  21092 & 51.5444 &  48.2216 &   8.489&   248624661649191296 &  &  55.3585 &  48.1450 &  10.213 \\
  441356921010671232 &  21091 & 51.5454 &  48.3839 &   7.525&   251450990648112128 & 232823 &  55.4209 &  51.2764 &   9.306 \\
  441486044909931136 &  & 51.5928 &  49.4270 &  11.097&   248376137662086784 &  23219 &  56.3649 &  47.6602 &   7.198 \\
  243096832581342848 &  21122 & 51.6362 &  47.2663 &   8.166&   244788843539087104 &  &  56.3727 &  45.7420 &  11.511 \\
  249180396057550848 &  & 51.6634 &  47.8822 &   9.494&   245189890404732160 &  &  56.4311 &  46.3011 &  11.820 \\
  441901694662492928 &  21117 & 51.6645 &  50.8464 &   7.633&   244596497721602176 &  23287 &  56.4719 &  45.5998 &   7.602 \\
  441383171850821888 &  & 51.6700 &  48.7768 &   9.596&   251154431747668096 &  23255 &  56.4783 &  50.4191 &   9.352 \\
  441780164270958464 &  & 51.6831 &  49.9094 &  10.672&   248480453827267456 &  &  56.6770 &  48.0868 &  10.413 \\
  441405917997620352 &  & 51.7089 &  48.7921 &  10.164&   250324850223865472 &  &  56.7326 &  49.6907 &  10.449 \\
  249180567856242048 &  21152 & 51.7095 &  47.9160 &   7.721&   251516652110388992 &  23452 &  56.9172 &  51.5287 &   7.320 \\
  441405780558667264 &  & 51.7637 &  48.7869 &   9.808&   251087224100861952 &  &  57.1336 &  50.0471 &  10.128 \\
  249197404128036480 &  21181 & 51.7717 &  48.2054 &   6.857&   249956685626486144 &  &  57.9157 &  48.3759 &   9.619 \\
  441901282346949760 & 232793 & 51.8090 &  50.8788 &  10.016&   246596887329673728 &  &  58.9503 &  47.0402 &  10.480 \\
\hline
\end{tabular}
}
\end{table*}
\begin{figure}[t]
\centering
\includegraphics[width=7cm]{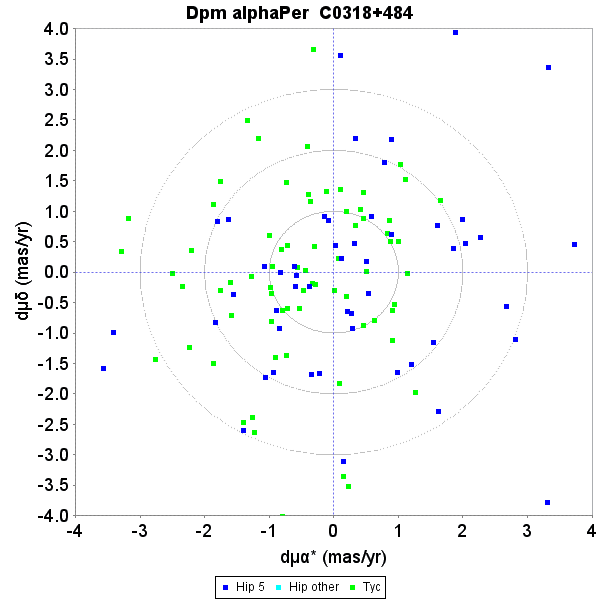}
\includegraphics[width=7cm]{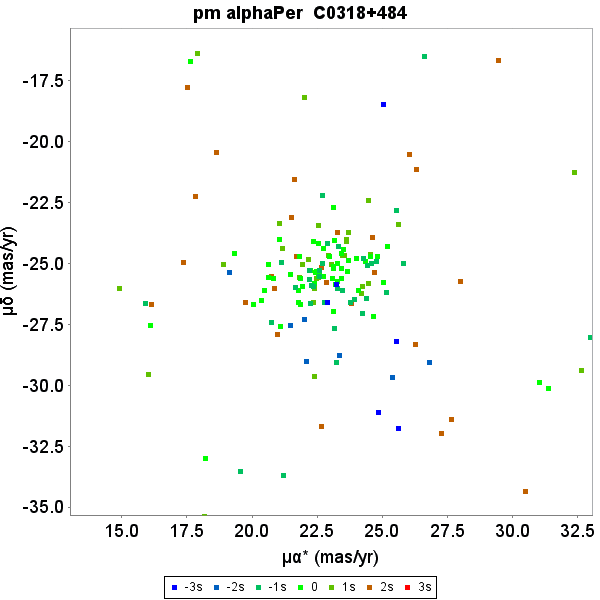}
\caption{Proper motion charts for the $\alpha$ Per cluster. Left: unit weight residual proper motions. Green dots have first epoch \Tycho data, the dark blue dots have \Hipparcos first epoch 5-parameter solutions. The concentric circles represent 1, 2, and 3$\sigma$ \su levels. Right: actual proper motion distribution, where the colour indicate the difference from the cluster parallax in \su units.}
\label{fig:alphaper}
\end{figure}
\begin{figure}[t]
\centering
\includegraphics[width=12cm]{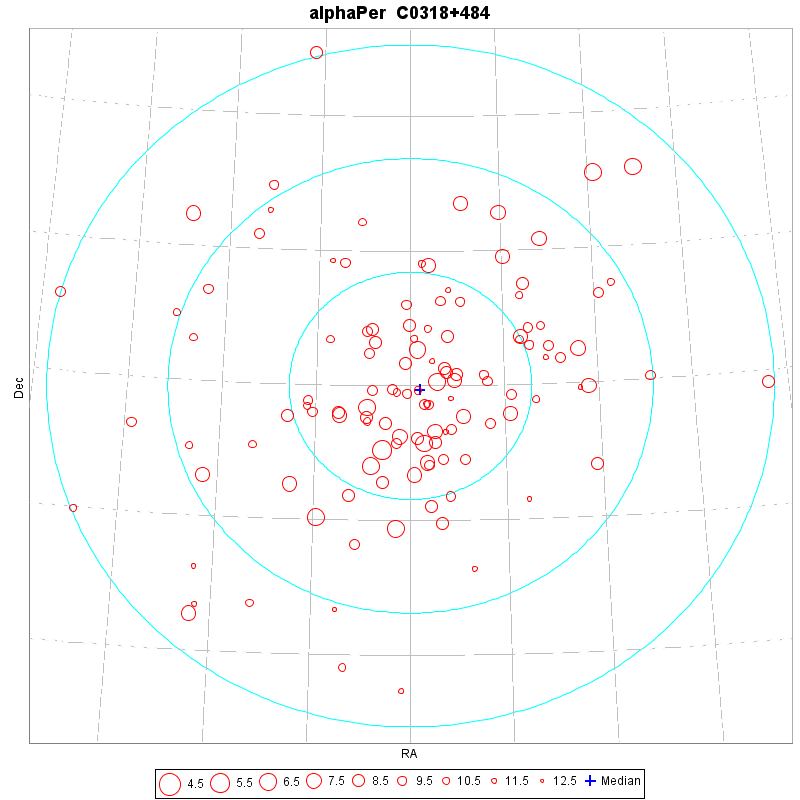}
\caption{A map of members of the $\alpha$ Per cluster as identified from the TGAS catalogue. The coordinate grid is at 2 degrees intervals, the three concentric circles are at 5, 10 and 15 pc from the cluster centre at the cluster distance.}
\label{fig:mapalphaper}
\end{figure}

\subsection{The cluster IC2391 \label{app:ic2391}}
\begin{table*}
\centering
\caption{Identifiers and positions for members of the cluster IC2391.}\label{tab:ic2391}
\scriptsize{
\begin{tabular}{rrrrr|rrrrr}
\hline\hline 
SourceId & HD & $\alpha$ (degr) & $\delta$ (degr) & G & SourceId & HD &  $\alpha$ (degr) & $\delta$ (degr) & G \\ \hline 
 5513364924200650752 &  68276 &122.3267 & -51.0123 &   7.620&  5317906150880166528 &  74374 & 130.3446 & -53.6358 &   9.367 \\
 5319716806015487488 &  70560 &125.0151 & -53.9216 &   9.253&  5318499818442087936 &  74516 & 130.5413 & -52.9676 &   7.395 \\
 5515719116036643072 &  &126.4811 & -47.5548 &  11.003&  5317887321743547264 &  74561 & 130.5757 & -53.9022 &   9.236 \\
 5515330266875244544 &  &127.0046 & -48.6839 &   9.923&  5317884435525524224 &  & 130.7515 & -53.9020 &  10.841 \\
 5321517668619002240 &  &127.1900 & -52.0907 &  10.247&  5318486074546754048 &  74678 & 130.7643 & -53.0779 &   7.664 \\
 5316373981428187904 &  72323 &127.3967 & -55.4168 &   7.703&  5318536995675325952 &  74714 & 130.8244 & -52.6030 &   9.014 \\
 5515501481453454208 &  72516 &127.8404 & -47.4435 &   8.627&  5318702128577954816 &  74734 & 130.8866 & -52.0037 &   7.881 \\
 5321225473403915136 &  &128.5853 & -52.8346 &  10.115&  5318316096918509568 &  & 130.9678 & -53.2332 &   9.579 \\
 5318069600154320384 &  &129.1009 & -54.0182 &  10.000&  5318512256667367040 &  & 131.0216 & -52.8880 &  10.546 \\
 5322839178517948800 &  73462 &129.1030 & -50.2554 &   9.325&  5316501971457325952 &  & 131.0555 & -57.2547 &  11.133 \\
 5321190151592875904 &  73777 &129.4459 & -52.8700 &   9.510&  5329039874346093056 &  & 131.0715 & -48.5083 &   9.301 \\
 5318077502894130944 &  &129.4647 & -53.7626 &  11.159&  5318521671232021632 &  & 131.1088 & -52.7089 &  11.214 \\
 5318077915210988928 &  73904 &129.5995 & -53.7217 &   7.658&  5318630900840369152 &  & 131.4129 & -52.4331 &   9.714 \\
 5321582711603786880 &  74044 &129.8187 & -52.3137 &   8.442&  5318296271349488640 &  & 131.4496 & -53.4306 &  10.048 \\
 5318113546259668864 &  74071 &129.8492 & -53.4397 &   5.539&  5318229269859699072 &  & 131.5635 & -53.7562 &  10.190 \\
 5321795845062877312 &  &129.9808 & -51.3655 &  10.042&  5317423293481147264 &  & 131.8926 & -54.4835 &  11.827 \\
 5318117325830884224 &  74169 &129.9972 & -53.2609 &   7.241&  5318412682141952512 &  75466 & 132.0007 & -52.8501 &   6.305 \\
 5318567953805446400 &  74145 &130.0067 & -52.7034 &   8.468&  5317551352220546048 &  76472 & 133.5550 & -53.3850 &   9.393 \\
 5318096125872352768 &  &130.0258 & -53.6352 &  10.255&  5325336375583055232 &  & 134.0447 & -49.4907 &  10.068 \\
 5318554106825052160 &  74275 &130.2020 & -52.8018 &   7.294&  5317261802706432000 &  76840 & 134.1002 & -54.3225 &   9.186 \\
 5318093239654329472 &  &130.2044 & -53.6292 &  10.773&  5303472346348134656 &  & 135.5162 & -58.1471 &  11.095 \\
 5317797574106967552 &  &130.2551 & -54.5172 &  10.279&&&&\\
\hline
\end{tabular}
}
\end{table*}
\begin{figure}[t]
\centering
\includegraphics[width=7cm]{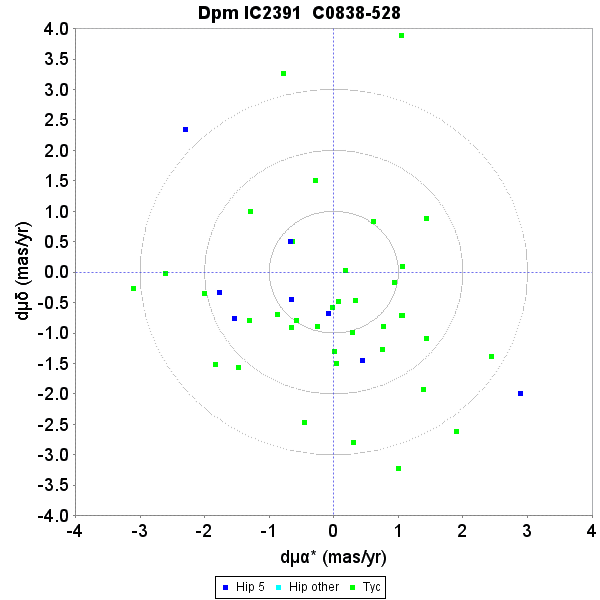}
\includegraphics[width=7cm]{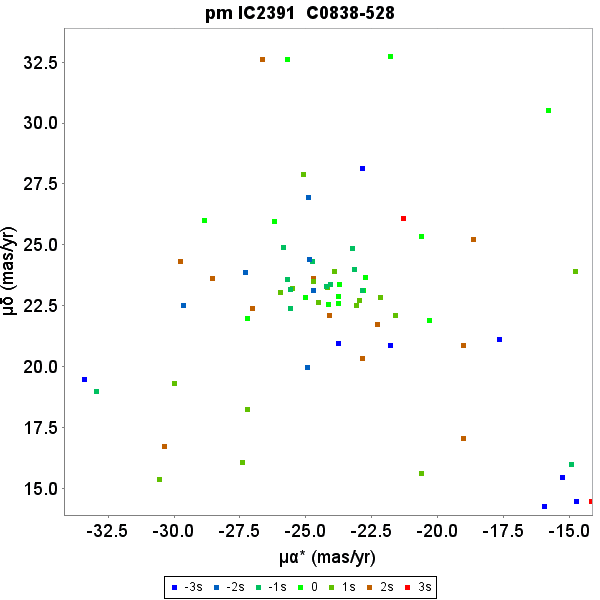}
\caption{Proper motion charts for the cluster IC2391. Left: unit weight residual proper motions. Green dots have first epoch \Tycho data, the dark blue dots have \Hipparcos first epoch 5-parameter solutions. The concentric circles represent 1, 2, and 3$\sigma$ \su levels. Right: actual proper motion distribution, where the colour indicate the difference from the cluster parallax in \su units.}
\label{fig:ic2391}
\end{figure}
\begin{figure}[t]
\centering
\includegraphics[width=12cm]{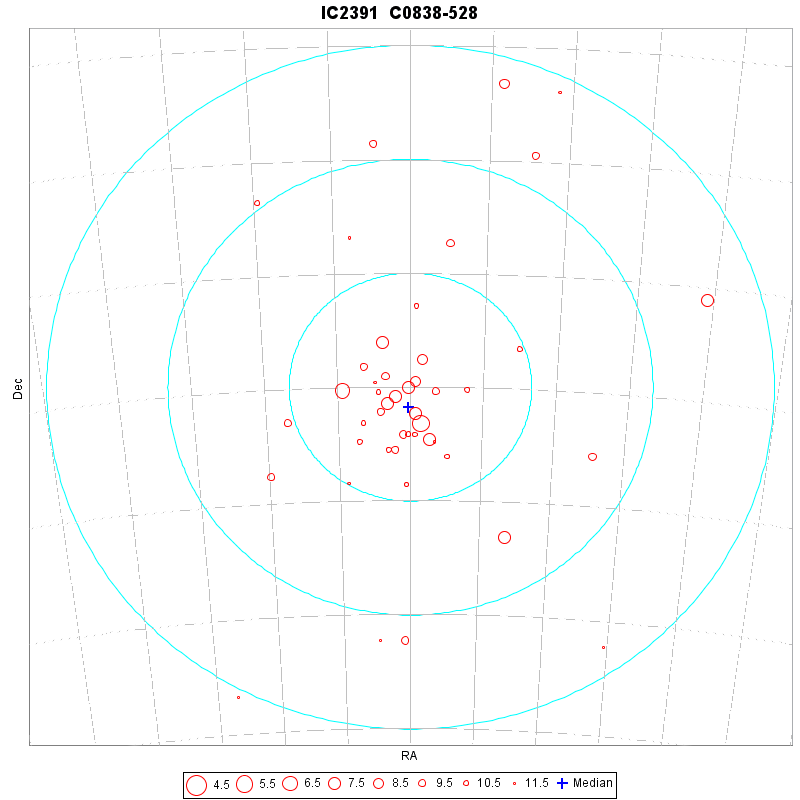}
\caption{A map of members of the cluster IC2391 as identified from the TGAS catalogue. The coordinate grid is at 2 degrees intervals, the three concentric circles are at 5, 10 and 15 pc from the cluster centre at the cluster distance.}
\label{fig:mapic2391}
\end{figure}

\subsection{The cluster IC2602 \label{app:ic2602}}
\begin{table*}
\centering
\caption{Identifiers and positions for members of the cluster IC2602.}\label{tab:ic2602}
\scriptsize{
\begin{tabular}{rrrrr|rrrrr}
\hline\hline 
SourceId & HD & $\alpha$ (degr) & $\delta$ (degr) & G & SourceId & HD & $\alpha$ (degr) & $\delta$ (degr) & G \\ \hline 
 5246562414765401472 &  87713 &151.1418 & -64.6477 &   6.947&  5239689092704584704 &  92664 & 160.0475 & -65.1002 &   5.568 \\
 5246170163993645056 &  88386 &152.2842 & -65.7982 &   9.002&  5239251727594949120 & 310144 & 160.1382 & -65.9278 &  10.629 \\
 5256461077433424896 &  88307 &152.2888 & -60.4162 &   8.073&  5239772689945600128 &  92715 & 160.1753 & -64.6531 &   6.834 \\
 5244883529229180928 &  &152.3408 & -67.6105 &  11.150&  5239869103371472000 &  92783 & 160.2766 & -64.4743 &   6.740 \\
 5252566023132496256 & 307557 &152.9157 & -63.2241 &   9.439&  5239883946781824256 &  92837 & 160.3962 & -64.1063 &   7.183 \\
 5252392884409210624 &  88980 &153.4413 & -63.8999 &   8.677&  5239823782876560512 &  92966 & 160.6047 & -64.3988 &   7.287 \\
 5245507295921159680 & 309933 &153.7489 & -65.8036 &   9.571&  5239725548384560128 &  92989 & 160.6599 & -64.6779 &   7.600 \\
 5245462112865217024 &  89903 &155.0783 & -66.0636 &   7.534&  5239660367966484736 &  & 160.7508 & -65.5049 &  10.958 \\
 5251495957802232448 &  90020 &155.3235 & -65.1956 &   7.333&  5239858726733859968 &  93098 & 160.8727 & -64.0684 &   7.605 \\
 5245270626044873344 &  90083 &155.4193 & -66.1127 &   8.219&  5241458584871666176 & 307912 & 161.0135 & -63.1726 &   9.555 \\
 5253053725257383424 &  &155.4318 & -62.2425 &  11.126&  5239701565287189120 & 307979 & 161.2482 & -65.0386 &  10.619 \\
 5232616277998058752 &  90456 &156.0607 & -68.1295 &   7.683&  5242010093028921856 &  & 161.3487 & -61.7751 &  11.182 \\
 5251663495883800448 &  90731 &156.6849 & -64.3517 &   7.448&  5239801689564803072 &  93424 & 161.3913 & -64.7038 &   8.105 \\
 5252084918078077952 &  90837 &156.8759 & -63.5232 &   8.280&  5239809557944885888 &  93517 & 161.5411 & -64.5955 &   7.844 \\
 5251470943909316096 &  &157.0373 & -64.5052 &  10.302&  5239849896277704960 &  & 161.5616 & -64.0494 &  10.484 \\
 5252077668171627648 &  &157.1300 & -63.7376 &  11.170&  5239498739751922432 & 310131 & 161.6374 & -65.4550 &  10.694 \\
 5253965770155144704 &  91042 &157.2239 & -61.1637 &   9.478&  5239843196128730368 &  93648 & 161.7879 & -64.2646 &   7.841 \\
 5251238225402532352 &  91144 &157.3580 & -65.8758 &   8.785&  5241357189281206656 &  93874 & 162.1746 & -63.8330 &   8.179 \\
 5251888517809705216 &  &157.3863 & -63.8209 &  11.157&  5239626420542800512 &  & 162.4515 & -64.7745 &  11.286 \\
 5233181908011057536 & 310053 &157.5378 & -66.5259 &  10.061&  5239637759256457472 &  94174 & 162.6916 & -64.4794 &   7.734 \\
 5251845946093877632 & 307802 &157.7811 & -64.1819 &   9.412&  5241909865679004288 &  & 162.7754 & -62.6092 &  10.616 \\
 5251991356506637568 &  91451 &157.9370 & -63.4894 &   9.095&  5241082311374967296 &  & 163.5722 & -64.6809 &  11.416 \\
 5251880511990665216 & 307793 &157.9753 & -63.8144 &  10.189&  5241132579670834816 &  94684 & 163.6177 & -63.9726 &   8.931 \\
 5251765681750027520 &  &158.4151 & -64.7811 &  11.404&  5239525196750143616 &  & 163.7080 & -65.4460 &  11.618 \\
 5251822100435455744 &  91906 &158.7351 & -64.1339 &   7.483&  5241066608973054080 & 308100 & 164.0480 & -64.8004 &  10.312 \\
 5239891677722958976 &  &159.5734 & -64.1351 &  10.393&  5241109696086432640 & 308094 & 164.3439 & -64.2762 &  10.039 \\
 5239789148260273792 &  92467 &159.7139 & -64.4980 &   7.195&  5241109867885123328 &  & 164.3560 & -64.2497 &  10.790 \\
 5239740804110817664 &  92478 &159.7602 & -64.9749 &   7.585&  5238698467088998400 &  95786 & 165.4778 & -66.0123 &   7.529 \\
 5239895629092870144 &  92536 &159.8450 & -64.1117 &   6.354&  5240861412615531136 &  95911 & 165.6875 & -63.4946 &   9.107 \\
 5239926827735288960 &  92535 &159.8539 & -63.7781 &   8.222&  5240380376282066816 &  96287 & 166.2079 & -64.6157 &   7.252 \\
 5239736199905884416 &  92570 &159.8777 & -65.0833 &   9.189&  5337081393150208768 & 308215 & 166.9832 & -62.0480 &  10.787 \\
 5239304538510338432 & 310113 &159.9163 & -65.3489 &  11.160&  5237676539747932032 &  98616 & 169.9780 & -63.7503 &   8.657 \\
 5253452057710889216 &  &160.0000 & -63.2530 &  10.509&  5237036555261723648 &  99149 & 170.9130 & -65.8344 &   8.127 \\
\hline
\end{tabular}
}
\end{table*}
\begin{figure}[t]
\centering
\includegraphics[width=7cm]{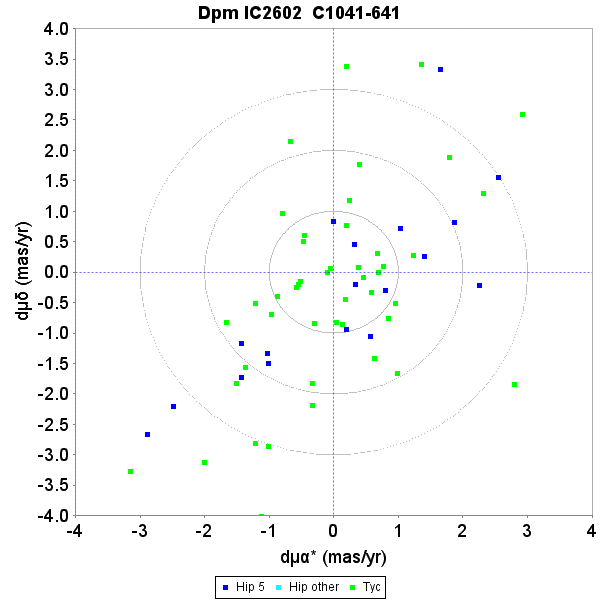}
\includegraphics[width=7cm]{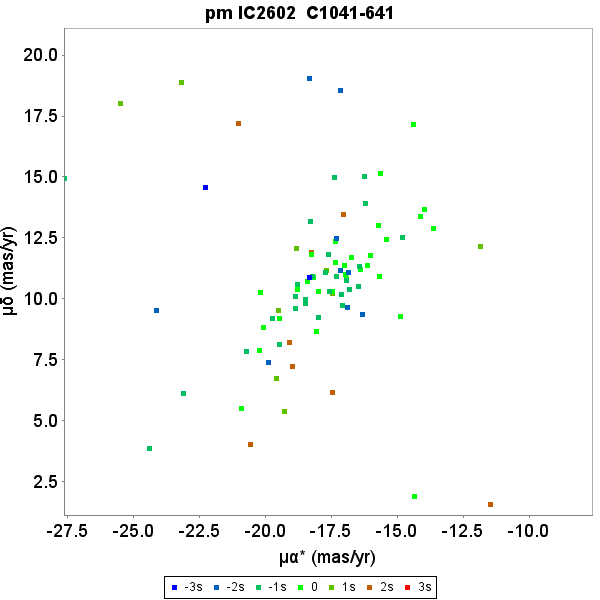}
\caption{Proper motion charts for the cluster IC2602. Left: unit weight residual proper motions. Green dots have first epoch \Tycho data, the dark blue dots have \Hipparcos first epoch 5-parameter solutions. The concentric circles represent 1, 2, and 3$\sigma$ \su levels. Right: actual proper motion distribution, where the colour indicate the difference from the cluster parallax in \su units.}
\label{fig:ic2602}
\end{figure}
\begin{figure}[t]
\centering
\includegraphics[width=12cm]{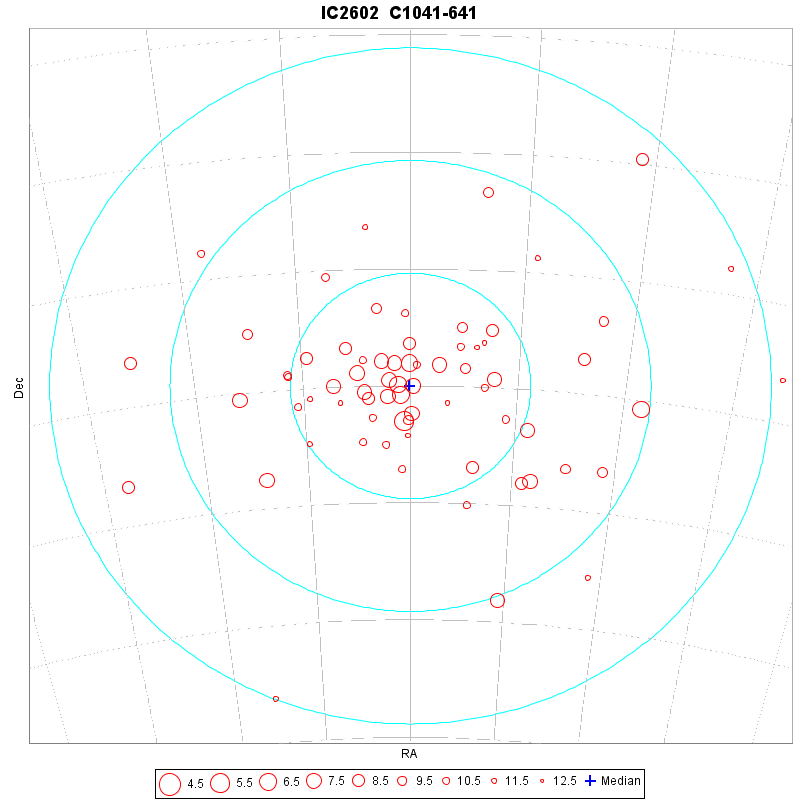}
\caption{A map of members of the cluster IC2602 as identified from the TGAS catalogue. The coordinate grid is at 2 degrees intervals, the three concentric circles are at 5, 10 and 15 pc from the cluster centre at the cluster distance.}
\label{fig:mapic2602}
\end{figure}

\subsection{The cluster Blanco 1 \label{app:blanco1}}
\begin{table*}
\centering
\caption{Identifiers and positions for members of the cluster Blanco 1.}\label{tab:Blanco 1}
\scriptsize{
\begin{tabular}{rrrrr|rrrrr}
\hline\hline 
SourceId & HD & $\alpha$ (degr) & $\delta$ (degr) & G & SourceId & HD & $\alpha$ (degr) & $\delta$ (degr) & G \\ \hline 
 2328281430196519680 &  &357.3891 & -27.8397 &  10.443&  2320877215816767360 & 225111 &   0.8830 & -29.7180 &  10.024 \\
 2327952401341889664 &  &357.9863 & -29.0023 &  12.040&  2320770185232173312 &  &   0.8901 & -30.4783 &  10.458 \\
 2334037751525146624 &  &359.2535 & -27.9233 &  10.312&  2320816227281177088 &  &   0.9591 & -30.0655 &  10.906 \\
 2326921471751885440 &  &359.4123 & -29.8351 &  10.144&  2320826432123463040 &  &   1.0782 & -29.8257 &  10.378 \\
 2326866942847101824 &  &359.4679 & -30.0883 &  11.908&  2320979848354547840 & 225206 &   1.0955 & -29.3816 &   7.797 \\
 2326849247581847296 &  &359.5839 & -30.2810 &   9.943&  2320798635095540736 &  &   1.1320 & -30.2449 &  11.508 \\
 2313104527601463296 &  &359.7039 & -33.5562 &  10.689&  2320926178443221888 & 225264 &   1.2119 & -29.6330 &   8.330 \\
 2326850415812950656 &  &359.7597 & -30.2214 &  10.718&  2320610721685102336 & 225282 &   1.2223 & -30.2566 &   8.335 \\
 2333096432132693632 &  &  0.2852 & -28.6157 &  11.436&  2320617524913298304 &  &   1.2453 & -30.1616 &  11.288 \\
 2314771146710725504 &  &  0.3521 & -30.6495 &  10.367&  2320933153470108032 &  &   1.3267 & -29.5016 &  11.129 \\
 2320833441510104960 &  &  0.3696 & -30.2058 &  10.690&  2320612748909666560 &     50 &   1.3616 & -30.2901 &   9.646 \\
 2320834747180161408 & 224948 &  0.4907 & -30.1580 &   9.866&  2320903844613139072 &  &   1.3790 & -29.8856 &  11.088 \\
 2320886218068219264 &  &  0.5488 & -29.6943 &  11.651&  2320902057906744704 &     91 &   1.4291 & -29.9606 &   9.823 \\
 2320793893451241344 &  &  0.5902 & -30.1393 &  10.570&  2321022626228808064 &  &   1.4851 & -28.9363 &  11.292 \\
 2314555230115494528 &  &  0.6372 & -30.9921 &  10.994&  2320916763874909568 &  &   1.4961 & -29.6513 &  11.450 \\
 2333013075407421824 &  &  0.6636 & -29.0752 &  10.097&  2321010325442474752 &    141 &   1.5379 & -29.1514 &   7.941 \\
 2320994588682307072 &  &  0.7772 & -29.3622 &  11.464&  2320709643372054656 &  &   1.5682 & -30.0992 &  11.512 \\
 2320790285678713984 &  &  0.7952 & -30.1803 &  11.147&  2313393493000760064 &  &   1.6470 & -32.8017 &  11.330 \\
 2320869897192498432 &  &  0.8360 & -29.8230 &  11.117&  2320645734258792064 &  &   1.8926 & -30.2872 &  10.421 \\
 2320869725393806336 &  &  0.8517 & -29.8137 &  10.498&  2320702187308925184 &    343 &   2.0080 & -30.0317 &   9.687 \\
 2320771216024323840 &  &  0.8544 & -30.4351 &  11.741&  2319380333814252544 &    704 &   2.8317 & -32.4039 &   8.436 \\
 2320776644862584960 &  &  0.8708 & -30.3242 &  10.468&  2319554915645229312 &  &   3.7501 & -31.2733 &  10.251 \\
\hline
\end{tabular}
}
\end{table*}
\begin{figure}[t]
\centering
\includegraphics[width=7cm]{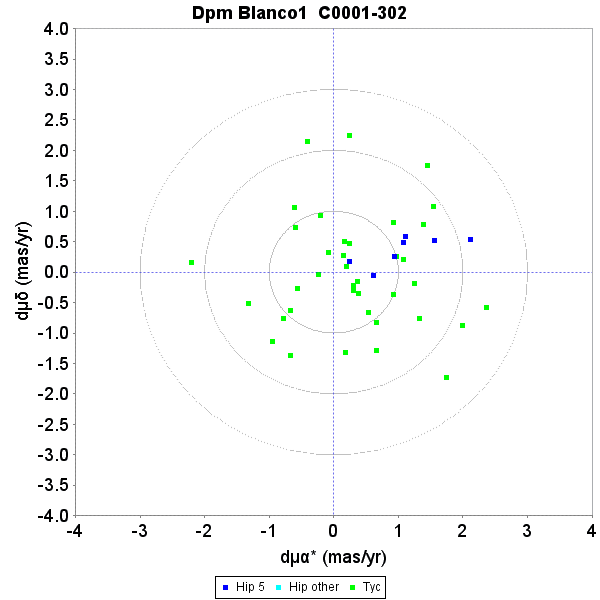}
\includegraphics[width=7cm]{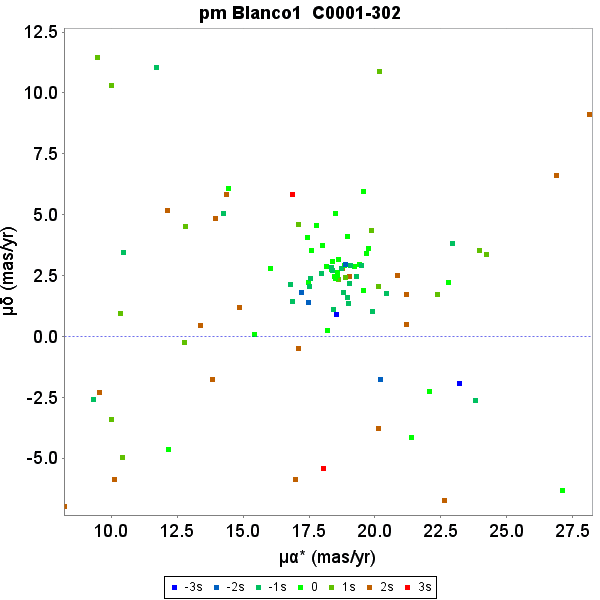}
\caption{Proper motion charts for the cluster Blanco 1. Left: unit weight residual proper motions. Green dots have first epoch \Tycho data, the dark blue dots have \Hipparcos first epoch 5-parameter solutions. The concentric circles represent 1, 2, and 3$\sigma$ \su levels. Right: actual proper motion distribution, where the colour indicate the difference from the cluster parallax in \su units.}
\label{fig:Blanco1}
\end{figure}
\begin{figure}[t]
\centering
\includegraphics[width=12cm]{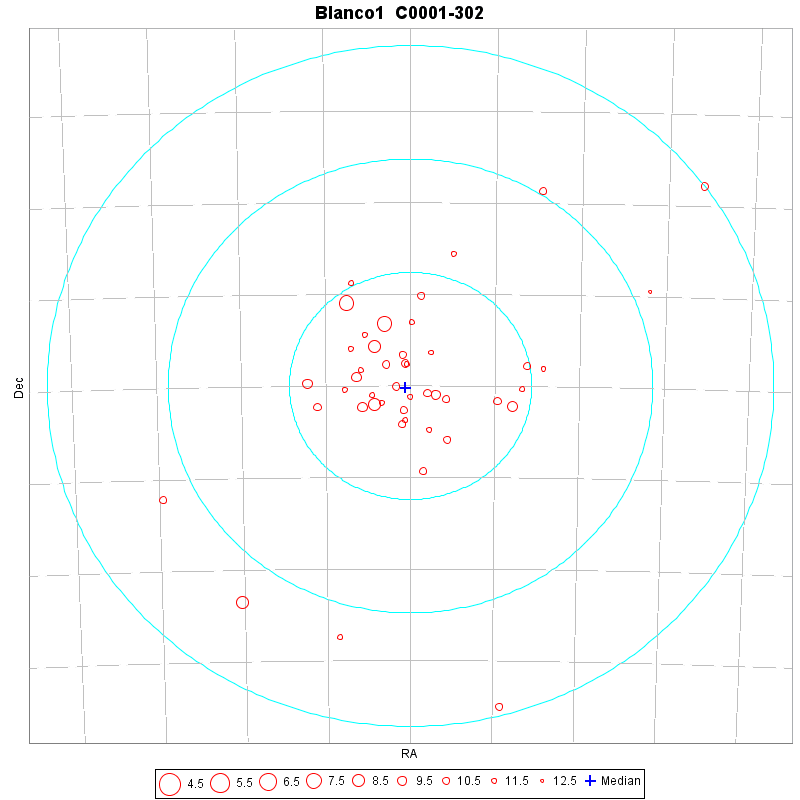}
\caption{A map of members of the cluster Blanco 1 as identified from the TGAS catalogue. The coordinate grid is at 2 degrees intervals, the three concentric circles are at 5, 10 and 15 pc from the cluster centre at the cluster distance.}
\label{fig:mapBlanco1}
\end{figure}

\subsection{The cluster NGC2451 \label{app:ngc2451}}
\begin{table*}
\centering
\caption{Identifiers and positions for members of the cluster NGC2451.}\label{tab:NGC2451}
\scriptsize{
\begin{tabular}{rrrrr|rrrrr}
\hline\hline 
SourceId & HD &  $\alpha$ (degr) & $\delta$ (degr) & G & SourceId & HD & $\alpha$ (degr) & $\delta$ (degr) & G \\ \hline 
 5585438770393517312 &  60330 &113.0701 & -38.1741 &   8.578&  5586861641519108352 &  62642 & 115.8727 & -37.7505 &   7.612 \\
 5585369844760452224 &  &113.5430 & -38.6319 &  11.380&  5538642456004614784 &  & 116.1002 & -38.5997 &  10.781 \\
 5585190555647523840 &  61375 &114.3182 & -39.0545 &   9.129&  5538816247560408448 &  62876 & 116.1170 & -37.9877 &   8.624 \\
 5588006714160268928 &  &114.4807 & -35.9994 &  10.942&  5538637336406254720 &  & 116.1170 & -38.7782 &   9.155 \\
 5587015401348292736 &  &114.4933 & -37.6214 &  11.132&  5538816831675959168 &  62893 & 116.1423 & -37.9429 &   5.925 \\
 5585278928890699264 &  &114.5179 & -38.4535 &   9.641&  5538612253797251456 &  & 116.1545 & -38.9079 &   9.369 \\
 5587001657452951680 &  &114.5298 & -37.7534 &  11.810&  5538598819139552768 &  62961 & 116.1828 & -39.0207 &   8.126 \\
 5537126538706711936 &  &114.8141 & -39.4657 &  11.702&  5538812227471020672 &  62938 & 116.1869 & -38.0537 &   7.616 \\
 5587037563379531008 &  &114.8219 & -37.3751 &  10.477&  5538386166717422080 &  63198 & 116.4983 & -39.4444 &   8.867 \\
 5536231123926763264 &  &114.8398 & -41.9947 &   9.718&  5538855555103296512 &  & 116.5371 & -37.3991 &  10.487 \\
 5586793471798169600 &  &114.9047 & -38.0427 &   9.854&  5538805183726897024 &  63215 & 116.5438 & -37.9336 &   5.906 \\
 5536717142426664704 &  &115.0181 & -40.6927 &   9.693&  5538853596598218112 &  & 116.5476 & -37.5307 &  11.055 \\
 5588530665810298880 &  &115.1928 & -34.6106 &   9.445&  5532490207410038528 &  & 116.7050 & -43.2293 &  11.497 \\
 5536727106748162816 &  &115.3105 & -40.6270 &  10.983&  5538534016672382464 &  63511 & 116.9133 & -38.6766 &   8.748 \\
 5586820925229156480 &  &115.3260 & -37.8968 &   9.842&  5538777695934246656 &  & 116.9339 & -37.9841 &  11.602 \\
 5586745643042408704 &  &115.3918 & -38.0832 &  10.843&  5537606750413835136 &  & 118.2994 & -40.0162 &   9.677 \\
 5586755676086015232 &  62479 &115.6242 & -38.1079 &   9.057&  5534089447072632320 &  & 118.7978 & -42.3349 &  11.899 \\
 5538660357428319360 &  &115.6419 & -38.8488 &  11.085&  5544009309698790400 &  & 120.5397 & -38.0250 &  11.414 \\
 5587582508831567488 &  62578 &115.7998 & -36.0500 &   5.672&&&&\\
\hline
\end{tabular}
}
\end{table*}
\begin{figure}[t]
\centering
\includegraphics[width=7cm]{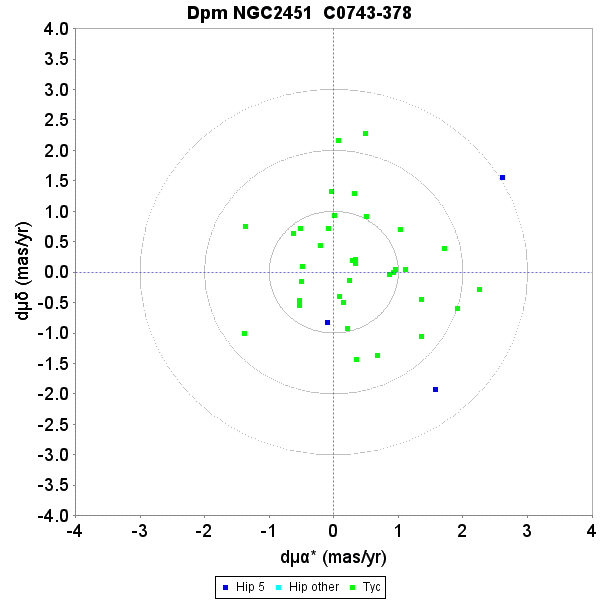}
\includegraphics[width=7cm]{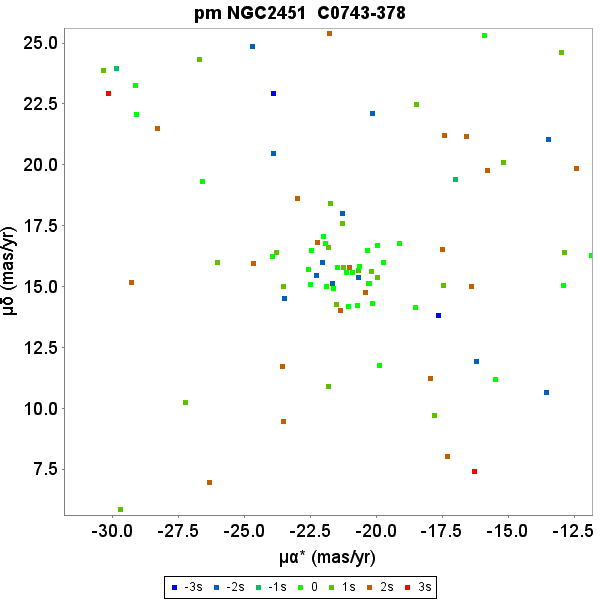}
\caption{Proper motion charts for the cluster NGC2451. Left: unit weight residual proper motions. Green dots have first epoch \Tycho data, the dark blue dots have \Hipparcos first epoch 5-parameter solutions. The concentric circles represent 1, 2, and 3$\sigma$ \su levels. Right: actual proper motion distribution, where the colour indicate the difference from the cluster parallax in \su units.}
\label{fig:NGC2451}
\end{figure}
\begin{figure}[t]
\centering
\includegraphics[width=12cm]{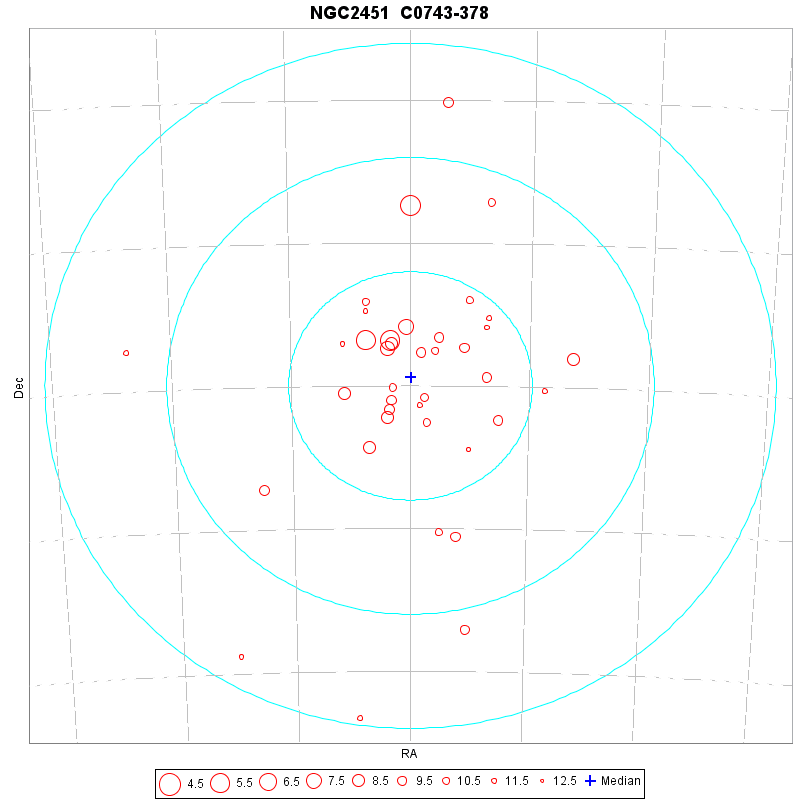}
\caption{A map of members of the cluster NGC2451 as identified from the TGAS catalogue. The coordinate grid is at 2 degrees intervals, the three concentric circles are at 5, 10 and 15 pc from the cluster centre at the cluster distance.}
\label{fig:mapNGC2451}
\end{figure}

\subsection{The cluster NGC6475}
\begin{table*}
\centering
\caption{Identifiers and positions for members of the cluster NGC6475.}\label{tab:NGC6475}
\scriptsize{
\begin{tabular}{rrrrr|rrrrr}
\hline\hline 
SourceId & HD & $\alpha$ (degr) & $\delta$ (degr) & G & SourceId & HD & $\alpha$ (degr) & $\delta$ (degr) & G \\ \hline 
 4041052505956833536 & 320665 &266.7198 & -35.6233 &  10.473&  4040637715192089472 & 162612 & 268.3694 & -35.5117 &   9.489 \\
 4040221824918071424 &  &266.8167 & -36.7436 &  11.605&  4040372492371770880 & 162613 & 268.3750 & -36.5086 &   7.984 \\
 4041744648522172288 & 161575 &266.9481 & -34.3107 &   6.947&  4040826521954932096 & 162780 & 268.5599 & -34.7277 &   6.876 \\
 4040279618003552768 & 161576 &266.9784 & -36.0220 &   9.178&  4040728596700624896 & 320885 & 268.6121 & -35.3338 &  10.014 \\
 4041422491620010368 & 161651 &267.0722 & -35.5376 &   9.027&  4040847756273228416 & 162817 & 268.6130 & -34.4667 &   6.082 \\
 4041554158132660096 & 161685 &267.1030 & -34.4658 &   9.148&  4040717532863953408 & 162839 & 268.6497 & -35.4995 &   8.443 \\
 4040291197235375616 & 161686 &267.1345 & -35.7515 &   8.940&  4043113952821300352 & 162874 & 268.6594 & -33.9563 &   7.821 \\
 4041428264056050816 & 161855 &267.3713 & -35.3781 &   7.342&  4043120068854728704 & 162873 & 268.6732 & -33.9055 &   7.673 \\
 4041458397546585600 & 320649 &267.4005 & -35.1666 &   9.876&  4040728287462977152 &  & 268.7205 & -35.2821 &  10.985 \\
 4041484510943000576 &  &267.4681 & -34.7804 &  11.088&  4040744436539998720 & 162891 & 268.7234 & -35.0893 &   8.022 \\
 4040702002264923136 &  &267.5717 & -35.3390 &  10.824&  4040346447690086784 & 162926 & 268.7835 & -36.4757 &   6.029 \\
 4041629852636717312 & 162016 &267.5731 & -34.4761 &   8.147&  4040754984979680512 & 162942 & 268.7853 & -35.1504 &   8.579 \\
 4041466197202462080 &  &267.6191 & -35.0089 &  11.077&  4042353503086066816 & 162980 & 268.8133 & -34.2356 &   7.494 \\
 4041633151171596928 & 320741 &267.6197 & -34.3410 &   9.694&  4042365219756839296 & 163001 & 268.8413 & -34.0165 &   9.322 \\
 4043474970596660224 & 318439 &267.6642 & -32.7352 &  10.025&  4042365013598409856 &  & 268.8461 & -34.0238 &  10.500 \\
 4041584807019715840 & 320745 &267.7199 & -34.4255 &   9.936&  4042333986754687104 & 320841 & 268.8799 & -34.4817 &  10.057 \\
 4041462245832553856 & 162144 &267.7589 & -35.0706 &   7.616&  4040561539657230464 & 163067 & 268.9467 & -35.3191 &   9.487 \\
 4041464101258425088 &  &267.8166 & -35.0052 &  11.982&  4042340171507583104 & 163109 & 268.9841 & -34.2866 &   7.977 \\
 4041579240742120832 & 162223 &267.8571 & -34.5600 &   8.907&  4040766529851750656 & 320952 & 269.0952 & -34.8332 &   9.523 \\
 4040713134816625792 & 162224 &267.8579 & -35.0344 &   9.046&  4042287635467629056 & 163193 & 269.0967 & -34.4978 &   8.671 \\
 4040603218018657280 &  &267.9507 & -35.8614 &  10.781&  4040778590119911424 & 163194 & 269.1027 & -34.7538 &   9.089 \\
 4040709355245406848 & 162286 &267.9562 & -35.1022 &   9.378&  4040519380251978624 &  & 269.1672 & -35.5569 &  10.844 \\
 4041573605745028096 & 162285 &267.9597 & -34.5482 &   8.643&  4042283168701645312 & 320946 & 269.1879 & -34.5919 &   9.428 \\
 4040603218018656640 &  &267.9611 & -35.8542 &  10.657&  4043357254123008640 & 163274 & 269.2085 & -32.6889 &   6.658 \\
 4040692347178449664 & 162287 &267.9650 & -35.3654 &   7.303&  4043274034836706560 & 318671 & 269.2626 & -33.1078 &   9.785 \\
 4041568795381660288 & 162348 &268.0284 & -34.6557 &   8.997&  4040760929214406528 & 320950 & 269.2863 & -34.9638 &   9.801 \\
 4040808964127026432 & 320764 &268.0413 & -34.8946 &   8.897&  4043303721650637696 &  & 269.4830 & -32.7873 &  11.081 \\
 4040803913245489024 & 162349 &268.0435 & -35.0018 &   8.430&  4037427587917137152 & 320999 & 269.5119 & -36.4534 &  10.696 \\
 4040804153763656576 & 320768 &268.0655 & -34.9699 &   9.131&  4042479603329702272 & 318778 & 269.5494 & -33.6397 &  10.282 \\
 4040804085044179968 & 162393 &268.0834 & -34.9745 &   8.074&  4042299489577363200 & 321043 & 269.6404 & -34.4211 &  10.758 \\
 4040805871752189568 & 162457 &268.1704 & -34.9294 &   8.281&  4042063678692927232 &  & 269.9128 & -34.9376 &  11.188 \\
 4040642972232475136 & 320776 &268.1921 & -35.3967 &   9.791&  4042197647312832896 & 321037 & 270.2051 & -34.3946 &  10.728 \\
 4041595699062347136 & 162513 &268.2352 & -34.4461 &   8.654&  4042036878101137152 &  & 270.2373 & -34.9650 &   9.851 \\
 4040786286701335168 & 162542 &268.2629 & -35.1648 &   8.984&  4042036878101137280 & 321058 & 270.2404 & -34.9665 &  11.110 \\
 4040814461686768640 & 320863 &268.2694 & -34.7612 &   8.880&  4042198197068643968 & 321036 & 270.2432 & -34.3489 &  10.476 \\
 4040814564765981824 & 320862 &268.2812 & -34.7478 &   9.166&  4041992313522568064 & 164108 & 270.2995 & -35.3291 &   8.739 \\
 4040737117915740672 & 320891 &268.2887 & -35.3490 &  10.510&  4039009785146773376 &  & 270.6158 & -35.4038 &  10.787 \\
 4040813259095930752 & 320864 &268.2984 & -34.8278 &   9.198&  4042169953363702528 &  & 271.1769 & -34.2802 &  11.362 \\
 4043179167604718336 & 162610 &268.3394 & -33.7643 &   9.129&  4042167616901494784 & 321249 & 271.3293 & -34.3027 &  10.132 \\
\hline
\end{tabular}
}
\end{table*}
\begin{figure}[t]
\centering
\includegraphics[width=7cm]{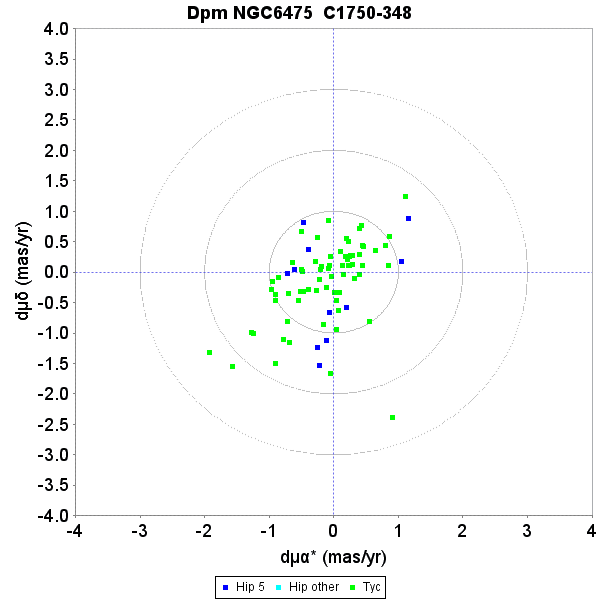}
\includegraphics[width=7cm]{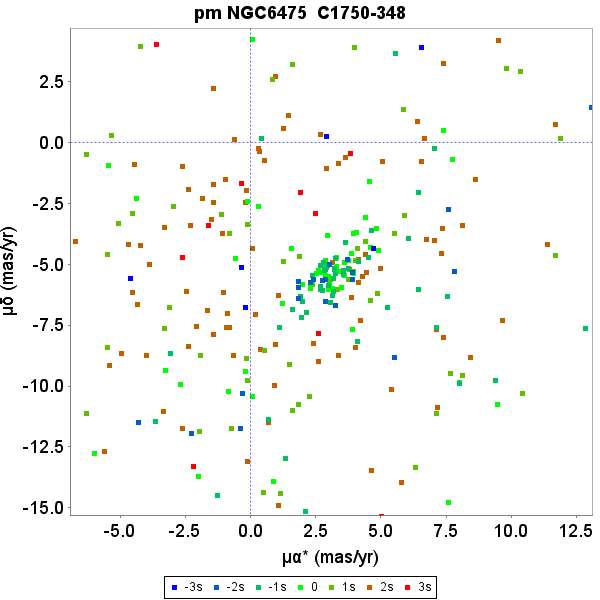}
\caption{Proper motion charts for the cluster NGC6475. Left: unit weight residual proper motions. Green dots have first epoch \Tycho data, the dark blue dots have \Hipparcos first epoch 5-parameter solutions. The concentric circles represent 1, 2, and 3$\sigma$ \su levels. Right: actual proper motion distribution, where the colour indicate the difference from the cluster parallax in \su units.}
\label{fig:NGC6475}
\end{figure}
\begin{figure}[t]
\centering
\includegraphics[width=12cm]{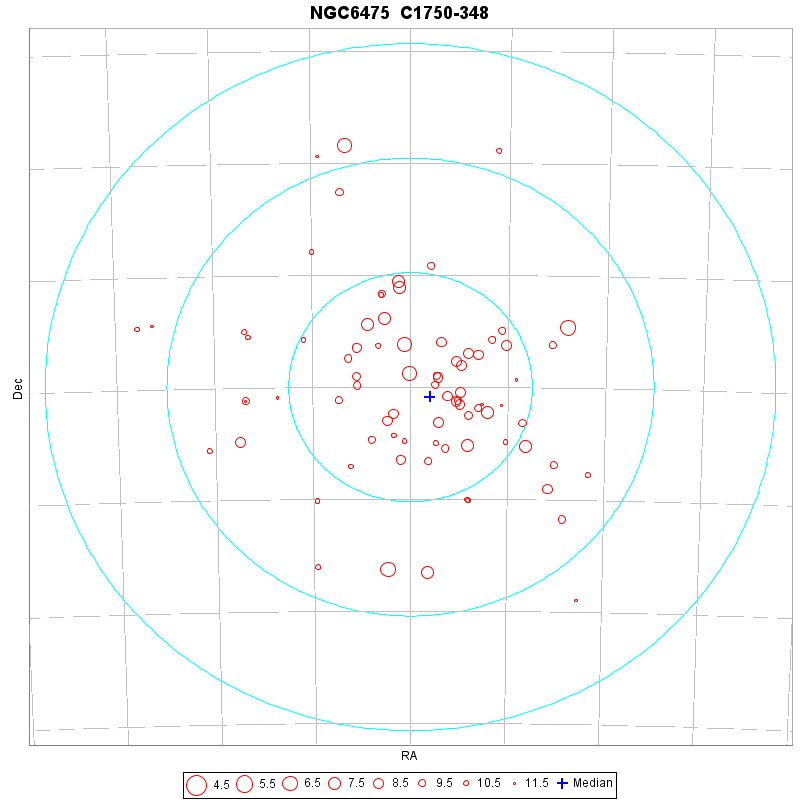}
\caption{A map of members of the cluster NGC6475 as identified from the TGAS catalogue. The coordinate grid is at 1 degrees intervals, the three concentric circles are at 5, 10 and 15 pc from the cluster centre at the cluster distance.}
\label{fig:mapNGC6475}
\end{figure}

\subsection{The cluster NGC7092}
\begin{table*}
\centering
\caption{Identifiers and positions for members of the cluster NGC7092.}\label{tab:NGC7092}
\scriptsize{
\begin{tabular}{rrrrr|rrrrr}
\hline\hline 
SourceId & HD & $\alpha$ (degr) & $\delta$ (degr) & G & SourceId & HD & $\alpha$ (degr) & $\delta$ (degr) & G \\ \hline 
 2170750757152762240 &  &321.6688 &  48.5799 &   9.716&  1978441947209204736 &  & 323.2545 &  48.2349 &   9.672 \\
 1978555643589899648 & 204917 &322.5789 &  48.3908 &   7.385&  1978443321598738560 & 205331 & 323.2921 &  48.3033 &   6.954 \\
 1978529633267976704 &  &322.6364 &  47.9997 &  11.698&  1978742904159340416 &  & 323.4489 &  49.1609 &  10.317 \\
 1978336187938996992 &  &322.7520 &  47.8002 &   9.741&  1978743350835937536 &  & 323.4864 &  49.1968 &   9.363 \\
 1978533137961284864 &  &322.7939 &  48.0704 &  10.097&  1978460467108321664 &  & 323.8055 &  48.4372 &   9.576 \\
 1978652744205824384 & 205116 &322.9266 &  48.5845 &   6.863&  1978484999961562624 &  & 323.8920 &  48.7588 &  10.381 \\
 1978647762043764608 & 205117 &322.9360 &  48.4843 &   7.689&  1978404495094544128 &  & 323.9485 &  48.1082 &   9.430 \\
 1978648483598270464 &  &322.9565 &  48.4817 &   8.874&  1977411326859661056 &  & 323.9917 &  46.5545 &  10.449 \\
 1978656214539396352 & 205198 &323.0610 &  48.6394 &   8.269&  1978464418478228352 &  & 324.2888 &  48.5571 &   9.685 \\
 1978636285891151744 & 205210 &323.0711 &  48.4437 &   6.593&  1978933291465578240 &  & 325.3494 &  49.3301 &  10.002 \\
 1978740430258165632 &  &323.1142 &  49.1810 &  11.285&  1977881058837817344 &  & 326.6370 &  48.1986 &   9.607 \\
 1978641852168765184 &  &323.1798 &  48.4831 &   9.053&&&&\\
\hline
\end{tabular}
}
\end{table*}
\begin{figure}[t]
\centering
\includegraphics[width=7cm]{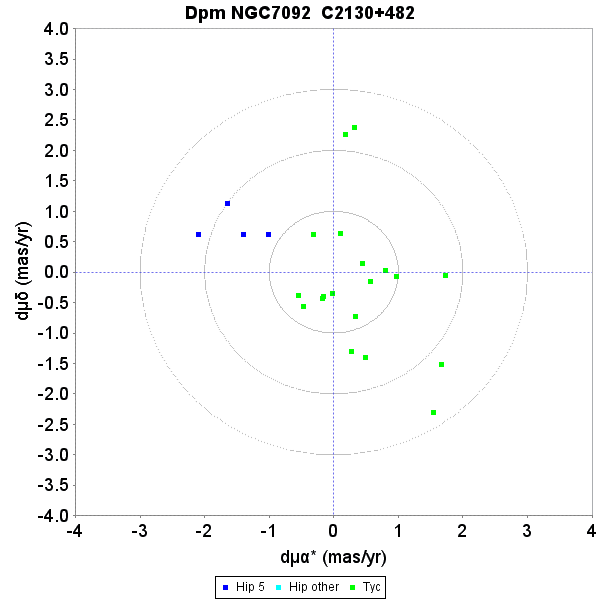}
\includegraphics[width=7cm]{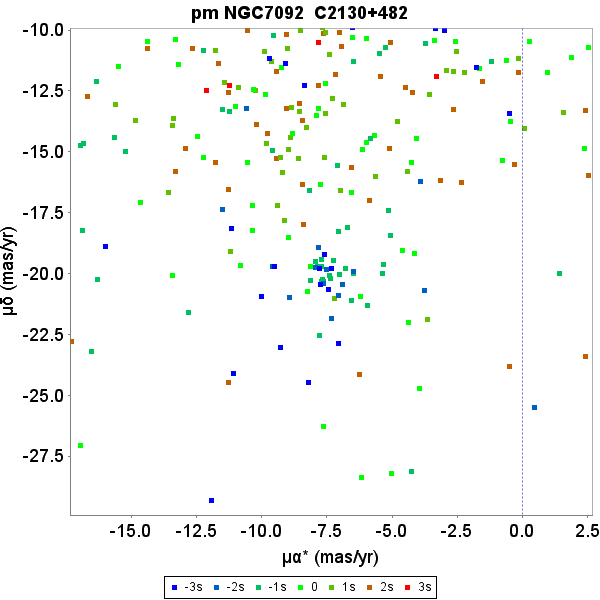}
\caption{Proper motion charts for the cluster NGC7092. Left: unit weight residual proper motions. Green dots have first epoch \Tycho data, the dark blue dots have \Hipparcos first epoch 5-parameter solutions. The concentric circles represent 1, 2, and 3$\sigma$ \su levels. Right: actual proper motion distribution, where the colour indicate the difference from the cluster parallax in \su units.}
\label{fig:NGC7092}
\end{figure}
\begin{figure}[t]
\centering
\includegraphics[width=12cm]{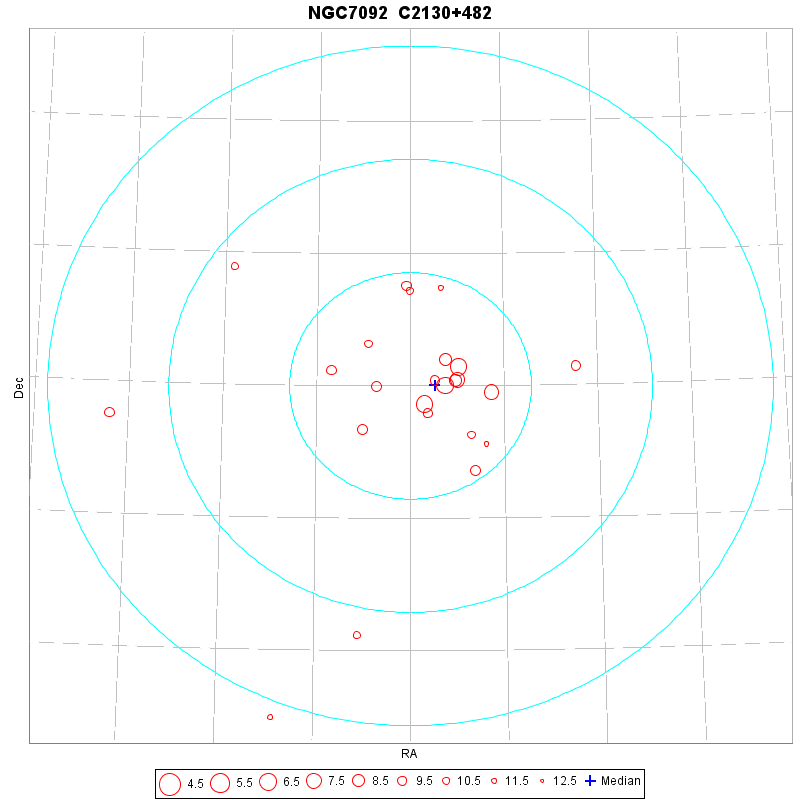}
\caption{A map of members of the cluster NGC7092 as identified from the TGAS catalogue. The coordinate grid is at 1 degrees intervals, the three concentric circles are at 5, 10 and 15 pc from the cluster centre at the cluster distance.}
\label{fig:mapNGC7092}
\end{figure}

\subsection{The cluster NGC2516}
\begin{table*}
\centering
\caption{Identifiers and positions for members of the cluster NGC2516.}\label{tab:NGC2516}
\scriptsize{
\begin{tabular}{rrrrr|rrrrr}
\hline\hline 
SourceId & HD & $\alpha$ (degr) & $\delta$ (degr) & G & SourceId & HD & $\alpha$ (degr) & $\delta$ (degr) & G \\ \hline 
 5292754032119550976 &  63011 &115.8099 & -59.8235 &   9.590&  5287997854056218880 &  66081 & 119.5204 & -63.2295 &   9.413 \\
 5289355166802297984 &      & 115.9907 & -61.6941 &  10.333&  5290673756119783168 &      &  119.5542 & -60.8741 &   8.203 \\
 5294285239500148480 &      & 116.1331 & -59.3370 &  11.772&  5291137853105909632 &  66029 & 119.5827 & -59.4363 &   9.707 \\
 5294282250202911104 &      & 116.3152 & -59.3482 &  10.441&  5290667949324010240 &      &  119.6589 & -61.0135 &   9.435 \\
 5292482074791467136 &      & 116.5823 & -60.4544 &  10.814&  5290767626925303936 &  66194 & 119.7106 & -60.8244 &   5.867 \\
 5292478948054155776 &      & 116.6608 & -60.5114 &   9.900&  5291055458454839296 &      &  119.7854 & -59.8373 &  10.590 \\
 5294345162885186176 &  63709 &116.7270 & -58.9067 &   8.182&  5290818582417293696 &  66259 & 119.8112 & -60.5871 &   8.382 \\
 5289457318302408960 &  63872 &116.8751 & -60.7841 &   9.648&  5290767386407134720 &  66318 & 119.8644 & -60.7964 &   9.609 \\
 5293999572636389760 &      & 116.9295 & -60.1684 &   9.964&  5290650700735361664 &      &  119.9147 & -61.1944 &  10.850 \\
 5288243216948851968 &      & 117.6310 & -62.8854 &  11.865&  5290847856916705152 &  66341 & 119.9170 & -60.2073 &   6.311 \\
 5294040838682140416 &  64507 &117.7367 & -59.6462 &   7.313&  5291062467841475456 &  66388 & 119.9597 & -59.9670 &   9.574 \\
 5290696124310372352 &  64644 &117.8000 & -61.2545 &   9.075&  5291542404667473152 &  66481 & 120.1364 & -58.8667 &   9.587 \\
 5288322587942769152 &      & 117.8623 & -62.4960 &  11.359&  5290772196770548096 &      &  120.1435 & -60.7155 &   9.882 \\
 5290968081639463552 &      & 117.8949 & -60.4124 &  10.887&  5290868816357122816 &      &  120.2115 & -60.1529 &  11.068 \\
 5291009519483916160 &      & 117.9648 & -60.1588 &  10.286&  5290747079801445248 &      &  120.2139 & -61.0192 &  10.270 \\
 5290968459596583296 &  64762 &117.9885 & -60.3843 &   9.251&  5289989688088614912 &      &  120.2572 & -61.2432 &  10.343 \\
 5294211640940573184 &  64743 &118.0035 & -58.9986 &   9.815&  5289976081632231552 &  66707 & 120.3084 & -61.3677 &   9.721 \\
 5290895513871203712 &      & 118.0578 & -60.8858 &  10.239&  5290850468254523904 &      &  120.4019 & -60.3519 &  10.772 \\
 5290918569256484480 &      & 118.0700 & -60.5472 &  11.210&  5289811945162807296 &      &  120.5536 & -61.7774 &  10.766 \\
 5294045099291234560 &  64831 &118.0928 & -59.4931 &   7.816&  5289977868338618240 &      &  120.5714 & -61.2769 &  11.748 \\
 5291011340551279872 &      & 118.2444 & -60.0532 &  10.324&  5290004668936594560 &      &  120.6922 & -60.8937 &  10.771 \\
 5290701931106148096 &      & 118.2496 & -61.0844 &  11.283&  5290780992863569408 &      &  120.7087 & -60.7159 &  10.840 \\
 5290682689652668416 &      & 118.2528 & -61.2178 &  11.066&  5290863799832998144 &  67107 & 120.7604 & -60.0696 &   9.566 \\
 5294047229593277184 &      & 118.2712 & -59.4145 &   9.326&  5290779755912990976 &      &  120.8441 & -60.7642 &  10.695 \\
 5290944098542081280 &      & 118.3249 & -60.3956 &  11.355&  5290859401786488704 &  67197 & 120.8576 & -60.1530 &   9.357 \\
 5290901286307238528 &      & 118.3321 & -60.7136 &  10.028&  5291237324551501184 &      &  121.0781 & -60.0178 &  11.165 \\
 5290901286307238656 &      & 118.3430 & -60.7150 &  10.738&  5290030335661147136 &      &  121.1402 & -60.7379 &  10.660 \\
 5290944132901818880 &      & 118.3445 & -60.3840 &  11.546&  5289934746866964352 &      &  121.1573 & -61.2504 &  11.313 \\
 5290943067749932544 &      & 118.3564 & -60.4320 &  10.442&  5290790441791605248 &      &  121.2040 & -60.4691 &  11.431 \\
 5290704301928090240 &      & 118.5403 & -60.9742 &  11.379&  5290006146403282432 &      &  121.2353 & -61.1145 &  11.219 \\
 5294443294297626368 &      & 118.5494 & -58.4124 &  10.888&  5290795183437025408 &  67515 & 121.2601 & -60.3879 &   7.683 \\
 5290940971807119104 &      & 118.6115 & -60.4304 &  10.960&  5289697492873552896 &      &  121.3343 & -61.9716 &  10.223 \\
 5290726910635935872 &      & 118.6749 & -60.9635 &  10.987&  5290020646214928640 &      &  121.4521 & -60.7742 &  11.528 \\
 5294246962752996736 &  65387 &118.8005 & -58.6090 &  10.263&  5289620664501064832 &      &  121.7018 & -62.5151 &  10.946 \\
 5289781674232532224 &  65623 &118.9673 & -62.3468 &   8.989&  5291290375984527104 &      &  121.7257 & -59.4200 &  11.302 \\
 5290616512796591744 &      & 119.1227 & -61.5126 &  11.340&  5289938285920024960 &  68058 & 121.8216 & -61.3661 &   9.257 \\
 5290722924905383808 &      & 119.1927 & -60.8161 &   8.783&  5289748345286306048 &      &  121.8795 & -61.5315 &  11.332 \\
 5290724196215701120 &      & 119.2482 & -60.7838 &   9.647&  5289950483627126016 &      &  121.8958 & -61.1153 &  10.596 \\
 5291032128192514176 &      & 119.2540 & -60.1960 &  11.254&  5289964708560878336 &      &  122.0436 & -60.9933 &  11.694 \\
 5291032609228851840 &      & 119.3943 & -60.2067 &  10.327&  5277548370424297344 &  68949 & 122.7451 & -62.8368 &   9.875 \\
 5290718733017309824 &      & 119.4077 & -60.9089 &   8.770&  5290112146196485888 &      &  122.8538 & -61.4296 &  10.624 \\
 5290725261370377600 &      & 119.4561 & -60.7212 &   9.666&  5291311885181828480 &      &  122.8821 & -59.5398 &  11.360 \\
 5290642694916327424 &      & 119.4840 & -61.3655 &  10.346&  5277627432182254208 &      &  122.9566 & -62.4664 &  10.367 \\
 5290720451004223616 &      & 119.4918 & -60.8459 &   8.992 & & & & & \\
\hline
\end{tabular}
}
\end{table*}
\begin{figure}[t]
\centering
\includegraphics[width=7cm]{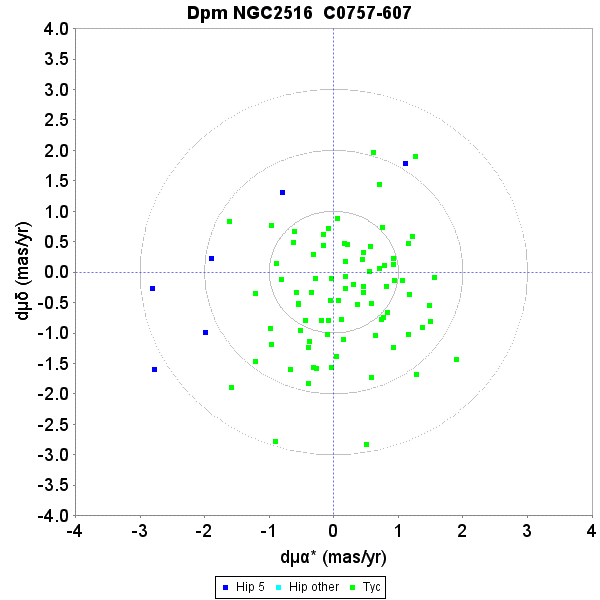}
\includegraphics[width=7cm]{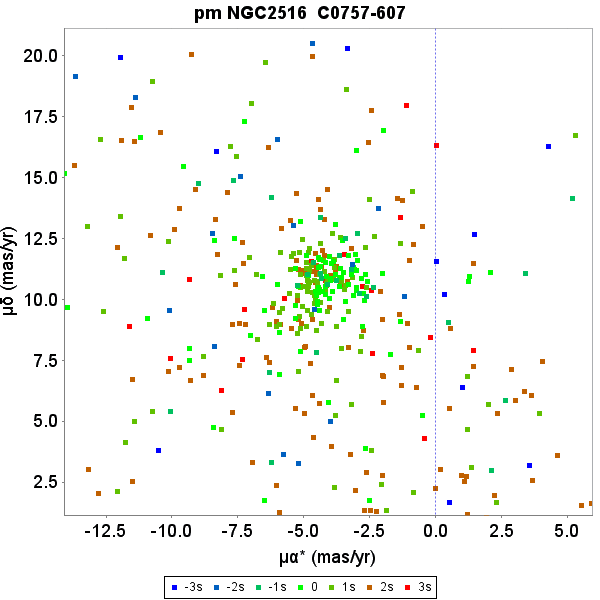}
\caption{Proper motion charts for the cluster NGC2516. Left: unit weight residual proper motions. Green dots have first epoch \Tycho data, the dark blue dots have \Hipparcos first epoch 5-parameter solutions. The concentric circles represent 1, 2, and 3$\sigma$ \su levels. Right: actual proper motion distribution, where the colour indicate the difference from the cluster parallax in \su units.}
\label{fig:NGC2516}
\end{figure}
\begin{figure}[t]
\centering
\includegraphics[width=12cm]{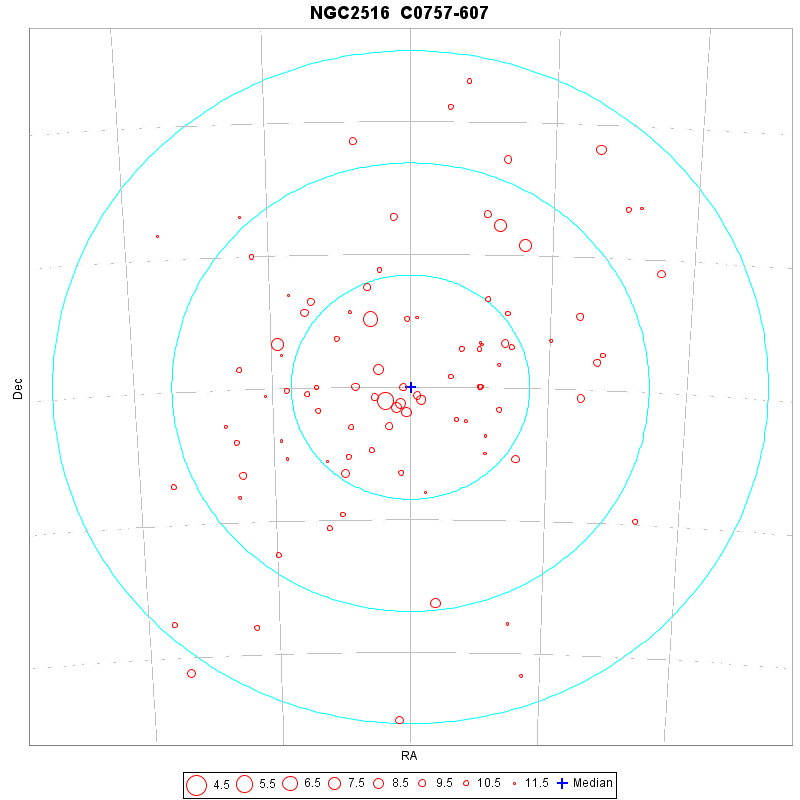}
\caption{A map of members of the cluster NGC2516 as identified from the TGAS catalogue. The coordinate grid is at 1 degrees intervals, the three concentric circles are at 5, 10 and 15 pc from the cluster centre at the cluster distance.}
\label{fig:mapNGC2516}
\end{figure}

\subsection{The cluster NGC2232}
\begin{table*}
\centering
\caption{Identifiers and positions for members of the cluster NGC2232.}\label{tab:NGC2232}
\scriptsize{
\begin{tabular}{rrrrr|rrrrr}
\hline\hline 
SourceId & HD & $\alpha$ (degr) & $\delta$ (degr) & G & SourceId & HD & $\alpha$ (degr) & $\delta$ (degr) & G \\ \hline 
 3116780972691989632 & 294906 & 94.9130 &  -3.9144 &  10.505&  3104529458223283584 &  &  96.9888 &  -4.7122 &  10.585 \\
 3116548975739569664 &  44702 & 95.7465 &  -4.1871 &   8.688&  3117056469074206080 &  45601 &  97.0893 &  -3.4611 &   8.665 \\
 3116554198419796992 &  & 95.9913 &  -4.0883 &  11.512&  3104226817645484672 &  45627 &  97.0962 &  -5.0340 &   8.915 \\
 3008314116254649600 &  & 96.0353 &  -5.3520 &  10.965&  3006948247933933568 &  &  97.2972 &  -7.1706 &  11.245 \\
 3008110637881857664 &  & 96.1198 &  -5.6776 &  10.219&  3104591649350135552 &  &  97.4085 &  -4.0665 &  11.700 \\
 3116951225195812736 & 295008 & 96.1369 &  -3.1461 &  10.685&  3105035336649025408 &  &  97.4321 &  -3.5793 &  11.306 \\
 3104421706081544320 & 295044 & 96.2830 &  -4.8583 &   9.687&  3104155315031782016 &  &  97.4766 &  -5.3775 &  11.375 \\
 3104401502557764224 &  & 96.4836 &  -5.0643 &  10.847&  3104158476127719680 &  45935 &  97.5686 &  -5.2604 &   9.819 \\
 3104454897588787840 &  45238 & 96.5333 &  -4.6282 &   8.439&  3104257500891889280 &  &  97.8864 &  -5.1211 &  10.543 \\
 3007037858130227200 &  45284 & 96.5549 &  -7.3614 &   7.416&  3006897086283522944 &  46282 &  98.0469 &  -7.3063 &   8.252 \\
 3104456031460151296 &  45321 & 96.6435 &  -4.5974 &   6.275&  3103225403071448320 &  &  98.7380 &  -5.7721 &  11.517 \\
 3104452629848438272 &  45399 & 96.7436 &  -4.6251 &   8.509&  3103153694297494400 &  &  99.0283 &  -6.0804 &   9.830 \\
 3104547325287231744 &  45434 & 96.7817 &  -4.5466 &   9.590&  3103128302450862464 &  47091 &  99.1701 &  -6.3635 &   9.711 \\
 3104193316900585856 &  & 96.8102 &  -5.3022 &  11.211&  3100016856344868992 &  &  99.1994 &  -6.7876 &  11.877 \\
 3007264082648616064 &  45547 & 96.9668 &  -6.2919 &   8.757&  3103618581557297536 &  47340 &  99.4843 &  -5.5335 &   9.142 \\
 3117085537415090688 & 295066 & 96.9790 &  -3.1366 &  10.052&&&&\\
\hline
\end{tabular}
}
\end{table*}
\begin{figure}[t]
\centering
\includegraphics[width=7cm]{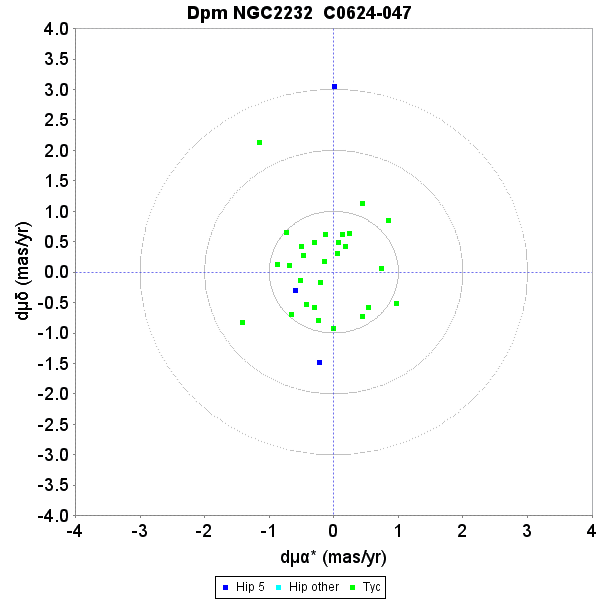}
\includegraphics[width=7cm]{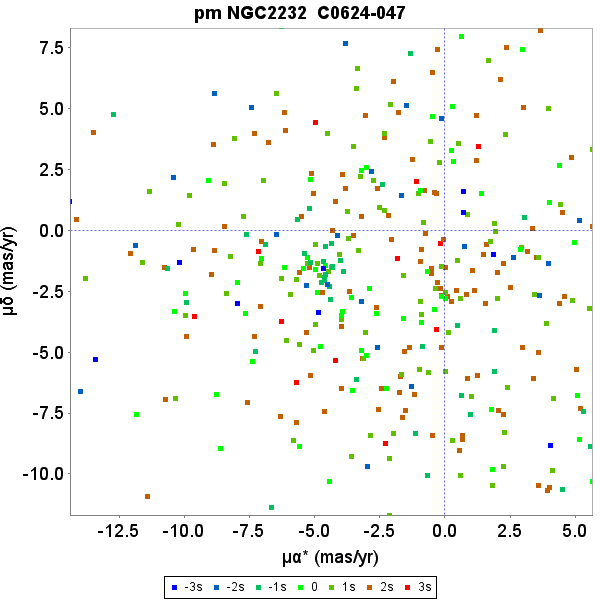}
\caption{Proper motion charts for the cluster NGC2232. Left: unit weight residual proper motions. Green dots have first epoch \Tycho data, the dark blue dots have \Hipparcos first epoch 5-parameter solutions. The concentric circles represent 1, 2, and 3$\sigma$ \su levels. Right: actual proper motion distribution, where the colour indicate the difference from the cluster parallax in \su units.}
\label{fig:NGC2232}
\end{figure}
\begin{figure}[t]
\centering
\includegraphics[width=12cm]{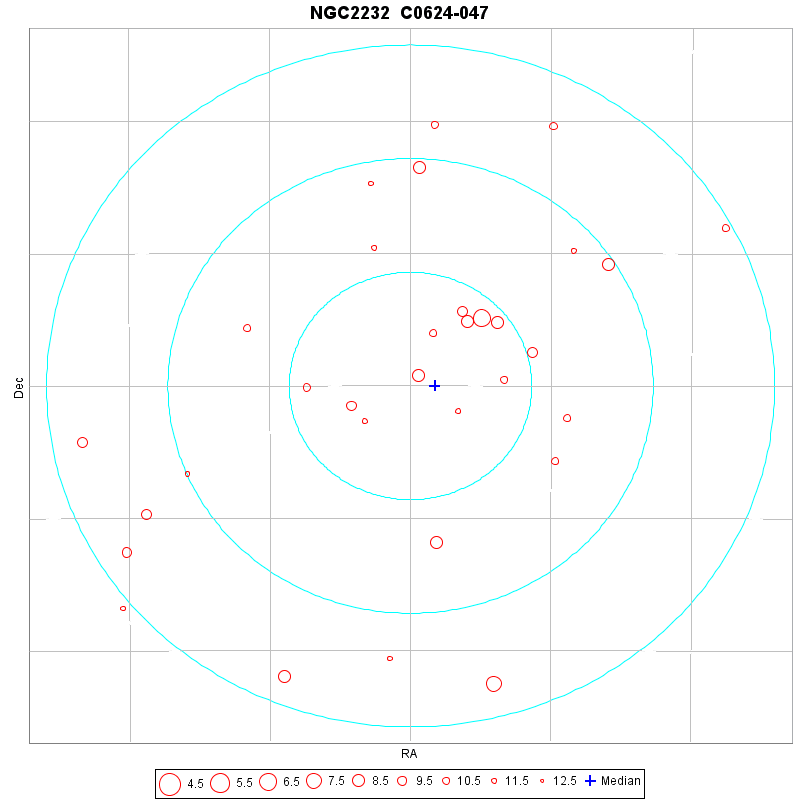}
\caption{A map of members of the cluster NGC2232 as identified from the TGAS catalogue. The coordinate grid is at 1 degrees intervals, the three concentric circles are at 5, 10 and 15 pc from the cluster centre at the cluster distance.}
\label{fig:mapNGC2232}
\end{figure}

\subsection{The cluster IC4665}
\begin{table*}
\centering
\caption{Identifiers and positions for members of the cluster IC4665.}\label{tab:IC4665}
\scriptsize{
\begin{tabular}{rrrrr|rrrrr}
\hline\hline 
SourceId & HD & $\alpha$ (degr) & $\delta$ (degr) & G & SourceId & HD & $\alpha$ (degr) & $\delta$ (degr) & G \\ \hline 
 4486156425152983680 & 161055 &265.7854 &   6.2153 &   9.977&  4474071143094987520 & 161572 & 266.4878 &   5.6944 &   7.602 \\
 4473740877289974528 & 161261 &266.0656 &   5.7143 &   8.283&  4474059082826822144 & 161603 & 266.5455 &   5.6582 &   7.348 \\
 4473363126327469184 &  &266.2138 &   4.4274 &  11.181&  4474061831605886080 & 161677 & 266.6710 &   5.7742 &   7.129 \\
 4473687173019891328 & 161370 &266.2537 &   5.5229 &   9.299&  4474057433559379072 & 161733 & 266.7590 &   5.6918 &   7.998 \\
 4486313139920906496 & 161425 &266.3149 &   6.8997 &   8.272&  4473838905626668800 &  & 266.7859 &   5.2253 &  10.627 \\
 4474064442946007040 & 161426 &266.3220 &   5.6676 &   9.042&  4473768811760404864 &  & 267.0638 &   4.9130 &  11.832 \\
 4474066504530306688 & 161480 &266.3902 &   5.7157 &   7.688&  4473992974693569280 &  & 267.4240 &   5.9250 &  10.610 \\
 4473670783424690816 & 161481 &266.3952 &   5.4265 &   9.014&  4472970772475428224 & 162954 & 268.3992 &   4.8589 &   7.710 \\
\hline
\end{tabular}
}
\end{table*}
\begin{figure}[t]
\centering
\includegraphics[width=7cm]{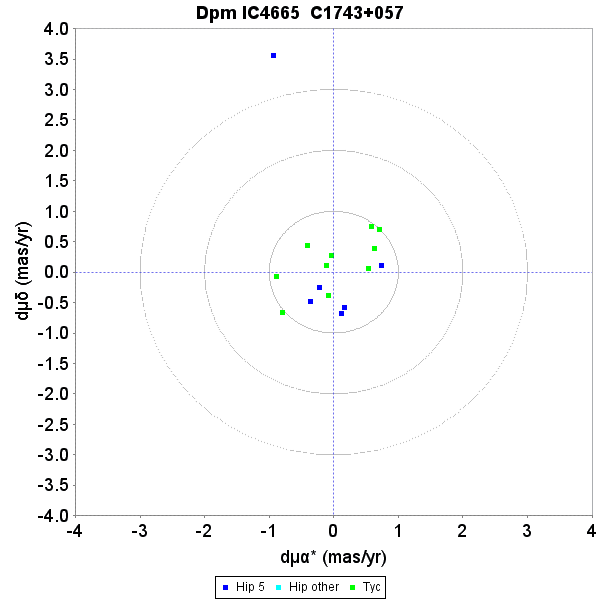}
\includegraphics[width=7cm]{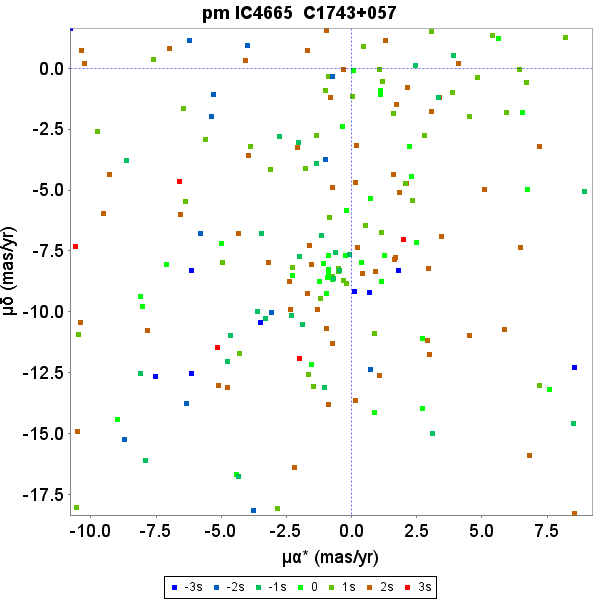}
\caption{Proper motion charts for the cluster IC4665. Left: unit weight residual proper motions. Green dots have first epoch \Tycho data, the dark blue dots have \Hipparcos first epoch 5-parameter solutions. The concentric circles represent 1, 2, and 3$\sigma$ \su levels. Right: actual proper motion distribution, where the colour indicate the difference from the cluster parallax in \su units.}
\label{fig:IC4665}
\end{figure}
\begin{figure}[t]
\centering
\includegraphics[width=12cm]{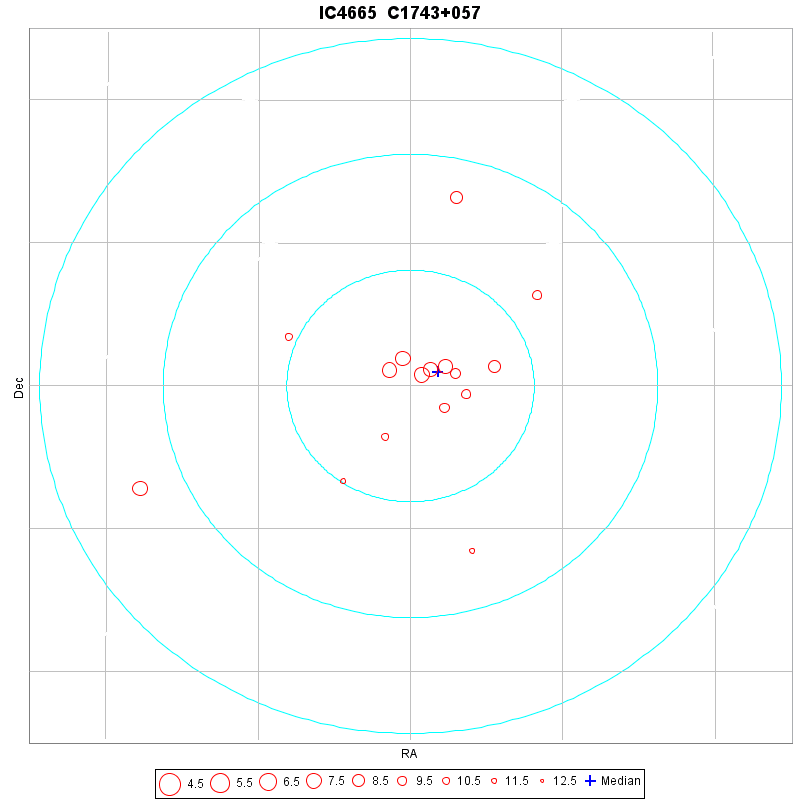}
\caption{A map of members of the cluster IC4665 as identified from the TGAS catalogue. The coordinate grid is at 1 degrees intervals, the three concentric circles are at 5, 10 and 15 pc from the cluster centre at the cluster distance.}
\label{fig:mapIC4665}
\end{figure}

\subsection{The cluster NGC6633}
\begin{table*}
\centering
\caption{Identifiers and positions for members of the cluster NGC6633.}\label{tab:NGC6633}
\scriptsize{
\begin{tabular}{rrrrr|rrrrr}
\hline\hline 
SourceId & HD & $\alpha$ (degr) & $\delta$ (degr) & G & SourceId & HD & $\alpha$ (degr) & $\delta$ (degr) & G \\ \hline 
 4476902900932589056 & 168699 &275.2333 &   5.9268 &   8.350&  4477222618306462848 & 170135 & 276.9130 &   6.5321 &   8.370 \\
 4478329620357361792 &      & 275.9270 &   7.8384 &  10.600&  4477173518240339968 & 170158 & 276.9487 &   6.4559 &   8.992 \\
 4476646955248355328 &      & 276.0366 &   5.6034 &  11.906&  4477266461332603776 &      &  276.9652 &   6.8313 &  10.045 \\
 4478439262280847744 & 169596 &276.2681 &   8.2389 &   9.651&  4477223374220701696 & 170174 & 276.9781 &   6.6001 &   7.896 \\
 4477430769601777024 &      & 276.3198 &   6.8252 &  11.738&  4284608247206265856 &      &  276.9947 &   5.8031 &  11.737 \\
 4477243234149473664 &      & 276.5428 &   6.6982 &   9.875&  4477172521807928448 &      &  276.9952 &   6.4303 &  11.265 \\
 4477231276960532224 &      & 276.6133 &   6.5270 &  10.670&  4477273264560797696 & 170231 & 277.0008 &   6.9143 &   8.232 \\
 4477213444256326528 &      & 276.6871 &   6.4050 &  10.042&  4477569926535247488 & 170271 & 277.0490 &   7.3553 &   8.859 \\
 4477212172946010368 &      & 276.6954 &   6.3490 &  10.255&  4477174892629871360 &      &  277.0601 &   6.5122 &   9.957 \\
 4477212413464176640 &      & 276.7201 &   6.3875 &   8.719&  4477249109664733568 & 170292 & 277.0957 &   6.7081 &   8.369 \\
 4477465198053904768 & 169984 &276.7218 &   7.1184 &   9.172&  4477170803821011328 & 170293 & 277.0958 &   6.4139 &   8.520 \\
 4477212001147318400 &      & 276.7293 &   6.3490 &   9.933&  4477374591421100800 &      &  277.1231 &   7.1732 &  10.493 \\
 4477214784286122112 &      & 276.7518 &   6.4172 &   9.406&  4477259761183622400 &      &  277.1373 &   6.8211 &  11.776 \\
 4477214028371878528 & 170011 &276.7585 &   6.4111 &   8.937&  4477256531368216960 & 170346 & 277.1625 &   6.7896 &   8.650 \\
 4477214818645859328 &      & 276.7649 &   6.4315 &   9.441&  4477256668798782080 &      &  277.1792 &   6.8132 &  11.278 \\
 4477268694715597440 &      & 276.7975 &   6.8371 &  11.258&  4477373079596895360 &      &  277.2584 &   7.1701 &   9.981 \\
 4477460387696238336 & 170053 &276.8095 &   7.0091 &   6.728&  4477373801151400064 & 170426 & 277.2733 &   7.2052 &   8.954 \\
 4477221656230795136 & 170054 &276.8112 &   6.5186 &   8.193&  4477258077548054528 & 170472 & 277.3325 &   6.8514 &   9.074 \\
 4477266152094957440 &      & 276.8323 &   6.8583 &  11.273&  4477385930139039744 &      &  277.4129 &   7.4038 &  10.047 \\
 4477221381355884672 &      & 276.8353 &   6.4942 &   9.597&  4477305665786045312 &      &  277.4593 &   6.9237 &  11.578 \\
 4477158571754162432 & 170079 &276.8502 &   6.1430 &   8.943&  4477412009180490624 & 170676 & 277.6046 &   7.5330 &   9.349 \\
 4477160427180032000 & 170095 &276.8599 &   6.2408 &   9.442&  4284914976584397440 &      &  277.6049 &   5.7124 &  11.818 \\
 4477223820897301248 & 170094 &276.8713 &   6.5890 &   9.242&  4285146251985356160 &      &  278.3180 &   6.7648 &  11.769 \\
 4477267629563706240 &      & 276.9017 &   6.8809 &  11.335 & & & & & \\
\hline
\end{tabular}
}
\end{table*}
\begin{figure}[t]
\centering
\includegraphics[width=7cm]{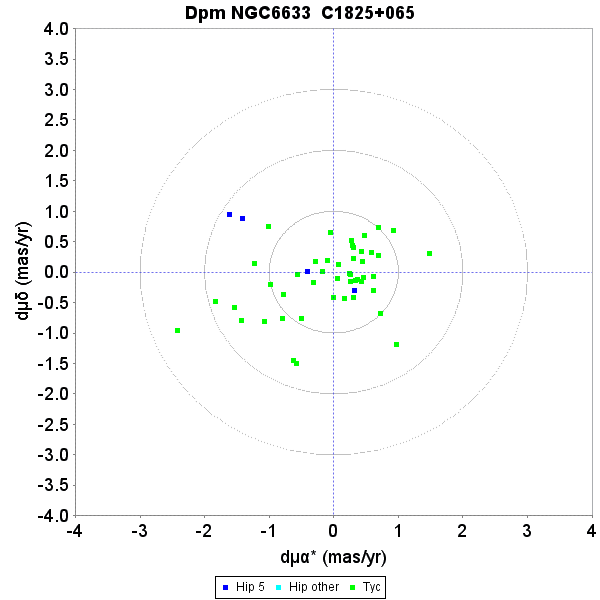}
\includegraphics[width=7cm]{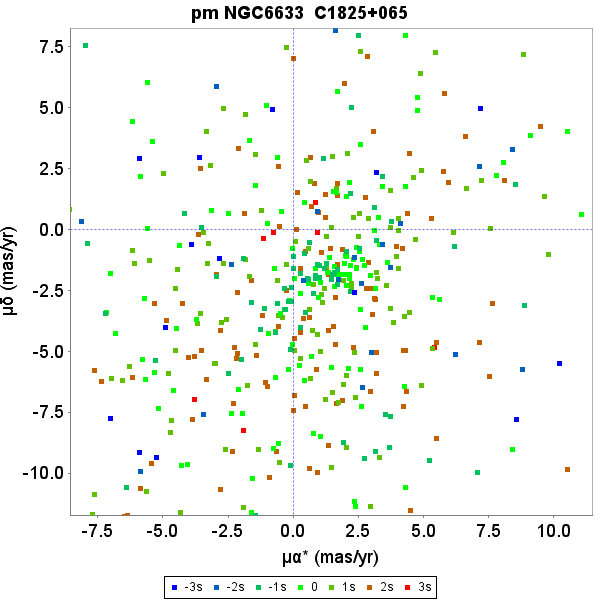}
\caption{Proper motion charts for the cluster NGC6633. Left: unit weight residual proper motions. Green dots have first epoch \Tycho data, the dark blue dots have \Hipparcos first epoch 5-parameter solutions. The concentric circles represent 1, 2, and 3$\sigma$ \su levels. Right: actual proper motion distribution, where the colour indicate the difference from the cluster parallax in \su units.}
\label{fig:NGC6633}
\end{figure}
\begin{figure}[t]
\centering
\includegraphics[width=12cm]{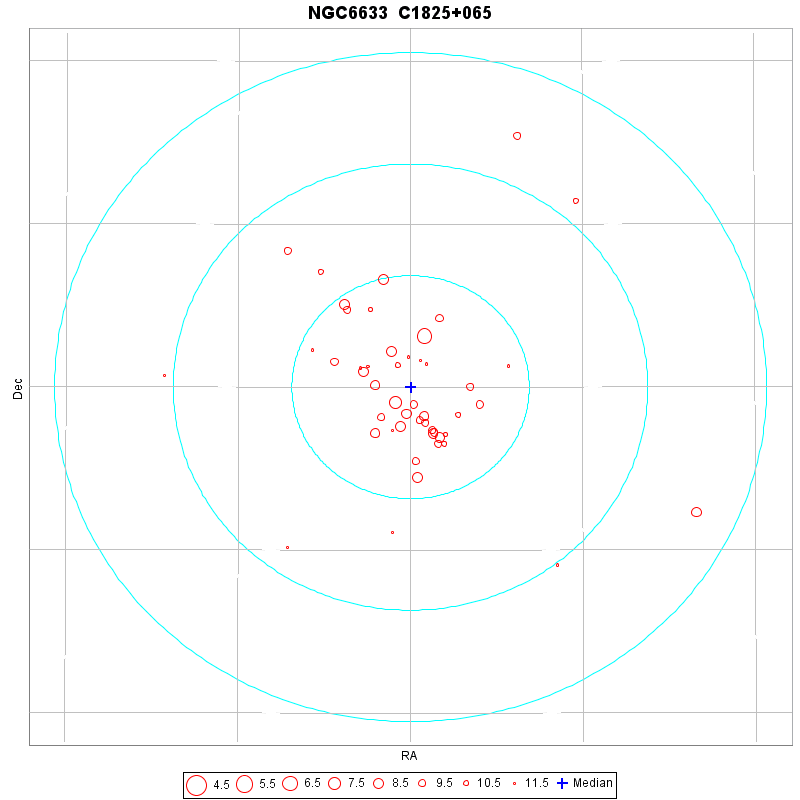}
\caption{A map of members of the cluster NGC6633 as identified from the TGAS catalogue. The coordinate grid is at 1 degrees intervals, the three concentric circles are at 5, 10 and 15 pc from the cluster centre at the cluster distance.}
\label{fig:mapNGC6633}
\end{figure}

\subsection{The cluster Coll140}
\begin{table*}
\centering
\caption{Identifiers and positions for members of the cluster Coll140.}\label{tab:Coll140}
\scriptsize{
\begin{tabular}{rrrrr|rrrrr}
\hline\hline 
SourceId & HD & $\alpha$ (degr) & $\delta$ (degr) & G & SourceId & HD & $\alpha$ (degr) & $\delta$ (degr) & G \\ \hline 
 5603486910168526464 &      & 108.0499 & -32.4072 &  11.302&  5592886106247453440 &  58534 & 111.1732 & -31.7827 &   7.674 \\
 5603471276488592000 &  55764 &108.2304 & -32.5091 &   9.309&  5605346734084513536 &      &  111.2547 & -30.7787 &  10.892 \\
 5603528622888474240 &      & 108.7214 & -31.8839 &  10.325&  5592828107009120896 &      &  111.2769 & -31.9735 &  11.109 \\
 5604862674090330624 &      & 109.7420 & -31.6362 &  11.579&  5592585939571471104 &      &  111.4850 & -32.6238 &  10.557 \\
 5591131663648556928 &      & 109.9207 & -33.6117 &  11.466&  5593263341813388032 &      &  111.4879 & -31.5568 &  10.675 \\
 5604944381551323520 &      & 110.1113 & -31.5505 &  10.629&  5591637370276050432 &      &  111.7394 & -33.7077 &  11.222 \\
 5592713963958910976 &      & 110.2371 & -32.6575 &  11.362&  5591558445957080576 &      &  111.9184 & -34.2274 &  10.868 \\
 5604778011699119232 &  57759 &110.3159 & -32.0267 &   9.346&  5593461838022243072 &  59572 & 112.3336 & -30.1935 &  10.062 \\
 5604898545660357888 &      & 110.4101 & -31.6288 &  11.195&  5593416723685787904 &      &  112.3633 & -30.5931 &  10.295 \\
 5592762548628967680 &  57912 &110.4892 & -32.1911 &   8.880&  5592911326293920128 &      &  112.5672 & -32.6002 &  10.284 \\
 5604923662629096960 &      & 110.5632 & -31.5046 &  11.761&  5591658089201524096 &      &  112.6466 & -33.7526 &   9.717 \\
 5592858343578910976 &      & 110.6150 & -32.1157 &   9.775&  5592948194293169920 &      &  112.7961 & -32.2772 &  11.616 \\
 5592893905908057728 &      & 110.9707 & -31.7265 &  10.158&  5591877991522153728 &  60498 & 113.3549 & -33.3999 &   7.381 \\
 5605735961203862144 &      & 110.9890 & -29.6577 &  10.135&  5591883317281598720 &      &  113.4522 & -33.3901 &  10.403 \\
 5592878650184244352 &  58395 &111.0086 & -31.9106 &   9.026&  5591869264148606848 &      &  113.6875 & -33.3917 &  10.787 \\
\hline
\end{tabular}
}
\end{table*}
\begin{figure}[t]
\centering
\includegraphics[width=7cm]{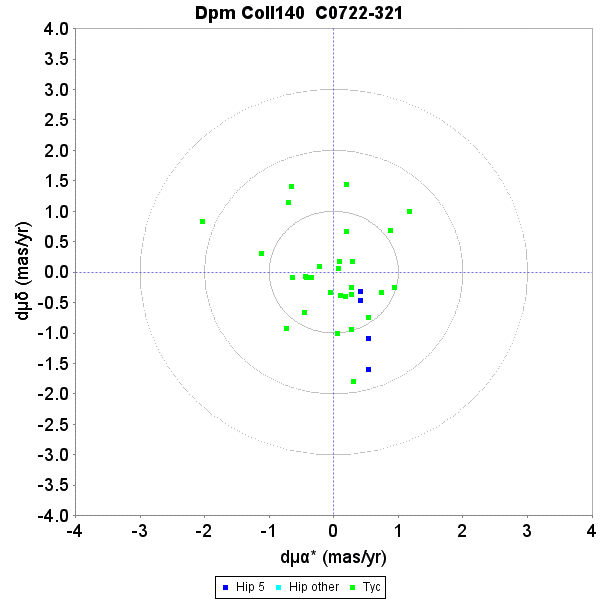}
\includegraphics[width=7cm]{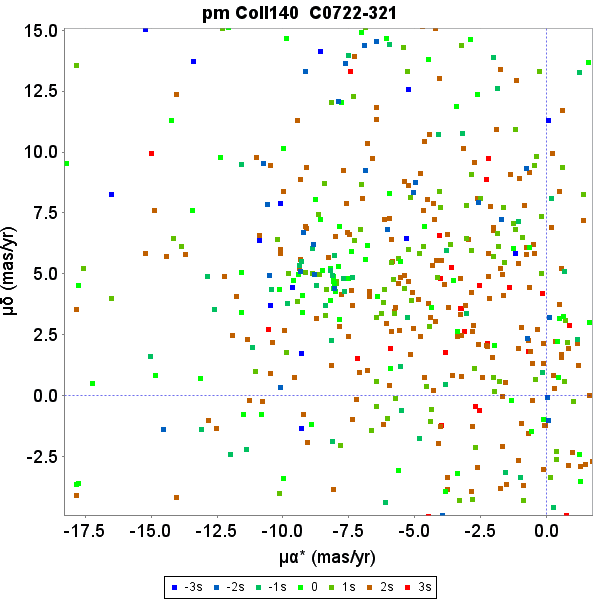}
\caption{Proper motion charts for the cluster Coll140. Left: unit weight residual proper motions. Green dots have first epoch \Tycho data, the dark blue dots have \Hipparcos first epoch 5-parameter solutions. The concentric circles represent 1, 2, and 3$\sigma$ \su levels. Right: actual proper motion distribution, where the colour indicate the difference from the cluster parallax in \su units.}
\label{fig:Coll140}
\end{figure}
\begin{figure}[t]
\centering
\includegraphics[width=12cm]{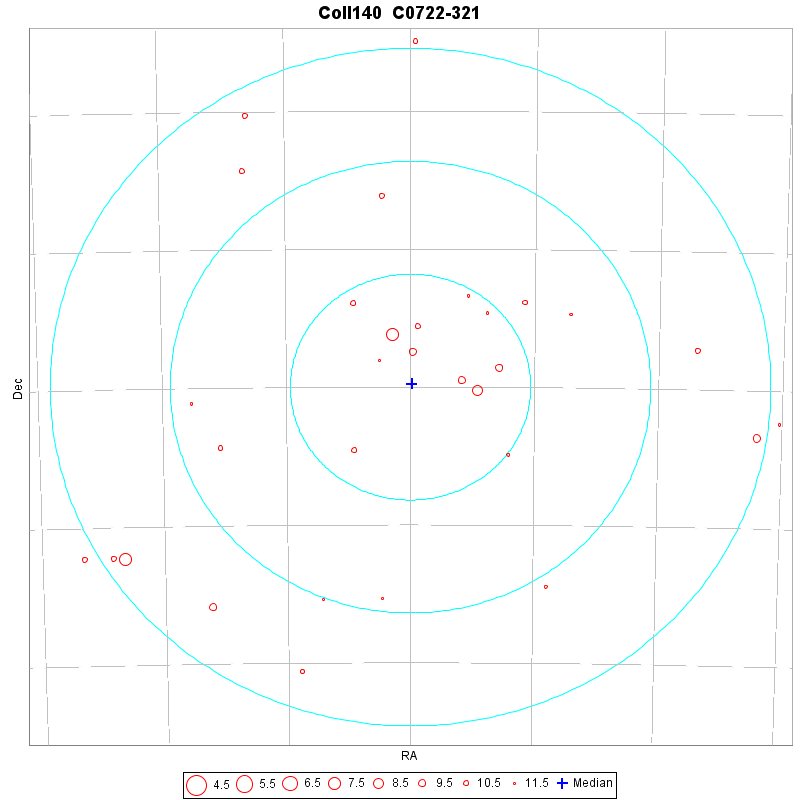}
\caption{A map of members of the cluster Coll140 as identified from the TGAS catalogue. The coordinate grid is at 1 degrees intervals, the three concentric circles are at 5, 10 and 15 pc from the cluster centre at the cluster distance.}
\label{fig:mapColl140}
\end{figure}

\subsection{The cluster NGC2422}
\begin{table*}
\centering
\caption{Identifiers and positions for members of the cluster NGC2422.}\label{tab:NGC2422}
\scriptsize{
\begin{tabular}{rrrrr|rrrrr}
\hline\hline 
SourceId & HD & $\alpha$ (degr) & $\delta$ (degr) & G & SourceId & HD & $\alpha$ (degr) & $\delta$ (degr) & G \\ \hline
 3033318763016101120 &      & 113.2187 & -13.6996 &  10.558&  3030228688664529408 &      &  114.1149 & -14.2268 &   9.951 \\
 3030085752152201600 &      & 113.2347 & -14.8403 &  10.584&  3030013253105462528 &  60999 & 114.1282 & -14.6655 &   8.796 \\
 3030085442914556800 &  60278 &113.2811 & -14.8684 &  10.154&  3030259956026959616 &      &  114.1306 & -14.0059 &  11.110 \\
 3028387796967333632 &  60279 &113.2828 & -15.1784 &   9.288&  3033778805551456512 &      &  114.1490 & -12.7578 &  10.580 \\
 3033430603965039104 &      & 113.3280 & -13.0412 &   9.991&  3030028886786394752 &      &  114.1512 & -14.4612 &   7.793 \\
 3033383634202689536 &  60476 &113.5347 & -13.0395 &   7.901&  3030231781041515520 &      &  114.1520 & -14.1438 &   9.830 \\
 3033837904304239232 &      & 113.6452 & -12.4722 &  10.327&  3029807678792327808 &      &  114.1908 & -15.4626 &  11.343 \\
 3030298370214451072 &  60597 &113.6975 & -13.9760 &  10.263&  3030024282581449088 &      &  114.4046 & -14.4237 &   8.915 \\
 3030243051035701632 &  60624 &113.7271 & -14.1628 &   7.590&  3030069259478942336 &      &  114.4132 & -14.0341 &  10.507 \\
 3033756746599448832 &      & 113.7453 & -12.9845 &  10.861&  3030067953808888448 &      &  114.4529 & -14.1026 &  10.144 \\
 3030250747617082752 &  60659 &113.7721 & -14.0448 &   9.891&  3030004525731877632 &      &  114.8204 & -14.3367 &   9.804 \\
 3029910448767038592 &      & 113.9117 & -14.8534 &  11.993&  3030681000261287552 &      &  114.8374 & -13.4457 &   9.730 \\
 3030247311643245696 &      & 113.9129 & -14.0425 &  11.371&  3029184393140868480 &      &  114.8589 & -15.0898 &  10.704 \\
 3029905191727079168 &      & 114.0572 & -14.9320 &  10.546&  3029983909887619328 &  61865 & 115.1938 & -14.5134 &   9.912 \\
 3030730306486641280 &      & 114.0748 & -13.3205 &   9.775&  3029232702927668096 &      &  115.2640 & -14.5687 &  10.140 \\
 3030025656971005696 &  60941 &114.0797 & -14.5923 &   9.138&  3029096878882496384 &  62051 & 115.3829 & -15.5418 &  10.175 \\
 3030729172615280128 &      & 114.1056 & -13.3717 &  11.339&  3030376641697207424 &      &  115.4815 & -14.0834 &   9.807 \\
\hline
\end{tabular}
}
\end{table*}
\begin{figure}[t]
\centering
\includegraphics[width=7cm]{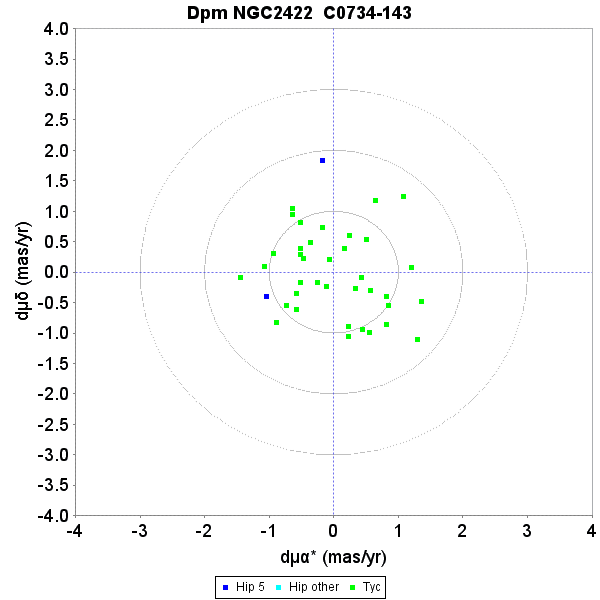}
\includegraphics[width=7cm]{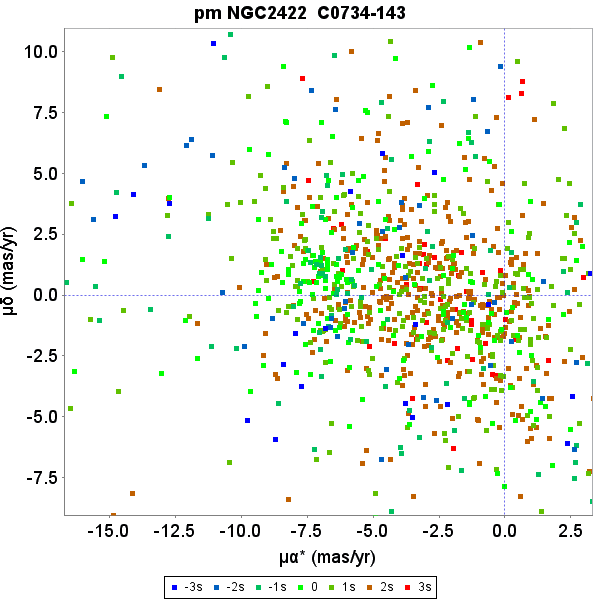}
\caption{Proper motion charts for the cluster NGC2422. Left: unit weight residual proper motions. Green dots have first epoch \Tycho data, the dark blue dots have \Hipparcos first epoch 5-parameter solutions. The concentric circles represent 1, 2, and 3$\sigma$ \su levels. Right: actual proper motion distribution, where the colour indicate the difference from the cluster parallax in \su units.}
\label{fig:NGC2422}
\end{figure}
\begin{figure}[t]
\centering
\includegraphics[width=12cm]{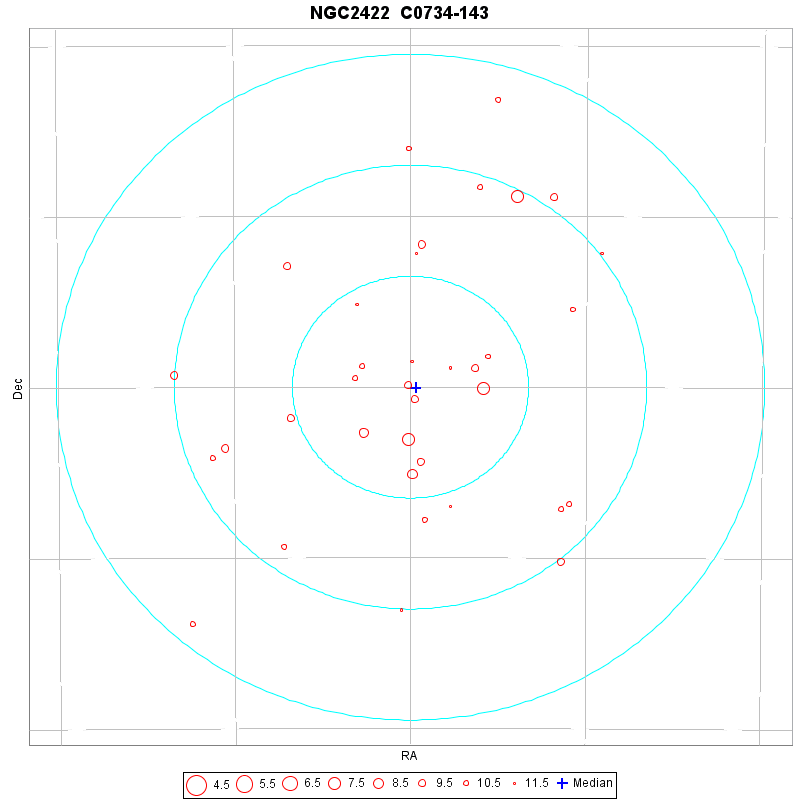}
\caption{A map of members of the cluster NGC2422 as identified from the TGAS catalogue. The coordinate grid is at 1 degrees intervals, the three concentric circles are at 5, 10 and 15 pc from the cluster centre at the cluster distance.}
\label{fig:mapNGC2422}
\end{figure}

\subsection{The cluster NGC3532}
\begin{table*}
\centering
\caption{Identifiers and positions for members of the cluster NGC3532.}\label{tab:NGC3532}
\scriptsize{
\begin{tabular}{rrrrr|rrrrr}
\hline\hline 
SourceId & HD & $\alpha$ (degr) & $\delta$ (degr) & G & SourceId & HD & $\alpha$ (degr) & $\delta$ (degr) & G \\ \hline
 5350433450039970176 & 303392 &162.3573 & -59.1874 &  10.466&  5338655344045619328 &  96306 & 166.3047 & -58.8424 &   9.261 \\
 5338425339955648384 & 303505 &163.1798 & -59.4636 &  10.152&  5338661425719301888 &  96305 & 166.3195 & -58.7836 &   8.580 \\
 5350988978295475200 &      & 163.3526 & -57.9345 &  11.464&  5340224862533206912 &  96324 & 166.3268 & -58.1997 &   9.093 \\
 5338429497484007168 &  94558 &163.4453 & -59.3828 &   8.852&  5338663693462017408 &  96304 & 166.3288 & -58.6767 &   9.561 \\
 5338803606315364096 &      & 164.1178 & -58.6939 &  10.323&  5338661803676414336 &      &  166.3869 & -58.7305 &   8.924 \\
 5340558461233008768 & 301224 &164.2416 & -56.8914 &  10.706&  5338658745659705344 &      &  166.4079 & -58.7636 &   9.652 \\
 5338742136743476864 &      & 164.3027 & -59.4441 &  11.273&  5338658092824688512 &  96388 & 166.4268 & -58.8362 &   8.823 \\
 5338910671260868352 & 303540 &164.4137 & -58.2177 &  10.062&  5338663865260709888 &      &  166.4400 & -58.6776 &   7.860 \\
 5338776427762345728 &  95163 &164.5684 & -59.1025 &   9.848&  5338660841603735680 &      &  166.4858 & -58.6925 &   9.269 \\
 5338738185373567488 &      & 164.6477 & -59.5534 &  11.354&  5338660669805043200 &      &  166.5159 & -58.6877 &   7.121 \\
 5338787457238337664 &      & 164.7054 & -58.8049 &  11.291&  5340162052934957440 &  96472 & 166.5311 & -58.6384 &   8.593 \\
 5338772132795055104 &  95290 &164.7736 & -59.1684 &   7.708&  5338651117797806976 &  96473 & 166.5326 & -58.8922 &   8.463 \\
 5338559205501201024 &  95291 &164.7977 & -59.3981 &   8.295&  5338657165111753728 &  96489 & 166.5547 & -58.8440 &   8.036 \\
 5338782578155498240 &      & 164.8872 & -58.9297 &  11.361&  5340176346582639232 &  96488 & 166.5562 & -58.3198 &   9.789 \\
 5338925514667835904 &  95412 &164.9703 & -58.1667 &   8.883&  5340171226981629312 &  96509 & 166.5725 & -58.4252 &   9.977 \\
 5338888715387278208 &      & 165.0102 & -58.5733 &  11.277&  5338659501573946112 &  96564 & 166.6493 & -58.7441 &   7.819 \\
 5338594733470670208 &  95495 &165.0972 & -58.9831 &   7.972&  5338636411829832576 &  96585 & 166.6672 & -59.0839 &   9.387 \\
 5338912011289889280 & 303643 &165.1418 & -58.4935 &  10.659&  5337712409748364800 &  96587 & 166.6921 & -60.2453 &   9.630 \\
 5338912389247010816 & 303641 &165.1947 & -58.4758 &  10.480&  5338647887982435968 &  96610 & 166.7060 & -58.7423 &   8.701 \\
 5340428684503196544 &  95599 &165.2757 & -57.9364 &   9.668&  5340149236752553216 &  96609 & 166.7134 & -58.7068 &   8.641 \\
 5338923590525853184 & 303627 &165.3136 & -58.1572 &  10.187&  5340306844870519552 &  96607 & 166.7228 & -57.7840 &  10.015 \\
 5338693380275987456 & 303659 &165.3185 & -58.8121 &  10.621&  5340170264908977408 &  96619 & 166.7378 & -58.4686 &  10.134 \\
 5338567383119554176 &      & 165.4894 & -59.4011 &  11.434&  5338646960269502080 &      &  166.7558 & -58.8314 &   9.783 \\
 5338720112152408320 &      & 165.5253 & -58.5794 &  11.357&  5340170299268715648 &  96636 & 166.7572 & -58.4637 &   9.012 \\
 5338729251842805376 &  95751 &165.5284 & -58.3540 &  10.151&  5338605281905483776 & 303837 & 166.7581 & -59.3781 &  10.709 \\
 5338692590002009344 & 303662 &165.5420 & -58.8341 &  10.174&  5340148996234385280 &  96653 & 166.7842 & -58.7163 &   8.368 \\
 5338694204909705472 &  95765 &165.5453 & -58.7911 &   9.299&  5340172085975106688 &  96651 & 166.7855 & -58.3902 &   8.954 \\
 5338695098262889728 & 303650 &165.5561 & -58.7064 &  10.528&  5340157929762905216 &  96652 & 166.7893 & -58.5270 &   9.211 \\
 5338674207541996032 &      & 165.5822 & -58.9441 &  11.076&  5338646651031856256 &  96667 & 166.7950 & -58.8158 &   9.625 \\
 5338720936786128768 &  95825 &165.6228 & -58.5752 &  10.000&  5338644486368342144 &  96668 & 166.7976 & -58.9074 &   8.308 \\
 5338730420073910144 &  95824 &165.6291 & -58.3731 &   9.630&  5338634041007885952 &  96703 & 166.8515 & -59.1028 &   9.612 \\
 5338721108584820352 &      & 165.6501 & -58.5692 &  10.789&  5340159819548513792 &  96714 & 166.8755 & -58.4736 &   9.434 \\
 5338679670740391808 & 303741 &165.7314 & -58.9080 &  10.840&  5338630845550826880 &  96732 & 166.9026 & -59.2184 &  10.210 \\
 5338709220115347968 &      & 165.7422 & -58.6047 &  11.521&  5340160506743277184 &      &  166.9128 & -58.3978 &  10.668 \\
 5340233727345705344 &  95910 &165.7557 & -58.2031 &   9.581&  5340159029274531712 &  96772 & 166.9472 & -58.4852 &   9.555 \\
 5338669122300733440 &  95931 &165.7701 & -59.1115 &   8.314&  5340159544670605824 &      &  166.9510 & -58.4488 &  10.327 \\
 5338680942050709760 &  95948 &165.8019 & -58.8920 &   9.295&  5340186207827570048 &  96771 & 166.9611 & -58.2864 &   9.838 \\
 5338709117036131712 &  95947 &165.8086 & -58.5965 &   9.939&  5338639469846546688 &  96791 & 166.9806 & -59.0646 &  10.235 \\
 5338085075467999744 &  95991 &165.8253 & -60.5955 &   9.422&  5340159613390081536 &      &  166.9997 & -58.4378 &  11.139 \\
 5338717191574643200 &  95968 &165.8386 & -58.4757 &   9.242&  5340158788756363648 &  96808 & 167.0134 & -58.5003 &   8.850 \\
 5340219571133507072 &  95967 &165.8436 & -58.3311 &  10.227&  5338640088321834496 &  96826 & 167.0215 & -59.0025 &   9.599 \\
 5340240564937995136 & 303723 &165.8814 & -58.0948 &  10.346&  5340290008597838208 &  96823 & 167.0266 & -57.9414 &   9.379 \\
 5338714717673483520 &  96011 &165.8883 & -58.5595 &   9.050&  5338641771949012096 &  96849 & 167.0621 & -58.9419 &   8.953 \\
 5338717019775948416 &      & 165.8939 & -58.4253 &  10.602&  5337851944644789632 &  96881 & 167.0902 & -59.5563 &   9.609 \\
 5338703997435135360 &  96059 &165.9719 & -58.7653 &   8.074&  5340155902538341632 &  96896 & 167.1518 & -58.5332 &   9.713 \\
 5340218231107182720 &  96058 &165.9729 & -58.3953 &   8.384&  5340152122970576000 &  96931 & 167.1798 & -58.7108 &   9.964 \\
 5338716641818829696 &      & 165.9794 & -58.4815 &  10.345&  5339394765619269760 &  96944 & 167.2231 & -58.8310 &   9.075 \\
 5340218024948753408 &      & 166.0102 & -58.4445 &   9.663&  5337850982572111872 & 303847 & 167.2508 & -59.4886 &  10.234 \\
 5338703104081939200 &  96137 &166.0802 & -58.7694 &   8.237&  5337875515425300992 & 303843 & 167.3467 & -59.3791 &  10.460 \\
 5338654347613209216 &  96157 &166.1259 & -58.8630 &   9.858&  5339402599637990528 &  97081 & 167.4173 & -58.6724 &   9.915 \\
 5338709632432217728 &  96174 &166.1409 & -58.6943 &   7.480&  5339402256040608512 &  97093 & 167.4369 & -58.7017 &   8.714 \\
 5340214107938580096 &      & 166.1489 & -58.4715 &  11.123&  5339438780442473216 &  97124 & 167.4662 & -58.2624 &   8.859 \\
 5338656649715663232 &  96175 &166.1498 & -58.7558 &   7.334&  5337861805889698688 &  97173 & 167.5687 & -59.5130 &   8.377 \\
 5338656684075397376 &      & 166.1549 & -58.7306 &   9.751&  5337872594847528960 &  97272 & 167.7078 & -59.1918 &   9.443 \\
 5338709666791955072 &  96191 &166.1634 & -58.6854 &   9.165&  5339422700084933248 &  97296 & 167.7704 & -58.6294 &   9.748 \\
 5338626756743327488 & 303755 &166.1715 & -59.1271 &  10.514&  5337872319969622784 &      &  167.7943 & -59.2249 &  11.298 \\
 5340226752318802304 & 303722 &166.1771 & -58.0930 &  10.770&  5337865963418035200 &  97396 & 167.9236 & -59.3914 &   8.059 \\
 5338662731389346048 &  96212 &166.1850 & -58.6958 &   8.652&  5337757180483341184 & 306155 & 168.3274 & -60.1784 &   9.928 \\
 5338662284712754048 &  96227 &166.2298 & -58.7496 &   8.226&  5339368617855163520 &  97656 & 168.3502 & -58.6031 &   8.402 \\
 5338663212425678848 &  96245 &166.2351 & -58.6659 &   8.367&  5339356935542014848 &  97669 & 168.3784 & -58.9111 &   8.518 \\
 5338655034807973248 &      & 166.2499 & -58.8314 &   9.882&  5339472418622698112 &  97747 & 168.4811 & -58.2820 &   9.909 \\
 5338662971907513344 &      & 166.2633 & -58.6856 &   9.542&  5339495405287654656 & 303970 & 168.9172 & -58.1581 &  10.434 \\
 5338628852687359360 &  96285 &166.2787 & -59.0167 &   9.022&  5339105284821839744 &  98420 & 169.6519 & -59.4973 &   8.866 \\
 5338661460079038976 &  96284 &166.2831 & -58.7793 &   9.339&  5339650676950667520 &  98833 & 170.4533 & -58.5297 &   9.874 \\
\hline
\end{tabular}
}
\end{table*}
\begin{figure}[t]
\centering
\includegraphics[width=7cm]{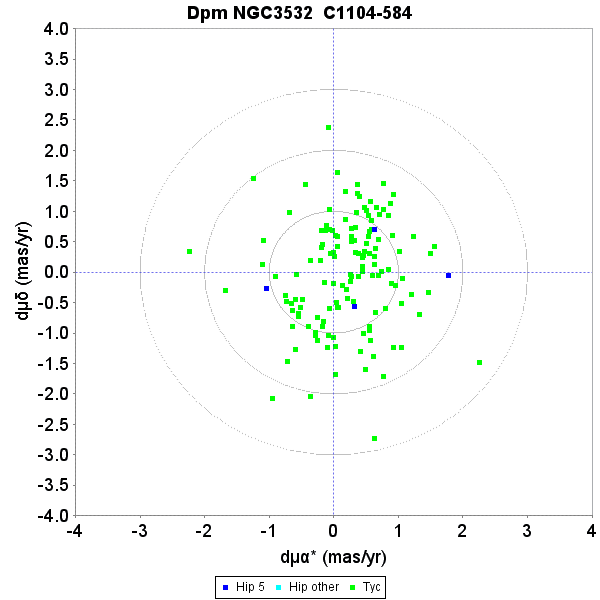}
\includegraphics[width=7cm]{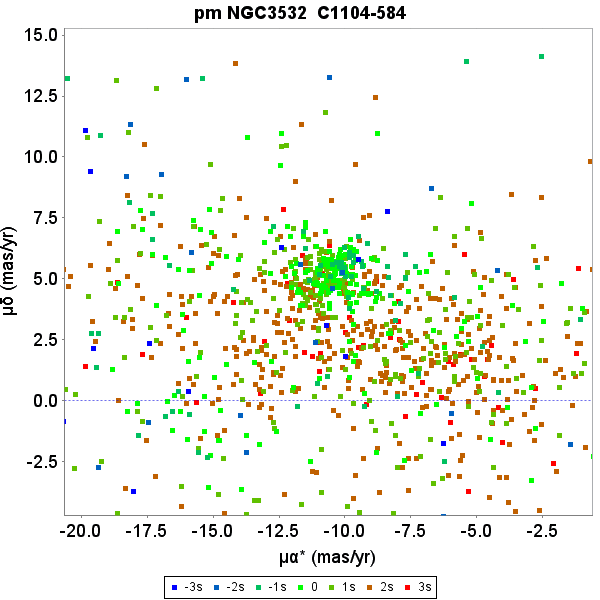}
\caption{Proper motion charts for the cluster NGC3532. Left: unit weight residual proper motions. Green dots have first epoch \Tycho data, the dark blue dots have \Hipparcos first epoch 5-parameter solutions. The concentric circles represent 1, 2, and 3$\sigma$ \su levels. Right: actual proper motion distribution, where the colour indicate the difference from the cluster parallax in \su units.}
\label{fig:NGC3532}
\end{figure}
\begin{figure}[t]
\centering
\includegraphics[width=12cm]{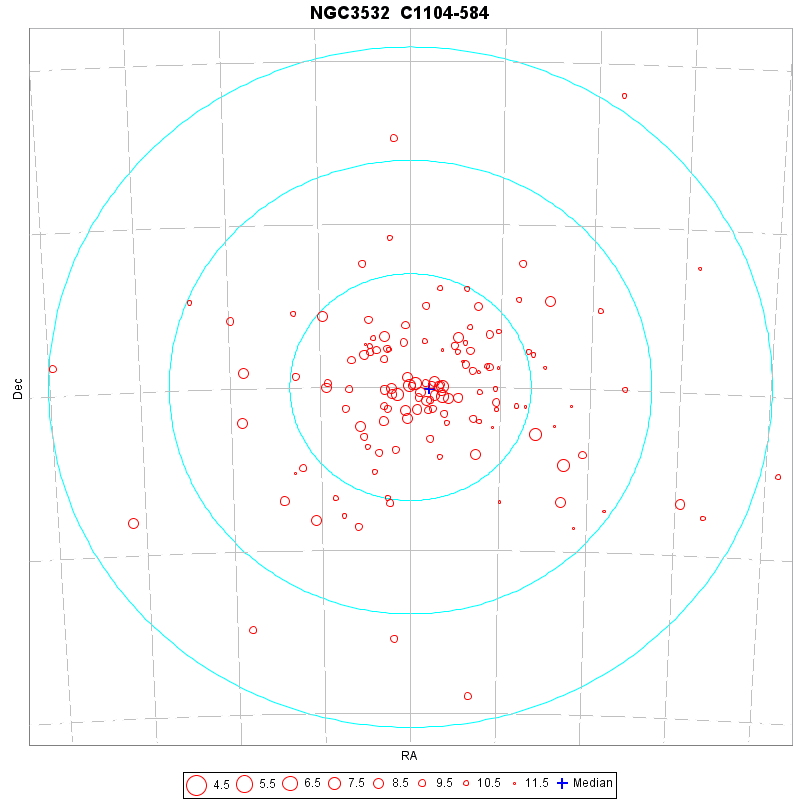}
\caption{A map of members of the cluster NGC3532 as identified from the TGAS catalogue. The coordinate grid is at 1.0 degrees intervals, the three concentric circles are at 5, 10 and 15 pc from the cluster centre at the cluster distance.}
\label{fig:mapNGC3532}
\end{figure}

\subsection{The cluster NGC2547}
\begin{table*}
\centering
\caption{Identifiers and positions for members of the cluster NGC2547.}\label{tab:NGC2547}
\scriptsize{
\begin{tabular}{rrrrr|rrrrr}
\hline\hline 
SourceId & HD & $\alpha$ (degr) & $\delta$ (degr) & G & SourceId & HD & $\alpha$ (degr) & $\delta$ (degr) & G \\ \hline 
 5517947345066590720 &      & 121.4272 & -47.2090 &  10.700&  5514372832766057216 &  68495 & 122.6310 & -49.1084 &   9.414 \\
 5514542123196527744 &  67612 &121.6349 & -49.2020 &   8.229&  5514343695707976064 &  68516 & 122.6408 & -49.5439 &   9.747 \\
 5514203507976717952 &      & 121.6431 & -50.2674 &  10.978&  5514563735473921280 &  68558 & 122.6763 & -48.9570 &   9.745 \\
 5517729435606473728 &  67610 &121.6506 & -47.8057 &   9.876&  5514356065213760768 &  68608 & 122.7481 & -49.2844 &   7.910 \\
 5517679991942975232 &      & 121.7884 & -47.9315 &   8.686&  5514374859990613888 &  68631 & 122.7858 & -49.0045 &   9.904 \\
 5514629534372854272 &  67867 &121.9613 & -48.6601 &  10.028&  5513434433952494592 &  69260 & 123.4664 & -50.5528 &   9.895 \\
 5514340568971771008 &      & 122.1904 & -49.3965 &  10.633&  5515810719098074112 &  69282 & 123.5370 & -49.2344 &   8.227 \\
 5514334762176002944 &      & 122.2064 & -49.5606 &  11.383&  5513454362600737280 &  69347 & 123.5841 & -50.3661 &   8.839 \\
 5514553427552434176 &  68114 &122.2075 & -49.2288 &   9.385&  5515896206125579392 &  69360 & 123.6207 & -48.6597 &  10.381 \\
 5514552705997928576 &  68115 &122.2212 & -49.2303 &   9.697&  5515011236705810816 &  69428 & 123.6768 & -49.9832 &   8.885 \\
 5514638570981291776 &      & 122.2895 & -48.4566 &  10.458&  5515823363481821952 &  69514 & 123.8152 & -49.1801 &   8.072 \\
 5514366957250806784 &      & 122.4639 & -49.1876 &   9.328&  5515039720928933376 &  69595 & 123.8727 & -49.4750 &  10.470 \\
 5514367060333996032 &      & 122.4675 & -49.1840 &   9.641&  5515049341655679104 &      &  124.1441 & -49.5485 &  10.487 \\
 5514362112527719168 &  68398 &122.4833 & -49.3250 &   8.624&  5514996702536508800 &  69911 & 124.2445 & -49.7457 &   9.712 \\
 5514373038924490752 &  68396 &122.4897 & -49.1389 &   8.919&  5515077757159293568 &      &  124.3191 & -49.0957 &  10.561 \\
 5514362799722478464 &  68397 &122.4967 & -49.2697 &   8.156&  5515955785911323136 &      &  124.4473 & -48.3585 &  10.863 \\
 5514362627923788032 &  68432 &122.5327 & -49.2766 &   8.416&  5515592466041607296 &  70950 & 125.6698 & -48.2100 &   7.937 \\
 5513099048544801664 &      & 122.5742 & -51.5853 &  11.979&  5515592466042082560 &      &  125.6757 & -48.1976 &   9.087 \\
 5514373932277680512 &  68452 &122.5867 & -49.0602 &   9.282&  5322644564963225088 &      &  125.7107 & -50.5029 &  11.662 \\
 5514369465511703936 &  68496 &122.6135 & -49.1641 &   7.947&  5514852048036348288 &  71288 & 126.0938 & -49.7885 &  10.334 \\
\hline
\end{tabular}
}
\end{table*}
\begin{figure}[t]
\centering
\includegraphics[width=7cm]{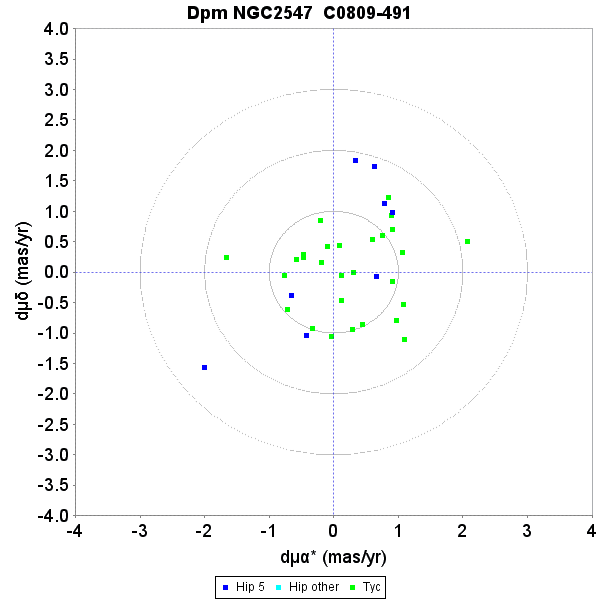}
\includegraphics[width=7cm]{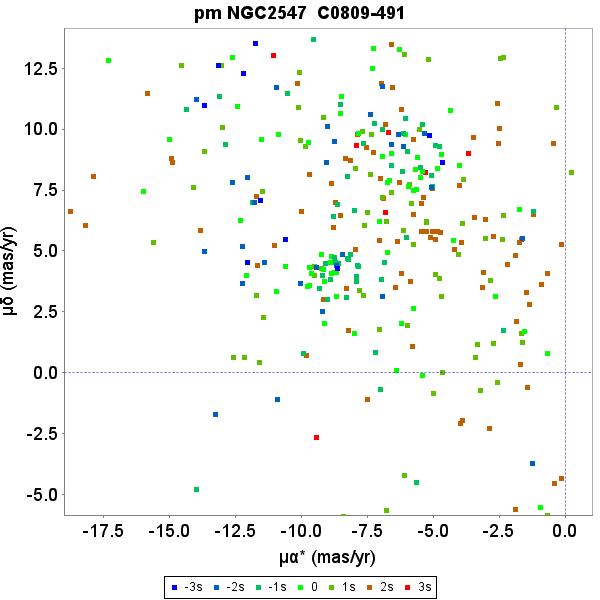}
\caption{Proper motion charts for the cluster NGC2547. Left: unit weight residual proper motions. Green dots have first epoch \Tycho data, the dark blue dots have \Hipparcos first epoch 5-parameter solutions. The concentric circles represent 1, 2, and 3$\sigma$ \su levels. Right: actual proper motion distribution, where the colour indicate the difference from the cluster parallax in \su units.}
\label{fig:NGC2547}
\end{figure}
\begin{figure}[t]
\centering
\includegraphics[width=12cm]{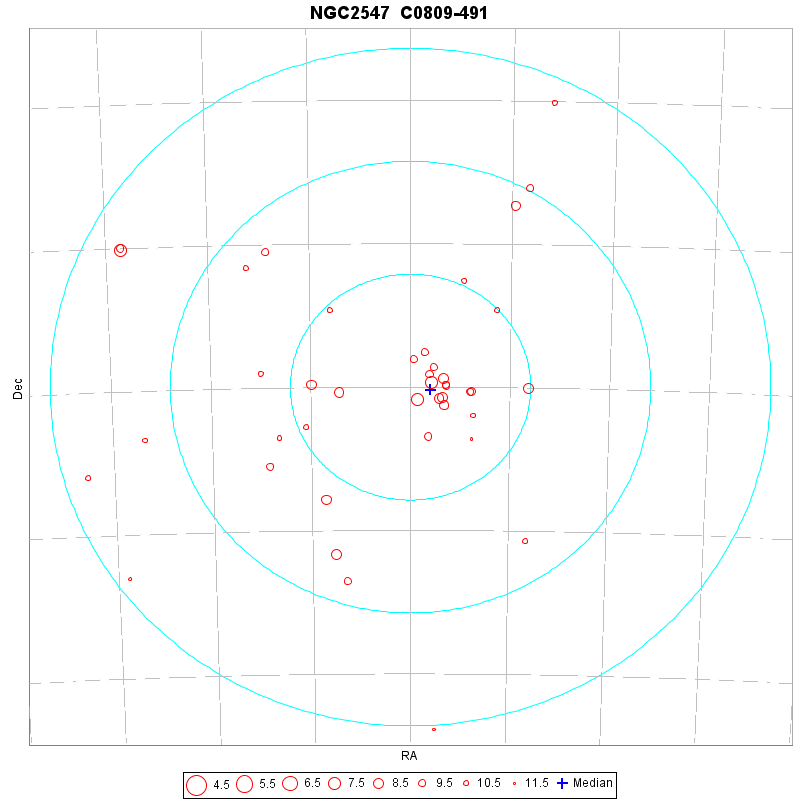}
\caption{A map of members of the cluster NGC2547 as identified from the TGAS catalogue. The coordinate grid is at 0.5 degrees intervals, the three concentric circles are at 5, 10 and 15 pc from the cluster centre at the cluster distance.}
\label{fig:mapNGC2547}
\end{figure}

\bibliographystyle{aa} 
\bibliography{refs,dpac} 

\end{document}